\documentclass[prd,aps,amsfonts,eqsecnum,nofootinbib,longbibliography,notitlepage,superscriptaddress]{revtex4-1}
\usepackage[dvipsnames]{xcolor}
\usepackage[colorlinks=true, urlcolor=violet, linkcolor=blue, citecolor=red, hyperindex=true, linktocpage=true]{hyperref}
\usepackage{comment}
\usepackage{microtype}
\usepackage{tabularx}
\usepackage{suffix}

\usepackage{amsmath,amssymb,graphics,amsthm,mathtools,dsfont}
\usepackage{array}

\usepackage{bbm}

\usepackage{xpatch}
\makeatletter
\patchcmd{\@ssect@ltx}
    {\addcontentsline{toc}{#1}{\protect\numberline{}#8}}
    {}
    {}
    {}
\makeatother

\newcolumntype{P}[1]{>{\centering\arraybackslash}p{#1}}
\allowdisplaybreaks

\newcommand{\figref}[1]{Fig.~\ref{#1}}

\newcommand{\secref}[1]{Sec.~\ref{#1}}
\newcommand{\appref}[1]{App.~\ref{#1}}

\numberwithin{thm}{section}

\renewcommand{\thesection}{\arabic{section}}
\renewcommand{\thesubsection}{\thesection.\arabic{subsection}}

\makeatletter
\renewcommand{\p@subsection}{}
\renewcommand{\p@subsubsection}{}
\makeatother

\newcommand{\vpd}[0]{\vphantom{\dagger}}

\newcommand{\vpp}[0]{\vphantom{\prime}}

\usepackage{physics}
\usepackage{booktabs}
\usepackage{verbatim}
\usepackage[inline]{enumitem}
\usepackage[caption=false]{subfig}

\newcommand{\ii}[0]{\mathrm{i}}
\newcommand{\pd}[1]{\partial^{\vpp}_{#1}}\WithSuffix\newcommand\pd*[1]{\partial^{\,}_{#1}}
\newcommand{\pdp}[2]{\partial^{#2}_{#1}}
\newcommand{\thed}[0]{\mathrm{d}}
\newcommand{\theD}[0]{\mathrm{D}}
\newcommand{\dby}[2]{\frac{\thed #1}{\thed #2}}

\newcommand{\DiracDelta}[1]{\delta \left( #1 \right)}
\newcommand{\dDelta}[2]{\delta^{#2} \left( #1 \right)}

\newcommand{\kron}[1]{\delta^{\,}_{#1}}

\newcommand{\Reals}{\mathbb{R}}

\newcommand{\Ints}{\mathbb{Z}}

\newcommand{\Orth}[1]{\mathsf{O} \left( #1 \right)}
\newcommand{\SOrth}[1]{\mathsf{SO} \left( #1 \right)}

\newcommand{\chempo}[1]{\mu^{\vpp}_{#1}}\WithSuffix\newcommand\chempo*[1]{\mu^{\,}_{#1}}

\newcommand{\inprod}[2]{ \left\langle #1 \middle| #2 \right\rangle}\WithSuffix\newcommand\inprod*[2]{\langle #1 | #2 \rangle}
\newcommand{\matel}[3]{\left\langle #1 \middle| #2 \middle| #3 \right\rangle}\WithSuffix\newcommand\matel*[3]{\langle #1 | #2 | #3 \rangle}
\newcommand{\BKop}[2]{\left| #1 \middle\rangle \hspace{-0.5mm} \middle\langle #2 \right|}\WithSuffix\newcommand\BKop*[2]{| #1 \rangle \hspace{-0.5mm} \langle #2 |}
\newcommand\mO{\mathcal{O}}

\newcommand\mL{\mathcal{L}}

\newcommand\mJ{\mathcal{J}}

\newcommand\e{\mathrm{e}}
\newcommand\ud{\mathrm{d}}

\newcommand{\im}{\mathrm{Im}~} 
\newcommand{\re}{\mathrm{Re}~}

\newcommand{\p}{\partial}

\newcommand{\antisym}{\mathrm{anti}}

\newcommand{\odd}{\mathrm{odd}}
\newcommand{\even}{\mathrm{even}}

\begin{document}
\title{Generalized time-reversal symmetry and effective theories for nonequilibrium matter}
%\title{Effective theories for stochastic dynamics and active matter}

\author{Xiaoyang Huang}
\email{xiaoyang.huang@colorado.edu}
\thanks{These authors contributed equally.}
\affiliation{Department of Physics and Center for Theory of Quantum Matter, University of Colorado, Boulder CO 80309, USA}

\author{Jack H. Farrell}
\email{jack.farrell@colorado.edu}
\thanks{These authors contributed equally.}
\affiliation{Department of Physics and Center for Theory of Quantum Matter, University of Colorado, Boulder CO 80309, USA}

\author{Aaron J. Friedman}
\affiliation{Department of Physics and Center for Theory of Quantum Matter, University of Colorado, Boulder CO 80309, USA}

\author{Isabella Zane}
\affiliation{Department of Physics and Center for Theory of Quantum Matter, University of Colorado, Boulder CO 80309, USA}

\author{Paolo Glorioso}
\affiliation{Department of Physics, Stanford University, Stanford CA 94305, USA}
\affiliation{Department of Physics \& Condensed Matter Theory Center,
University of Maryland, College Park, Maryland 20740, USA}

\author{Andrew Lucas}
\email{andrew.j.lucas@colorado.edu}
\affiliation{Department of Physics and Center for Theory of Quantum Matter, University of Colorado, Boulder CO 80309, USA}

\date{\today}

\begin{abstract}
The past decade has witnessed the development of systematic effective theories for dissipative \emph{thermal} systems.  Here, we describe an analogous effective theory framework that applies to the classical stochastic dynamics of \emph{nonequilibrium} systems. We illustrate this approach using a range of examples, including nonreciprocal (predator-prey) dynamics, dissipative and driven rigid-body motion, and active chiral fluids and solids. Many of these systems  exhibit a generalized time-reversal symmetry, which plays a crucial role within our formalism, and in many cases can be implemented within the Martin-Siggia-Rose path integral. This effective theory formalism yields generalizations of the fluctuation-dissipation theorem and second law of thermodynamics valid out of equilibrium. By stipulating a stationary distribution and a set of symmetries---rather than postulating the 
stochastic equations of motion directly---this formalism provides an alternative route to building phenomenological models of driven and active  matter. We hope that this approach facilitates a systematic 
investigation of the universality classes of active matter, and provides a common language for 
nonequilibrium many-body physics from high energy to condensed matter.
\end{abstract}

\maketitle

\tableofcontents

\section{Introduction}
\label{sec:intro}

A hallmark achievement of 20th-century physics was the development of the  Wilsonian renormalization group (RG), and with it, effective field theory (EFT)  \cite{WilsonPRBI, WilsonPRB2, hohenberghalperin, WeinbergLagrangians, altland_simons_2010}. This framework provides a systematic approach to predicting the emergent macroscopic behavior of \emph{generic} microscopic systems based on coarse graining and symmetries. The EFT framework was initially developed in the contexts of ground-state 
physics in quantum field theories and the finite-temperature thermal physics of statistical field theories. In this paper, we describe an EFT framework for general dissipative and stochastic classical systems, both in and out of equilibrium.

\subsection{Review of Wilson's paradigm}
\label{subsec:Wilson overview}

In the context of classical systems, EFTs were first systematically developed as statistical field theories describing thermal equilibrium. Starting from a microscopic model, one can use the principles of RG \cite{WilsonPRBI, WilsonPRB2, hohenberghalperin, WeinbergLagrangians, altland_simons_2010} to recover a simple effective theory (i.e., a free energy) that captures all universal properties. 

As a concrete example, consider the statistcal theory of the paramagnet-to-ferromagnet phase transition. In 3D, an appropriate \emph{microscopic} model that realizes this transition is the Ising model. Denoting by $S_i = \pm 1$ a binary classical spin on site $i$ of a 3D lattice, the partition function for the Ising model is
\begin{equation} \label{eq:Zising}
    Z_{\rm Ising} \, = \, \e^{-\beta H} = \sum_{\{S_i\}} \exp\left[\beta \sum_{\expval{i,j}} S_i S_j\right] \, ,~~
\end{equation}
where $\expval{i,j}$ denotes neighboring spins $i$ and $j$. The model described by Eq.~\eqref{eq:Zising} hosts a phase transition at a critical temperature corresponding $\beta=\beta_c = 1/T_c$. At small $\beta< \beta_c$ (high temperature), a typical configuration drawn with probability $Z^{-1} \exp(-\beta H)$ is exponentially likely to have roughly half of each $S=\pm 1$; at large $\beta>\beta_c$ (low temperature), the configurations are exponentially more likely have a clear majority of $S$ aligned. While the microscopic theory captured by Eq.~\eqref{eq:Zising} is useful for numerical simulations of the transition (e.g., via Monte Carlo), it
gives little insight into the long-distance behavior one can expect in each phase (or at the critical point), nor whether various \emph{emergent} macroscopic properties are \emph{universal} (i.e., insensitive to fine-tuned microscopic details).

The \emph{philosophy} of Wilsonian EFT differs substantially from the foregoing approach. Instead of a microscopic lattice model, one instead starts by considering the symmetries of the system and the natural IR modes that should appear in the EFT. In the ferromagnetic phase of the Ising model, one expects macroscopic ordering, corresponding to all spins aligning as either $S_i=+1$ or $S_i=-1$. We then coarse grain the lattice model into a continuum EFT by defining, e.g.,
\begin{equation}
    \phi(x) \, = \, \sum_i S_i \frac{\e^{-(x-x_i)^2/2\sigma^2}}{(2\pi \sigma^2)^{3/2}} \, ,~~
\end{equation}
so that when $\sigma \gg a$ is much  larger than the lattice spacing, we expect $\phi(x)$ to be reasonably smooth. We then postulate that the partition function \eqref{eq:Zising} is well approximated by 
\begin{equation}
\label{eq:isingeft} 
Z_{\rm Ising} \, \sim \, \exp\left[-\int \thed^3x \, \left(A \, \abs{\nabla \phi}^2 + B \, \phi^2 + C \phi^4 + D \phi^6  +  \cdots \right)\right] \, ,~~  
\end{equation}
up to normalization and higher-order terms in $\phi$, where the ($\beta$-dependent) phenomenological coefficients $A,B,C,D,\cdots$ are not calculated from Eq.~\eqref{eq:Zising}, but instead determined from experimental data. The universal physics of the microscopic theory \eqref{eq:Zising} is encoded in the effective theory \eqref{eq:isingeft} via the parameter $B$, which like $\beta$ in the microscopic theory, controls whether the system realizes a paramagnet or ferromagnet as one tunes through the critical value $B_c$.

We comment that the \emph{effective} theory \eqref{eq:isingeft} follows from the microscopic theory \eqref{eq:Zising} under RG. Importantly,  RGs  organize degrees of freedom by their ``relevance'' to the system's long-distance and late-time behavior, and systematically integrate out the short-distance  degrees of freedom to recover a ``coarse-grained'' theory. In the thermodynamic limit, the result is an effective field theory \eqref{eq:isingeft} that is insensitive to the precise details of the microscopic theory \eqref{eq:Zising}. 

However, one can also write down an effective theory \eqref{eq:isingeft} \emph{without} coarse graining a microscopic model. The crucial features of  Eq.~\ref{eq:isingeft} are that the effective free energy is (\emph{i}) spatially local and (\emph{ii}) has the same $\Ints_2$ symmetry as the partition function \eqref{eq:Zising}, where $S_i \to - S_i$ becomes $\phi \to -\phi$. This symmetry reflects the fact that the ferromagnetic phase is equally likely to have all spins up versus down. To recover the effective theory, one need only write down all terms---organized by their order in the field $\phi$ and number of derivatives $\p$---that are local and compatible with the anticipated symmetry of the model (in this case, $\Ints_2$). In the vicinity of the critical point $B=B_c$, the EFT captured by Eq.~\eqref{eq:isingeft} accurately predicts the same critical exponents and universal properties as the discrete Ising model \eqref{eq:Zising}. Such an EFT describes \emph{any} system with the same universal properties as the Ising model described by Eq.~\eqref{eq:Zising}, independent of microscopic details. In the remainder of this paper, our perspective is that one may safely construct EFTs in the form of an action (Lagrangian) or free energy by writing down all terms compatible with locality and any desired symmetries in an organized fashion.

\subsection{The challenge of dissipative, noisy dynamics} \label{sec:challenge}

The EFT paradigm provides a systematic way to analyze and classify the universal physics of equilibrium systems.  We seek an analogue of this approach for nonequilibrium systems and \emph{active} matter\footnote{While there does not seem to be a formal definition for active matter, a heuristic definition is that active matter corresponds to dynamics with ``self-propelled" particles.  In general, such systems cannot be described by Hamiltonian mechanics---at least, not straightforwardly---and thus, there is no notion of thermal equilibrium or conventional thermodynamics.}. For concreteness, we adopt the perspective herein that ``nonequilibrium'' systems are those in which detailed balance---or a more general ``reversibility'' (i.e.,  the stochastic notion of microscopic time-reversal symmetry)---are violated. 

Because nonequilibrium systems do not generally obey Hamiltonian dynamics---and may seem to violate a second law of thermodynamics and a fluctuation-dissipation theorem---it is unclear \emph{a priori} what aspects of the EFT paradigm can be applied in such a setting.  Some aspects of EFT---including the philosophy of working with coarse-grained degrees of freedom and performing an RG analysis---have been applied to nonequilibrium systems in the literature. Still, we argue that the conventional approach to studying nonequilibrium matter has some \emph{philosophical} differences  compared to equilibrium EFTs.  These differences are summarized by the following fact: \textit{Effective theories for dynamical systems typically start from equations of motion}. For example,  Hohenberg and Halperin endeavored to classify all possible (stochastic) equations of motion that could realize dissipative relaxation to thermal equilibrium \cite{hohenberghalperin}. The path-integral framework of Martin, Siggia, and Rose (MSR) then assigns to such stochastic equations of motion a corresponding \emph{Lagrangian} \cite{MSR}; one can then apply RG to that Lagrangian, but \emph{only after postulating the equations of motion} that went into the MSR path integral. The MSR framework also features in Toner and Tu's phenomenological model of \emph{flocking} \cite{TonerTu}, which takes the form of a coarse-grained hydrodynamic equation of motion that, roughly speaking, extends the Navier-Stokes equation to birds. While these efforts have indeed identified new universality classes beyond the traditional equilibrium setting, starting from equations of motion does make it more challenging to recover \emph{all} of the benefits of the equilibrium EFTs:  (\emph{i}) The symmetries that the EFT should obey are less clear than in equilibrium; (\emph{ii}) dynamical stability constraints are not manifest; and (\emph{iii}) it is not obvious whether the equations of motion might contain terms that, while apparently allowed by symmetry, are in fact forbidden in any theory that arose from coarse graining microscopic degrees of freedom.

This paper is organized around the goal of building an effective theory framework for nonequilibrium dynamics that (\emph{i}) takes the same philosophical approach as the Wilsonian EFTs, (\emph{ii}) avoids the potential pitfalls mentioned above, and (\emph{iii}) generalizes as much of equilibrium physics as possible to nonequilibrium systems.  Our motivation is twofold. On a \emph{practical} level, it is easier to construct effective theories in terms of Lagrangians than equations of motion, because the latter transform \emph{covariantly} under symmetries, while the former are invariant, and thereby easier to classify. On a \emph{fundamental} level, an EFT approach has the advantage that stability of the resulting theory and invariance under appropriate symmetries become manifest.

We now give a brief overview of several examples of nonequilibrium systems for which starting from microscopic equations of motion---even coarse-grained ones---can fail to recognize constraints that a systematic ``top-down'' approach (starting from a Lagrangian) would naturally incorporate. 
\begin{enumerate}
    \item Consider a continuous active medium, such as an odd elastic solid \cite{odd_elasticity_2020, VV_odd_review}, which can be microscopically modeled by an active mechanical system consisting of Hookean springs with nonreciprocal couplings. The equations of motion for this microscopic medium often lead to instabilities, which in turn drive the medium out of the regime of validity of Hooke's Law, which assumes that each spring is only weakly perturbed out of equilibrium.  
    
    An EFT aims not to describe a system whose microscopic equations are known, but rather, the coarse-grained dynamics of a system whose microscopic dynamics are too complex to deduce \emph{ab initio}. We would not observe in nature an odd elastic solid, with a linear stress-strain relation, unless the phase was stable. Hence, weseek to incorporate this stability criterion directly into our EFT.  This stability criterion is often related to a generalized notion of time-reversal symmetry, which is challenging to see starting from equations of motion.  An EFT in which the time-reversal transformation is more transparent thus makes it easier to understand the physical origin of any constraints on coarse-grained stochastic equations of motion. 
    \item In fluids that intrinsically break rotational invariance---i.e., in which $\Orth{d}$ is broken to a discrete subgroup---there are numerous terms that can be added to the hydrodynamic equations of motion compared to the Navier-Stokes equations describing isotropic fluids  \cite{huang_2022_discrete, Friedman:2022qbu}. In \emph{thermal equilibrium}, many such terms are forbidden by the requirement that the theory remains consistent when coupled to curved spacetime; this is easiest to diagnose in a (Lagrangian) effective field theory \cite{huang_2022_discrete}. That EFT also elucidates the connection between such anisotropic terms and \emph{anomalies} \cite{lucaprl, Qi:2022vyu}, which place further constraints on the functional form of these anisotropies. Diagnosing whether these constraints carry over to nonequilibrium and active systems---and if so, their fundamental origin---likely requires  ``a Lagrangian EFT'' for nonequilibrium systems as well.
    \item In kinetically constrained fluids \cite{gromovfrachydro, knap2020, morningstar, hart2021hidden, Glorioso:2021bif, Grosvenor:2021rrt, Burchards:2022lqr}, there are terms that are are forbidden from appearing in the stochastic equations of motion despite being compatible with all known symmetries. Adding such terms changes the universal dynamics (i.e., the critical exponents), realizing a \emph{different} theory! Such terms cannot arise from the coarse-graining procedure that recovers an EFT from a microscopic model, and this is known to hold even out of equilibrium \cite{Guo2022}, as we discuss in Sec.~\ref{sec:fracton}. Hence, knowledge of all symmetries is not sufficient to adequately constrain the equations of motion.
\end{enumerate}

Until recently, the phenomenology of dissipative and stochastic dynamics---even those that relax to thermal equilibrium---was developed primarily through a ``bottom-up'' approach starting from microscopic equations of motion. Over the past decade, high-energy theorists have developed systematic ``top-down'' \emph{quantum} EFTs starting from Lagrangians to describe the hydrodynamics of dissipative thermal fluids \cite{harder2015thermal, eft1, eft2, haehl2016fluid, jensen2018dissipative}. Through careful consideration of the system's symmetries, these efforts have derived features---such as the usual fluctuation-dissipation theorem (FDT)---of the corresponding hydrodynamics that previously had to be inserted by hand (e.g., into the MSR path integral). Crucially, the FDT is guaranteed by the Kubo-Martin-Schwinger (KMS) symmetry \cite{Kubo_1966, martinschwinger, ArakiKMS78} that generalizes time-reversal symmetry to dissipative thermal systems. A more careful analysis within EFT has also derived an \emph{entropy} current with nonnegative divergence realizing a second law of thermodynamics in such dissipative systems \cite{Glorioso:2016gsa}. Hence, the celebrated phenomenology of hydrodynamics in thermal equilibrium has been carefully derived from dissipative effective field theories based on Lagrangians, rather than equations of motion.

\subsection{The purpose of this paper}

This paper addresses the question: Can the EFT for dissipative thermal systems can be extended to (certain) nonthermal systems?  The goal of our approach is to systematically obtain a stochastic (Langevin) equation for a set of variables $\vb{q} = \{q_a\}$, such that the resulting equations avoid all of the potential pitfalls above.  Here $q_a$ represent slow degrees of freedom that survive after coarse-graining out complicated microscopic dynamics. For much of the paper $q_a$ will be a discrete set of variables, but in Sec.~\ref{sec:continuum} we discuss the natural extension of our methods to field theories.  

As we will see, it is easiest to systematically build stochastic equations by constructing a \emph{Fokker-Planck equation} for the probability distribution $P(\vb q,t)$ of finding any configuration at time $t$.  The Fokker-Planck equation will be constructed using the philosophy of the Wilsonian EFT paradigm, where it will be easier to impose the desired symmetries and positivity constraints.  More concretely, we proceed as follows:
\begin{itemize}
    \item The \emph{starting point} of our effective theory is to determine---or, more often, to \emph{postulate}, \`a la Eq.~\eqref{eq:isingeft}---the stationary probability distribution $\exp (-\Phi)$ for the stochastic dynamics.  In other words, we begin by proposing the structure of the equal-time correlation functions of the nonequilibrium fixed point\footnote{In contrast, $\exp(-\Phi)$ can at best be calculated perturbatively in other approaches that instead postulate equations of motion (in the absence of detailed balance). See, e.g., Ref.~\citenum{wijland}. On the other hand, there could exist local equations of motion that lead to nonlocal stationary distributions $\Phi$, and this is challenging to describe with our framework.}.
    \item Once the stationary distribution is known, we write down the most general (Markovian) stochastic dynamics consistent with this stationary state. The algorithmic approach to effective field theory also accounts for all symmetries, conservation laws, and constraints present in the dynamics.
    \item Stability criteria arise naturally from two places: the stability of the proposed phase in $\Phi$, and the positivity of noise variance. A generalized fluctuation-dissipation theorem (Tab.~\ref{tab:summary}) and second law of thermodynamics~\eqref{eq:2nd law} can be formally stated.  The Mermin-Wagner theorem \cite{merminwagner} and other constraints on equilibrium statistical physics extend neatly to our nonequilibrium formalism.
    \item With knowledge of $\Phi$, identifying terms that are ``even" and ``odd" under detailed balance~\eqref{eq:FPE balance condition} (a stochastic notion of time-reversal symmetry) is transparent.  In particular, it is straightforward to impose generalized time-reversal symmetries, labeled as T, and to understand generalized time-reversal transformations within the Martin-Siggia-Rose (MSR) path integral (see Sec.~\ref{sec:MSR}).
\end{itemize}

Philosophically, our approach can be thought of as blending the MSR approach to nonequilibrium matter with the KMS-invariant effective field theories for dissipative thermal systems, which were organized systematically in terms of symmetry-invariant building blocks, as in Wilsonian equilibrium EFTs.\footnote{Another approach to try and extend the KMS-invariant EFT for a thermal system to active matter was recently presented in Ref. \cite{landry23}.} However, while our approach is inspired by EFTs for thermal systems, we emphasize that it is not simply recasting the same EFT used for thermal systems in a ``nonequilibrium setting".   In a thermal system, the Hamiltonian $H$ plays two crucial roles: first, it generates time evolution, and second, the probability of finding the system in any given state $q$ is $\e^{-\beta H(q)}$.  In our EFT for more general stochastic systems, we will see that the $\Phi$ introduced above need not directly relate to the dissipationless dynamics.  Indeed, we will see that many active systems are characterized by ``equations of motion" that are not directly related to the $\Phi$ which characterizes the steady state of the stochastic dynamics.  Indeed, part of this paper will provide an algorithm to systematically study the most general possible active or stochastic theory compatible with the desired stationary state.

\begin{figure}[t]
    \
    \centering
    \resizebox{0.8\textwidth}{!}{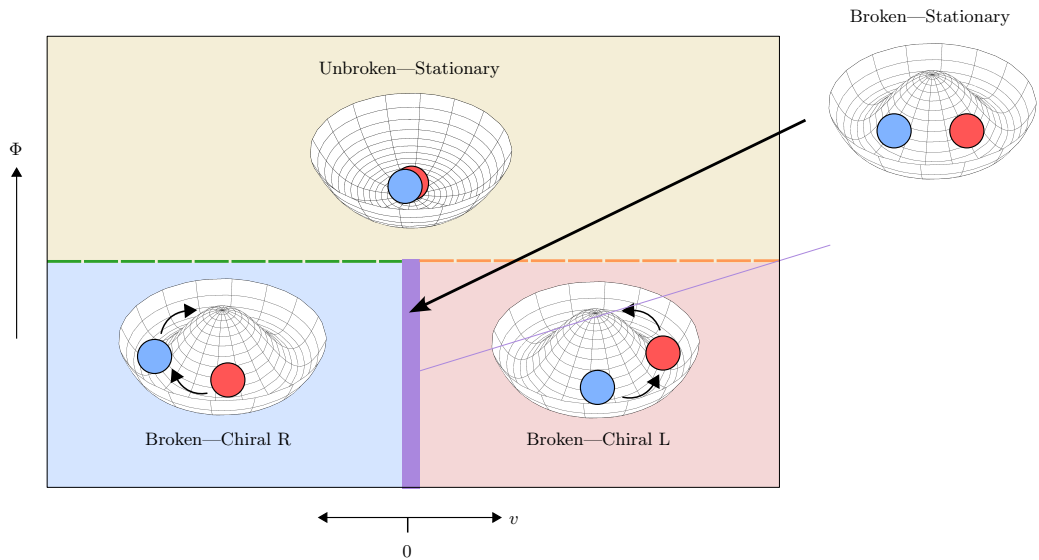}
    \caption{A schematic ``phase diagram'' that summarizes our formalism, including T-breaking phase transitions.  Starting with $v=0$, we may tune the stationary distribution $\Phi$ so that it has degenerate minima---this control over $\Phi$ is represented by the vertical axis of the figure.  Phase transitions from the disordered phase into the stationary broken phase are controlled by noise strength and the minima of $\Phi$ according to statistical mechanics.  But separately, we may tune the dissipationless and T-breaking $v$ from positive to negative, driving motion in one of two directions along the degenerate minima of $\Phi$ and creating a T-broken phase.  Control over $v$ is represented by the horizontal axis, orthogonal to control over $\Phi$ (vertical axis). }\label{fig:summaryofapproach}
\end{figure}

A key advantage of taking this perspective is that the phase diagram of the models that we build will always ``decompose" along two orthogonal (sets of) ``axes".  Firstly, the phase diagram will contain all \emph{thermodynamic} phase transitions captured by $\Phi$: for example, tuning the overall prefactor of $\Phi$ could result in an ordering transition from a paramagnet to a ferromagnet.  Once we have found the phases captured by $\Phi$, we can then systematically determine all of the possible dissipationless stochastic dynamics: intuitively, these capture how the system can circulate around the phase space without modifying $\Phi$.  When such dissipationless terms are deterministic, we can completely classify them (see e.g. Sec.~\ref{sec:classify SSB}).  Additional phase transitions can further then arise as a consequence of these dissipationless dynamics.  A simple  example of this phenomenon is the chirality-breaking phase transition found in nonreciprocal predator-prey dynamics (Sec. \ref{sec:kuramoto}).  Putting these two ingredients together, we obtain a schematic phase diagram illustrated in Fig.~\ref{fig:summaryofapproach}.

Some phases of active matter, such as flocking \cite{TonerTu} or motility-induced phase separation \cite{mips}, are not easy to describe within our framework.   As one example, one must sacrifice strict spatial locality in $\Phi$ to realize the Toner-Tu fixed point, as one finds nontrivial scaling exponents in spatial correlators.  As incorporating this directly into $\Phi$ may feel less motivated within the context of our formalism, we have elected to focus on other recent theories of active matter including active solids \cite{odd_elasticity_2020}, fluids \cite{VV_2021_fluctuating}, and nonreciprocal dynamics \cite{nonreciprocalPhase}, where we (\emph{i}) reproduce many of the nontrivial active phases and phenomena observed previously in the literature and (\emph{ii}) provide generalizations of these models.

\subsection{Overdamped oscillator: a minimal example of our framework}\label{sec:introT}

Having outlined our approach to model building in words, we now provide an example (with equations) describing the simplest possible stochastic dynamical system: the overdamped harmonic oscillator. We gloss over a few technical subtleties, which we instead be address more carefully in Sec. \ref{sec:EFT}. 

We first describe how one might approach this system by starting with equations of motion directly, as is common in the literature.  We start by considering the dissipative system
\begin{equation}
    \dot{x} = -\gamma x \, ,~~ \label{eq:newton}
\end{equation}
and, for simplicity, we assume that $x$ is the coordinate of a macroscopic rigid body that is attached to a spring while suspended in a very viscous fluid. The coefficient $\gamma$ denotes the damping of the body's motion towards the equilibrium position $x=0$, where the spring is unstretched. Note that Eq.~\eqref{eq:newton} explicitly breaks time-reversal symmetry: The arrow of time corresponds to the pull of the particle towards its equilibrium position at $x=0$. The condition $\gamma>0$ is required because the system flows towards, rather than away, from equilibrium.  Indeed, the solutions to Eq.~\eqref{eq:newton} take the form 
 \begin{equation}
    x(t) \, = \, A\,  \e^{-\gamma t}\,, \label{eq:introinit}
\end{equation}
which is clearly not invariant under $t\to-t$.

Microscopically, collisions between the molecules of the fluid and the rigid body impart small stochastic forces (thermal fluctuations) that modify Eq.~\eqref{eq:newton} according to 
\eqref{eq:newton} to 
\begin{equation}
     \dot{x} = -\gamma x + \xi\,,  \label{eq:newton2}
\end{equation} 
where the noise term above satisfies 
\begin{equation}
    \overline{\xi(t)\xi(t^\prime)} = 2 D \delta(t-t^\prime)\, ,~~
\end{equation}
and we require that $D\geq0$, as $\xi(t)$ is a real-valued random variable. Since the trajectory $x(t)$ is \emph{stochastic}, we can calculate the probability $P(x_0,t)$ to find $x(t)=x_0$ given some initial condition, which obeys a Fokker-Planck equation \cite{Fokker, Planck, Kolmogorov-backward, Kramers, Moyal, PawulaFPEK, PawulaBoltzmann, KadanoffBook}:
\begin{equation}
    \partial_t P = \partial_x \left(\gamma x P \right) + D \partial_x^2 P\,. \label{eq:introFP1}
\end{equation}
We deduce that, at late times $t\to \infty$, the system relaxes to the stationary distribution 
\begin{equation}
    P(x) = \e^{-\Phi(x)} = \e^{-\gamma x^2/2D}\,.
\end{equation}
Importantly, the justification above for the noisy Fokker-Planck description \eqref{eq:introFP1} of the overdamped oscillator began with an equation of motion, subsequently incorporated noise, and ultimately deduced that, unsurprisingly, the system is most likely to be found near $x=0$.   

We now outline the implementation of our approach for the simple example above. The starting point is to write down the general form of a master equation that governs the evolution for the probability distribution $P(x,t)$ \eqref{eq:introFP1}, without knowing anything about the microscopic details of the underlying system. As is common in macroscopic approaches, the natural next step is to write the master equation in terms of a derivative expansion:
\begin{equation}
    \partial_t P(x,t) = \partial_x \left[ A(x) \left(\partial_x + B(x) \right) P(x,t)\right]+\cdots\,, \label{eq:introFP20}
\end{equation}
where the dots denote higher-derivative terms that encode non-Gaussian noise. Higher derivatives become progressively less important as far as the macroscopic length scales of interest are much longer compared to lengths that are intrinsic to the microscopic processes of the system.  Keeping only second derivatives in Eq.~\eqref{eq:introFP20} implies that the noise is effectively Gaussian because, in our example, the object on a spring is heavy compared to the microscopic particles whose collisions with it lead to damping forces.  Note that all terms on the right-hand side of Eq.~\eqref{eq:introFP20} are total derivatives as required by conservation of probability. The next step is to postulate the existence of a steady state $\exp (-\Phi(x))$. The choice of $\Phi(x)$ depends on the of phenomenology of interest; in this example, we assume that $\Phi$ has a minimum at $x=0$, and we further restrict to the linear regime, so that we can approximate
 \begin{equation}
 \Phi(x) = \frac{1}{2}\beta k x^2 + \cdots \, ,~~ \label{eq:introkx2}
\end{equation}
where the dots stand for higher powers of $x$. In the above, we wrote the prefactor as $\beta k$ for convenience, and we assume $\beta k \geq 0$ to ensure stability at $x=0$. Requiring that $\Phi$ is a steady state for \eqref{eq:introFP20}, leads to the requirement that \begin{equation}
    B(x) = \pd{x} \Phi(x)
\end{equation}
and hence Fokker-Planck equation
\begin{equation}
    \pd{t} P = \pd{x} \left[ A(x) \left(\pd{x} + \beta k x \right) P\right]\, , ~~  \label{eq:introFP2}
\end{equation}
where we additionally require that $A(x)\geq 0$ to ensure that the system relaxes \emph{towards} the steady state. Then, Eq. \eqref{eq:introFP2} is the most general Fokker-Planck equation consistent with our requirements. Upon identifying
\begin{equation}\label{eq:ident1}
    A = D, \;\;\;\;\; \;\;\; A\beta k =\gamma\,,
\end{equation}
we precisely recovered Eq.~\eqref{eq:introFP1}. We thus uniquely determined the dynamics of this stochastic system solely in terms of a minimal set of general assumptions, without knowing any detail about the underlying dynamics.

When $A(x)=D$ is a constant, the approach above can also be elegantly summarized using the MSR path integral \cite{MSR}.  We do not introduce the notation here, although expect it to be familiar enough for many readers ( for details, see Sec.~\ref{sec:MSR}). Letting the noise field be $\pi$, the MSR Lagrangian is 
\begin{equation}
L = \pi \dot x + \ii \pi A \left(\pi -\ii kx\right) \, ,~~    \label{eq:introLMSR} 
\end{equation}
which can intuitively be read off from Fokker-Planck equation \eqref{eq:introFP20} upon replacing $\partial_x \to -\ii \pi$.   Conventionally, if one started with a stochastic equation \eqref{eq:newton2}, \eqref{eq:introLMSR} can be written down using standard rules.  However, we will see that the form of (\ref{eq:introLMSR}) is completely fixed by demanding stationarity with $\Phi$ given by \eqref{eq:introkx2}; this constraint can be viewed as implying that, upon transforming \begin{equation}
    \pi \rightarrow -\pi + \ii k x \, ,~~
\end{equation}
the transformed $L$ is again a valid MSR Lagrangian (all terms have at least one power of $\pi$).  In fact, in this particular theory, $L$ is \emph{invariant} under this transformation.

The invariance under this transformation turns out to imply that this simple one-dimensional model is compatible with  microscopic \emph{time-reversal symmetry}.  It is often desirable to demand this as a feature of a dissipative EFT: Since $F=ma$ is time-reversal symmetric, this is a symmetry that must manifest in the spring-fluid system, so long as the noise merely arises from ``integrating out" the bath degrees of freedom. In a dissipative system, this integrating-out  procedure implies that the stochastic system obeys the \emph{detailed balance} condition:
\begin{equation}
   \mathbb{P}\left[x(t_1)=x_1 \to x(t_2)=x_2 \right] \e^{-\Phi(x_1)} = \mathbb{P}\left[x(t_1)=x_2 \to x(t_2)=x_1 \right] \e^{-\Phi(x_2)} \, ,~~ \label{eq:introdb}
\end{equation}
and we refer to this property as \emph{reversibility}, which we denote by ``T" going forward.

As we show in Sec.~\ref{sec:EFT}, reversibility manifests in the requirement that the generator of the Fokker-Planck equation is invariant under a simple transformation, even as we generalize to systems with more degrees of freedom and multiplicative and/or non-Gaussian noise.  Therefore, reversibility can prove to be a very powerful symmetry.  And even if reversibility is not a symmetry, we still find it extremely helpful to keep track of the reversibility transformation: for example, it may often be a reasonable physical postulate that the reversed theory is also a local theory.  We will see that this requirement is equivalent to the postulate that $\Phi$ is local.  So long as $\Phi$ is local, the formalism that we develop in this paper provides non-trivial constraints on the stochastic effective theory.

Notice that the definition of reversibility given in Eq.~\eqref{eq:introdb} depends on $\Phi$.   One alternative definition of this symmetry that avoids introducing $\Phi$ is 
\begin{equation}
    \mathbb{P}[x(0) = a, x(t) = b, x(2t) = c, x(3t)=a] = \mathbb{P}[x(0)=a, x(t) = c, x(2t) = b, x(3t)=a].
\end{equation} Unfortunately, it is not easy to implement such a symmetry manifestly within effective theories, while -- in contrast -- (\ref{eq:introdb}) is.  It is for this reason that, in our framework,  we first deduce  $\Phi$ before writing down the equations of motion.

\subsection{Organization of the paper}
The rest of this paper is organized as follows. 

Sec.~\ref{sec:EFT} contains a pedagogical mathematical tour through the Fokker-Planck equations, and emphasizes how to properly implement reversibility and detailed balance (and classify dissipative and dissipationless coefficients) for any system (thermal or nonthermal).   At the end of the section, we explain how in certain limits, our approach relates to the MSR path integral and the thermal EFTs \cite{eft1,eft2} which were recently developed.

Sec.~\ref{sec:not active} details the application of our formalism to thermal systems, whose dissipationless dynamics are described by Hamiltonian mechanics.  In doing so, we provide elegant and algorithmic derivations of a few hallmark results, such as the minimal Landau-Lifshitz-Gilbert damping for a classical spin.    

Sec.~\ref{sec:active} extends our formalism to active matter, beginning with a few simple examples.   In Sec.~\ref{sec:general_T_breaking}, we then describe a systematic classification of all possible phases in which a generalized time-reversal symmetry is spontaneously broken.  This discussion gives a more systematic way of understanding the possible phases of nonreciprocal matter, and includes a number of interesting extensions of the predator-prey-like models of \cite{nonreciprocalPhase}.   The example of active (or even simply damped) rigid body motion is presented in Sec.~\ref{sec:rigidbody}; the nontrivial classical phase space of the rigid body renders our formalism especially well-suited to the careful analysis of this system.

Sec.~\ref{sec:continuum} then describes the generalization of our methods to continuum models and field theories.  We explain how to understand active solids and fluids in our framework, and connect to some of the puzzles we raised in Sec.~\ref{sec:challenge}.  A general method for incorporating desired active terms into our effective theories, in the simplifying limit of lattice models, is described briefly in Sec.~\ref{sec:phase separation}, in the context of active phase separation.

\section{Effective (field) theory for stochastic dynamics}
\label{sec:EFT}

Here we develop the building blocks of the effective field theory (EFT) %formalism 
framework that describes thermal, dissipative, and active classical matter. We make precise the intuitive arguments of Sec.~\ref{sec:introT} on the role of reversibility symmetry T---also known as \emph{detailed balance} in this context. As noted therein, an effective theory for dissipative dynamics \emph{must} incorporate noise; this is most naturally captured using an inherently \emph{stochastic} formalism. For convenience, we represent the configurations of a system as elements of a linear vector space, in analogy to quantum mechanics, and as prescribed by the Doi-Peliti formulation \cite{Doi1, Doi2, Peliti1, Peliti2}. %As a warm up, we 
We first consider the stochastic dynamics of classical systems with a discrete configuration space, with dynamics generated by a continuous-time Markov process. The continuum analogue corresponds to a stochastic differential equation---e.g., the Langevin equation \cite{Langevin, LangevinTranslate}. However, we find it most transparent to consider the Fokker-Planck equation (FPE) \cite{Fokker, Planck, Kolmogorov-backward, Kramers, Moyal, PawulaFPEK, PawulaBoltzmann, KadanoffBook} instead. Working in the FPE language, we derive powerful constraints from the existence of a stationary state (before connecting to T), a generalization of the fluctuation-dissipation theorem \cite{Kubo_1966, ThoulessStochastic, FDTPRL}, and finally, the Martin-Siggia-Rose (MSR) path-integral approach \cite{MSR}.

\subsection{Warm up: Discrete state spaces}
\label{subsec:discrete states}

As a warm up to the more general Fokker-Planck dynamics of systems with continuous state spaces, we first consider the simpler scenario of stochastic classical systems with a \emph{discrete} configuration space. We consider a single particle hopping between vertices $v \in V$ connected by edges $e \in E$ of a graph $G=(V,E)$. The allowed configurations are associated with vertices $v \in V$, and may realize physical positions, generic single-particle configurations (so that $G$ is analogous to a ``Hilbert space''), or even many-body configurations (in which case $G$ is analogous to a ``Fock space''), while the allowed transitions are associated with edges $e \in E$ \cite{Doi1, Doi2, Peliti1, Peliti2}. 

Hence, the  single-particle hopping problem captures generic classical systems with discrete state spaces. The configurations (i.e., vertices  $v \in V$) correspond to positions, while the transitions (i.e., edges $e\in E$) correspond to hops. We denote by $\ket{x}$ the state of the system in which the particle is at the position (i.e., vertex) $x$; these kets form an orthonormal basis for the space $\mathcal{H}$ of $N=\abs{V}$ total allowed states given by
\begin{equation}
\label{eq:discrete state orthonormal}
    \mathcal{H} \, = \, \Reals^{N} \, = \,  \text{span} \, \left\{ \, \ket{v} \, \middle| \,v \in V \, \right\} ~~~\text{with}~~~\inprod{x}{y}\,  = \,  \kron{x,y} ~~\forall \, x,y \in V \, , ~~
\end{equation}
where the state vectors $\ket{x} \in \mathcal{H}$ are real valued and the basis states (or classical configurations) $\{ \ket{x} \}$ are orthonormal and complete. The utility of this basis is that it facilitates the construction of a probability distribution
\begin{equation}
    \label{eq:discrete probability dist}
    \ket{P} \, = \, \sum\limits_{x \in V} \, p_x \, \ket{x}~~~\text{with}~~~p_x \, =\,  \inprod{x}{P} \, , ~~
\end{equation}
where $p_x$ is the probability to find the particle at position $x \in V$ when sampled from the distribution $\ket{P}$. The relation $p_x = \inprod{x}{P}$ follows from completeness of the $\ket{x}$ basis, with $p_x$ the overlap of the probability distribution $\ket{P}$ with the configuration $\ket{x}$. Because $p_x$ is a probability, we must have $\sum_x \, p_x = 1$. We define
\begin{equation}
    \label{eq:discrete uniform}
    \ket{u} \, = \, \sum\limits_{x \in V} \, \ket{x} ~~~\text{such~that}~~~\inprod{u}{P} \, = \, \sum\limits_{x \in V} \, p_x \equiv 1 \, ,~~
\end{equation}
where we refer to $\ket{u}$ as the \emph{uniform distribution}. We also note that the normalization of a state is unimportant, so long as the state is time independent. Time evolution (i.e., dynamics) of $\ket{P}$ \eqref{eq:discrete probability dist} follow from the master equation
\begin{equation}
\label{eq:discrete FPE}
    \pd{t}  \ket{P} \, = \, -W \ket{P} \, , ~~
\end{equation}
where $W$ is the generator of infinitesimal translations in time (much like a Hamiltonian; we use the symbol $W$ to denote the generator of stochastic Markovian dynamics). Using the fact that probability is conserved (i.e.,  $\inprod{u}{P}=1$ for any probability distribution $\ket{P}$), taking the inner product of Eq.~\eqref{eq:discrete FPE} with $\bra{u}$ gives
\begin{equation}
    \label{eq:discrete FPE left condition}
    \matel{u}{\pd{t}}{P} \, = \, \pd{t} \inprod{u}{P} \, = \, \pd{t} 1 \, =\,  0 \, = \, - \matel*{u}{W}{P}~~\forall \, \ket{P} ~~~\implies ~~~\bra{u} W \, = \, 0 \, . ~~
\end{equation}
If the initial probability distribution $p^{\,}_x (0)$ is known,  the probability distribution $p^{\,}_x (t)$ at any time $t$ recovers from
\begin{equation}
    \label{eq:discrete prob evo gen}
    \ket{P_t} \, = \, \mathbb{T} \, \e^{- \int_0^t \, \thed t' \, W (t') }  \ket{P_0} \, = \, \mathcal{U} (t,0) \ket{P_0} \,, ~~
\end{equation}
where $\mathbb{T}$ is the ``time-ordering operator'' (which ensures that the integral in the exponential is applied to the distribution $\ket{P_0}$ in chronological order), and $\mathcal{U} (t,0)$ is the \emph{propagator} from time $0$ to time $t$.  Relatedly, we define the probability that the system transitions from the configuration $x$ at time $0$ to the configuration $y$ at time $t$ under $W$ via
\begin{equation}
    \label{eq:discrete prob x to y}
    \mathbb{P} \left[ x,0 \to y,t ; W\right] = \matel*{y}{\e^{- W\, t}}{x} \, = \, \mathcal{U}^{\vpd}_{yx} (t,0) \, ,~~
\end{equation}
e.g., in the case where $W$ \eqref{eq:discrete FPE} is  time independent (i.e., static); above, $\mathcal{U}^{\,}_{yx}$ denotes the matrix elements of the propagator $\mathcal{U}$ \eqref{eq:discrete prob evo gen}. A crucial ingredient in our framework is the existence of a stationary distribution; we define
\begin{equation}
    \label{eq:discrete stationary dist}
    S \equiv \sum\limits_{x \in V} \, \e^{- \Phi_x} \, \BKop{x}{x} ~~~\text{and}~~~ \ket{P_{\rm ss}} \equiv S \, \ket{u} = \sum\limits_{x \in V} \, \e^{- \Phi_x} \, \ket{x} \, ,~~
\end{equation}
as the stationary operator and stationary state (distribution), respectively, where  $P^{\rm ss}_x = \inprod{x}{P^{\rm ss}} = \exp (- \Phi_x)$. 

Using the  definitions above, the condition of stationarity is equivalent to
\begin{equation}
    \label{eq:discrete stationarity condition}
    \dv{}{t}  \ket{P_{\rm ss}} = -W  \ket{P_{\rm ss}} = 0 \, , ~~
\end{equation}
so that the stationary state is unchanged by the dynamics generated by $W$, leading to $\mathcal{U} (t) \ket{P_{\rm ss}} = \ket{P_{\rm ss}}$. If $W$ generates stochastic dynamics via Eq.~\eqref{eq:discrete FPE} and has a stationary state \eqref{eq:discrete stationary dist} such that Eq.~\eqref{eq:discrete stationarity condition} holds, then the operator
\begin{equation}
\label{eq:discrete reverse W}
    \widetilde{W} \equiv S \, W^{T} \, S^{-1} \, , ~~
\end{equation}
generates stochastic dynamics and satisfies Eqs.~\eqref{eq:discrete FPE left condition} and \eqref{eq:discrete stationarity condition} for the \emph{same} $P^{\,}_{\rm ss}$ \eqref{eq:discrete stationary dist} as $W$. In this sense, $\widetilde{W}$ is the ``time-reversed'' partner to $W$, in that it generates the same sequence of configurations as $W$, albeit in reverse chronological order. Accordingly, we refer to the $\Ints^{\,}_2$ transformation \eqref{eq:discrete reverse W} as a \emph{reversibility} operation T. Numerous properties of the time-reversed generator \eqref{eq:discrete reverse W} are discussed in  Sec.~\ref{subsec:FPE TR}, in the context of continuous state spaces; here, we briefly state several key results for discrete state spaces.

For a particle hopping on the  graph $G=(V,E)$ with vertex set $V$ and (directed) edge set $E$, the most general form of the generator $W$ \eqref{eq:discrete FPE} of time evolution compatible with a stationary operator $S$ \eqref{eq:discrete stationary dist} takes the form
\begin{equation}
\label{eq:Discrete FPE particle W}
    W = -\sum\limits_{e \in E} \, Q_e \, \BKop{e}{e} \, S^{-1} + \sum\limits_{C=\text{cyc}(e_1,\dots, e_n)} \hspace{-3mm}  \alpha_C ~ \sum\limits_{i=1}^n \left(\BKop{e_i}{e_{i-1}} - \BKop{e_{i-1}}{e_i} \right) \, S^{-1} \, , ~~
\end{equation}
where we define a directed edge configuration $\ket{e_{xy}}\equiv \ket{x}-\ket{y}$ (ignoring normalization) if the edge $e_{xy} \in E$ corresponds to a hop from $x$ to $y$ (for vertices $x,y \in V$), the ordered set of $n$ directed edges $\{e_i\}$ forms a \emph{closed} loop in $V$, and $C=\mathrm{cyc}(e_1, \dots, e_n)$ denotes a cyclic permutation around that closed loop (with weight $\alpha_C$). The statement that $e_1, \dots e_n$ forms a closed loop on the graph $G$ is equivalent to the condition
\begin{equation}
    \sum\limits_{i=1}^n \, \ket{e_i} = 0 \, , ~~
\end{equation} 
and the sum over $C$ in Eq.~\eqref{eq:Discrete FPE particle W} corresponds to all possible cycles in the directed graph. 

Because $W$ \eqref{eq:Discrete FPE particle W} is a Markov ``rate'' matrix, it must satisfy the conditions $\matel{i}{W}{j} \leq 0$ for all $i \neq j$, $\matel{i}{W}{i} \geq 0$ for all $i$, and each row must sum to zero (where, e.g., $i$ and $j$ correspond to vertices).  Under time reversal T \eqref{eq:discrete reverse W}, the $Q^{\,}_e$ terms remain invariant, while $\alpha_C \to -\alpha_C$ switches sign.  Hence, the two terms in Eq.~\eqref{eq:Discrete FPE particle W} correspond to the most general possible T-even and T-odd terms allowed, respectively.

To see this, first consider the case in which the stationary distribution $P_{\rm ss}$ \eqref{eq:discrete stationary dist} is \emph{uniform}. Since all configurations are equiprobable (i.e.,  $\Phi = \log N$ is a constant), the operator $W$ \eqref{eq:Discrete FPE particle W} must be \emph{doubly} stochastic, which means that both the rows and columns of $W$ sum to zero. This implies that the uniform distribution $\ket{u}$ \eqref{eq:discrete uniform} is both a left and right null vector of $W$. Crucially, we note that every configuration in the ``edge basis'' $\{\ket{e}\}$ is orthogonal to the uniform distribution, meaning that diagonal operators $\BKop{e}{e}$ are compatible with $W$ being doubly stochastic, and realize the T-even terms in Eq.~\eqref{eq:Discrete FPE particle W}. Having constrained the symmetric terms in $W$, it is easy to check that nonsymmetric terms must act on at least three vertices to be compatible with a uniform stationary state. This is satisfied by considering off-diagonal terms in the edge basis, so long as the edges are nearest neighbors \emph{and} form a closed loop, as in Eq.~\eqref{eq:Discrete FPE particle W}. This is most transparent if one considers a system with three configurations, such that $W$ \eqref{eq:Discrete FPE particle W} realizes ``flow'' around a triangle; the loop condition simply extends this example to arbitrary graphs. Finally, to capture generic, nonuniform stationary distributions $\ket{S}=S \ket{u}$, we simply take $W = W_u\,  S^{-1}$, where $W^{\,}_u$ has a uniform stationary distribution proportional to $\ket{u}$), which ensures that $W$ obeys $W \, \ket{P_{\rm ss}} = W_u \, S^{-1} \, S \, \ket{u} = W_u \, \ket{u} = 0$ as required by Eq.~\eqref{eq:discrete stationarity condition}.

\subsection{From stochastic differential equations to the Fokker-Planck equation}
\label{subsec:SDE to FPE}
We now extend the discussion of Sec.~\ref{subsec:discrete states} to the more general context in which there is a continuum of allowed configurations. Consider a set of $M$ continuous variables $\vb{q} = (q_1, \dots, q_M)$, which may describe, e.g., the positions (and momenta) of $M$ particles, conserved hydrodynamic modes, the phase-space coordinates of a dynamical system, or other continuous classical degrees of freedom.   In general, each component $q_a$ is a continuous, real-valued variable, and a function of time $t$.  While we do not consider this possibility here, in principle, one may also consider systems with \emph{both} discrete variables $q_b$ and continuous variables $q_a$.  We also remark that one can write down a stochastic \emph{field theory} using analogous tools to the discrete particle mechanics we discuss here. By taking a standard continuum limit, in Sec.~\ref{sec:MSR} we recover the equivalent Martin-Siggia-Rose (MSR) \cite{MSR} path-integral representation. However, there are a number of subtleties in the MSR formalism: While the formalism is useful in a number of field theoretic (i.e., continuum) settings, our perspective is that the operator approach developed herein is much more conceptually transparent, and especially with respect to reversibility (which plays a crucial role in this work).

A stochastic differential equation (SDE) for $\vb{q}$ is realized by a Langevin equation of the form 
\begin{equation}
    \pd{t} q_a  \, =  \, f_a (\vb{q}) + \xi_a (t) \, , ~~
    \label{eq:SDE main}
\end{equation}
where $\vb{f}$ is an $M$-component force that depends on $\vb{q}$, and $\vb*{\xi}$ is a time-dependent noise source (specified below).  Unfortunately, Eq.~\eqref{eq:SDE main} is not well defined without stipulating a particular discretization convention for derivatives.  Multiple prescriptions (e.g., the Stratonovich and It\^o formulations of stochastic calculus) have been proposed \cite{lubensky}. For this reason, we argue that an algorithmic effective theory approach to Eq.~\eqref{eq:SDE main} requires a completely unambiguous framework. Moreover, it can be conceptually advantageous to consider the evolution of \emph{all} possible trajectories in a statistical sense, analogous to studying the evolution of \emph{distributions} of random variables, rather than the stochastic evolution of individual random variables themselves. These \emph{probability distributions} $P( \vb{q})$ obey the Fokker-Planck equation (FPE), realized by the standard partial differential equation of Eq.~\eqref{eq:FPE main}, which does \emph{not} suffer from the complications of stochastic calculus, and is thus unambiguous.  

As is common in the literature on stochastic dynamics \cite{Fokker, Planck, Doi1, Doi2, Peliti1, Peliti2}---and reviewed below---it is convenient to associate the allowed configurations of a classical system with a vector space, analogous to Hilbert spaces from quantum mechanics. One can then interpret the probability distribution $P( \vb{q})$ as a mathematical \emph{analogy} to the wavefunction $\psi (x) $ in quantum mechanics, and apply the intuition and methods from quantum dynamics to the underlying stochastic dynamics of Eq.~\eqref{eq:SDE main} via $P( \vb{q})$. Formally, distinct configurations of the system are labelled by ``kets'' of the form $\ket{\vb{q}} = \ket{q_1, \dots, q_M}$. We define the inner product between two configurations $\ket{\vb{q}}$ and $\ket{\vb{q'}}$ as
\begin{equation}
    \inprod{\vb{q}}{\vb{q'}} \, = \,  \dDelta{\vb{q}-\vb{q'}}{M} \, = \,  \prod_{n=1}^M \,\DiracDelta{q^{\vpp}_n-q^{\prime}_n} \, , ~~
    \label{eq:Bra-Ket inner product}
\end{equation}
in analogy to the inner product(s) for finite- and infinite-dimensional Hilbert spaces.

The utility of this bra-ket formulation is that one can identify a probability distribution $P (\vb{q})$ for a stochastic classical system analogous to the wavefunction $\psi (x)$ of a deterministic quantum system. The probability density function $P(\vb{q})$ can be expressed as an overlap with the distribution vector $\ket{P}$, given by
\begin{equation}
    \label{eq:Probability}
    P( \vb{q}) \, = \,  \inprod{\vb{q}}{P} \, ,~~
\end{equation}
where the effective theory we develop %describes 
captures the dynamics of $P(\vb{q})$.

In particular, the dynamics associated with the SDE  \eqref{eq:SDE main} is naturally and unambiguously described by the  Fokker-Planck equation (FPE), given in the It\^o formulation (where all derivatives are placed on the \emph{left}) by
\begin{equation}
    \label{eq:FPE main}
    \pd{t} P \, = \, - \pd{a} \left( f_a \, P - \frac{1}{2} \pd{b} (Q^{\vpp}_{ab} \, P )+ \dots \right) \, ,~~
\end{equation}
where $\pd*{a}$ is a shorthand for $\pdv*{}{q_a}$, and  the matrix $Q$ %is associated with 
relates to the variance of %the \emph{Gaussian} part of the 
noise term $\vb*{\xi}$ in Eq.~\eqref{eq:SDE main} via
\begin{equation}
    \label{eq:Q noise def}
    Q^{\vpp}_{ab} \delta(t-t') \, \equiv  \, \expval{\xi_a (t) \xi_b (t') } \, \to \, \Delta \, \kron{a,b} \, \DiracDelta{t-t'}  \,, ~~
\end{equation}
where the latter relation applies to additive Gaussian noise---for which  $\vb*{\xi}$ is independent of $\vb{q}$ and only the second moment \eqref{eq:Q noise def} is nonzero---with strength (i.e., variance) $\Delta$, which we take to be independent of $a$ for simplicity. 

For more general noise sources $\vb*{\xi} (\vb{q})$ in Eq.~\eqref{eq:SDE main}, we always require that $\expval{\xi_a(t)}=0$, while nonvanishing higher-order cumulants of $\vb*{\xi}$ contribute to the $\dots$ terms in Eq.~\eqref{eq:FPE main}. For clarity of presentation, many of the examples we consider in later sections involve additive, uncorrelated Gaussian noise, which satisfies Eq.~\eqref{eq:Q noise def}; however, our formalism extends to more generic, non-Gaussian and/or multiplicative noise sources, as we describe in Sec.~\ref{subsec:FDT}.

Importantly, Eq.~\eqref{eq:FPE main} can be recast in the form of a \emph{master equation},
\begin{equation}
    \label{eq:FPE kets}
    \dby{}{t} \ket{P(t)}  \, = \, - W \ket{P(t)} \, ,~~
\end{equation}
also called the ``quantum Hamiltonian'' description \cite{Schutz_SSEP} due to its resemblance of the Schr\"odinger equation (up to being real valued; see also \cite{Doi1, Doi2, Peliti1, Peliti2}). The \emph{differential} operator $W$ generates time evolution, and is known as the ``rate matrix'' in the context of continuous-time Markov chains. We infer its form from Eq.~\eqref{eq:FPE main},
\begin{equation}
    \label{eq:W form}
    W \, \equiv \, \pd{a} f_a - \frac{1}{2} \,\pd{a} \pd{b} Q^{\vpp}_{ab} \, , ~~
\end{equation}
where, because $W$ is an operator, the partial derivatives above act not only on $F$ and $Q$, but also on any object to which $W$ is applied (as in quantum mechanics). Regarding Eq.~\eqref{eq:FPE kets}, the probability distribution $P (\vb{q})$ \eqref{eq:Probability} at time $t$ can be expressed in terms of initial conditions using the \emph{propagator} ${\cal U}$, which is implicitly defined via
\begin{equation}
    \ket{P(t)} \, = \, \mathbb{T} \, \e^{- \int_0^t \, \thed t' \, W (t') } \, \ket{P (0) }\, = \, \mathcal{U}(t,0) \, \ket{ P(0) } \,, ~~
    \label{eq:FPE propagator}
\end{equation}
where the operator $\mathbb{T}$ enforces time ordering of the integral in the exponential,  and $\mathcal{U}(t,0)$ is the propagator from time 0 to time $t$ (we simply write ${\cal U}(t)$ when $W$ is time independent). Additionally, the probability that the initial configuration $\vb{q}_i$ at time $0$ evolves under $W$ into the configuration $\vb{q}_f$ at time $t$ is given by the transition amplitude 
\begin{align}
    \mathbb{P}[\vb{q}_i,0 \to \vb{q}_f,t; W] \, = \, \matel{ \vb{q}_f}{\exp ( - \int_0^{t} \thed t' \, W (t') )}{\vb{q}_i} \, \equiv \, \mathcal{U}^{\vpd}_{fi} (t,0) \, ,~~
    \label{eq:FPE transition prob}
\end{align}
where $\mathcal{U}^{\,}_{fi}(t,0)$ are matrix elements of the propagator \eqref{eq:FPE propagator}, which simplifies when $W$ is time independent. 

Note that Eq.~\eqref{eq:FPE transition prob} resembles a quantum \emph{amplitude} for some ``path'' $i \to f$, suggesting %the possibility of 
a formulation \`a la the Feynman path-integral representation of the evolution of quantum states. Indeed, such a prescription is well established by the Martin-Siggia-Rose (MSR) path integral \cite{MSR}, which we consider in Sec.~\ref{sec:MSR}.  However, for reasons that will become apparent in the path-integral context, it is actually more elegant (and straightforward) to implement reversibility T in terms of the operator $W$ \eqref{eq:W form}. This %formalism 
represents a rare instance in which the effective theory is \emph{more difficult} to implement systematically (to all orders in nonlinearities) in a Lagrangian (path-integral) formalism.

Finally, our interest lies in Fokker-Planck dynamics that admit a \emph{stationary} state $\ket{P_{\rm ss}}$ satisfying
\begin{equation}
    \label{eq:FPE stationary state}
    W \, \ket{P_{\rm ss}} = -\dby{}{t} \, \ket{P_{\rm ss}} = 0 \, , ~~
\end{equation}
or equivalently, $\mathcal{U} \, \ket{P_{\rm ss}} = \ket{P_{\rm ss}}$, so that $\ket{P_{\rm ss}}$ is invariant under the dynamics generated by $W$. Note that if $W$ is time independent, at least one choice of $\ket{P_{\rm ss}}$ is guaranteed to exist, but it may not be unique. For simplicity, throughout this work, we assume that $\ket{P^{\,}_{\rm ss}}$ exists and is nonsingular, so that a stationary operator $S$  \eqref{eq:discrete stationary dist} exists and is \emph{invertible}; we also explain how to incorporate conservation laws in our EFT framework (a physically relevant scenario in which $\ket{P_{\rm ss}}$ may fail to be unique) in Sec.~\ref{sec:noether}.  Under the foregoing assumptions---and without loss of generality---we define the stationary state $\ket{P_{\rm ss}}$ according to
\begin{equation}
    \label{eq:FPE stationary distribution}
    P_{\rm ss} (\vb{q}) \, \equiv \, \inprod{\vb{q}}{P_{\rm ss}} \, = \, \e^{- \Phi (\vb{q})} \, \Longleftrightarrow \, \ket{P_{\rm ss}} = \int \thed %^M
    \vb{q} \, \e^{- \Phi (\vb{q})} \, \ket{\vb{q}} \, ,~~
\end{equation}
which is a simple Ansatz given Eq.~\eqref{eq:FPE stationary state} and the requirement that $P_{\rm ss} (\vb{q}) = \inprod{\vb{q}}{P_{\rm ss}}$ be positive. We need not worry about overall normalization, since we only consider linear operations on $\ket{P_{\rm ss}}$.

\subsection{Stationarity constraints on the Fokker-Planck equation}
\label{subsec:FPE TR}

We now discuss several features of the dynamics generated by the FPE \eqref{eq:FPE main} that follow from stationarity, including the important relations to reversibility T. First, it will prove useful to define several quantities. For example, related to the stationary \emph{state} $\ket{P_{\rm ss}}$ \eqref{eq:FPE stationary state} with distribution $\Phi$ \eqref{eq:FPE stationary distribution} is the stationary \emph{operator} given by
\begin{equation}
    \label{eq:Stationary operator}
    S  \, = \, \int \thed %^M 
    \vb{q} \, \e^{-\Phi(\vb{q})} \, \BKop{\vb{q}}{\vb{q}} ~~~ \implies ~~~ \ket{P_{\rm ss}} \, = \, S \, \ket{u}\,,~~
\end{equation}
in analogy to Eq.~\eqref{eq:discrete stationary dist} in Sec.~\ref{subsec:discrete states}, where $\ket{u}$ is the \emph{uniform} distribution
\begin{equation}
    \label{eq:uniform state}
    \ket{u} \, \equiv \, \int \thed %^M 
    \vb{q} \, \ket{\vb{q}} ~~~\implies ~~~u (\vb{q}) \, = \, 1 \, , ~~
\end{equation}
up to normalization, in analogy to Eq.~\eqref{eq:discrete uniform} for discrete spaces in Sec.~\ref{subsec:discrete states}. 

As before, numerous useful properties follow from $\ket{u}$ \eqref{eq:uniform state}. For example, conservation of probability mandates that $\inprod{u}{P(t)} = \int \thed \vb{q} \, P (\vb{q}) = 1$. Noting that $u$ \eqref{eq:uniform state} is independent of time, applying $\bra{u}$ to Eq.~\eqref{eq:FPE kets} leads to
\begin{equation}
\label{eq:stoch1}
   \bra{u} \dby{}{t} \ket{P} \, = \, \dby{}{t} \, \inprod{u}{P} \, = \, \dby{}{t} 1  \, =\,  0 ~~~\implies ~~~\bra{u} \, W  \, = \,  0  \,  ,~~
\end{equation}
for any $\ket{P}$, where $0$ is technically the null vector. In other words, $W^T$ annihilates the uniform distribution $\ket{u}$ \eqref{eq:uniform state}.

The assumption of stationarity implies that the probability distribution $\ket{P_{\rm ss}} \equiv S\ket{u}$ \eqref{eq:FPE stationary distribution} is a time-independent (i.e., stationary) solution to the FPE \eqref{eq:FPE main}, and thus
\begin{equation}
\label{eq:stoch2} 
    W \,\ket{P_{\rm ss}} = W \,  S \ket{u} =  0  \, ,~~
\end{equation}
as in Eq.~\eqref{eq:FPE stationary state}, while $\mathcal{U} \, \ket{P_{\rm ss}}  = \ket{P_{\rm ss}} $. The two preceding expressions constrain the action of $W$ to the left and right.

We now consider the crucial transformation $W \to \widetilde{W}$, corresponding to a $\Ints^{\,}_2$ \emph{reversibility} operation T.  In analogy to Eq.~\eqref{eq:discrete reverse W} in the case of discrete state spaces, the ``time-reversed'' evolution generator $\widetilde{W}$ is defined in terms of the original generator $W$ \eqref{eq:W form} and the stationary operator $S$ \eqref{eq:Stationary operator} via
\begin{equation}
\label{eq:W reversed}
    % \widetilde{W}^T \, \equiv \, S^{-1} W S \, . ~~
    \widetilde{W} \, \equiv \,  S \,  W^T \, S^{-1} \, , ~~
\end{equation} 
which exists whenever Kolmogorov's criterion is satisfied, and which has the \emph{same} stationary distribution as $W$ by Kelly's Lemma \cite{Kolmogorov-backward, KellyStochastic}. The above can be rewritten component wise as
\begin{align}
    \label{eq:backward markov rule}
    (\, \widetilde{W}^{\,T} \, )^{\vpp}_{xy} = \frac{P^{\vpp}_{\rm ss}(x) \,W^{\vpp}_{xy}}{P^{\vpp}_{\rm ss}(y)} \, , ~~
\end{align}
where there is no sum over $y$ or $x$. The relation in Eq.~\eqref{eq:backward markov rule} is well known in the theory of Markov chains  \cite{KellyStochastic}.

We now motivate this definition and the interpretation of $\widetilde{W}$ \eqref{eq:W reversed} as the time-reversed counterpart of $W$ \eqref{eq:W form}.  Consider the probability that the system starts in the state $\ket*{\vb{q}_i}$ and ends in the state $\ket*{\vb{q}_f}$. Note that 
\begin{equation}
    \matel{ \vb{q}_f}{\exp (-W \, t) }{\vb{q}_i} \, = \,  \matel{ \vb{q}_i}{\exp ( -W^T \, t)}{\vb{q}_f} \, , ~~
\end{equation}
yet $W^T$ does not, in general, generate valid Fokker-Planck dynamics captured by Eq.~\eqref{eq:FPE kets}, since it obeys neither Eq.~\eqref{eq:stoch1} nor Eq.~\eqref{eq:stoch2}.  However, as in the case of discrete state spaces considered in Sec.~\ref{subsec:discrete states},  the operator $\widetilde{W}$ \eqref{eq:W reversed} \emph{is} a proper generator of Fokker-Planck dynamics with the \emph{same stationary distribution} $\ket{S}$ \eqref{eq:FPE stationary distribution} as $W$ \cite{KellyStochastic}. 
To see this, it is sufficient to check that Eqs.~\eqref{eq:stoch1} and \eqref{eq:stoch2} hold for  $\widetilde{W}$, i.e.,
\begin{subequations}
\label{eq:TR W stoch}
    \begin{align}
    \bra{u} \widetilde{W} \, &= \, \bra{u} \left(S^{-1} \, W \, S \right)^T \, =\,  \bra{u} S^T W^T (S^{-1})^T \, =\, 0 \label{eq:TR stoch 1}\\
    \widetilde{W} S \ket{u} \, &= \,  \left(S^{-1}WS\right)^T S \ket{u} \, = \,  S \, W^T  S^{-1}S \ket{u} \, = \, S \left(\bra{u} W\right)^T \, =\,  S %\ket{0} = \ket{0} . ~~
    \, (0) = 0\label{eq:TR stoch 2} \, , ~~
    \end{align}
\end{subequations}
so that $\widetilde{W}$ is also a stochastic matrix with stationary distribution $P_{\rm ss} = \exp(-\Phi)$  \eqref{eq:FPE stationary distribution}. 

Using the results above, we identify the \emph{global balance condition}
\begin{equation}
   \mathbb{P}[\vb{q}_i(0) \to \vb{q}_f(t)] \, = \, \matel*{ \vb{q}_f}{\e^{-W \, t}}{\vb{q}_i} \, = \, \matel*{ \vb{q}_i}{S^{-1}\e^{-\widetilde{W} \, t} S}{\vb{q}_f}\,  =\,  \e^{\Phi_i - \Phi_f} \, \widetilde{\mathbb{P}}[\vb{q}_f(0) \to \vb{q}_i(t)] \, , ~~\label{eq:FPE balance condition}
\end{equation}
given the stationary distribution $P^{\rm ss}=\e^{-\Phi}$ \eqref{eq:FPE stationary distribution}. Lastly, the reversibility T is an involution:
\begin{equation}
    \widetilde{\widetilde{W}} \, = \, S \, \widetilde{W}^T \, S^{-1} \, = \, S \, \left( S \,  W^T \, S^{-1} \right)^T \, S^{-1} \, = S \, S^{-1} \, \left(W^T \right)^T \, S \, S^{-1} \, = W \, , ~~
\end{equation}
and is therefore $\Ints^{\,}_2$, as required.  Applying T twice returns the same operator  $W\to \widetilde{W} \to W$.

If the Fokker-Planck generator $W$ \eqref{eq:W form} obeys $W=\widetilde{W}$, the system is said to obey \emph{detailed balance}, while all $W$s with a stationary state $P_{\rm ss}$ \eqref{eq:FPE stationary distribution} obey \emph{global balance} \eqref{eq:backward markov rule}. In the Markov-chain literature, $\widetilde{W}$ \eqref{eq:W reversed} is known as the ``backward''---or \emph{time-reversed}---Markov generator. In the setting of stochastic effective theories, we adopt the perspective that the transformation $W \to \widetilde{W}$ \eqref{eq:W reversed} corresponds to a \emph{reversibility} transformation; accordingly, we associate detailed balance with reversibility T. Theories with $W=\widetilde{W}$ obey detailed balance and are ``T even''; theories with $W \neq \widetilde{W}$ break detailed balance, and thus T; in particular, T-breaking theories with $W=-\widetilde{W}$  are ``T odd.'' The connection between the transformation of Eq.~\eqref{eq:W reversed} and time reversal is more transparent when accompanied by the familiar, \emph{microscopic} notion of time reversal (e.g., $t \to -t$, $p \to -p$, etc.); we discuss such details in later sections, where we consider how T \eqref{eq:W reversed} acts on quantities other than the generator $W$ \eqref{eq:W form}. The formalism we develop herein is a powerful theoretical tool, especially for systems in which reversibility T plays an important role and is a symmetry of $W$, either on its own or in the generalized sense of a gT symmetry (i.e., in combination with another $\Ints^{\,}_2$ transformation). Even if time reversal T (or gT) is not preserved, there is still great value in separating  the T-even versus T-odd contributions to $W$.  As we explain in Sec.~\ref{subsec:FDT}, the T-even and T-odd contributions are related to dissipative versus dissipationless dynamics, and identifying the corresponding terms leads to generalizations of the second law of thermodynamics in Eq.~\eqref{eq:KL rate} and a fluctuation-dissipation theorem in Eq.~\eqref{eq:FPE force FDT}.

\subsection{Construction of the Fokker-Planck generators}
\label{subsec:FDT}

Inspired by Wilsonian effective theory \cite{WilsonPRBI, WilsonPRB2, hohenberghalperin, WeinbergLagrangians, altland_simons_2010}, we now %expound on 
provide a general construction for the differential operator $W$ \eqref{eq:W form}, which must be compatible with (\emph{i}) the existence of a stationary distribution $P_{\rm ss} \propto \exp(-\Phi)$ \eqref{eq:FPE stationary distribution}, (\emph{ii}) normalization of the probability distribution $P(\vb{q},t)$ \eqref{eq:Probability}, and (\emph{iii}) the existence of a time-reversed evolution operator $\widetilde{W}$ \eqref{eq:W reversed} that fulfils the same conditions as $W$ itself. While massaging this FPE-based formalism into a more conventional path-integral EFT involves a few subtleties (which we discuss in Sec.~\ref{sec:MSR}), the FPE approach nonetheless captures the philosophical underpinnings of Wilson's EFT formalism in the context of stochastic dynamics \eqref{eq:SDE main}.

For simplicity, we assume herein that the reversibility operation T \eqref{eq:W reversed} that maps $W \to \widetilde{W}$ also maps the trajectory $q_a (t) \to q_a (-t)$. Later, in Sec.~\ref{subsec:Hamiltonian setup}, we provide an examples in the context of Hamiltonian mechanics that involve momentum variables $q_a = p_a $, for which $q_a (t) \to q_a (-t) = - q_a (t)$ under T. We comment further on the treatment of T---and modifications to the results recovered herein---in the presence of T-odd variables $q_a$.

\subsubsection{General form of \texorpdfstring{$W$}{W}}
\label{subsub:Gen W}
We begin by defining the generalized chemical potentials---from which various aspects of time-reversal symmetry T and the EFT framework follow---in terms of the stationary distribution $\Phi( \vb{q} ) $ \eqref{eq:FPE stationary distribution} via
\begin{equation}
    \label{eq:Chemical Potential Def}
    \chempo{a} \, \equiv \, \frac{\partial \Phi}{\partial q_a} \, ,~~
\end{equation}
and we now introduce a generic---and particularly convenient---expression for the evolution operator $W$ \eqref{eq:W form}. The conditions of Eqs.~\eqref{eq:stoch1} and \eqref{eq:stoch2} suggest that the evolution operator $W$ \eqref{eq:W form} may be written
\begin{equation}
    W \, = - \frac{1}{2} \pd{a} \mathcal{Q}_{ab}(\pd{b} + \chempo{b}) \,,~~
    \label{eq:W nice}
\end{equation}
where the factor of $\pd*{a}$ on the left ensures satisfaction of Eq.~\eqref{eq:stoch1} by annihilating the uniform state $\bra{u}$ \eqref{eq:uniform state} acting to the left, the factor of $( \pd*{b} + \chempo*{b} )$ on the right ensures satisfaction of Eq.~\eqref{eq:stoch2} by annihilating the stationary operator $S$ \eqref{eq:Stationary operator} acting to the right, and $\mathcal{Q}$ is a matrix that we expect to be related to $Q$ \eqref{eq:Q noise def}. However, compared to $Q$, the object $\mathcal{Q}$ in Eq.~\eqref{eq:W nice} may be nonlinear in the coordinates $\vb{q}$, include further derivatives of arbitrary order, etc. 

It can be helpful to introduce the notation 
\begin{equation}
\label{eq:FPE pi}
    \pi^{\vpp}_a \, = -\ii \, \pd{a} ~~~\text{and}~~~\mathcal{H} \, = - \ii \, W \, ,~~
\end{equation}
so that, in analogy with quantum mechanics, $\pi_a$ is the ``conjugate operator'' to $q_a$ and $\ii \, \mathcal{H}$ is the infinitesimal generator of time translations. Note that $\cal H$ resembles a quantum Hamiltonian \cite{Doi1, Doi2, Peliti1, Peliti2}; we reserve $H$ for Hamiltonians mechanics, where $H$ relates to both $W$ and $\Phi$ \eqref{eq:FPE stationary distribution}. This notation provides a useful shorthand in certain scenarios, especially when we discuss genuine stochastic PDEs and field theories.  In fact, the field $\pi_a$  naturally arises in the path integral construction discussed in Sec.~\ref{sec:MSR}. Of course, the notation of Eq.~\eqref{eq:FPE pi} awkwardly introduces complex numbers into a formalism that is otherwise real valued, and so we generally utilize the FPE notation (i.e., $\pd*{a}$ and $W$), until our discussion of continuum field theories (where one requires functional Fokker-Planck equations) in Sec.~\ref{sec:continuum}.

It will also prove useful to derive an expression for the time-reversed operator $\widetilde{W}$ \eqref{eq:W reversed} analogous to Eq.~\eqref{eq:W nice}. We note that the derivative operator $\pd*{a}$ is antisymmetric under T, and that
\begin{equation}
    S \pd{a} S^{-1} \, = \, \pd{a} + [\pd{a} \Phi] \, = \, \pd{a} + \chempo{a} \, ,~~
\end{equation} 
where the derivative in $[\pd*{a} \Phi]$ acts \emph{only} on $\Phi$, and not to the right (in general, derivatives inside square braces act only within those braces). We use the relation above to recover the expression
\begin{align}
    \widetilde{W} \, &= \, -\frac{1}{2} \, S \, W^T S^{-1} \, = \, -\frac{1}{2} \, S \, \left( \pd{a} \, \mathcal{Q}^{\vpd}_{ab} \, \left( \pd{b} + \chempo{b} \right) \right)^T \, S^{-1} \, = \, -\frac{1}{2} \, S(-\pd{b} + \mu^{\vpp}_b) \, \mathcal{Q}^T_{ab} (-\pd{a}) S^{-1} \notag \\
    &= \, -\frac{1}{2} \, S(\pd{b} - \mu^{\vpp}_b) S^{-1} S \mathcal{Q}^T_{ab} S^{-1} S (\pd{a}) S^{-1} \, =\,  -\frac{1}{2} \, \pd{b} \left( S\mathcal{Q}^T_{ab} S^{-1}\right) (\pd{a} + \chempo{a}) \,,~~ \label{eq:W reverse nice}
\end{align}
which resembles Eq.~\eqref{eq:W nice} for $W$, up to a transpose on the matrix $\mathcal{Q}$.

We now consider $W$ \eqref{eq:W nice} in more detail. We first note that Eq.~\eqref{eq:W nice} is generally  complicated, and sensitive to the various details contained in $\mathcal{Q}$. The matrix $\mathcal{Q}$ \eqref{eq:nice Q def} may be nonlinear in $\vb{q}$ and capture non-Gaussian and/or multiplicative noise, indicated by the $\dots$ in Eq.~\eqref{eq:FPE main}. As is standard practice both in EFTs and in the treatment of stochastic (FPE) dynamics, we consider a \emph{derivative expansion} of $\mathcal{Q}$ in powers of $\pd*{a}$, organizing terms by their derivative order; in all cases of interest in this paper, the low-order terms are sufficient to capture the relevant physics.

\subsubsection{Gaussian noise and the fluctuation-dissipation theorem}
\label{subsub:Gaussian+FDT}

For concreteness, we begin by considering the  derivative-free terms in $\mathcal{Q}$. These capture the \emph{only} nontrivial contribution in the presence of purely Gaussian noise, for which moments higher than the variance vanish. Also note that derivative-free matrices $\mathcal{Q}$ commute with the stationary operator $S$ \eqref{eq:Stationary operator}.  In general, Gaussian noise still allows for $\mathcal{Q}$ to be a functional $\mathcal{Q} (\vb{q})$ of the coordinates $\vb{q}$ (i.e., multiplicative noise). We decompose $\mathcal{Q} (\vb{q})$ according to
\begin{equation}
    \mathcal{Q}_{ab}\,  =\,  Q_{ab} -V_{ab}\, , ~~
    \label{eq:nice Q def}
\end{equation}
where, on the right-hand side, $Q$ and $V$ represent undetermined symmetric and antisymmetric matrices, respectively. 

This notation is intentional: The matrix $Q$ in Eq.~\eqref{eq:nice Q def} is the (symmetric) noise-variance matrix $Q$ \eqref{eq:Q noise def}---i.e., the \emph{same} object $Q$ that appears in Eq.~\eqref{eq:W form}. Subtleties relating to the ordering of derivatives may arise at the level of the SDE \eqref{eq:SDE main}, but are unimportant at the level of the FPE \eqref{eq:FPE main}. For higher-order (non-Gaussian) noise \eqref{eq:higher derivatives}, this matching between the SDE and FPE is unambiguous at the highest derivative order. Then, applying the time-reversal operation T \eqref{eq:W reversed} to the derivative-free part of $\mathcal{Q}$ \eqref{eq:nice Q def} gives
\begin{align}
\label{eq:TR on nice Q}
    S\, \Big(Q - V \Big)^T\, S^{-1} \, = \, S\, \Big( Q^T-V^T \Big) \, S^{-1}\,=\, S \, \big(Q + V \big) \, S^{-1}\, = \, Q+V \,,~~
\end{align}
meaning that the symmetric term $Q$ in $\mathcal{Q}$ \eqref{eq:nice Q def} is invariant under time reversal T \eqref{eq:W reversed}, while the antisymmetric term $V$ flips sign. So when $\mathcal{Q}$ contains no derivatives, we will see that the symmetric part $Q$ of $\mathcal{Q}$ with \emph{dissipative} (T-even) dynamics and the antisymmetric part $V$ with dissipation\emph{less} (T-odd) dynamics.  

In the simplest scenario---in which $\mathcal{Q}$ \eqref{eq:nice Q def} contains no derivatives---we expand $W$ \eqref{eq:W nice} using Eq.~\eqref{eq:nice Q def} to find 
\begin{align}
\label{eq:W nice no deriv}
    W \, &=\,  -\frac{1}{2} \, \pd{a} \, \mathcal{Q}_{ab} \, ( \pd{b} + \chempo{b} )\,  = \, -\frac{1}{2} \,  \pd{a} \, ( Q_{ab}-V_{ab}  \,  ) \, ( \pd{b} + \chempo{b} )  \notag \\
    &= \, \frac{1}{2} \, \pd{a} \big( \chempo{b} + \pd{b} \big) \big( V_{ab} - Q_{ab} \big) + \frac{1}{2} \, \pd{a} \Big( \big[ \pd{b} Q_{ab} \big] - \big[ \pd{b} V_{ab} \big] \Big) \, , ~~
\end{align}
where the divergences in square braces $[ \pd*{b} Q_{ab}]$ are \emph{functions} (the derivative does not continue acting to the right), and in the lower line, we place $\pd*{b}$ to the left of the constant matrix $Q_{ab}$ to match the It\^o formulation of Eqs.~\eqref{eq:SDE main}, \eqref{eq:Q noise def}, and \eqref{eq:W form}. Comparing Eq.~\eqref{eq:W nice no deriv} to Eq.~\eqref{eq:W form}, the final term on the right-hand side of in Eq.~\eqref{eq:W nice no deriv} corresponds to the dissipative contribution from the \emph{noise} via $Q$ \eqref{eq:Q noise def}, while the other terms correspond to the force 
\begin{equation}
    2 \, f_a = - Q_{ab} \, \chempo{b} +  V_{ab} \, \chempo{b} +  \big[ \pd{b} Q_{ab} \big] - \big[ \pd{b} V_{ab} \big]  \, , ~~
    \label{eq:FPE force FDT}
\end{equation}
in Eq.~\eqref{eq:W form}, which contains \emph{both} dissipative and  dissipationless terms.  In particular, if the symmetric part $Q$ of $\mathcal Q$ is independent of $\vb{q}$ and free of derivatives---as in Eq.~\eqref{eq:W nice no deriv} above---then $Q$ is simply the variance of white noise \eqref{eq:Q noise def}. 

The constrained form of the dissipative part of the force \eqref{eq:FPE force FDT} is a manifestation of a \emph{fluctuation-dissipation theorem} (FDT) \cite{Kubo_1966, ThoulessStochastic, FDTPRL}. The fact that the dissipative force is $-Q_{ab}\, \chempo*{b}/2$---where $Q_{ab}$ is the noise variance \eqref{eq:Q noise def}---is essential to preserving stationarity. Note that $\Phi$ \eqref{eq:FPE stationary distribution} shows up in the FDT \eqref{eq:FPE force FDT} via $\chempo*{}$ \eqref{eq:Chemical Potential Def}.

We now recast the FDT in a more familiar form \cite{FDTPRL}. Consider the correlation function $\expval{ q_a(t) q_b(0)}$,  where $\expval{\cdot}$ denotes an average with respect to the stationary distribution $S$ \eqref{eq:FPE stationary distribution}. This correlation function can be rewritten as
\begin{equation}
\label{eq:corr fdt 0}
    \expval{ q_a(t) \, q_b(0)} \, = \, \dby{}{\epsilon} \, \left. \int \thed \vb{q} \, \e^{-\Phi + \epsilon q_b} \, q_a(t) \right|^{\vpp}_{\epsilon=0} \, ,~~
\end{equation}
where the right-hand side essentially captures a ``shift'' of the chemical potential $\chempo*{b} \to \chempo*{b} + \epsilon$ \eqref{eq:Chemical Potential Def} for $t<0$. Defining
\begin{equation}
    \chi^{\vpp}_{ab}(t)\, \equiv\, \frac{\delta \expval{ q_a(t)}}{\delta \chempo{b}(0)} \, ,~~
\end{equation}
we can rewrite the correlation function of Eq.~\eqref{eq:corr fdt 0} in the familiar form
\begin{equation}
\label{eq:corr fdt}
    \expval{ q_a(t) \, q_b(0)} \, = \, \int\limits_t^\infty  \thed s \, \chi^{\vpp}_{ab}(s) \, ~~
\end{equation}
where $\chi$ is a generalized susceptibility (the type depends on $q_a$).

In the context of active matter, the lore is that there is no FDT \cite{Kubo_1966, ThoulessStochastic, FDTPRL}. However, if a stationary distribution $P_{\rm ss}=\exp(-\Phi)$ \eqref{eq:FPE stationary distribution} is known to exist (for a given $W$)---or, if $W$ \eqref{eq:W nice} is constructed to be compatible with some $\Phi$---then the FDT \eqref{eq:corr fdt} is a direct mathematical consequence of $P_{\rm ss}$ \eqref{eq:FPE stationary distribution}, which emerges naturally as a constraint on the force $f_a$ \eqref{eq:FPE force FDT}. The difference between our perspective and that of the active-matter literature is more precisely stated as follows: If a specific active system has local equations of motion, but a nonlocal $\Phi$, the FDT \eqref{eq:corr fdt} relates physics in a local theory to physics in a nonlocal theory.  The emergence (or lack thereof) of an FDT can then be traced to whether or not a ``simple'' stationary state of the form $P_{\rm ss} \propto \exp(-\Phi)$ \eqref{eq:FPE stationary distribution} can be identified.

\subsubsection{Non-Gaussian noise}
\label{subsub:Non Gaussian}
We now extend this construction from the derivative-free example above to more general operators $\mathcal{Q}$. To this end, we would like to organize the derivative expansion of $\mathcal{Q}_{ab}$ such that the symmetry of $W$ under T \eqref{eq:W reversed} is related to the symmetry of $\mathcal{Q}_{ab}$ under $a \leftrightarrow b$.  To handle non-Gaussian noise (corresponding to more derivatives in $\mathcal{Q}$), we make use of the following useful (albeit intimidating) derivative expansion \`a la Kramers-Moyal \cite{Kramers, Moyal, PawulaFPEK, PawulaBoltzmann}.  We define
\begin{equation}
    \label{eq:pm derivatives}
    \pdp{a}{-} \, \equiv \, \pd{a} \, = \, \pdv{}{q_a} ~~~\text{and}~~~\pdp{a}{+} \, \equiv \, \pd{a} + \chempo{a} \, ,~~
\end{equation}
using which we expand the matrix $\mathcal{Q}$ \eqref{eq:W nice} order by order according to
\begin{align}
\label{eq:higher derivatives}
    \mathcal{Q}^{\vpd}_{ab} \, &= \, \mathcal{Q}_{ab}^{(0)} + (\pdp{c}{-}  \mathcal{Q}^{(1)}_{ab,c} + \mathcal{Q}^{(1)}_{ab,c}\pdp{c}{+})  + \cdots 
    \, = \, \sum\limits_{n=0}^{\infty} \left( \pdp{c_1}{-} \cdots \pdp{c_n}{-} \mathcal{Q}^{(n)}_{ab, c_1\dots c_n}\, + \mathcal{Q}^{(n)}_{ab, c_n\ldots c_1} \pdp{c_n}{+}\cdots \pdp{c_1}{+}\right) \, ,~~
\end{align}
where $Q^{(n)}_{abc_1\ldots c_n}$ are $\vb{q}$-dependent coefficients that are related to cumulants of the noise, and summation over repeated $c_k$ indices is implicit. In Eq.~\eqref{eq:higher derivatives}, the rightmost sum contains all orderings: Pulling some factors of $\pdp{}{\pm}$ through a $\mathcal{Q}^{(n)}$ only leads to a multiplicative factor, which can simply be absorbed into another term $\mathcal{Q}^{(n)}$ at lower order $n$. Changing any $\pdp{}{+} \to \pdp{}{-}$ (or vice versa) similarly leads to a multiplicative factor, allowing us to arrange  for all terms to contain \emph{only} $\pdp{}{+}$ or $\pdp{}{-}$ at each order, acting on the right or left of $\mathcal{Q}$, respectively, without loss of generality.

The motivation for the Kramers-Moyal-esque expansion \cite{Kramers,Moyal,PawulaFPEK,PawulaBoltzmann} of $\mathcal{Q}$ \eqref{eq:higher derivatives} is that each term in the sum transforms ``nicely'' under the time-reversal transformation T \eqref{eq:W reversed}.  % In other words, whether a given term in the sum \eqref{eq:higher derivatives} %is % OR: generates, results in, leads to, yields, gives rise to
Whether a term in the sum in Eq.~\eqref{eq:higher derivatives} corresponds to dissipative or dissipationless dynamics depends only on the symmetry (or antisymmetry) of the corresponding matrix $\mathcal{Q}^{(n)}$ in the $a,b$ indices!  In particular, for even $n$, the sign of $W$ under time reversal  is equivalent to the sign of $\mathcal{Q}_{ab}$ under interchanging $a \leftrightarrow b$, while for odd $n$, the sign of $W$ under time reversal is opposite the sign of $\mathcal{Q}_{ab}$ under interchanging $a \leftrightarrow b$. In this way, the expansion in Eq.~\eqref{eq:higher derivatives} %allows for %%OR: facilitates
provides for the identification of dissipative and dissipationless terms that appear in the presence of higher-derivative (i.e., non-Gaussian) noise.

Formally, the derivative expansion in Eq.~\eqref{eq:higher derivatives} provides a convenient means by which to organize the various terms that may contribute to $\mathcal{Q}$ \eqref{eq:nice Q def}. 
More physically, this derivative expansion can be viewed on a similar footing as, e.g., those of hydrodynamic theories \cite{harder2015thermal, eft1, eft2, haehl2016fluid, jensen2018dissipative, Glorioso:2016gsa}, where $\pd*{a}$ is associated with a small length scale $\ell \ll 1$. In the FPE context, the intuition is that the length scale of variation of a given probability distribution $P(t,\vb{q})$  is much larger than any microscopic length scales (e.g., the mean free path of the elementary constituents) that appear in the coefficients of higher-derivative terms. This implies that higher derivatives in Eq.~\eqref{eq:higher derivatives}---which correspond to higher-order, non-Gaussian correlations of the noise $\xi(t)$ in Eq.~\eqref{eq:Q noise def}---become progressively less important.

\subsubsection{Generating all possible T-breaking, noise-free terms}
\label{subsec:Make T-Break}

Having established the general form of $W$ \eqref{eq:W nice} compatible with the existence of a stationary state \eqref{eq:FPE stationary state} and global---but not necessarily \emph{detailed}---balance \eqref{eq:W reversed}, we now discuss the generalized T-breaking (i.e., dissipationless) contributions to the right-hand side of Eq.~\eqref{eq:W nice}. For simplicity, we ignore the effects of noise for the time being. Up to a subtlety described below, this simply amounts to choosing the most general from of the antisymmetric term $V_{ab}$ in Eq.~\eqref{eq:nice Q def}. Up to second order in derivatives, taking $\mathcal{Q}_{ab} = -V_{ab} = V_{ba}$ we find, from Eq.~\eqref{eq:W nice no deriv},
\begin{equation}
\label{eq:dissipationless-second-order}
    W \, = \, \frac{1}{2} \, \pd{a} V^{\vpd}_{ab} (\pd{b} + \chempo{b}) \, %=\, \frac{1}{2} \, \pd{a} (\pd{b} + \chempo{b}) V^{\vpd}_{ab} - \frac{1}{2} \, \pd{a} [\pd{b} V^{\vpd}_{ab}]\,  
    = \, \frac{1}{2} \, \pd{a} \left(\pd{b} V^{\vpd}_{ab} + \chempo{b} V^{\vpd}_{ab}-[\pd{b} V^{\vpd}_{ab}] \right) \, , ~~
\end{equation}
where all derivatives are on the left, as in Eq.~\eqref{eq:W nice no deriv}, so that the rightmost expression above realizes the standard It\^o formulation.  Here and below, any derivative terms in square brackets indicate that the derivative operators act only within those brackets. Hence, the term \eqref{eq:dissipationless-second-order} is a purely deterministic contribution to the FPE \eqref{eq:FPE main}. Accordingly, the dissipationless part $v_a$ of the It\^o force $f_a$ introduced in Eq.~\eqref{eq:SDE main} has components given by
\begin{align}
\label{eq:v_anti}
    v_a \, = - \frac{1}{2}\,[ \pd{b} V_{ab} ]+ \, \frac{1}{2}\chempo{b} \, V_{ab} \, , ~~
\end{align}
so that $v_a$ leads to a nonvanishing probability current $J_a$, even when evaluate on a steady state. To see this, we rewrite the Fokker-Planck equation \eqref{eq:FPE main} as a \emph{continuity equation},
\begin{align}
    \pd{t} P\,  = - \pd{a} J^{\vpp}_a \, ,~~
\end{align}
where $J_a$ is the probability current; for the stationary state $P^{\rm ss}= \exp(-\Phi)$, the corresponding probability current is
\begin{align}
\label{eq:SS current}
    J_a^{\,\rm ss} \, = \,  v_a \, \e^{-\Phi} \, , ~~
\end{align}
and since $P^{\rm ss}$ is \emph{stationary}, the probability current must satisfy $\pd*{a}\,  J_a^{\,\mathrm{ss}} = 0$---i.e., $J_a^{\,\mathrm{ss}}$ is divergenceless. Taking the divergence of Eq.~\eqref{eq:SS current} leads to the following  partial differential equation for $v_a$,
\begin{align}
\label{eq:dissipationless_terms}
    \pd{a} v_a - \chempo{a} \, v_a  \, = \, 0 \, ,~~
\end{align}
which is compatible with Eq.~\eqref{eq:v_anti} \cite{pingao}. Thus,  Eq.~\eqref{eq:v_anti} is most the general solution to Eq.~\eqref{eq:dissipationless_terms}, except when $q_a$ lives on a topologically nontrivial manifold, such as the torus.  This is explained in App.~\ref{sec:topoappendix}. Restoring the noise, we have
\begin{equation}
    \label{eq:W v_a}
    W \, = \, -\frac{1}{2} \pd{a} Q_{ab} \left( \pd{b}-\chempo{b}\right) + \frac{1}{2} \pd{a} v_a \, ,~~
\end{equation}
where it is instructive to compare to Eq.~\eqref{eq:W form}.
\begin{table}[t!]
    \centering
    \begin{tabular}{l@{\hskip 24pt}c@{\hskip 24pt}c}
    \toprule
    & Dissipationless terms & Dissipative terms\\
    \toprule
    Fokker-Planck operator ($W$): & $\p_a v_a$ & $-\frac12 \p_a Q_{ab} (\p_b + \mu_b)$ \\
    Constraint on It\^o ``force" ($f_a$): & $(\pd{a} v_a) - \chempo{a} v_a = 0$ & $g_a = -\frac{1}{2} Q_{ab}\chempo{b} + \frac12 \pd{b} Q_{ab}$\\
    It\^o Langevin equation: & $\pd{t} q_a = v_a + \xi_a(t)$ & $\pd{t} q_a = g_a + \xi_a(t)$\\
    Relation to noise ($\xi_a$): & N/A & $\expval{\xi_a(t)\xi_b(t')} = Q_{ab}\delta(t-t')$  \\
    \bottomrule
    \end{tabular}
    \caption{Summary of constraints from stationarity of $\Phi$ on the dissipationless and dissipative parts of the Fokker-Planck generator $W$ \eqref{eq:W nice} and the corresponding Langevin equation \eqref{eq:SDE main} in  the It\^o formulation.} 
    \label{tab:summary}
\end{table}

\subsection{Entropy and the second law of thermodynamics} \label{sec:2ndlaw}

In dissipative theories of \emph{thermal} systems, the existence of well-defined entropy currents---and an associated \emph{second law of thermodynamics}---lead to powerful constraints on the resulting EFTs \cite{Glorioso:2016gsa}. Thus, it is natural to ask whether the EFT formalism we have developed for \emph{generic} stochastic dynamics---which may relax to \emph{nonequilibrium} stationary distributions $P^{\rm ss}$ \eqref{eq:FPE stationary distribution}---admits a notion of entropy $S (\vb{q},t)$, and a second law of thermodynamics, 
\begin{align}
\label{eq:2nd law}
    \dv{S(\vb q, t)}{t} \, \geq  \, 0 \, .~~
\end{align}
In fact, there is a significant body of literature (see, e.g., \cite{schnakenberg, oliveira}) concerning the generalization of entropy (currents)---and the second law \eqref{eq:2nd law}---to generic, nonequilibrium stochastic systems (e.g., with nonthermal stationary states). It is also natural to expect that, if such notions exist, they should lead to useful constraints on nonequilibrium EFTs, including those describing \emph{active} matter. The present discussion is further motivated by the fact that, in the literature, entropy production is often used as a measure of the extent to which a system is ``active'' \cite{oliveira}.

There is a formal notion of entropy density $s[P(\vb{q},t)]$ which leads to a second law of thermodynamics \eqref{eq:2nd law} in our EFT framework. Moreover, when $P(\vb{q},t)$ is close to the stationary distribution $P^{\rm ss} \propto \exp(-\Phi(\vb{q}))$ \eqref{eq:FPE stationary distribution}, these notions of entropy and the second law \eqref{eq:2nd law} closely resemble those of dissipative EFTs for thermal fluids \cite{Glorioso:2016gsa}. For this reason---and others discussed below---we believe this definition of entropy to be a logical starting point for generic, stochastic EFTs. However, we caution that the definition of entropy we recover may differ from that typically associated with entropy production in active and nonequilibrium dynamics \cite{schnakenberg, oliveira}. Additionally, we restrict the derivation below to Gaussian noise (either additive or multiplicative) for simplicity. We conclude this discussion by commenting on the extension of this notion of entropy for nonequilibrium EFTs beyond Gaussian noise.

We can work out the ``correct'' choice of the entropy $D[P]$ for a probability distribution $P (\vb{q},t)$ \eqref{eq:Probability} in a pedagogical way. The most important property that $D[P]$ must satisfy is that $D_{\rm max} = D [ P^{\rm ss}]$---i.e., that the stationary distribution $P^{\rm ss}$ \eqref{eq:FPE stationary distribution} \emph{uniquely} maximizes the entropy $D$. A natural guess for the entropy density $s (\vb{q},t)$ for the probability distribution $P (\vb{q},t)$ \eqref{eq:Probability} is the standard one $s (\vb{q},t) = - P (\vb{q},t) \, \log P (\vb{q},t)$, such that the global entropy 
\begin{equation}
\label{eq:Shannon ent}
    D_{\rm Shannon} (t) \, = \, - \int \thed \vb{q} \, P (\vb{q},t) \, \log P (\vb{q},t) \, , ~~
\end{equation}
is the \emph{Shannon entropy} for $P (\vb{q},t)$. However, it is straightforward to see that this is not the correct choice. Importantly, the Shannon entropy \eqref{eq:Shannon ent} for the \emph{uniform distribution} $P^{\rm uni} = {\cal V}^{-1}  \int \thed \vb{q}$ \eqref{eq:uniform state} is $D_{\rm uni} = \log {\cal V}$, where ${\cal V} = \int \thed \vb{q}$. Meanwhile, the Shannon entropy \eqref{eq:Shannon ent} for the stationary distribution $P^{\rm ss} = {\cal Z}^{-1} \, \int \thed \vb{q} \, \exp ( - \Phi (\vb{q}) ) $ \eqref{eq:FPE stationary distribution}---where  ${\cal Z} =  \int \thed \vb{q} \, \exp ( - \Phi (\vb{q}) ) $ is analogous to a partition function for $P^{\rm ss}$---is given by
\begin{align}
    D_{\rm Shannon} \left[ P^{\rm ss} \right] \, &=  - \int \thed \vb{q} \frac{e^{-\Phi(\vb{q})}}{\cal Z} \, \log \frac{e^{-\Phi(\vb{q})}}{\cal Z}  \, = \,\log {\cal Z} +  \frac{1}{\cal Z} \int \thed \vb{q} \,  \Phi (\vb{q}) \, e^{-\Phi(\vb{q})} \, \leq \, D_{\rm Shannon} \left[ P^{\rm uni} \right] \, , ~~
\end{align}
where the final relation follows from Gibbs' inequality, which says that $- \int \thed \vb{q} \, P^{\rm ss} (\vb{q}) \log P^{\rm ss} (\vb{q}) \leq - \int \thed \vb{q} \, P^{\rm ss} (\vb{q}) \log P^{\rm uni} = \log {\cal V}$. In other words, Gibbs' inequality guarantees that the Shannon entropy \eqref{eq:Shannon ent} for the stationary distribution $P^{\rm ss} $ \eqref{eq:FPE stationary distribution} saturates $\log {\cal V}$ (i.e., is maximal) if and only if $P^{\rm ss} $ is uniform. Hence, the Shannon entropy \eqref{eq:Shannon ent} is only a valid definition of the entropy for stochastic EFTs in which the stationary distribution $P^{\rm ss}$ \eqref{eq:FPE stationary distribution} is uniform \eqref{eq:uniform state}, which is not generally the case for the systems of interest (e.g., active matter).

For generic stationary distributions $P^{\rm ss} $ \eqref{eq:FPE stationary distribution}, we require a functional of $D[P(\vb{q},t)]$ that is maximal when $P(\vb{q},t)=P^{\rm ss}$ . This suggests that the correct entropy $D[P]$ should be of the form of a \emph{relative entropy} of the true probability distribution $P(\vb{q},t)$ and the stationary distribution $P^{\rm ss} $ \eqref{eq:FPE stationary distribution}. A natural choice---which follows from Gibbs' inequality---is the Kullback-Leibler (KL) divergence (or relative entropy), $D_{\rm KL} \left[ P^{\rm ss} (\vb{q}) \middle\| P(\vb{q},t) \right] =  \int \thed \vb{q} \, P^{\rm ss} (\vb{q}) \, \log \left[  \, P^{\rm ss} (\vb{q}) / P(\vb{q},t) \, \right] \leq 0$. However, this shows that the KL divergence is \emph{minimized} when $P(\vb{q},t)=P^{\rm ss}$, suggesting that the correct notion of entropy in generic stochastic EFTs is given by \emph{minus} the KL divergence,
\begin{equation}
\label{eq:EFT entropy def}
    D \left[ P(\vb{q},t)  \right] \,  \equiv \, \int \thed \vb{q} \, \e^{-\Phi(\vb{q})} \, \log \frac{P(\vb{q},t)}{\e^{-\Phi(\vb{q})}} \, ,~~
\end{equation}
where we have ignored the normalization of $P^{\rm ss}$ and $P(\vb{q},t)$, since this merely amounts to an unimportant rescaling and overall constant shift of $D$ \eqref{eq:EFT entropy def}. We now explore this quantity in more detail.

First, we prove that a second law of thermodynamics \eqref{eq:2nd law} holds for $D(t)$ \eqref{eq:EFT entropy def}. Recall that, by Gibbs' inequality, $D(t)$ is maximal only when $P \propto \exp(-\Phi)$. For convenience, we write $P (\vb{q},t)$ in the form
\begin{equation}
    P(\vb{q},t) \, = \, \e^{-\Phi(\vb{q}) - \Psi(\vb{q},t)} \, ,~~
\end{equation} 
where $\Psi$ is small when $P$ is close to the stationary distribution $P^{\rm ss}$ \eqref{eq:FPE stationary distribution}. We then find that
\begin{align}
    \dv{D}{t} \, &=\, \int \thed \vb{q} \, \e^{-\Phi} \, \dv{\,}{t} \log P \, = \, \int \thed  \vb{q} \, \e^{-\Phi} \, \frac{1}{P} \, \dv{P}{t}  \, = \, \int \thed \vb{q} \, \e^{-\Phi} \, \e^{\Phi + \Psi} \, \left( - W \, P \right) \notag \\
    %
    %&= - \int \thed \vb{q} \, \e^{\Psi} \, W \, P \, = 
    &= \,  \frac{1}{2} \int \thed \vb{q} \, \e^{\Psi} \, \pd{a} \, \mathcal{Q}_{ab} \, \left( \pd{b}+\chempo{b} \right) \e^{-\Phi-\Psi} \, = \,  -\frac{1}{2} \int \thed \vb{q} \, \e^{\Psi} \, \pd{a} \, \mathcal{Q}_{ab} \, \e^{-\Phi- \Psi} \, \left( \pd{b} \Psi  \right)\notag \\
    &= \,  \frac{1}{2} \int \thed \vb{q} \, \left( \pd{a} \Psi\right) \, \e^{\Psi} \, \mathcal{Q}_{ab} \, \e^{-\Phi - \Psi} \, \left( \pd{b} \Psi \right) \,  , ~~\label{eq:dD/dt main}
\end{align}
where, in the last line, we used integration by parts and the fact that $\e^{\Psi} \mathcal{Q}_{ab} \, \e^{-\Phi-\Psi} \left( \pd*{b} \Psi \right)$ vanishes at the boundary of phase space. For simple noise distributions---in which $\mathcal{Q}_{ab}$ depends only on $\vb{q}$ and not $\vb*{\p}$---we observe that
\begin{equation}\label{eq:KL rate}
    \dv{D}{t} \, \to \, \frac{1}{2} \int \thed \vb{q} \, (\pd{a} \psi) \, \mathcal{Q}_{ab} \, \e^{-\Phi} \, (\pd{b} \psi) \, \geq \, 0 \, ,~~
\end{equation}
by the fact that $\mathcal{Q}_{ab} \sim Q_{ab}$ \eqref{eq:nice Q def} is positive semidefinite.  Thus, if we define $D$ \eqref{eq:EFT entropy def} to be the entropy of the probability distribution $P(\vb{q},t)$ for a system that relaxes to the stationary distribution $P^{\rm ss} (\vb{q}) = \exp(-\Phi(\vb{q}))$ \eqref{eq:FPE stationary distribution}, and for noise distribution in which $\mathcal{Q}$ \eqref{eq:nice Q def} contains no derivatives, we obtain a second law of thermodynamics, captured by Eq.~\eqref{eq:KL rate}. %This fact is  well known within the statistics community. 
This fact is well understood in, e.g., the statistics literature.

For non-Gaussian noise, which contributes higher-derivative corrections to the Fokker-Planck generator $W$ \eqref{eq:W nice} via Eq.~\eqref{eq:higher derivatives}, it is no longer obvious that $D(t)$ \eqref{eq:EFT entropy def} must be a nondecreasing function of $t$. However, $D(t)$ should increase due to the data-processing inequality \cite{cover2012elements}.  This suggests that our generalized notion of a second law of thermodynamics provides nontrivial constraints on any effective theory of non-Gaussian stochastic dynamics, which cannot be deduced by simply writing down all invariant building blocks in a derivative expansion in $W$.  It would be interesting to more systematically analyze these constraints, but we expect that they are related to the fact that in the presence of non-Gaussian noise, the moments of the noise distribution are not independent and must be compatible with the Cauchy-Schwarz inequality.  It would be interesting to explicitly check this by using the relations between microscopic noise and higher-derivative terms in the FPE, following Ref.~\citenum{dubkov2005generalized}.

The formal notion of entropy that we have defined depends \emph{nonlinearly} on the solution to the Fokker-Planck equation \eqref{eq:FPE main}.  Thus, it is not defined for \emph{single} stochastic trajectories---in fact, it is not possible for such an entropy to be well defined. For example, in the overdamped oscillator discussed in Sec.~\ref{sec:intro}, such an entropy would have to be maximal when $x=0$, but a system initialized with $x=0$ can reach a state with $x\neq 0$, meaning that such an entropy function would decrease on this trajectory. To define a suitable entropy function $S(q_a)$ on phase space, the best one can hope for is that this function is nondecreasing \emph{on average}. For linear stochastic processes, such an entropy function always exists (see App.~\ref{app:entropy} for details). For nonlinear processes, we have not systematically found such a function, but we anticipate that combining the construction above with the arguments of Ref.~\citenum{bankslucas} could imply the existence of such an entropy function in certain models (e.g., in hydrodynamic systems). 

\subsection{Symmetries and conservation laws}
\label{sec:noether}

Having developed the basic EFT formalism for stochastic dynamics \eqref{eq:FPE main} in terms of the stationary distribution $\Phi$ \eqref{eq:FPE stationary distribution}, we now discuss the role of symmetries and conservation laws. As we will see, even though symmetries may lead to nonunique stationary distributions (e.g., depending on the ``symmetry sector'' to which the conserved charge belongs), they are elegantly imposed within our framework, using an analogue of Noether's theorem for stochastic dynamics in Eq.~\eqref{eq:noether}. Note that there are two subtly different ways by which symmetry can enter our EFT formalism, corresponding to symmetries of the dynamics versus symmetries of the stationary distribution $\Phi$ \eqref{eq:FPE stationary distribution}. The former case admits two types of conserved quantities, corresponding to ``strong'' and ``weak'' symmetries: A \emph{strong} symmetry holds for each individual, random (stochastic) trajectory, while a \emph{weak} symmetry only holds upon averaging over trajectories. However, there also exist situations where a symmetry of the stationary distribution $P^{\rm ss}\propto \exp(\Phi)$ is not a symmetry of the dynamics (even on average). We address each case in turn, and detail how such symmetries constrain the evolution operator $W$ \eqref{eq:W nice}.

\subsubsection{Strong and weak symmetries of the dynamics}
\label{subsub:strong v weak}

We first consider the case of \emph{strong} symmetries, in which a quantity $F(\vb{q})$ is conserved (i.e., $\dv*{F}{t} = 0$) on \emph{all} trajectories, and derive the consequences of such a conservation law on $W$ \eqref{eq:W nice}.  Since $F$ is constant in time by construction, multiplying and dividing $W$ by (any function of) thereof should not change $W$. In particular,
\begin{align}
\label{eq:W F condition stronk}
    \e^{-F}\, W\, \e^{F} \, = \, W \, , ~~
\end{align}
must hold for any such $F$. Since $W$ is generically a differential operator, we must account for the action of $W$ on the left-hand side of Eq.~\eqref{eq:W F condition stronk} on the function $\exp(F)$ to its right. We then find that, for  $F$ to be conserved,  $W$ must be invariant under the following shift of the differential operator  $\pd*{a}$,
\begin{align}
\label{eq:noether}
    W(\pd{a}, q_a) \, = \, W(\pd{a} + [ \pdv*{F}{q_a}] ,q_a) \, , ~~
\end{align}
where the quantity $[\pdv*{F}{q_a}]$ in square brackets is a function---not a differential operator. Thus, the form of $W$ \eqref{eq:W nice} is further constrained by \emph{each} conserved quantity $F$; this also extends to the case of vector conserved quantities, in which each component of $F$ is conserved. Examples of strong symmetries appear in Secs.~\ref{sec:spinchain} and \ref{sec:rigidbody}.

We now consider the case of \emph{weak} symmetries, in which case,  instead of $\dv*{F}{t}=0$, we instead have
\begin{align}
    \dv{t} \expval{F(t)}  \, =\,  \dv{t} \int\dd{\vb{q}} F(\vb{q}) P(\vb{q}, t) \, = \, 0 \, ,~~
\end{align}
where $P(\vb{q}, t)$ is the probability distribution at time $t$, and $\expval{F}$ denotes the average of $F$ over random trajectories (since $F$ is only a conserved quantity on average, and not along individual trajectories).  Note that $F(\vb{q})$ has no explicit time dependence (it depends on time only implicitly via its dependence on $\vb{q}$); as a result,
\begin{align}
    0 \, = \, \int \dd{\vb{q}} \, F(\vb{q})\,  \pd{t} P(\vb{q}) \, =\, \int \dd{\vb{q}} \, F(\vb{q})\,  [-W \, P(\vb{q})]\,  =\, \int \dd{\vb q} \,  P(\vb q) [-W^T F(\vb{q})]\, ,~~
\end{align}
and, since $W$ \eqref{eq:W nice} always has a derivative operator on the left (to preserve normalization in the FPE), we find
\begin{align}
\label{eq:W F condition weak}
    W^T F(\vb{q}) \, = \,\frac{1}{2} \,  (\pd{a} - \chempo{a} ) \mathcal{Q}_{ba} \, \pd{b} F  \, = \, 0 \, ,~~
\end{align}
so that, in many cases, we may simply enforce the condition. Thus, both strong and weak symmetries in stochastic dynamics may be enforced via constraints on $W$ \eqref{eq:W nice}.  

\subsubsection{Symmetries of the stationary distribution}
\label{subsub:sym+stationary}

Alternatively, there may also exist ``symmetries'' of the stationary distribution function $\Phi$ \eqref{eq:FPE stationary distribution} that do \emph{not} correspond to quantities $F$ that are conserved by the dynamics, even on average. Nonetheless, there remains a sense in which a particular invariance of $\Phi$ corresponds to a genuine symmetry; moreover, these nondynamical symmetries are relevant to our EFT formalism and have implications for the time-evolution operator $W$ \eqref{eq:W nice}. 

In particular, such symmetries directly facilitate the  construction of certain dissipationless terms in $W$.  For instance, suppose that a given stationary distribution $\Phi[\vb{q}]$ \eqref{eq:FPE stationary distribution} is invariant under coordinate shifts
\begin{equation}
    \label{eq:dist sym general}
    q_a \, \to \,  q_a + \epsilon F_a(\vb{q}) \, , ~~
\end{equation}
which implies that the generator $\vb{F}$ and chemical potential $\vb*{\mu}$ are orthogonal, i.e.,
\begin{equation}
\label{eq:dist sym mu F ortho}
    \vb{F} \cdot \vb*{\mu} \, = \, F_a \,\chempo{a} \, = \, \pdv{q_a}{\epsilon} \, \pdv{\Phi}{q_a} \, = \, \dv{\Phi}{\epsilon} \, = \, 0 \, , ~~
\end{equation}
and we also take $F$ to be divergenceless (i.e., $\pd*{a} F_a = 0$). In that case, a dissipationless term 
\begin{align}
    v_a \, \sim \, F_a \, ,~~
\end{align}
as defined in Eq.~\eqref{eq:v_anti} automatically satisfies the constraint of Eq.~\eqref{eq:dissipationless_terms} on all dissipationless contributions to the Fokker-Planck force $\vb{f}$ \eqref{eq:FPE force FDT}, and is therefore allowed.  More generally, dissipationless terms $v^{\,}_a$ are always allowed to move the system along \emph{level sets} \eqref{eq:dist sym general} of the stationary distribution function $\Phi$ \eqref{eq:FPE stationary distribution}.

A particularly important application of this idea arises when the distribution $\Phi$ is invariant under some group $G$. This provides for the inclusion of nontrivial (and quite exotic) dissipationless terms, even though there may be no corresponding dynamically conserved quantity. We treat this more general case in Sec.~\ref{sec:general_T_breaking}, where we also discuss the resulting possibility of spontaneous breaking of generalized time-reversal symmetry gT.

We note that symmetries of the distribution function $\Phi$ \eqref{eq:FPE stationary distribution} also offer a means by which to generate additional terms in $W$ \eqref{eq:W nice}, starting from some ``minimal'' choice of $W$  compatible with $\Phi$. To see this, note that a continuous symmetry of $\Phi$ generated via Eq.~\eqref{eq:dist sym general} acts on a generic probability distribution $P$ \eqref{eq:Probability} via
\begin{align}
\label{eq:symmetryPhi}
    P \, \to \, P' \, = \, \e^{\epsilon F_a \pd{a}} P \,, ~~
\end{align}
where invariance of $\Phi$ \eqref{eq:FPE stationary distribution} under the transformation of Eq.~\eqref{eq:dist sym general} generated by $\vb{F}$ requires that $\Phi' = \Phi$ (which recovers from taking $P=P^{\vpp}_{\rm ss}$ and $P'=P_{\rm ss}'$). Applying  $\exp(\epsilon F_a \pd*{a})$ to both sides of $\pd*{t} P = -W P$ \eqref{eq:FPE kets} leads to
\begin{align}
    \pd{t} P \, = \, -W P \implies \e^{-\epsilon F_a \pd*{a}} \pd{t} P' \, = \, -W \e^{-\epsilon F_a \pd*{a}} P' ~~\implies ~~\pd{t} P'\,  = \, - \e^{+\epsilon F_a \pd*{a}} W \e^{-\epsilon F_a \pd*{a}} P' \, \equiv \,  - W' P' \, , ~~
\end{align}
and, since $P_{\rm ss}^{\vpp} = P_{\rm ss}'$  \eqref{eq:FPE stationary distribution}, we have found that $W' \equiv \e^{\epsilon F_a \pd*{a}} W \e^{-\epsilon F_a \pd*{a}}$ is a valid Fokker-Planck generator with the \emph{same} stationary distribution $P_{\rm ss} = \exp (\Phi)$ \eqref{eq:FPE stationary distribution} as $W$! In particular, for infinitesimal transformations, we find
\begin{align}
\label{eq:Wshift}
    W' \, = \, W + \epsilon \left[F_a \pd{a}, W\right] \, ,~~
\end{align}
where $[A,B]=AB-BA$ is the commutator.  The shift captured by Eq.~\eqref{eq:Wshift} shows how additional terms (consistent with the same stationary distribution) may be added to a given Fokker-Planck generator $W$.

\subsection{Connection to the Martin-Siggia-Rose path integral}
\label{sec:MSR}

%One of the key insights provided by the path-integral formalism of dissipative systems relates to the  the KMS transformation is closely related to the time-reversal operation T in such thermal systems \cite{eft1, eft2}.  In the MSR formalism, generalized time-reversal symmetry T appears, at least superficially, of KMS

We now connect the foregoing operator formalism to the path-integral description of stochastic Langevin dynamics \eqref{eq:SDE main} and FPEs \eqref{eq:FPE main} due to Martin, Siggia, and Rose (MSR) \cite{MSR}. The latter affords a Lagrangian description of dissipative systems, so that familiar path-integral methods can be used to study active and dissipative dynamics not only of point-particle systems, but continuum systems (i.e., field theories). As with the Fokker-Planck operator formalism developed thus far, the mathematical implementation and physical role of the time-reversal operation T are deeply important to the MSR description of the stochastic EFT. The MSR implementation of T appears---at least superficially---quite similar to that of the Kubo-Martin-Schwinger (KMS) invariance of systems both in thermal equilibrium \cite{ArakiKMS78, martinschwinger} and beyond \cite{Guo2022, Qi:2022vyu}. However, we note that there are several %subtle discrepancies that we emphasize later on. 
subtleties; accordingly, we view the corresponding $\Ints_2$ operation as the path-integral implementation of T. 

%The EFTs we develop based on the existence of stationary states $P^{\,}_{\rm ss} \propto \exp(-\Phi)$ \eqref{eq:FPE stationary distribution} can be viewed as extending this analogy between time-reversal symmetry and KMS invariance, thereby generalizing familiar results for systems in thermal equilibrium to dissipative and active classical matter.

We also note that T is often cumbersome to implement in the MSR framework---at least in full generality---due to subtleties related to the regularization  (i.e., It\^o versus Stratonovich) of SDEs such as Eq.~\eqref{eq:SDE main}. We relegate the appertaining technical details to App.~\ref{app:PI}; other subtleties related  to the regularization are discussed in Ref.~\citenum{lubensky}. In most of this work, we instead formulate EFTs in terms of the FPE operator $W$ \eqref{eq:W nice}, in which T is straightforward. However, we note that a direct connection between the FPE and MSR implementations of T exists in certain simple limits. Accordingly---and owing to the utility of the MSR formalism in describing continuum models, and its relevance to recent developments concerning active and dissipative systems---we now review the MSR formalism, the implementation of T therein, and its relation to the operator-based Fokker-Planck approach discussed previously.

%In most of the paper, we will therefore use the operator formalism where T is manifestly implementable. However, because time-reversal symmetry is easily implemented within the MSR formalism in certain limits, and relating our operator-based approach to the problem to the MSR framework provides strong connections to other recent literature, we now 

\subsubsection{Details of the MSR formalism}
\label{sec:MSR main}

The MSR path integral reproduces the transition probability \eqref{eq:FPE transition prob} according to
\begin{align}
    \label{eq:MSR amplitude}
    \mathbb{P}[\vb{q}_i(0) \, \to \, \vb{q}_f(t);\mathcal{H}]  \, = \, \matel*{\vb{q}_f(t)}{\e^{-\int_0^{t} \thed t' \, W (\vb{q}(t'))}}{\vb{q}_i(0)} \, = \, \int\limits_{\vb{q}(0)=\vb{q}_i}^{\vb{q}(t)=\vb{q}_f}  \theD \vb{q} \, \theD \vb*{\pi} ~\exp \left[ \ii \, \int\limits^t_0 \thed t' \, ( \pi_a \pd{t} q_a - \mathcal{H}(\vb{q},\vb*{\pi})) \right] \, ,~~
\end{align}
where %, in the MSR formulation on the RHS, we dispense with the bold-faced vector notation (facilitating generalization to fields), and 
the conjugate momenta $\pi_{a} \sim - \ii \pd*{a}$ and Hamiltonian $\mathcal{H} \sim - \ii \, W$ are defined in Eq.~\eqref{eq:FPE pi} for FPE dynamics. As a reminder, $\cal H$ is generally \emph{not} the microscopic Hamiltonian (which appears in Sec.~\ref{sec:not active}), and the coordinates $\vb{q}$ include all phase-space variables describing the dynamical system---for $N$ point particles in $d$ spatial dimensions, $\vb{q}$ includes all $dN$ canonical positions $x_a$ \emph{and} momenta $p_b$. The fields $\pi_{a}$ are the conjugate momenta to $q_a$ in the sense of the EFT formalism, and are often easier to work with than $\pd*{a}$ in the context of continuum dynamics.
%We find that $H$ and $\pi$---and the MSR formalism in general---are more useful in field-theoretic (i.e., continuum) contexts, as the operators $\pd*{a}$ and $W$ are notationally unwieldy (owing, e.g., to functional differentiation). Moreover, in the absence of multiplicative or non-Gaussian noise, there is no advantage to the operator formalism. 
% Still, certain concepts---like the central role of time-reversal symmetry---are more transparent in the operator language; 

Before explaining the details of the MSR formalism and its  connection to the FPE operator language, we first introduce a slightly modified notation to incorporate \emph{multiplicative} noise. A stochastic process with random noise source $\xi_a(t)$---where $\expval{\xi_a (t)}=0$ and $\expval{\, \xi_a(t) \, \xi_b(t') \, } = \Delta \, \kron{a,b} \, \delta (t-t')$---can be extended to account for multiplicative noise via $\eta_a (\vb{q},t) = M_{ab}  (\vb{q}) \, \xi_b(t)$. Essentially, the bare noise $\xi_b$ is modulated by $M_{ab}(\vb{q})$, so that 
\begin{equation}
    \label{eq:multiplicative noise matrix}
    Q_{ab}(\vb{q}) \, =\,  \int \thed t' \, \expval{\eta_a(\vb{q},t) \eta_b (\vb{q},t')} \,  = \,  M_{ac} (\vb{q}) \, M_{bd} (\vb{q}) \int \thed t' \, \expval{\xi_c (t) \xi_{d}(t')} \, = \,  M_{ac}(\vb{q}) \, M_{bc} (\vb{q}) \, ,~~ 
\end{equation}
is the symmetric, positive-semidefinite noise matrix \eqref{eq:Q noise def}. In the case $Q_{ab}(\vb{q})=\Delta \, \kron{a,b}$, the multiplicative noise reduces to Gaussian additive noise, described by the noise matrix \eqref{eq:Q noise def} with variance $\Delta$. 

Now, consider a stochastic process captured by a Langevin SDE  \eqref{eq:SDE main} and subject to multiplicative noise $\eta_a$ with noise matrix $Q$ \eqref{eq:multiplicative noise matrix}. The corresponding MSR path integral (with boundary conditions suppressed) is given by 
\begin{align}\label{eq:zxi}
    Z[\vb*{\xi}] \, = \, \int \theD \vb{q} \, \mJ[\vb{q}] ~\DiracDelta{\pd{t} q_a - f_a(\vb{q}) - M_{ab}(\vb{q})\, \xi_b(t)}\, =\, 1 \, ,~~
\end{align}
where $\mJ[\vb{q}]$ is the Jacobian of the operators inside the delta function (see App.~\ref{app:PI} for technical details), and  $\vb*{\xi}$ appears in Eq.~\eqref{eq:zxi} because $Z$ is not yet averaged over noise. Note that Eq.~\eqref{eq:zxi} is equal to one because the integral is indefinite, and realizes the transition amplitude of Eq.~\eqref{eq:MSR amplitude} when appropriate boundary conditions are imposed. For convenience, we consider the It\^o regularization, for which $\mathcal{J}[\vb{q}]=1$.  The noise-averaged partition function \eqref{eq:zxi} is 
\begin{align}
    Z \, & = \, \int \theD  \vb*{\xi} ~ Z[\vb*{\xi}] \, \exp \left( -\frac{1}{2}\int \thed t \, \xi_a^2 \right) \notag \\
    &= \,   \int \theD \vb*{\xi} \, \theD \vb{q} \, \theD \vb*{\pi} ~\exp \left( \ii \int \thed t \, \pi_a  (\pd{t} q_a - f_a (\vb{q})- M_{ab} (\vb{q})\, \xi_b) \right)  ~\exp \left( -\frac{1}{2} \int \thed t \, \xi_a^2 \right) \, ,~~\label{eq:L before averaging}\\
\intertext{and integrating out the noise (i.e., evaluating the integral over $\vb*{\xi}$) leads to }
    Z \, &= \,  \int  \theD \vb{q} \, \theD \vb*{\pi} ~\exp \left( \ii \int \thed t \,  (\pi_a \,\pd{t} q_a - \pi_a f_a (\vb{q}) +\ii\frac{1}{2} \pi_a Q_{ab}\pi_b ) \right) \notag\\
    &= \, \int  \theD  \vb{q} \, \theD \vb*{\pi} ~\exp \left( \ii \int \thed t \, L(\vb{q},\vb*{\pi}) \right) \equiv \int  \theD \vb{q} \, \theD \vb*{\pi} ~\exp \left( \ii \int \thed t \, (\pi_a \pd{t} q_a - \mathcal{H}(\vb{q},\vb*{\pi})) \right)  \, , ~~\label{eq:L after averaging}
\end{align}
which implicitly defines the MSR \emph{Lagrangian}
\begin{equation}
    \label{eq:MSR Lagrangian}
    L (\vb{q}, \vb*{\pi}) \,  = \, \pi_a \pd{t} q_a - \mathcal{H}(\vb{q},\vb*{\pi}) \, = \, \pi_a \pd{t} q_a - \pi_a f_a (\vb{q}) + \frac{\ii}{2} \pi_a \, Q_{ab} (\vb{q}) \, \pi_b \, ,~~
\end{equation}
where we have written both Eqs.~\eqref{eq:L after averaging} and \eqref{eq:MSR Lagrangian} such that replacing $\pi_a \to- \ii \, \pd*{a}$ \eqref{eq:FPE pi} recovers $W = \ii \, \mathcal{H}$ \eqref{eq:W nice}. 

\subsubsection{Time-reversal symmetry}
\label{sec:KMSvsT}
% \label{sec:MSR EO}

In developing the operator-based Wilsonian EFT for stochastic dynamics, we have %thus far 
emphasized the simple nature of the ``time-reversal'' transformation T acting on the Fokker-Planck generator $W$ \eqref{eq:W reversed}. There is also a natural definition of T in the MSR formulation of the same stochastic effective theories. Importantly, the latter implementation of T is (\emph{i}) generally more convenient in the context of continuum systems described by field theories and (\emph{ii}) closely connected to the KMS transformation \cite{ArakiKMS78, martinschwinger}, around which the thermal EFTs of, e.g., hydrodynamic systems are organized \cite{Liulec, Guo2022, Qi:2022vyu,Mullins:2023ott}. We first present the path-integral implementation of T \eqref{eq:MSR T pi relation} in a simple (yet physically relevant) limit in which it is directly related to its FPE-operator analogue \eqref{eq:W reversed}. We then justify in the path-integral framework that the corresponding transformation \eqref{eq:MSR T pi relation} is the correct MSR implementation of T. Finally, we comment on the connection between these two implementations of T beyond the simple limit above.

%T transformation of Eq.~\eqref{eq:MSR T pi relation} and the KMS transformation that features prominently in thermal EFTs \cite{ArakiKMS78, martinschwinger, Liulec, Qi:2022vyu, Guo2022}. 

%At the same time, we note that the development of hydrodynamic EFTs for thermal fluids is fundamentally organized around the closely related KMS transformation. Importantly, while time reversal T is straightforward in the Fokker-Planck operator formalism, the definition of the KMS transformation is not immediately clear in that approach. By contrast, in the MSR path-integral approach, the KMS transformation is straightforward, while time reversal T is less transparent. We now discuss the relationship between KMS and T in detail; while we work in the MSR language,  the statements below can be translated into the Fokker-Planck operator language via, e.g., \eqref{eq:FPE pi}.

%We now provide an affirmative answer to this question by 
Consider a T-invariant stochastic model (i.e., with ${\cal W} = \widetilde{W} \implies {\cal H} = \widetilde{\cal H}$) subject to a Gaussian noise source with constant variance matrix $Q_{ab}$ \eqref{eq:Q noise def}.  Ignoring for the moment the dissipationless term $v_a$ \eqref{eq:v_anti} in $\vb{f}$ \eqref{eq:FPE force FDT}, the MSR Lagrangian for such a stochastic system is given by 
\begin{equation}
    L \, = \, \pi_a \pd{t} q_a + \frac{\ii}{2} \pi_a \, \Delta_{ab} \, \left(\pi_b-\ii \chempo{b} \right)  \, , ~~ \label{eq:MSR_L_Teven}
\end{equation}
which follows from Eqs.~\eqref{eq:Q noise def}, \eqref{eq:FPE force FDT}, and \eqref{eq:MSR Lagrangian}, where the elements $Q_{ab}=\Delta_{ab}$ \eqref{eq:Q noise def} are independent of both $\vb{q}$ and $\vb*{\p}$. We consider the inclusion of dissipationless terms---along with multiplicative and higher-derivative noise---momentarily. 

The corresponding \emph{MSR implementation of T} in the limit of Gaussian noise and T-even $W$ \eqref{eq:W nice} is given by % the symmetry %% on it's own this isn't always a symmetry?
\begin{align}
    t \, \to-t ~,~~q_a \to \pm q_a ~,~~\text{and}~~\pi_a \to \mp \left( \pi_a - \ii \chempo{a} \right) \, ,~~
    \label{eq:MSR T pi relation}
\end{align}
where $\pm$ denotes whether $q_a$ is even ($+$) or odd ($-$) under T (i.e., $x \to x$ and $p \to -p$), and we find that $\chempo*{a} \to \pm \chempo*{a}$ under T \eqref{eq:MSR T pi relation}. We note that the transformation in Eq.~\eqref{eq:MSR T pi relation} is mathematically equivalent to %previous: mathematically similar to
the implementation of KMS in the path-integral (EFT) description of thermal dissipative systems \cite{ArakiKMS78, martinschwinger, Liulec, Qi:2022vyu, Guo2022}.

By applying the transformation T \eqref{eq:MSR T pi relation} to $\mathbb{P} [ \vb{q}_i (0) \to \vb{q}_f (t) ]$ \eqref{eq:MSR amplitude}, we can compare the probability to go from configuration $\vb{q}_i$ to $\vb{q}_f$ in time $t$ to the probability to go from $\vb{q}_f$ to $\vb{q}_i$ in time $t$ under the time-reversed process \eqref{eq:W reversed},
\begin{align}
    \widetilde{\mathbb{P}} \, [\vb{q}_i(0) \to \vb{q}_f(t);\mathcal{H}] \, &= \,  \int_{\vb{q}(0)=\vb{q}_f}^{\vb{q}(t) = \vb{q}_i}  \theD \vb{q} \, \theD \vb*{\pi} ~\exp \left( \ii \int_0^t \thed t' \, \left[ (\pi_a-\ii\chempo{a}) \pd{t} q_a - \widetilde{\mathcal{H}}(\vb{q},\vb*{\pi}) \right] \right) \label{eq:MSR reverse prob} \, ,~~ \\
\intertext{where $\widetilde{\cal H}= - \ii \widetilde{W}$ \eqref{eq:W reversed} is the  time-reversed MSR Hamiltonian. Using the standard chain rule, we note that $\chempo*{a} \pd*{t} q^{\,}_a= (\pdv*{\Phi}{q_a} ) \, \pd*{t} q_a = \dv*{\Phi}{t}$, so that the probability amplitude above becomes}
    \widetilde{\mathbb{P}} \, [\vb{q}_i(0) \to \vb{q}_f(t);\mathcal{H}] \,  &= \, \e^{\Phi(\vb{q}_i)-\Phi (\vb{q}_f )} \, \int_{\vb{q}(0)=\vb{q}_f}^{\vb{q}(t) = \vb{q}_i}   \theD \vb{q} \, \theD \vb*{\pi} ~\exp \left( \ii \int_0^t \thed t' \, \left[ \pi_a\pd{t} q_a - \widetilde{\mathcal{H}}(q,\pi) \right] \right) \notag \\
    &= \, \e^{\Phi(\vb{q}_i)-\Phi (\vb{q}_f )}  ~\mathbb{P}[\vb{q}_f(0) \to \vb{q}_i(t);\widetilde{\mathcal{H}}] \, ,~~\label{eq:MSR balance condition}
\end{align}
which is the probability to transition from %the \emph{initial} configuration
$\vb{q}_f$ to %the \emph{final} configuration 
$\vb{q}_i$ in time $t$ under $\widetilde{\cal H}$ \eqref{eq:W reversed}---the time-reversed partner to ${\cal H}= - \ii W$---in analogy to Eq.~\eqref{eq:FPE balance condition}. Note that we have restricted to models in which ${\cal H}$ is invariant under T, and thus obeys \emph{detailed balance} \eqref{eq:introdb}. When $Q_{ab}$ \eqref{eq:Q noise def} corresponds to Gaussian noise (i.e., contains no derivatives $\pd*{a}$), we find that 
\begin{align}
    {\cal H} \, = \, - \frac{\ii }{2} \pi_a \, Q_{ab} \, \left( \pi_a - \ii \, \chempo{a}\right) \, &\to \, \widetilde{\cal H} \, %= - \frac{\ii \, \Delta}{2} \left( \mp \right) \left( \pi^{\vpp}_a - \ii \, \chempo{a} \right) \left( \mp ( \pi^{\vpp}_a  - \ii \chempo{a} ) \mp \ii \, \chempo{a} \right) \notag \\
    %&=
    =  - \frac{\ii}{2} \left( \pi_a - \ii \, \chempo{a} \right) \, Q_{ab} \,  \pi_a \, = \,  \,{\cal H} ,~~
\end{align}
meaning that ${\cal H} = \widetilde{\cal H}$, as required. Note that the transformation T \eqref{eq:MSR T pi relation}---and the connection to its FPE counterpart \eqref{eq:W reversed}---are complicated in the presence of non-Gaussian noise, where the derivatives $\pd*{a}$ in the expansion for ${\cal Q}_{ab}$ \eqref{eq:higher derivatives} spoil the derivations above. For such noise sources, the connection between the two implementations of T is more elusive. However for Gaussian noise, invariance under the transformation of Eq.~\eqref{eq:MSR T pi relation} implies that
\begin{equation}
\label{eq:LKMSeven}
    W_{\rm MSR-even} \, =   - \frac{1}{2} \pd{a} \left( \pd{b} + \chempo{b} \right) Q_{ab}  \, = \, \ii \, {\cal H}_{\rm MSR-even} \, = \, \ii \, \widetilde{\cal H} \,  = \,  \widetilde{W} \, ,~~
\end{equation}
in the FPE operator language, since the Fokker-Planck generator $W$ \eqref{eq:W nice} is straightforwardly related to the MSR Hamiltonian $\cal{H}$ \eqref{eq:MSR Lagrangian} via Eq.~\eqref{eq:FPE pi} when written in the It\^o formalism (with all derivatives to the left).

In the presence of \emph{multiplicative} noise, Eq.~\eqref{eq:LKMSeven} fails, as $Q_{ab}$ depends on $\vb{q}$. Moreover, even in the case of dissipation toward equilibrium, the connection to thermal KMS \cite{Liulec} is complicated in the presence of multiplicative noise. In App.~\ref{app:PI}, we show that Eq.~\eqref{eq:MSR T pi relation} does \emph{not} preserve the It\^o regularization\footnote{A similar issue is also known to arise in standard quantum mechanical path integrals for particles for which $H \neq p^2/2m + V(q)$ \cite{kleinert}.}; thus, the correct implementation of T involves supplementing Eq.~\eqref{eq:MSR T pi relation} with an appropriate change in the regularization itself\footnote{In previous work \cite{ghostbusters}, the KMS transformation was implemented using a regularization stated in the \emph{frequency} domain, in which case it is not straightforward to cast the dynamics in terms of a master equation.}. In App.~\ref{app:PI}, in the case of multiplicative Gaussian noise, we explain how to implement the path integral \eqref{eq:MSR amplitude} with a suitable regularization such that T \eqref{eq:MSR T pi relation} is manifestly a symmetry. We relegate a study of more general, non-Gaussian noise terms to future work.

%to carry out the path integral in a correct regularization, where time reversal is manifest as a symmetry, for multiplicative Gaussian noise; we leave the more general case to future work.%We relegate the consideration T in the presence of more complicated noise sources---and its formal connection to the KMS transformation for thermal systems---to future work. 

However, we also note that, from a field-theoretic \emph{renormalization-group} perspective,  multiplicative noise is typically \emph{irrelevant}. Thus, any potential discrepancy between the MSR implementation of T \eqref{eq:MSR T pi relation} and both the FPE analogue \eqref{eq:W reversed} and the KMS transformation for thermal  systems \cite{ArakiKMS78, martinschwinger} may  ultimately be unimportant to the universal properties and classification of active and dissipative systems.

% In the simple limit in which all of the coordinates $\vb{q}$ transform as $\vb{q} \to \pm \vb{q}$ under \emph{microscopic} time reversal (e.g., for position coordinates), then $v_a \to \pm v_a$ under T \eqref{eq:MSR T pi relation}. As a result, 

We now consider more general models, which may include \emph{dissipationless} contributions $v_a$ \eqref{eq:v_anti} to the Fokker-Planck force $\vb{f}$ \eqref{eq:FPE force FDT}. In this case, we adjust the (T-even) MSR Lagrangian \eqref{eq:MSR_L_Teven} as 
\begin{equation}
\label{eq:MSR L even with v}
    L \, = \, L_{\rm MSR-even} + \pi_a \, v_a \, = \, \pi_a  \pd{t} q_a + \frac{\ii}{2} \pi_a  \, Q_{ab} (\pi_b - \ii \chempo{b} ) + \pi_a v_a \, ,~~
\end{equation}
where $v_a$ \eqref{eq:v_anti} is generically a function of $\vb{q}$. Suppose that $v_a \to \widetilde{v}_a$ under T, then for each label $a$, the $\pi_a v_a$ term in the MSR Lagrangian \eqref{eq:MSR L even with v} transforms as
\begin{equation}
\label{eq:MSR nice v under T}
    \pi_a v_a \, \to \,  \mp (\pi_a - \ii \chempo{a}) \widetilde{v}_a \, ,~~
\end{equation}
and if $\widetilde{v}_a = v_a$, the $V$ term in Eq.~\eqref{eq:nice Q def}---corresponding to $\pi_a v_a$ in $L$ \eqref{eq:MSR L even with v}---flips sign under T, as expected. On the other hand, if $\widetilde{v}_a = -v_a$, the $\pi_a v_a$ in $L$ \eqref{eq:MSR L even with v} is invariant under T (such dissipationless terms obey detailed balance, and are compatible with Hamiltonian dynamics). We now observe an important complication: The extra $\ii \chempo*{a} v_a$ term in Eq.~\eqref{eq:MSR nice v under T} that is ``spawned'' when we apply T \eqref{eq:MSR T pi relation} to $L$ \eqref{eq:MSR L even with v}. For the Lagrangian $L$ \eqref{eq:MSR L even with v} to transform under T into another valid MSR Lagrangian in the It\^o regularization, we must have $\chempo*{a}v_a=0$ since all terms must have at least one power of $\pi_a$.  However, this is \emph{not} the constraint on T-odd terms derived in Eq.~\eqref{eq:dissipationless_terms} unless $\pd*{b} V_{ab}=0$. This is guaranteed so long as $V_{ab}$ is constant, but may not hold in general. The origin of this complication is the subtlety surrounding regularizations noted below Eq.~\eqref{eq:LKMSeven}. The takeaway is that there are extra constraints on the dissipationless terms $\pi_a v_a$ in the MSR Lagrangian due to stationarity: Dynamics compatible with a stationary distribution $\Phi$ \eqref{eq:FPE stationary distribution} have a time-reversed evolution operator $\widetilde{W}$ \eqref{eq:W reversed} that is also a valid FPE generator; however, if $\chempo*{a}v_a \neq 0$, then the time-reversed MSR Lagrangian does not generate valid Fokker-Planck dynamics, as all terms in the MSR Lagrangian \eqref{eq:MSR Lagrangian} must contain at least one power of $\pi_a$.  

The subtleties discussed above concerning multiplicative noise and dissipationless contributions arise when $\p_b Q_{ab}$ or $\p_b V_{ab}$ are nonzero. However, for continuum field theories, such as hydrodynamics, we show explicitly in  App.~\ref{app:diffusion} that it is possible to choose a spatial discretization of $Q_{ab}$ and $V_{ab}$ such that their divergences vanish. This discretization, which can be thought of as a regularization of the model, leads to an MSR realization of time reversal \eqref{eq:MSR T pi relation} that is \emph{exactly} equivalent to the Fokker-Planck realization of T \eqref{eq:W reversed}. The crucial assumption is that the expansion in spatial gradients is treated perturbatively.  This observation further suggests that precision tests \cite{Delacretaz:2023ypv} of nonlinear fluctuating hydrodynamics would \emph{not} detect any discrepancy between reversibility \eqref{eq:W reversed} and KMS invariance \cite{ArakiKMS78, martinschwinger}; meaning that such discrepancies can only arise in theories where higher-gradients are not perturbative.

\subsubsection{Time-reversal symmetry without integrating out noise} \label{subsub:MSR TR pre noise}

The MSR framework also makes transparent how a noisy SDE \eqref{eq:SDE main} is \emph{microscopically} T invariant, and thus obeys detailed balance.  For simplicity, consider additive Gaussian noise with $Q_{ab}=\Delta \, \kron{a,b}$ \eqref{eq:Q noise def}, so that T \eqref{eq:MSR T pi relation} acts as
\begin{equation}
    \label{eq:MSR TR}
    t \, \to - t ~,~~q_a \, \to \pm q_a ~,~~ \pi_a \, \to \mp \left( \pi_a - \ii \chempo{a} \right) ~,~~\xi_a \, \to \mp \left( \xi_a - \Delta \chempo{a} \right) \, , ~~
\end{equation}
where $\pm$ refers to whether $q_a$ is T even ($+$) or T odd ($-$). It is straightforward to verify that the MSR Lagrangian $L$ \eqref{eq:L before averaging} prior to integrating out the noise $\xi_a$ is invariant under the  transformation T \eqref{eq:MSR TR}. 

In other words, Eq.~\eqref{eq:MSR TR} is the transformation in the path-integral EFT corresponding to T (i.e., reversibility). Considering the damped harmonic oscillator described by Eq.~\eqref{eq:newton} of Sec.~\ref{sec:intro}, from our perspective, it makes sense to say that $\dot{x} = -\gamma x+\xi$, with $\gamma = \Delta \mu/2 x$ as in Eq.~\eqref{eq:newton2}, is time-reversal symmetric (or T even).  Under time reversal T, the noise variable $\xi$ shifts so as to undo precisely the na\"ive sign change to the dissipative term. This does not happen by chance, but is a direct consequence of the existence of a stationary state.

%\section{Adding dissipation to Hamiltonian dynamics}
% \section{Dissipative generalizations of Hamiltonian dynamics for inactive matter}
\section{Dissipative generalizations of Hamiltonian dynamics}
\label{sec:not active}

Before we consider the application of the effective field theory (EFT) framework of Sec.~\ref{sec:EFT} to nonequilibrium \emph{active} matter, we first consider EFTs for dissipative thermal systems. The formalism of Sec.~\ref{sec:EFT} provides a simple theoretical description of Hamiltonian systems in thermal equilibrium, and a natural prescription for adding dissipation to any such model. Such systems evolve toward a particular stationary distribution, corresponding to
\begin{equation}
    \label{eq:phibetaH} 
    \Phi \, = \, \beta \, H \, ,~~
\end{equation} 
meaning that stationary distribution $\Phi$ is simply the Boltzmann distribution for the underlying Hamiltonian $H$ at inverse temperature $\beta$. Such a distribution naturally describes thermal equilibrium of a system with its environment, where the Hamiltonian $H$ is also responsible for time evolution.  We now describe several simplifying aspects of Hamiltonian dynamics, along with a natural procedure for coupling Hamiltonian systems to dissipative baths. While the treatment of equilibrium systems herein largely reproduces well-known results, it does so in a more systematic manner using a simple algorithm compared to existing literature, and helps to build intuition for the EFT formalism in anticipation of its later application to active matter in the rest of the paper. 

We first review of Hamiltonian mechanics in Sec.~\ref{subsec:Hamiltonian setup}, using the EFT formalism of Sec.~\ref{sec:EFT}. In particular, given a stationary state $\Phi = \beta H$ \eqref{eq:phibetaH}, we relate the Hamiltonian $H$ to a Fokker-Planck generator $W$ compatible with $\Phi$; we note that this does not require the use of the canonical phase-space variables, and can incorporate symmetries of the dynamics and stationary distribution $\Phi$. In Sec.~\ref{sec:whitenoise}, we prescribe how to add dissipation to generic Hamiltonian systems (including those with particular symmetries) in equilibrium with a thermal bath; this is remarkably straightforward in the EFT framework of Sec.~\ref{sec:EFT}. Finally, in the remaining subsections, we survey a variety dissipative generalizations of paradigmatic Hamiltonian theories, including an example featuring nontrivial Poisson brackets (in Sec.~\ref{sec:spinchain}). 

\subsection{Hamiltonian dynamics and Poisson brackets}
\label{subsec:Hamiltonian setup}

Here we briefly review the important features of Hamiltonian dynamics in the absence of dissipation (i.e., noise), and explain the application of the effective theory formalism developed in Sec.~\ref{sec:EFT} to this setting.

\subsubsection{Hamiltonian mechanics with canonical phase space variables}

Numerous important aspects of Hamiltonian mechanics are captured by the \emph{Poisson bracket} $\{ \cdot , \cdot \}$, which is defined for arbitrary functions of the phase-space variables. We frequently refer to the simple example of $N$ particles in $d$ spatial dimensions, described by $Nd$ positions $x_a$ and $Nd$ momenta $p_a$, which satisfy the canonical Poisson brackets
\begin{subequations}\label{eq:31pb}
\begin{align}
    \qty{x_a,x_b} \, &= \, \qty{p_a,p_b} \, = \, 0, \\
    \qty{x_a,p_b} \, &= -\qty{p_b,x_a} \, = \, \kron{a,b} \,,~~
\end{align}
\end{subequations}
where the label $a$ combines the particle and spatial-axis indices (with a total of $Nd$ values). The Poisson bracket $\{ \cdot , \cdot \}$ is defined for \emph{all} classical dynamical systems, including those with noncanonical phase-space variables; it also extends to generic functions of the phase-space variables via the chain rule of calculus. 

As a reminder, in the EFT formalism of Sec.~\ref{sec:EFT}, the coordinate vector $\vb{q}$ contains \emph{all} phase-space variables. In the example above, $\vb{q}$ is a length-$2Nd$ vector of all positions $x_a$ \emph{and} momenta $p_b$, such that Hamilton's equations may be written simply as $\pd*{t} q_a = \qty{q_a,H}$, with $H$ the Hamiltonian. For %noiseless 
dissipationless Hamiltonian dynamics, we have
\begin{equation}
    W \, = \, T \, \pd{a} \qty{ q_a,q_b } \chempo{b} \, ,~~
    \label{eq:W therm}
\end{equation}
where $T \equiv \beta^{-1}$ is the temperature (with $k_B = 1$). Because the stationary distribution $\Phi$ \eqref{eq:phibetaH} is proportional to the Hamiltonian $H$, the ``chemical potentials'' $\chempo*{b}$ in Eq.~\eqref{eq:W therm} are the canonical ones---the Legendre transforms (or thermodynamic conjugates) of the corresponding coordinate $q_b$. For example, for a single nonrelativistic particle in one dimension  subject to the force $F(x)=V'(x)$ (i.e., $H=p^2/2m + V(x)$), we have
\begin{equation}
    \label{eq:mu therm}
    \chempo{a} \, =\, \pdv{[\beta H]}{q_a} ~~\implies~ ~\chempo{x}\, =\, \frac{1}{T} \pdv{V}{x} \, = - \beta F (x) ~,~~\chempo{p} \, =\,  \frac{p}{m T} \, = \, \beta \dot{x} \, , ~~
\end{equation}
and thus, for dissipationless Hamiltonian mechanics, knowledge of the Poisson bracket \eqref{eq:31pb} and the Hamiltonian $H$, directly implies the form of the Fokker-Planck generator $W$ \eqref{eq:W_therm2} of time evolution.

It is also straightforward to recover the corresponding MSR Lagrangian $L(\vb{q},\dot{\vb{q}}, \vb*\pi)$ \eqref{eq:MSR Lagrangian},
\begin{align}
\label{eq:thermalLa}
    L \, =  \, \pi_a \pd{t} q_a - \pi_a\qty{q_a, H} \, ,~~
\end{align}
where $\pi_a$ is conjugate to $q_a$ in the sense of Sec.~\ref{sec:MSR main}, the Hamiltonian $H$ both generates time evolution via $\pd*{t} q_a = \qty{q_a,H}$ and determines the stationary state $\Phi=\beta H$ \eqref{eq:phibetaH}, and Eqs.~\eqref{eq:W therm} and \eqref{eq:thermalLa} lead to the same equations of motion.

It is conventional to identify the Poisson bracket \eqref{eq:31pb} with the inverse of the \emph{symplectic form} $\omega$,
\begin{equation}
\label{eq:symp form}
    \omega\,  = \, \frac{1}{2} \omega^{\vpp}_{ab}\thed q_a\wedge \thed q_b ~~~ \implies ~~~  \omega^{-1}_{ab} \, \equiv \, \omega^{ab}  \, = \, \qty{q_a,q_b} \, ,~~
\end{equation}
where the \emph{inverse} of the symplectic form is identified by \emph{raised} indices. We can rewrite Eq.~\eqref{eq:W therm} as
\begin{align}
\label{eq:W_therm2}
    W \, =\,  T\, \pd{a}\, \omega^{ab}_{\,} \, \chempo{b} \, ,~~
\end{align}
in terms of the inverse symplectic form $\omega^{-1}$, which is more convenient than Eq.~\eqref{eq:W therm} when generalizing to systems in which the phase-space variables satisfy noncanonical Poisson brackets (as in Sec.~\ref{sec:spinchain}).

\subsubsection{Time-reversal symmetry}

It is natural to ask how the symplectic form $\omega$ \eqref{eq:symp form} and the Fokker-Planck generator $W$ \eqref{eq:W_therm2} behave under time reversal T~ \eqref{eq:W reversed}. Microscopically, time reversal transforms the phase-space coordinates as, e.g., $(x,p)\to (x,-p)$; accordingly, the symplectic form $\omega$ \eqref{eq:symp form} transforms as $\omega \to -\omega$ (and likewise for its inverse). Implementing T \eqref{eq:W reversed} in the operator language, this implies that $W$ \eqref{eq:W therm} is manifestly time-reversal symmetric for Hamiltonian dynamics (i.e., it obeys detailed balance); taking a simple example with $N=d=1$ so that $\vb{q}=(x,p)$, we find that
\begin{align}
    \widetilde{W} \, &=\, S \,  \left( \pd{x} \chempo{p} - \pd{p} \chempo{x} \right) \, S^{-1} \, = \, \e^{-\beta H} \, \left( (-\chempo{p})(-\pd{x})  -  (\chempo{x}) (\pd{p})\right) \e^{\beta H} \notag \\
    &= \, \left( \pd{x} \chempo{p} - [\pd{x}\chempo{p}] + \chempo{p} [-\chempo{x}] - \pd{p} \chempo{x} + [\pd{p} \chempo{x}] + \chempo{x} [-\chempo{p}] \right) \notag \\
    &= \, \pd{x} \chempo{p} - \pd{p} \chempo{x} \, =\,  W \, ,~~
\end{align}
where we used the chain rule and the fact that, e.g., $\chempo*{x}=\beta \pd{x}H$. Importantly, T symmetry is expected for Hamiltonian dynamics that give rise to the thermal equilibrium distribution $\Phi$ \eqref{eq:phibetaH}. We also observe that $\omega^{ab}\chempo*{b}$ is simply a special case of the term $[\pd*{b} V_{ab}] - V_{ab}\chempo*{b}$ in $v_a$~\eqref{eq:v_anti}, where the first term vanishes for constant $\omega$ \eqref{eq:symp form}. We also stress that the inverse symplectic form $\omega^{ab}$ is T even: Considering the case of $N$ particles in $d$ dimensions with canonical coordinates, the nonzero components are given by $\omega^{ab} = \qty{x_a,p_b} = \kron{a,b}$; under T, we have
\begin{equation}
    \omega^{ab} \, \to \, S \, ( \qty{x_a,-p_b} )^T \, S^{-1} \, = \, S \, \qty{x_a,p_b} \, S^{-1} \, = S \,\kron{a,b} \, S^{-1} \, = \, \kron{a,b} \, = \, \omega^{ab} \,, ~~
\end{equation}
and the generalization to additional degrees of freedom ($N\geq 1$) and spatial dimension $d \geq 1$ is straightforward.

\subsubsection{Symmetries and Noether's theorem}

We now consider how conservation laws of Hamiltonian systems manifest in the EFT formalism developed in Sec.~\ref{sec:EFT}. In particular, we translate the discussion of Sec.~\ref{sec:noether} for stochastic EFTs into the more conventional treatment of conservation laws in Hamiltonian mechanics. According to Eqs.~\eqref{eq:W therm} and \eqref{eq:thermalLa}, the Fokker-Planck generator \eqref{eq:W form} is
\begin{align}
    W \, = \, \pd{a} \qty{q_a, H} \, ,~~
\end{align}
and consequently, for a \emph{strong} symmetry with conserved quantity $F$, the symmetry condition in Eq.~\eqref{eq:noether} leads to
\begin{align}
    W \, \to \,  W+ \pdv{F}{q_a}\qty{q_a, H} \, = \, W + \qty{F, H} \, ,~~
\end{align}
from which we infer that, for the phase-space function $F$ to be conserved, the condition in %that $W$ be invariant under shifting $\pd*{a}$ by $[\pd*{a}F]$ 
Eq.~\eqref{eq:noether} is equivalent to the familiar requirement that $F$ must have a vanishing Poisson bracket with the Hamiltonian $H$, i.e., 
\begin{align}
\label{eq:Hamiltonian symmetry}
    \qty{F, H} \, = \, 0 \,. ~~
\end{align}
Additionally, in the absence of noise, strong  and weak symmetries are \emph{equivalent}.  Since $\mathcal{Q}_{ab} = \omega^{-1}_{ab}$ for dissipationless Hamiltonian mechanics, the analogy to Eq.~\eqref{eq:W F condition weak} for a ``weak'' symmetry $F$ is
\begin{align}
    0 \, = \, \pd{a} \omega^{ab}(\pd{b} + \chempo{b}) F \, =\,  (\pd{b} F) \omega^{ab}\chempo{b} \, = \, \beta \{F, H\} ~~~\implies ~~~\{F, H\} = 0 \, ,~~
\end{align}
which is precisely the same as the condition in Eq.~\eqref{eq:Hamiltonian symmetry} for a strong symmetry of dissipationless Hamiltonian dynamics. In such systems, the statement that $F$ is a conserved quantity is equivalent to the condition $\{F, H\} = 0$.

\subsubsection{Noncanonical symplectic forms}

The discussion above illustrates the utility of the EFT formalism we develop in Sec.~\ref{sec:EFT} for Hamiltonian systems where both time evolution and the stationary distribution $\Phi$ \eqref{eq:phibetaH} are generated by the Hamiltonian $H$. In fact, the EFT description of systems with ordinary, canonical phase-space variables is especially simple. However, the EFT construction also holds for \emph{nontrivial} Poisson brackets $\qty{q_a,q_b}$---or, equivalently, a nontrivial (inverse) symplectic form $\omega^{ab}_{\,}$ \eqref{eq:symp form}---which are not of the form, e.g., $\qty{x_i,p_j}=\kron{i,j}$. Such nontrivial Poisson brackets are realized by, e.g., systems in which the dynamics are most directly captured by angular-momentum variables $\vb{L}$, the components of which satisfy $\qty{L_i,L_j} = \epsilon_{ijk} \, L_k$. The implementation of the EFT formalism for such nontrivial Poisson brackets requires extra care; moreover, since the Poisson bracket is antisymmetric (i.e., $\qty{q_a,q_b} = -\qty{q_b,q_a}$, one must be careful regarding the ordering of variables in Eq.~\eqref{eq:thermalLa}. Additionally, when the symplectic form $\omega$ is a nontrivial function of the phase-space coordinates, its Pfaffian $\sqrt{\omega} \equiv \sqrt{\det \omega}$ (borrowing the convention of general relativity) is the prefactor of the volume form on phase space\footnote{The volume of a region $R$ of phase space is $\int_R \omega^{\wedge Nd} = \int_R \sqrt{\omega} \, \thed q_1\wedge \cdots \wedge \thed q_{2 N d}$.}. In such scenarios, the stationary distribution \eqref{eq:phibetaH} is modified according to
\begin{equation}
    \Phi \to \beta \, H - \log \sqrt{\omega} \, , ~~ \label{eq:Phiham}
\end{equation}
and the ``chemical potential'' \eqref{eq:Chemical Potential Def} changes to $\chempo*{a} = \beta \pd*{a} H - \pd*{a} \log\sqrt{\omega}$. However, we find that the $\log \sqrt{\omega}$ contribution to $\Phi$---which affects the chemical potential $\chempo*{a} = \pd{a}\Phi$---is precisely cancelled by another term, so that the Fokker-Planck generator remains in the form of Eq.~\eqref{eq:W_therm2} where $\chempo*{a} = \beta \pd{a}H$. To see this, note that
\begin{align}
    W \, &= \, T \, \pd{a} \, \omega^{ab}  \, (\pd{b} + \chempo{b} ) \, = \, T \, \left( \pd{a} \pd{b} \omega^{ab}- \pd{a} \left[ \pd{b} \, \omega^{ab} \right] \right)  + T\,  \pd{a} \omega^{ab} \, \left[ \pd{b} \, \Phi \right]  \notag \\
    &= \, \pd{a} \, \omega^{ab} \, \left[ \pd{b} \, H \right] - T \, \pd{a}\left(  \left[ \pd{b} \, \omega^{ab}\right]  +\omega^{ab} \left[ \pd{b} \log \sqrt{\omega} \right] \right) \notag \\
    &= \, \pd{a} \, \omega^{ab} \, \left[ \pd{b} \, H \right] - T \, \pd{a} \frac{1}{\sqrt{\omega}} \left[ \sqrt{\omega} \, \pd{b} \omega^{ab} + \omega^{ab} \,\frac{1}{2 \, \sqrt{ \omega}} \pd{b} \det\omega \right] \notag \\
    &= \, \pd{a} \, \omega^{ab} \, \left[ \pd{b} \, H \right] - T \, \pd{a} \frac{1}{\sqrt{\omega}}  \left[ \pd{b} \sqrt{\omega} \, \omega^{ab} \right] \,= \, \, \pd{a} \, \omega^{ab} \, \left[ \pd{b} \, H \right] \, ,~~\label{eq:W_therm2 nontriv}
\end{align}
where we used the fact that $\pd*{b} \left( \sqrt{\omega} \, \omega^{ab} \right) = 0$ because the symplectic form $\omega$ is closed. Note that Eq.~\eqref{eq:W_therm2 nontriv} precisely reproduces Eq.~\eqref{eq:W_therm2} if the contribution of the $\omega$ term in Eq.~\eqref{eq:Phiham} to $\chempo*{b} = \pd*{b} \Phi$ is ignored. Hence, we generically allow $\omega^{ab}$ to be nontrivial while taking $\chempo*{b} = \beta \, [\pd*{b} H]$. Thus, our EFT framework is compatible with nontrivial symplectic forms $\omega$, even when they depend nontrivially on the coordinates  $\vb{q}$.  Moreover, the T-odd contribution $v_a= -\omega^{ab}\pd{b} H$ to the force $f_a$ \eqref{eq:FPE force FDT} satisfies the constraint of Eq.~\eqref{eq:dissipationless_terms}, as required.

We make one final remark for readers familiar with the path-integral formulation of Sec.~\ref{sec:MSR}. Consider a T-invariant MSR Lagrangian \eqref{eq:LKMSeven} describing Hamiltonian dynamics; neglecting dissipation, one would write $L = \pi_a (\pd{t} q_a - \omega^{ab} \pd*{b} H)$. However, the MSR implementation of T  \eqref{eq:MSR T pi relation} is \emph{not} the same as the T operation in Eq.~\eqref{eq:W reversed} in the operator language, because $\chempo*{a} \neq \beta \pd*{a} H$, \emph{even in Hamiltonian mechanics without dissipation} when $\omega$ \eqref{eq:symp form} is not constant! The correct T-invariant path integral is far more complicated, and discussed in App.~\ref{app:PI}.

\subsection{Dissipation towards a thermal state}
\label{sec:whitenoise}

Having explained how noise-free Hamiltonian mechanics %may be described with our formalism, 
is captured by the EFT formalism of Sec.~\ref{sec:EFT}, 
we now demonstrate how to add dissipation (i.e., noise sources) to the dynamics in a way that preserves both the thermal stationary state $\Phi = \beta H$ \eqref{eq:phibetaH} toward which the system relaxes and any symmetries. While dissipative relaxation to thermal equilibrium is well understood, we find that such dynamics are especially transparent in the EFT formalism of Sec.~\ref{sec:EFT} using Fokker-Planck operators. A key benefit of  understanding dissipative Hamiltonian dynamics through the EFT lens is that this builds intuition for the application to \emph{active} systems in, e.g., Sec.~\ref{sec:active}.

A key observation of Sec.~\ref{subsec:Hamiltonian setup} is that the operator $W$ \eqref{eq:W therm} corresponding to noiseless Hamiltonian dynamics is antisymmetric, yet T even; at the same time, Sec.~\ref{sec:EFT} establishes that the symmetric part of $W$ \eqref{eq:W nice} corresponds to \emph{dissipationless} dynamics, and is generally T even. In this sense, the EFT formalism prescribes that adding \emph{dissipative} effects on top of the dynamics generated by Hamiltonian $H$---where the stationary distribution is given by $\Phi = \beta H$ \eqref{eq:phibetaH}---merely requires adding terms $W'$ to $W$ \eqref{eq:W therm} corresponding to \emph{noise}, which generally preserve both T invariance and the stationary state $\Phi$. The allowed additions to $W$ \eqref{eq:W therm} are of the form
\begin{align}
    W' \, = \, -\pd{a} Q_{ab} \qty(\pd{b} + \chempo{b}) \, ,~~ \label{eq:33diss}
\end{align}
where $Q_{ab}(\vb{q})$ \eqref{eq:nice Q def} is a symmetric matrix that may, in general, depend on the variables $\vb{q}$, and must be positive definite for thermodynamic stability, and respects any residual symmetries (conservation laws) as needed. 

The simplest such choice of $Q$---and the most common in the literature---is given by
\begin{equation} 
\label{eq:white noise Q}
    Q_{ab} \, = \, \Delta \, \kron{a,b} \, ,~~
\end{equation}
for some constant $\Delta>0$, which corresponds to additive Gaussian noise in Eq.~\eqref{eq:Q noise def}. In this case, the Langevin equation \eqref{eq:SDE main} may be interpreted in the It\^o formalism as
\begin{equation}
\label{eq:33eom}
    \pd{t} q_a \, = \, \qty{ q_a, H} - \gamma \pd{a} H + \xi_a(t) \, ,~~ 
\end{equation}
where $\gamma$ is defined in terms of the noise-variance matrix $Q$ \eqref{eq:white noise Q} via
\begin{equation}
\label{eq:33noise}
    \expval{\xi_a(t) \xi_b(t')} \, = \, Q_{ab} \DiracDelta{t-t'} \, = \, 2\gamma T \kron{a,b} \DiracDelta{t-t'} \, , ~~
\end{equation}
with $T=1/\beta$ and $\gamma = \beta \Delta / 2$. The relation between the noise strength $\Delta$ in Eq.~\eqref{eq:33noise} and the damping term in $\gamma$ in Eq.~\eqref{eq:33eom} is the \emph{classical fluctuation-dissipation theorem at finite temperature $T$}.  This celebrated result recovers immediately and transparently upon applying the EFT framework of Sec.~\ref{sec:EFT} to Hamiltonian dynamics with dissipation.

The foregoing results also extend to systems with \emph{multiplicative} noise \eqref{eq:multiplicative noise matrix}. We caution that such noise sources require care when dealing with the ordering of operators in, e.g., the Fokker-Planck equation \eqref{eq:FPE main}. Obtaining a fluctuation-dissipation theorem \eqref{eq:33noise}---even for thermal stationary states $\Phi = \beta H $ \eqref{eq:phibetaH}---requires that one first write down the dissipative correction $W'$ \eqref{eq:33diss} in the form of Fokker-Planck generator $W$ \eqref{eq:W nice}, and then, if desired, rewrites that same generator $W'$ in the It\^o (or other preferred) formulation of stochastic calculus (see also Sec.~\ref{sec:KMSvsT}).

\subsection{Classical harmonic oscillator}\label{sec:HO}

We now consider several pedagogical examples of classical dynamical systems in the presence of dissipation, using the EFT formalism of Sec.~\ref{sec:EFT}. In particular, we showcase how that formalism streamlines the derivation of several well-known models of dissipative dynamics (an apparently novel example appears in Sec.~\ref{sec:rigidbody}).

We first apply the EFT formalism described above to one of the simplest classical systems imaginable---the harmonic oscillator in $d$ spatial dimensions with frequency $\Omega=\sqrt{\kappa/m}$. This system is described by the canonical phase-space coordinates $\vb{q}=(x_1,\dots,x_d,p_1,\dots,p_d)$, and we couple this system to a general thermal bath, which induces dissipation. For convenience, we define the differential operators $\pd*{a} \equiv \pdv*{}{x_a}$ and $D_a \equiv \pdv*{}{p_a}$, and let $a \in \{1,2,\dots,d\}$.

\subsubsection{Overdamped limit}
\label{sec:HOoverdamped}
We first consider the overdamped limit, which corresponds to the model discussed in Sec.~\ref{sec:introT} of the introduction. In this limit, the kinetic energy of the oscillator is negligible, so the stationary distribution \eqref{eq:phibetaH} can be written
\begin{equation}
\label{eq:H0Over Phi}
    \Phi (\vb{q}) \, \to \, \beta V(\vb*{x}) \, = \, \frac{1}{2} \beta m \Omega^2 x^2_a \, , ~~
\end{equation}
where summation over $a$ is implicit, and for clarity, we assume an isotropic potential. Restricting to Gaussian white noise \eqref{eq:white noise Q}, the generator $W$ \eqref{eq:W therm} corresponding to T-even relaxation to $\Phi$ \eqref{eq:H0Over Phi} is given by
\begin{equation}
    W \, =\,  -T\gamma'\, \pd{a} \left(\pd{a} + \chempo{a}\right) \, ,~~
\end{equation}
which can be straightforwardly recast as an SDE \eqref{eq:SDE main} in the  It\^{o} prescription,
\begin{align}\label{eq:HOlangevin}
    \pd{t} x_a \, = \, - \gamma' m\Omega^2 x_a  +\xi_a(t) \, ,~~
\end{align}
where the noise term $\xi_a (t)$ obeys Eq.~\eqref{eq:white noise Q} and the FDT \eqref{eq:33noise}, and $\gamma' = \beta \Delta/2$ with $\Delta$ the noise strength. 
The result is the natural generalization of the overdamped harmonic oscillator previewed in Sec.~\ref{sec:intro}.

\subsubsection{Finite-mass limit}
\label{sec:SHO finite mass}
We next consider the limit of one or more $d$-dimensional oscillators in which the kinetic energy is no longer negligible (the overdamped limit can be viewed as taking $m \to \infty$). The corresponding stationary distribution $\Phi$ \eqref{eq:phibetaH} is
\begin{equation}\label{eq:HO phi}
    \Phi \, = \, \beta \left[\frac{p^2_a}{2m}+\frac{1}{2} m\Omega^2 x^2_a \right] \, , ~~
\end{equation} 
where summation over $a$ is implied, the coordinates satisfy the Poisson brackets $\qty{x_a,p_b} = \kron{a,b}$, and we define 
\begin{equation}\label{eq:HO var def}
    \chempo{a} \, \equiv \, \pdv{\Phi}{x^{\,}_a}\,  = \, \beta m \Omega^2 x_a~~~\text{and}~~~\nu^{\vpp}_a \, \equiv \, \pdv{\Phi}{p_a} \, = \, \beta p_a/m \, , ~~
\end{equation}
as the two flavors of chemical potentials. Assuming that only the force driving the oscillator is random (i.e., that noise only couples directly to the momentum $\vb*{p}$),
the Fokker-Planck generator \eqref{eq:W therm} is given by
\begin{equation}
\label{eq:SHO finite m W}
    W \, = \, T\left[\pd{a} \nu^{\vpp}_a - D_a \chempo{a}  \right] - T\,\gamma\, D_a \left(D_a +\nu^{\vpp}_a\right) \, , ~~
\end{equation}
and we note that $W$ \eqref{eq:SHO finite m W} is T symmetric and of the form of Eq.~\eqref{eq:W nice}.
% \begin{subequations}
%     \begin{align}
%         x_a(t) &\rightarrow x_a(-t), \\
%         p_a(t) &\rightarrow -p_a(-t), \\
%         \pi_a(t) &\rightarrow -\pi_a(-t) + \ii\mu_a(-t), \\
%         \sigma_a(t) &\rightarrow \sigma_a(t) - \ii\nu_a(-t).
%     \end{align}
% \end{subequations}
The equations of motion for the positions and momenta recover straightforwardly from $W$ \eqref{eq:SHO finite m W}; in the It\^o SDE form, they are
\begin{align}
\label{eq:finite-mass SHO EOM}
    \pd{t} x_a \, = \, \frac{p_a}{m} ~~~~\text{and}~~~~
    \pd{t} p_a \, = \, - m\Omega^2 x_a - \frac{\gamma}{m} p_a + \xi_a(t) \, ,~~
\end{align}
where the noise $\xi_a$ obeys Eq.~\eqref{eq:33noise}, and we note that $\gamma$ has different units than in the overdamped case in Sec.~\ref{sec:HOoverdamped}.

\subsection{Vortex dynamics in superfluid thin films}\label{sec:vortex}

We now consider a nontrivial example corresponding to the dynamics of point vortices in superfluid thin films \cite{Onsager1949, Newton_2011, pinesTheoryQuantumLiquids1998}.  Such systems are described by Hamiltonian mechanics in which the dynamical variables are the positions $\vb{r}_i = (x_i,y_i)$ of the vortices, which have quantized circulations $\Gamma_i = n_i \hbar / m$, and the coordinates $x_i$ and $y_i$ are \emph{canonically conjugate} \cite{reeves2022, Newton_2011}. The kinetic energy of the vortices is given in terms of the coordinates $\vb{r}_i$ via
\begin{align}
\label{eq:vortex Ham}
    H \, = -\sum_{i < j} \frac{\rho^{\,}_0 \Gamma_i \Gamma_j}{4\pi} \log (\abs{ \vb{r}_i-\vb{r}_j } / R)
\end{align}
where $\rho^{\,}_0$ is the mass density of the 2D superfluid and $R$ is an arbitrary constant with units of length. We note that the Hamiltonian \eqref{eq:vortex Ham} has the form of the Coulomb interaction in two spatial dimensions.  

The equations of motion follow from the following Poisson bracket for the canonical coordinates $x_i$ and $y_i$,
\begin{align}
    \{x_i, y_j\} \, =\,  \Gamma_i^{-1} \kron{i,j} \, ,~~
\end{align}
and writing $\vb{q}=(x_1,x_2,\dots, x_N, y_1, y_2, \dots , y_N)$, the canonical dissipationless equations of motion are
\begin{equation}
    \pd{t} q_a \, = \, \omega^{ab} \chempo{b} \, \equiv \, u_a \, , ~~ \label{eq:muaua}
\end{equation}
where $\omega$ is the inverse symplectic form \eqref{eq:symp form} coupling $x$ and $y$ coordinates and $u_a$ corresponds to the $x$ or $y$ component of the superfluid velocity field at vortex $i$'s location.  This leads to the well known Hamiltonian equations of motion
\begin{equation}
\label{eq:vortex Hamilton EOM}
    \Gamma_i \dot{x}_i \, = \, \pdv{H}{y_i} ~~~~ {\rm and} ~~~~
    \Gamma_i \dot{y}_i \, = \, -\pdv{H}{x_i} \, ,~~
\end{equation}
where there is no sum over $i$. We now introduce dissipation, so that the noisy equations take the form
\begin{align}
    \dot{q}_a \, = \, u_a + G_a + \xi_a(t) \, ,~~
\end{align}
where $G_a$ is the dissipative part of the force $f_a$ \eqref{eq:FPE force FDT}, and we take $\xi_a (t)$ to be Gaussian white noise with strength $\Delta$ \eqref{eq:white noise Q}. Then the only dissipative term $G_a$ we may include is of the form
\begin{align}
\label{eq:vortex dissip force}
    G_a \, = -\frac{1}{2} Q_{ab}\chempo{b} \, = - \Delta \chempo{a} \, = -\beta \Delta \Gamma_a \epsilon^{ab} u_b \, ,~~
\end{align}
where there is no sum over the label $a$, and $\mu_a$ and $u_a$ are defined in (\ref{eq:muaua}). We next define the \emph{mutual friction coefficient} $\gamma = \beta \Delta$. Including the dissipative correction $G_a$ \eqref{eq:vortex dissip force}, the equations of motion \eqref{eq:vortex Hamilton EOM} for the coordinate $\vb{r}_i = (x_i, y_i)$ becomes
\begin{equation} \label{eq:vortexdiss}
    \dot{\vb{r}}_i \, = \, \vb{u}_i -\gamma \,  \vu{z} \times  \Gamma_i  \, \vb{u}_i + \xi_i (t) \, , ~~
\end{equation}
where $\vb{u}_i$ is the superfluid velocity at the point $\vb{r}_i$. The second term above corresponds to the \emph{mutual friction} that is common to the literature on point-vortex models~\cite{ambegaokar}. 

It is illustrative to compare the derivation above to %the one which was presented in the classic paper
that of Ref.~\citenum{ambegaokar}, which argued %directly 
that the equations of motion for a point vortex should be local in the superfluid velocity $\vb{u}_i$ and reduce to $\dot{\vb{r}}_i = \vb{u}_i$ in the absence of dissipation. The simplest such equation of motion is Eq.~\eqref{eq:vortexdiss}, where the dependence on $\Gamma^{\,}_i$ in the dissipative term is required  for the two terms in Eq.~\eqref{eq:vortexdiss} to transform equivalently under parity.  While the derivation in  Ref.~\citenum{ambegaokar} is quicker than the one above, it has two potential sources of confusion. First, the correct choice of sign $\gamma>0$ in Eq.~\eqref{eq:vortexdiss} follows straightforwardly from positive semi-definiteness of  $Q$ \eqref{eq:Q noise def}, but it is not obvious  starting from Eq.~\eqref{eq:vortexdiss} why we must have $\gamma>0$. Second, in the presence of additive Gaussian noise, it is not clear from Eq.~\eqref{eq:vortexdiss} alone how the resulting dynamics are consistent with a thermal distribution at finite temperature $T$ with the Hamiltonian \eqref{eq:vortex Ham}. However, both of these properties are built into the stochastic EFT formalism of Sec.~\ref{sec:EFT}.

In principle, our framework could also apply to the more complicated mutual friction arising in the dynamics of three-dimensional vortex rings, using the field-theoretic generalization of our EFT framework described in Sec.~\ref{sec:continuum}, or to constrained vortex dynamics \cite{Qi:2023zfk}.

\subsection{Landau-Lifshitz-Gilbert spin dynamics}\label{sec:spinchain}

Another interesting application of our formalism is to the classical Heisenberg spin chain \cite{Heisenberg1928, girvin_yang_2019, altland_simons_2010}. This model also provides a nontrivial application of the generalized Noether's theorem \eqref{eq:noether} in Sec.~\ref{sec:noether}. For simplicity, consider a single classical spin $\vb{S}$ interacting with a magnetic field $\vb{B}$ in $d=3$ spatial dimensions; the corresponding Hamiltonian is
\begin{align}
\label{eq:LLG H}
    H \, = \, \gamma \, S_i \, B_i \, ,~~
\end{align}
where $\gamma= g \chempo*{B} / \hbar$ is the gyromagnetic ratio of the spin, with $g$ the Land\'e $g$-factor for an electron  and $\chempo*{B} = e \hbar / 2 m_e$ the Bohr magneton. Again, we take $\Phi = \beta H$ \eqref{eq:phibetaH}, and note that each spin component $S_i$ corresponds to the chemical potential $\chempo*{i}= \beta \pdv*{H}{S_i} = \beta \, \gamma B_i$ \eqref{eq:Chemical Potential Def}. Additionally, the spin components satisfy the nontrivial Poisson brackets
\begin{align}
\label{eq:spin Poisson}
    \qty{S_i,S_j} \, = \, \epsilon^{ijk}\,  S_k \, ,~~
\end{align}
so that dissipationless (Hamiltonian) evolution is generated by the Fokker-Planck operator %\eqref{eq:W nice} (in particular the form )
\eqref{eq:W therm},
\begin{equation}
\label{eq:LLG W from H}
    %W = T \pd{i}  \qty{S^{\vpp}_i,S^{\vpp}_j} \, \chempo{j} = - \gamma \, \pd{i}  \epsilon^{ijk}_{\vphantom{j}} \,  S^{\vpp}_k \, B^{\vpp}_j
   W \, = -\gamma \pd{i} \epsilon^{ijk}\, S_j \, B_k \, , ~~
\end{equation}
where, as a reminder, $W$ is a differential operator, so  $\pd*{i}$ continues to act to the right of $W$.

We now add dissipation to the dynamics generated by $W$ \eqref{eq:LLG W from H} while preserving a notion of time-reversal invariance. Although $W$ \eqref{eq:LLG W from H} itself is invariant under T, we only require that the \emph{dissipative} terms $W'$ we add thereto preserve a \emph{generalized} $\Ints^{\,}_2$ symmetry gT corresponding to the \emph{combination} of time reversal $\cal T$ (or T) and parity $\cal P$, as one expects of dynamics in the presence of a magnetic field $\vb{B}$. We note that the gT transformation $\cal PT$ is itself a  symmetry of the original $W$ \eqref{eq:LLG W from H}. The simplest dissipative correction $W'$ that respects this choice of gT is %Lagrangian must be:
\begin{align}
    W' \, = \, \pd{i} Q_{ij} (\pd{j} + \chempo{j}) \, ,~~
\end{align}
where $Q_{ij}$ \eqref{eq:Q noise def} is the (symmetric) noise matrix. However, we must further impose the conservation of the norm of the spin %, so that $M = S_iS_i$ is a conserved quantity.
$M = \vb{S} \cdot \vb{S} =S_i S_i$. Following  Eq.~\eqref{eq:noether} of Sec.~\ref{sec:noether}, we find that $W$ must %then 
be invariant  under the transformation
\begin{align}
    \pd{i} \, \to\,  \pd{i} + \pdv{M}{S_i} \, = \, \pd{i} + 2S_i \, ,~~
\end{align}
%This condition 
which is \emph{not} satisfied by a constant matrix $Q_{ij}$. However, this can be corrected by noticing that $\epsilon^{ijk} \pd{j} S_k$  is invariant under the shift $\pd*{i} \to \pd*{i} + S_i$. Hence, we should build the matrix $Q_{ij}$ \eqref{eq:Q noise def} out of these invariant objects. We then find that the simplest dissipative term that may be added to $W$ \eqref{eq:LLG W from H} corresponds to 
\begin{align}\label{eq:dissipative_spin}
    W' \, = -\Delta \, \pd{i} \epsilon^{ijk} \,  S_j \epsilon^{klm} \, S_l (\pd{m} + \chempo{m}) \, , ~~
\end{align}
for some constant $\Delta>0$ such that $Q_{im}$ is positive semidefinite. To gain some intuition, consider the contribution of the $\Delta$ term in Eq.~\eqref{eq:dissipative_spin} to the equation of motion for the spin,
\begin{align}
\label{eq:dissipative spin EOM 1}
    \pd{t} S_i \, = \, \gamma \, \epsilon^{ijk} \, S_j \, B_k + \Delta \, \epsilon^{ijk} \, \epsilon^{klm}  \, S_j S_l \, \chempo{m} + \eta_i(\vb{S},t) \, ,~~
\end{align}
where $\eta_i$ is a multiplicative noise source with variance matrix  \eqref{eq:dissipative_spin}
\begin{align}
\label{eq:dissip spin noise Q}
    Q_{ij}(\vb{S}) \, = \, \Delta \, \epsilon^{imk}\, S_j\, \epsilon^{klj}\, S_l \, ,~~
\end{align}
which is somewhat complicated, but required for$\abs{\vb{S}}^2$ to be conserved as a strong symmetry (i.e., for \emph{every} realization of the noise). Finally, we define  $\lambda = \beta \, \Delta \, \gamma$ and recast Eq.~\eqref{eq:dissipative spin EOM 1} into the vector form
\begin{align}\label{eq:llg}
    \pd{t} \vb{S} \, = \, -\gamma \vb{S} \times \vb{B} - \lambda \, \vb{S} \times (\vb{S} \times \vb{B}) + \vb*{\xi}(\vb{S}, t) \, ,~~
\end{align}
known as the the \emph{Landau-Lifshitz} equation, where the term with coefficient $\lambda$ is the \emph{Landau-Lifshitz-Gilbert} damping term \cite{landau1935theory, gilbert1955phenomenological}. Hence, this celebrated nonlinear phenomenological model follows straightforwardly from the EFT ``cookbook" principles outlined in Sec.~\ref{sec:EFT} using only knowledge of the stationary state \eqref{eq:LLG H} and symmetries.

Again, it is instructive to compare our derivation with the usual approach starting from the phenomenological equation of motion \eqref{eq:llg}. One would first note that $\pd*{t} \vb{S}\cdot\vb{S}=0$ is guaranteed if
\begin{equation}
    \pd{t} \vb{S} \, = \, \vb{S} \times \vb{g} \,
\end{equation}
for some vector $\mathbf{g}$,  which may include both determinstic and stochastic contributions.  For the system to be in equilibrium when $\vb{S}$ is parallel to $\vb{B}$, we must choose $\vb{g}$ such that $\vb{S}\times \vb{g}$ vanishes when $\vb{S}\propto \vb{B}$ are parallel. Writing 
\begin{equation}
    \vb{g} \, = \, \gamma \vb{B} + \lambda \vb{S}\times \vb{B} + \cdots \, ,~~
\end{equation}  
reproduces Eq.~\eqref{eq:llg}. The FDT \eqref{eq:33noise} can then be recovered by linearizing Eq.~\eqref{eq:llg} around a configuration where $\vb{S} \propto \vb{B}$ are parallel. In our view, the advantage of the Wilsonian EFT framework for stochastic systems that we develop in Sec.~\ref{sec:EFT} is that it can be applied more \emph{algorithmically}.  Using more traditional methods instead requires starting with a phenomenological equation of motion and then working through a sequence of physical conditions and assumptions to constrain Eq.~\eqref{eq:llg}.  In the EFT formalism, one merely specifies a stationary distribution \eqref{eq:LLG H} and a set of symmetries, from which all allowed terms recover using the prescription of Sec.~\ref{sec:EFT}.

\section{From thermal to active matter}
\label{sec:active}

We now consider additional examples and applications of the EFT formalism presented in Sec.~\ref{sec:EFT}, with an eye toward the description of \emph{active} matter. In particular, we consider systems for which the dynamics does \emph{not} result from Hamiltonian mechanics, in contrast to the examples of Sec.~\ref{sec:not active}

\subsection{An ``odd'' two-dimensional harmonic oscillator}
\label{sec:Odd 2d HO}

The simple harmonic oscillator is a useful toy model of equilibrium (thermal) physics, and is thus a useful starting point for examining physics \emph{near} equilibrium. We note that, in $d=1$, there is only one nontrivial choice of the dissipationless term $V$ in Eq.~\eqref{eq:nice Q def}; hence, we look for simple examples of non-Hamiltonian dynamics in a two-dimensional oscillator. While the standard approach in studying nonthermal systems is to start by writing down an equation of notion, we note that there is no guarantee of either a stationary state nor stable solutions. Instead, we use the EFT framework developed in Sec.~\ref{sec:EFT} and start by specifying the stationary distribution $\Phi$ \eqref{eq:FPE stationary distribution}, which guarantees stability, and derive constraints on the equations of motion compatible with $\Phi$.

As a reminder, a harmonic oscillator (in any number of spatial dimensions) that relaxes to the equilibrium stationary state $\Phi=\beta H$ \eqref{eq:phibetaH} with $H = (\vb{p}^2 + m^2 \Omega^2 \vb{x}^2)/2m$ \eqref{eq:HO phi} is captured by the Fokker-Planck generator
\begin{equation*}
\tag{\ref{eq:SHO finite m W}}
    W \, = \, T\left[\pd{a} \nu^{\vpp}_a - D_a \chempo{a}  \right] - T\gamma D_a \left(D_a +\nu^{\vpp}_a\right) \, , ~~
\end{equation*}
%where $(\pd*{a},D_a)=(\pdv*{}{x^{\,}_a},\pdv*{}{p^{\,}_a})$, $\nu^{\,}_a = \pdv*{\Phi}{p^{\,}_a}$  \eqref{eq:HO var def}, and $\gamma = \beta \Delta/2$ captures the noise strength. 
and there are a number of terms that one could add to the generator $W$ \eqref{eq:SHO finite m W} to make the system appear ``active.'' 

Here, we consider adding terms such that the equation of motion for $\pd*{t} x_a$ \eqref{eq:finite-mass SHO EOM} is a function not only of the momentum $p_a$, but the position $x_a$. We also add correlations in the noise sources for $x_a$ and $p_a$ in $Q$ \eqref{eq:Q noise def}. We add to the thermal (dissipative) Fokker-Planck generator $W$ \eqref{eq:SHO finite m W} the following ``active'' terms,
\begin{align}
\label{eq:W SHO active terms}
    W' \, &= \, - T\, \lambda \, \pd{a}(\pd{a}+ \chempo{a})- T\, \kappa \, \epsilon^{ab}  \, (2 \, \pd{a} D_b + \pd{a} \nu^{\vpp}_b - D_a \chempo{b})  -  V \, \epsilon^{ab}\,  \pd{a} \chempo{b} - V' \epsilon^{ab}\,  D_a \nu^{\vpp}_b \, ,~~
\end{align}
where the parameters $\gamma$, $\lambda$, and $\kappa$ capture correlations of noise sources $\vb*{\xi}$ and $\vb*{\zeta}$ via
\begin{subequations}
\label{eq:odd HO noise corr}
\begin{align}
    \expval{ \zeta_a(t) \zeta_b(t') } \, &= \, 2 \, T \, \gamma \, \kron{a,b}\, \DiracDelta{t-t'} \, \\
    \expval{ \xi_a(t) \xi_b(t') } \,  &= \, 2\, T\, \lambda\,  \kron{a,b}\, \DiracDelta{t-t'} \\
    \expval{ \xi_a(t)\zeta_b(t') } \, &= \, 2\, T\,  \kappa \, \epsilon^{\,}_{a,b} \, \DiracDelta{t-t'}\, , ~~
\end{align}
\end{subequations}
and since the total variance matrix $Q$ \eqref{eq:Q noise def} must be positive semidefinite, we must have 
\begin{align}\label{eq:constraintHO}
    \lambda \, > \, 0,~~\gamma \, > \, 0,~~~{\rm and}~~~\gamma\lambda \, > \, \kappa^2 \, ,~~
\end{align}
and we note that the couplings $V$ and $V'$ are  not constrained. We also note that, \emph{a priori}, starting from the equation of motion \eqref{eq:eomoddHO} \emph{without} noise, the constraint of Eq.~\eqref{eq:constraintHO} is ambiguous. 

The Fokker-Planck generator $W+W'$ gives rise to the following SDE \eqref{eq:SDE main} in the It\^o prescription,
\begin{subequations}
\label{eq:eomoddHO}
\begin{align}
    m \, \pd{t} x_a \, &= \, p_a -\lambda \, m^2 \, \Omega^2 \, x_a -  \beta\, m^2 \, \Omega^2 \, V \epsilon^{ab} \,  x_b-\kappa \, \epsilon^{ab}\, p_b+\xi_a(t) \label{eq:eomoddHO X} \\
    \pd{t} p_a \, &= \, -m\, \Omega^2 \, x_a- \frac{\gamma}{m} \, p_a - \frac{\beta}{m} \, V' \, \epsilon^{ab}\, p_b+ \kappa \, m \, \Omega^2\, \epsilon^{ab} \, x_b+\zeta_a(t) \, ,~~\label{eq:eomoddHO P}
\end{align}
\end{subequations}
which describes a system with a \emph{generalized} time-reversal symmetry gT corresponding to $\cal P  T$, where the parity operation $\cal P$ sends $\epsilon_{ab}\to -\epsilon_{ab}$.  The generator $W'$ \eqref{eq:W SHO active terms} includes T-odd terms (with coefficients $V$ and $V'$) that lead to \emph{dissipationless} rotations in position and momentum space. Notice that it is possible to obtain both $V,V'$ terms  in \emph{thermal} systems in which a modified Poisson bracket due to the presence of a magnetic field and Berry curvature, which correspond to $V'$ and $V$, respectively \cite{niu_review}. 
% ---they mix with the standard terms proportional to $\kappa$ in \eqref{eq:eomoddHO X}. 
On the other hand, the \emph{dissipative} T-odd term proportional to $\kappa$ in Eq.~\eqref{eq:eomoddHO P} realizes the ``odd'' force that has recently been investigated in active systems  \cite{VitelliRobot}.

To map the model described above to that of \cite{VitelliRobot}, we set $V=V'=0$ for simplicity, so that Eq.~\eqref{eq:eomoddHO} takes the form of the following second-order differential equation for $x_a$,
\begin{align}
\label{eq:odd HO robot EOM}
   \big(\pd{t} + \frac{\gamma}{m}\big)^2 x_a \, = -\big(1 - \kappa^2 \big)\, \Omega^2 \, x_a  - \big(\lambda \, m \, \Omega^2 - \frac{\gamma}{m}\big) \big(\pd{t} + \frac{\gamma}{m}\big) \, x_a +2\, \kappa \, \Omega^2 \, \epsilon^{ab} \, x_b \, ,~~
\end{align}
where the final term (proportional to $\kappa$) realizes the ``odd'' force, as in Ref.~\citenum{VitelliRobot}. %We note that positivity of the noise matrix \eqref{eq:odd HO noise corr} constrains the magnitude $\kappa$ of this odd force to obey $\kappa^2<\gamma\lambda$ \eqref{eq:constraintHO}. 
The normal modes %for the equation of motion \eqref{eq:odd HO robot EOM} correspond to
are given by
\begin{align}
    \omega \, = \, -\ii \frac{\lambda\, m^2 \, \Omega^2 + \gamma}{2\, m} \pm \frac{1}{2\, m} \, \sqrt{4(1-\kappa^2) \, m^2 \, \Omega^2 - (\lambda \, m^2 \, \Omega^2 - \gamma)^2  \pm \ii \, 8\, \kappa\, m^2 \, \Omega^2} \, ,~~
\end{align}
whose imaginary part is given by
\begin{align}
    \im \omega \, = \, - \frac{\lambda \, m^2 \,  \Omega^2 + \gamma}{2 \, m} +\frac{1}{2\, m} \, \re \, \sqrt{(\lambda m^2 \Omega^2 - \gamma)^2 - 4(1-\kappa^2)\, m^2 \, \Omega^2 + \ii\,  8 \, \kappa\, m^2 \, \Omega^2}
\end{align}
and is a monotonically increasing function of $\kappa$; by the constraint of Eq.~\eqref{eq:constraintHO}, we have that
\begin{align}
    \im \omega \, \leq \, - \frac{\lambda \,  m^2   \, \Omega^2 + \gamma}{2 \, m} + \frac{1}{2 \, m} \, \re \, \sqrt{(\lambda \,  m^2 \, \Omega^2 + \gamma)^2 - 4\, m^2 \, \Omega^2 + \ii \, 8 \, \sqrt{\gamma\lambda} \, m^2 \, \Omega^2} \, ,~~
\end{align}
where the right-hand side is maximized when $\lambda \, m^2 \, \Omega^2 = \gamma$, leading to
\begin{align}
    \im\omega \, \leq \, -\lambda \, m \, \Omega^2 + \re \, \sqrt{(\lambda m\, \Omega^2 +\ii \Omega)^2} \, = \, 0 \, ,~~
\end{align}
so that the constraints of Eq.~\eqref{eq:constraintHO} imposed by positivity of the noise-variance matrix $Q$ \eqref{eq:odd HO noise corr} imply \emph{stability} of the normal modes (i.e., $\im \omega \leq 0$). The dynamics described by Eq.~\eqref{eq:odd HO robot EOM} correspond to the decaying mode in Ref.~\citenum{VitelliRobot}.

We next define the dimensionless parameter
\begin{equation}
    \label{eq:odd HO robot parameter}
    \theta \, \equiv \, 2\, m \, \kappa \, \Omega/(\lambda \, m^2 \, \Omega^2 + \gamma) \, ,~~
\end{equation}
which diagnoses whether Eq.~\eqref{eq:odd HO robot EOM} realizes decaying motion ($\abs{\theta}\leq 1$) or limit-cycle motion ($\abs{\theta}>1$) in Ref.~\citenum{VitelliRobot}. When $\abs{\theta}\leq 1$, the dissipation dominates, so that the system relaxes to a stationary configuration; when $\abs{\theta} > 1$, the combination of dissipation and activity lead to nonlinear dynamics, resulting in periodic, limit-cycle behavior.  Crucially, we find that the constraint $\kappa^2 \leq \gamma \, \lambda$ \eqref{eq:constraintHO} from positive semidefiniteness of $Q$ \eqref{eq:odd HO noise corr} implies that 
\begin{align}
    \theta^2 \, \leq \, 1 - \left(\frac{\lambda \, m^2 \, \Omega^2 - \gamma}{\lambda \, m^2 \, \Omega^2 + \gamma}\right)^2 \, \leq 1 \, ,~~
\end{align}
which further implies that $\abs{\theta} \leq 1$ for the equation of motion \eqref{eq:odd HO robot EOM} for valid sources of noise \eqref{eq:odd HO noise corr}, which must obey Eq.~\eqref{eq:constraintHO}. In other words, the EFT perspective naturally gives rise \emph{only} to the decaying motion observed in Ref.~\citenum{VitelliRobot}. Realizing the limit-cycle motion of Ref.~\citenum{VitelliRobot} in an EFT requires identifying a stationary distribution $\Phi$ other than that of Eq.~\eqref{eq:HO phi}; in particular, that $\Phi$ must contain cubic (or higher) terms, so that Eqs.~\eqref{eq:eomoddHO} are \emph{nonlinear}.

\subsection{Nonreciprocal Kuramoto model}\label{sec:kuramoto}

For a more interesting realization of \emph{active} matter, we now consider the \emph{nonreciprocal Kuramoto model}~\cite{nonreciprocalPhase}, corresponding to classical spins on a lattice coupled via ``nonreciprocal'' interactions. This model has a rich phenomenology and phase structure, and even hosts a spinning chiral phase that spontaneously breaks a \emph{generalized} time-reversal symmetry gT~\cite{nonreciprocalPhase}. Using the EFT framework developed in Sec.~\ref{sec:EFT}, we first review the equilibrium Kuramoto model of \emph{synchronization}~\cite{Kuramoto_RMP} before incorporating nonreciprocal interactions to realize an \emph{active} analogue~\cite{nonreciprocalPhase}.  Indeed, it is this section precisely which motivates the schematic summary of our approach in Fig.~\ref{fig:summaryofapproach}. %%% AARON: never put parantheses around \ref...use \eqref{} for equations and \ref for everything else. 

\subsubsection{The equilibrium (reciprocal) Kuramoto model}
\label{sec:eq kuramoto}
The phenomenon of \emph{synchronization} appears %arises, manifests, etc.
in various areas of biology, chemistry, and even the social sciences, and is distinguished % characterized, heralded, indicated, signalled, marked, identified, etc. hallmark,
by the alignment of the periodic phase variables describing a large number of agents. A simple model that realizes this phenomenon is the noisy, equilibrium \emph{Kuramoto model} \cite{kuramoto1975, kuramoto1984chemical, strogatz2000kuramoto, Kuramoto_RMP}: The synchronizing agents correspond to stationary rotors described by an angle $\theta^{\,}_a$. We now consider the simple implementation of the Kuramoto model in which the angles $\{\theta^{\,}_a\}$ are described by a nonlinear SDE \eqref{eq:SDE main} of the form
\begin{equation}
\label{eq:eq Kuramoto SDE}
    \pd{t} \theta_a \, = \, \sum\limits_{a,b} J_{ab} \sin(\theta_a - \theta_b) + \xi_a (t) \, ,~~
\end{equation}
where $J$ is a matrix of coupling strengths $J_{ab}$ between distinct rotors $\theta_a$ and $\theta_b$, and $\vb*{\xi} (t)$ is a noise source. 

The \emph{equilibrium} variant of the stochastic model in Eq.~\eqref{eq:eq Kuramoto SDE} has been studied extensively in the literature, and corresponds to a \emph{symmetric} coupling matrix $J_{ab}=J_{ba}$. This model hosts two stationary phases, analogous to those of the classical XY model: A symmetry-breaking ``synchronized'' phase at weak noise $\Delta < \Delta_c \sim 1/N$, where $N$ is the number of rotors (analogous to a low-temperature ferromagnet for $\beta < \beta_c$), and a ``disordered'' phase at strong noise $\Delta > \Delta_c$ (analogous to a high-temperature paramagnet for $\beta > \beta_c$)~\cite{strogatz2000kuramoto}. We now review the phenomenology of the equilibrium Kuramoto model from the perspective of the EFT formalism developed in Sec.~\ref{sec:EFT}.

%To begin, we should imagine a stationary $\Phi$ consistent with the phenomenology of the Kuramoto model.  In analogy with the Hamiltonian for interacting spins on a lattice, the natural guess is:

The starting point of the EFT is the stationary distribution $\Phi$; a natural guess is the XY-like ``Hamiltonian,''
\begin{align}
\label{eq:stationaryKuramoto}
    \Phi \, = -\frac{1}{2\, \Delta} \sum\limits_{a,b=1}^N J_{ab} \,  \cos(\theta_a - \theta_b) \, ,~~
\end{align}
where $J$ is a symmetric matrix and $\Delta$ captures the noise strength. This is the simplest form of $\Phi$ consistent with rotational invariance, and is essentially a classical $2d$ XY model .  For convenience, we take the noise  $\vb*{\xi}$ to be additive and Gaussian so that $Q_{ab} = \Delta \kron{a,b}$ \eqref{eq:white noise Q}.
%we have explicitly introduced the noise strength $\Delta$ (the reason why will become clear---it is to make an analogy with $k_B T$ in a thermal system!). 
The chemical potential $\chempo*{a}$ corresponding to $\theta^{\,}_a$ is
\begin{align}
\label{eq:eq Kuramoto chempo}
    \chempo{a} \, = \, \frac{1}{\Delta}\sum\limits_{b=1}^N J_{ab}\, \sin(\theta_a - \theta_b) \, ,~~
\end{align}
where the factor of $1/2$ is cancelled by the sums over all $a,b$ in Eq.~\eqref{eq:stationaryKuramoto}. We then recover the Fokker-Planck generator $W$ \eqref{eq:W form}, which we choose to be invariant (i.e., even) under T \eqref{eq:W reversed}, giving
\begin{align}
\label{eq:eq Kuramoto simple W}
    W \, =  -\frac{1}{2} \, \pd{a} \, {\cal Q}_{ab} \, \left( \pd{b}+\chempo{b} \right) \, = -\frac{\Delta}{2} \, \pd{a}  \left( \pd{a} + \chempo{a} \right) \, ,~~
\end{align}
so that the resulting equation of motion is \eqref{eq:eq Kuramoto SDE}~\cite{Kuramoto_RMP}. The model is ``reciprocal'' in that $J$ is symmetric in $a \leftrightarrow b$.

From our EFT perspective, this variant of the Kuramoto model---with stationary state \eqref{eq:stationaryKuramoto} and equation of motion \eqref{eq:eq Kuramoto SDE}---is an \emph{equilibrium} model that realizes nonthermal physics. The model is ``equilibrium'' in that it admits the stationary state $\Phi$ \eqref{eq:stationaryKuramoto}, and the corresponding Fokker-Planck generator $W$ \eqref{eq:eq Kuramoto simple W} is T even (i.e., it obeys detailed balance). However, the physics realized by this model---i.e., the dynamics and phase structure---are \emph{not} captured by a stationary distribution $\Phi = \beta H$ \eqref{eq:phibetaH} at temperature $T = 1/\beta$, where $H$ is related to $W$. In other words, the system is not captured by Hamiltonian evolution as described in Sec.~\ref{sec:not active}, and the stationary state $\Phi$ \eqref{eq:stationaryKuramoto} does not correspond to a thermal distribution $\Phi= - \ii \beta W$ for the corresponding generator $W$ \eqref{eq:eq Kuramoto simple W}.

Instead, assuming sufficiently long-ranged interactions, the equilibrium Kuramoto model \eqref{eq:eq Kuramoto simple W} generically displays a nonthermal phase transition at some critical noise strength $\Delta_c$. To see this, we first change variables via
\begin{align}
    \vb{S}_a \, \equiv \, \begin{pmatrix}\cos \theta_a\\\sin \theta_a\end{pmatrix} ~~~~\implies~~~~ \Phi \, = - \frac{1}{2\, \Delta} \sum_{a,b=1}^N \, J_{ab} \, \vb{S}_a \cdot \vb{S}_b \, ,~~
\end{align}
where $\vb{S}_a$ is the two-component classical spin for the $a$th rotor. In this language, the \emph{order parameter} that captures alignment of the rotors in the ``low-temperature'' phase can be written simply as
\begin{equation}
\label{eq:Kuramoto mag}
     \vb{m} \, \equiv \, \frac{1}{N} \sum_{a=1}^N \, \expval{\vb{S}_a} \, , ~~
\end{equation}
where $\vb{m}$ is analogous to the \emph{magnetization} in an XY model, and its magnitude is $m =\abs{\vb{m}}$.

For a concrete example, consider the all-to-all Kuramoto model, where  $J_{ab} = J/N$, connects all $N$ rotors with the same strength.  Using mean field theory---which is exact in this effectively infinite-dimensional system---we write $\vb{S}_a = \expval{\vb{S}_a} + \delta \vb{S}_a$, where $\expval{\vb{S}_a}$ is the mean-field (equilibrium) value, and we discard all terms that are nonlinear in the fluctuations $\delta \vb{S}_a = \vb{S}_a - \expval{\vb{S}_a}$. The corresponding mean-field stationary state $\Phi$, at linear order in fluctuations, is
\begin{align}
\label{eq:eq Kuramoto MFT Phi 1}
    \Phi_{\rm MFT} \, = -\frac{J}{2 N \Delta} \sum_{a,b} \, \expval{\vb{S}_a} \cdot \left( \expval{\vb{S}_b} + 2\,  \delta \vb{S}_b \right) \, , ~~
\end{align}
and assuming a spatially uniform state with $ \expval{\vb{S}_a} = m \, \vu{x} = \vb{S}_0$ for all rotors $a$, we seek a self-consistency condition for a nonzero mean-field magnetization $m$ \eqref{eq:Kuramoto mag}. The corresponding mean-field stationary state \eqref{eq:eq Kuramoto MFT Phi 1} is simply
\begin{align}
    \Phi_{\rm MFT} \, = - \frac{J}{2\Delta} \sum\limits_{a,b} \, m \, \left( m \, + 2 \left[ \cos \theta_a - m \right] \right) \, = \, \frac{m^2 \, J \, N}{2 \Delta} - \frac{J \, m}{\Delta} \sum_a \, \cos \theta_a  \, , ~~
    \label{eq:eq Kuramoto uniform Phi}
\end{align}
at lowest order in the fluctuations. The self-consistency condition follows from evaluating $m = \vb{m}_x = \sum_a \expval{\cos \theta_a}/N$ in the uniform stationary state \eqref{eq:eq Kuramoto uniform Phi}; defining $\lambda = m J/\Delta$, we note that $m = \expval{\cos \theta_a}$ for all rotors $a$, giving
\begin{align} \label{eq:mft_consistency}
    m \, = \, \frac{\int \dd{\theta} \cos(\theta) \, \e^{\lambda \cos \theta}}{\int \dd{\theta} \e^{\lambda  \cos \theta}} \,  =   \frac{I_1(\lambda)}{I_0(\lambda)} \,  \approx \, \frac{J}{2\, \Delta} m - \frac{J^3}{16\, \Delta^3}  m^3 \, ,~~
\end{align}
to cubic order in $m$, where $I_n(\lambda)$ is a modified Bessel function of the first kind. This expression admits a solution
\begin{equation}
    m \, = \, \frac{2 \, \Delta}{J} \, \left(\frac{4 \, \Delta}{J} - 2 \right)^{1/2} > 0 ~~~~\mathrm{for}~~~~\Delta \, < \, \Delta_c \, = \,J/2 \, ,~~
\end{equation}
so that a nontrivial order---in which all rotors are aligned, captured by $m>0$ \eqref{eq:Kuramoto mag}---realizes for $\Delta < \Delta_c$. 

Essentially, the critical noise strength $\Delta_c$ the rotors align in a spontaneously chosen direction so long as $J>0$.  Above $\Delta_c$, the system is disordered, with average angle $\expval{\theta_a}=0$. These features are built into $\Phi$ \eqref{eq:stationaryKuramoto}, corresponds to the XY Hamiltonian with all-to-all interactions, which has precisely this ordering transition in equilibrium statistical mechanics. Since $\Phi \propto 1/\Delta$, the noise $\Delta$ is analogous to the temperature $T$, and the low-noise order for $\Delta < \Delta_c$ is analogous to a low-temperature XY ferromagnet, with $\exp(-\Phi)$ maximized on the manifold in which all rotors are aligned. Even away from the all-to-all limit considered above, the thermodynamics of this Kuramoto model is the same as that of a classical XY model in thermal equilibrium, with $\Delta$ analogous to $T$.

\subsubsection{Nonreciprocity and generalized time-reversal symmetry}
 \label{sec:NR Kuramoto SSB}
The equilibrium Kuramoto model described by Eq.~\eqref{eq:eq Kuramoto SDE} can also be generalized to the \emph{nonreciprocal} case \cite{nonreciprocalPhase}, via
\begin{equation}\label{eq:NR Kuramoto SDE}
    \pd{t} \theta_a \, = \, \sum\limits_{b} K_{ab} \, \sin(\theta_a - \theta_b) + \xi_a(t) \, ,~~
\end{equation}
where $K_{ab} \neq K_{ba}$ is nonsymmetric. Compared to the equilibrium version \eqref{eq:eq Kuramoto SDE}, the SDE \eqref{eq:NR Kuramoto SDE} does \emph{not} recover from a Hamiltonian stationary distribution $\Phi=\beta H$ \eqref{eq:phibetaH}, and thus cannot be realized via Hamiltonian dynamics (see Sec.~\ref{sec:not active}). Compared to the equilibrium phase structure~\cite{Kuramoto_RMP}, the nonreciprocal Kuramoto model \cite{nonreciprocalPhase} admits an additional, \emph{chiral} phase in which all rotors $\theta_a$ spin at a fixed rate about a spontaneously chosen axis.

From our EFT perspective, the difference between the nonreciprocal and equilibrium Kuramoto models relates to their  respective \emph{symmetries}. In particular, the nonreciprocal Kuramoto model \eqref{eq:NR Kuramoto SDE} breaks detailed balance (i.e., is not T even) while respecting a \emph{generalized} time-reversal symmetry gT: the product of the reversibility transformation T \eqref{eq:W reversed} and a $\Ints^{\,}_2$ ``chirality'' transformation $P$ that swaps the sign of the interaction term in $W$ or \eqref{eq:NR Kuramoto SDE}.

We now explain how to reproduce the phase transition to the chiral phase of Ref.~\citenum{nonreciprocalPhase} within our EFT formalism. We note that this formalism does not reproduce exactly the equations of motion of Ref.~\citenum{nonreciprocalPhase}. This is due to a lack of analytic control: The distribution $\Phi$ \eqref{eq:FPE stationary distribution} that leads to Eq.~\eqref{eq:NR Kuramoto SDE} is not known, and may be arbitrarily complicated. Instead, we start with a plausible guess for $\Phi$ and derive the allowed T-odd terms and resulting equations of motion that reproduce the desired phase transition.

We first sketch a choice of $\Phi$ that leads to nonreciprocal dynamics qualitatively similar to that of Eq.~\eqref{eq:NR Kuramoto SDE}. This $\Phi$ admits a phase with spontaneous symmetry breaking (SSB) of gT; in the SSB phase, the configuration that minimizes $\Phi$ itself breaks gT (i.e., $PT$), so that the dynamics of particular trajectories does not explore the entire low-energy manifold, and applying the gT operation to a particular trajectory $\vb*{\theta}(t)$ leads to a trajectory $\vb*{\theta}' (t)$ that has negligible overlap with $\vb*{\theta}(t)$. Fortunately, it is straightforward to concoct such a $\Phi$. Consider the \emph{equilibrium} stationary distribution $\Phi$ \eqref{eq:stationaryKuramoto}, but suppose that there are two, distinguishable species of rotors labeled $A$ and $B$. We modify $\Phi$ \eqref{eq:stationaryKuramoto} by including a term that is quadratic in $\cos (\theta_a - \theta_b)$ for $a \in A$ and $b \in B$, leading to 
\begin{equation}
\label{eq:weirdPhi}
    \Phi \, = - \frac{J}{2\Delta N} \sum\limits_{i,j} \cos \left( \theta^{\vpp}_i - \theta^{\vpp}_j \right) + \frac{J'}{2\Delta N} \sum\limits_{a\in A, b\in B}\cos^2 \left(\theta^{\vpp}_a - \theta^{\vpp}_b \right)\, ,~~
\end{equation}
where $i,j$ run over all rotors in both sets $A$ and $B$, and $J'/J$ is the parameter of interest. The new term (with coefficient $J'>0$) competes with the ordinary Kuramoto attraction (with coefficient $J>0$), and makes configurations with fixed angular separations between the $A$ and $B$ rotors $\theta_a$ and $\theta_b$ statistically favorable. We emphasize that the nonequilibirum distribution \eqref{eq:weirdPhi} is a natural modification of the equilibrium analogue \eqref{eq:stationaryKuramoto}---the nonequilibrium version simply incorporates one higher harmonic of the rotor variables $\theta^{\,}_i$ compared to Eq.~\eqref{eq:stationaryKuramoto}.

For simplicity, we again consider the \emph{mean-field} dynamics of systems that relax to the stationary distribution $\Phi$ \eqref{eq:weirdPhi}. %After a long time, 
At late times, we expect all rotors to align \emph{within} their species ($A$ or $B$), due to the first term in Eq.~\eqref{eq:weirdPhi}; the second term suggests that alignment between the two species is undesirable. Accordingly, we take  $\expval{\theta_{a/b}}=\theta_{A/B}$, so that all rotors in $a \in A$ ($b \in B$) have the average orientation $\theta_A$ ($\theta_B$); we further define $\delta \theta \equiv \theta_A - \theta_B$ as the average angular separation between $A$ and $B$ rotors. The mean-field stationary distribution \eqref{eq:weirdPhi} is then 
%so that the $A$ and $B$ species may be described respectively by only two degrees of freedom, $\theta_A$ and $\theta_B$.  If we define $\delta \theta \equiv \theta_A - \theta_B$, then the mean field 2D system has a potential given by:
\begin{align}\label{eq:effectiveWeirdPhi}
    \Phi_{\rm MFT} \, = -\frac{\alpha \, J \, N}{\Delta} \, \cos(\delta\theta)+\frac{\alpha \, \bar{J}' \, N}{2 \, \Delta} \, \cos^2(\delta \theta) \, , ~~
    \, ,~~
\end{align}
where we have ignored the constant term in $\Phi$ as well as fluctuations from $\theta_{a/b}=\theta_{A/B}$ and  $\alpha = N_A N_B / N^2$. Because  $\Phi$ \eqref{eq:weirdPhi} is fully connected, and effectively infinite dimensional, mean-field theory correctly describes the coarse-grained physics, even when we neglect fluctuations, as in Eq.~\eqref{eq:effectiveWeirdPhi}. At the mean-field level, the distinct phases of the rotors realized by $\Phi_{\rm MFT}$ \eqref{eq:effectiveWeirdPhi} are identified by its two minima: When $J \gg J'$, $\Phi_{\rm MFT}$ is minimized (and entropy and probability maximized) when \emph{all} rotors are aligned, so that $\delta \theta = 0$; in the opposing limit $J \ll J'$, $\Phi_{\rm MFT}$ has a \emph{pair} of degenerate minima corresponding to $\delta \theta = \pm \, \arccos(J/J')$, and the system spontaneously chooses one of the two minima with fixed, average angular separation $\expval{\theta_a - \theta_b} = \pm \arccos (J/J')$ between all $A$ and $B$ rotors.

We now consider the two phases of $\Phi_{\rm MFT}$ \eqref{eq:effectiveWeirdPhi}. In particular, we consider whether the rotors are stationary in the SSB phase (with two degenerate minima), or if they spin in a manner that violates gT. Such dynamical details do not follow from $\Phi$ alone, but depend on whether the resulting equations of motion are ``nonreciprocal.'' For convenience, we assume additive Gaussian noise with variance $\Delta$. First, consider the reciprocal case, which can be captured by purely dissipative dynamics, and whose Fokker-Planck generator $W$ is invariant under $P$ and T individually, 
\begin{equation}
    \label{eq:dissipative_NRK}
    W \, = \,-\frac{\Delta}{2} \pd{i} (\pd{i} + \chempo{i}) \, ,~~
\end{equation}
where the sum over $i \in A \cup B$ is implicit. The chemical potentials are 
\begin{equation}
    \chempo{A} \, = -\chempo{B} \, = \,\frac{\alpha \, N}{4 \, \Delta} \left( J \sin (\delta \theta) - J' \, \cos (\delta \theta) \sin (\delta \theta) \right) \, , ~~
\end{equation}
and the corresponding mean-field equations of motion for the $A$ and $B$ rotors are, respectively,
\begin{subequations}
\label{eq:Kuramoto MF EOM AB}
\begin{align}
    \pd{t} \theta_A \, &=  - \frac{\alpha\, N}{8} \left( J - J' \, \cos (\theta_A - \theta_B ) \right) \, \sin (\theta_A - \theta_B )  + \xi^{\vpp}_A (t)  \label{eq:Kuramoto MF EOM A}\\
    \pd{t} \theta_B \, &=  + \frac{\alpha\, N}{8} \left( J - J' \, \cos (\theta_A - \theta_B ) \right) \, \sin (\theta_A - \theta_B )  + \xi^{\vpp}_B (t)  , ~~
    \label{eq:Kuramoto MF EOM B}
\end{align}
\end{subequations}
where the only necessary contribution to the stochastic force $f_{A/B}$ is $- \Delta \chempo*{A/B}/2$, corresponding to dissipative, T-even dynamics. Note that Eq.~\eqref{eq:Kuramoto MF EOM AB} is manifestly invariant under $P : \theta_A \leftrightarrow \theta_B$. At late times, the average over trajectories should reproduce the distirubtion $\Phi_{\rm MFT}$ \eqref{eq:effectiveWeirdPhi}. In the symmetric (reciprocal) phase, we have $\theta_A=\theta_B$, so that the sine terms in Eq.~\eqref{eq:Kuramoto MF EOM AB} vanish, and $\expval{\pd*{t}\theta_{A/B}}=0$, meaning that the rotors are \emph{static}. However, in the putative SSB phase with $\theta_A-\theta_B = \arccos (J/J')$, we would find that $\expval{\pd*{t}\theta_{A/B}}= 0$, since the parenthetical terms in Eq.~\eqref{eq:Kuramoto MF EOM AB} vanish. In other words, this purely dissipative case \eqref{eq:Kuramoto MF EOM AB} leads to all rotors aligning in some stationary (spinless) configuration. On average, $\theta_A$ and $\theta_B$ are static at late times, with $\theta_A - \theta_B$ determined by the phase of $\Phi$ \eqref{eq:effectiveWeirdPhi}.

Realizing a spinning phase at late times requires adding dissipationless terms to the Fokker-Planck generator $W$ \eqref{eq:dissipative_NRK}, and corresponds to the nonreciprocal phase in which $\Phi_{\rm MFT}$ \eqref{eq:effectiveWeirdPhi} has two degenerate minima with finite angular separation between all $A$ and $B$ rotors. The new dissipationless terms explicitly break the T invariance of $W$ \eqref{eq:dissipative_NRK}; however, the target nonreciprocal stationary state $P_{\rm ss} = \exp (- \Phi_{\rm MFT})$ with spinning rotors also breaks T. Instead, both $P_{\rm ss}$ and the corresponding generator $W$ are invariant under a gT symmetry corresponding to the combination of T \eqref{eq:W reversed} and a $\Ints_2$ operation $P \, : \, \theta_A \leftrightarrow \theta_B$ that swaps the roles of the $A$ and $B$ species of rotors\footnote{The operation $P$ is only meaningful in the nonreciprocal case in which the two species are distinguishable.}. The identification of a generalized T symmetry gT provides analytic control over the theory (i.e., $W$, which is gT even), and the results of Sec.~\ref{sec:EFT} apply straightforwardly. In particular, the results of Sec.~\ref{subsec:FDT} guarantee that $W$ \eqref{eq:dissipative_NRK} is modified to
\begin{equation}
\label{eq:nonrecipWkuramoto}
    W \, = \, \pd{i} v_i - \frac{\Delta}{2} \pd{i} \, (\pd{i} + \chempo{i}) \, ,~~
\end{equation}
where $i \in {A,B}$ in the mean-field limit described by the pair of angles $\theta_A$ and $\theta_B$. The new, dissipationless contribution $v_i$ \eqref{eq:v_anti} to the Fokker-Planck force \eqref{eq:FPE force FDT} satisfies Eq.~\eqref{eq:dissipationless_terms}, which can be fulfilled by taking 
\begin{equation}
\label{eq:kuramoto_v}
    v_A \, = \, v_B \, = \, v(\delta \theta) \, ,~~
\end{equation}
for some as-yet undetermined function $v(\cdot)$, which may be constrained by symmetries. In particular,  for $W$ \eqref{eq:nonrecipWkuramoto} to be symmetric under gT (i.e., $P \, T$), $v$ must be an \emph{odd} function of $\delta \theta =\theta_A - \theta_B$. The resulting  equations of motion are
\begin{subequations}
\label{eq:NRK fun EOM}
\begin{align}
    \pd{t} \theta_A \, &=  \, v (\theta_A - \theta_B) - \frac{\alpha\, N}{8} \left( J - J' \, \cos (\theta_A - \theta_B ) \right) \, \sin (\theta_A - \theta_B )  + \xi^{\vpp}_A (t)  \label{eq:NRK fun EOM A}\\
    \pd{t} \theta_B \, &= \, v (\theta_A - \theta_B)   + \frac{\alpha\, N}{8} \left( J - J' \, \cos (\theta_A - \theta_B ) \right) \, \sin (\theta_A - \theta_B )  + \xi^{\vpp}_B (t)  , ~~
    \label{eq:NRK fun EOM B}
\end{align}
\end{subequations}
where $v (\delta \theta)$ need not vanish at the degenerate minima $\delta \theta = \pm \arccos (J/J')$, though it must vanish at $\delta \theta =0$. As before, all of the terms on the right-hand side of  Eq.~\eqref{eq:NRK fun EOM} vanish, on average, \emph{except} the new $v$ term. In the phase with $J \gg J'$ so that $\delta \theta = 0$, $v=0$, and all rotors are static and aligned. However, in the phase with $J \ll J'$, we find that all $A$ rotors have orientation $\theta_a = \theta_A$, while all $B$ rotors have orientation $\theta_b = \theta_B$, with fixed separation $\delta \theta = \theta_A - \theta_B = \pm \arccos (J/J')$. Crucially, all rotors \emph{spin} at the constant rate $\Omega = \expval{\pd*{t}\theta_{A/B}} = \pm v(\arccos (J/J'))$, where the direction $\pm$ (clockwise versus counterclockwise) is dictated by which of the two minima of $\Phi_{\rm MFT}$ \eqref{eq:effectiveWeirdPhi} is spontaneously chosen. While the stationary state $\Phi$ \eqref{eq:effectiveWeirdPhi}, the generator $W$ \eqref{eq:nonrecipWkuramoto}, and the equations of motion \eqref{eq:NRK fun EOM} respect gT (i.e., the combination of $P$ and T), the fact that $\Omega_{A/B} = \expval{ \pd*{t}\theta_{A/B} } \neq 0$ reflects spontaneous breaking of gT, where gT sends $\Omega_{A/B} \to - \Omega_{A/B}$. In other words, the two minima of $\Phi_{\rm MFT}$ \eqref{eq:effectiveWeirdPhi}, corresponding to all rotors spinning clockwise or counterclockwise at rate $\Omega_{A/B} = \expval{\pd*{t}\theta_{A/B}} = \pm v(\theta_A - \theta_B)$, are interchanged under gT.

It is instructive to compare the nonreciprocal case above the the reciprocal, equilibrium example of Sec.~\ref{sec:eq kuramoto}. In both cases, the transformation $P$---a symmetry of $\Phi$ that is not a symmetry of the equations of motion---is spontaneously broken. However, in the nonreciprocal case, the operation T---which is also a symmetry of $\Phi$ in both cases---is not a symmetry of the \emph{nonreciprocal} equations of motion, whereas the equilibrium variant inherently obeys detailed balance, and its equations of motion are T even. In the nonreciprocal case, the product of T and $P$ (i.e., gT) is preserved by $\Phi$ \eqref{eq:effectiveWeirdPhi} and the equations of motion \eqref{eq:NRK fun EOM}, but is spontaneously broken by solutions to the latter (corresponding to particular minima of $\Phi$). This spontaneous breaking of gT is due to the dissipationless term $v$ \eqref{eq:kuramoto_v} that appears in \eqref{eq:NRK fun EOM}, and is indicated by a nonzero expectation value of the order parameter
\begin{equation}
\label{eq:NRK MF order parameter}
    \Omega \,  = \, \expval{\pd{t}\theta_A + \pd{t} \theta_B)} \, = \, \pm 2 \, v (\arccos (J/J'))\, ,~~
\end{equation}
which changes sign under T but is symmetric under $\theta_A \leftrightarrow \theta_B$ ($P$), and hence breaks gT; the rightmost expression is evaluated in the stationary state $\Phi$ \eqref{eq:effectiveWeirdPhi} for $J \ll J'$, where $\pm$ corresponds to the two minima of $\Phi$. The nonreciprocal Kuramoto model \cite{nonreciprocalPhase} realizes the simplest example of spontaneous breaking of gT ($P$T), captured by the order parameter $\Omega \neq 0$ \eqref{eq:NRK MF order parameter}; we consider other examples of spontaneous gT breaking  in Sec.~\ref{sec:gT}. 

%In this case $P$---a symmetry of $\Phi$ that is not a symmetry of the equations of motion---is spontaneously broken, just as it was in the $T$-invariant case of the ordinary (reciprocal) Kuiramoto model.  The difference in the nonreciprocal case is that $T$---another symmetry of $\Phi$---is \emph{also} broken in the equations of motion.  Introducing an order parameter $\omega = \dot \theta_A + \dot \theta_B$, we find that $\omega$ changes sign under $T$, and therefore trivially also changes sign under $PT$.  Since it is $PT$ that is really a symmetry of our theory $W$ (and also $\Phi$), it is therefore most accurate to understand the chiral phase as spontaneously breaking $gT = PT$ symmetry.  This is the simplest example of the $gT$-breaking that we study more generally in Sec.~\ref{sec:gT}. To summarize, the nonreciprocal interactions in the equations of motion, which are follow from the dissipationless contribution to $W$, allow for a spontaneous breaking of $PT$-symmetry and a corresponding chiral phase, as reported in Ref.~\citenum{nonreciprocalPhase}.

We emphasize that the EFT presented above is \emph{not} precisely the nonreciprocal Kuramoto model \eqref{eq:NR Kuramoto SDE} described in Ref.~\citenum{nonreciprocalPhase}, as can be seen by comparing those equations of motion \eqref{eq:NR Kuramoto SDE} to those that result from the effective theory  \eqref{eq:NRK fun EOM}. Rather, our perspective is that the stationary distribution $\Phi$ that gives rise to the phenomenological Kuramoto equation of motion \eqref{eq:NR Kuramoto SDE} is likely quite complicated, involving many terms, some of which may be nonlocal. Instead, we consider a simpler stationary state $\Phi$ \eqref{eq:weirdPhi}, at the cost of complicating the resulting equations of motion. However, those equations are simple in the mean-field limit \eqref{eq:NRK fun EOM}. Still, the topology of the resulting phase diagram of $\Phi_{\rm MFT}$ \eqref{eq:effectiveWeirdPhi} agrees in part with that of the microscopic nonreciprocal Kuramoto model~\cite{nonreciprocalPhase}, and we expect that the phase transition from a static to chiral (i.e., spinning) phase is the same in each model. We relegate calculation (or simulation) of any dynamical critical exponents of these models to future work.

\subsection{Dynamics of wealth}
\label{sec:money}

As a final toy example of nonthermal dynamics, we apply the EFT formalism of Sec.~\ref{sec:EFT} to a cartoon model from the social science describing the dynamics of \emph{wealth} (or income) \cite{kalecki, angle, ispolatov, yakovenko}. Although real economic systems do not evolve toward a stationary distribution $\Phi$ \eqref{eq:FPE stationary distribution}, due to the ever-changing nature of society, technology, and markets, this approximation is useful from the perspective of modeling short-time dynamics.  Denoting by $x>0$ the wealth of an individual, it is well established in the literature \cite{kalecki, angle, ispolatov, yakovenko} that the appropriate stationary distribution takes the form $\exp(-\Phi) \sim x^{-\alpha}$ when $x \gg x^{\,}_0$ is large, while $\exp (-\Phi) \sim x \, \exp(-x)$ when $x \ll x^{\,}_0$ is small. 

A simple choice of $\Phi$ that realizes these two limits is given by
\begin{equation}
\label{eq:money stationary}
    \Phi \, = \, (1+\alpha)\,  x^{\vpp}_0 \, \log\left(1+\frac{x}{x^{\vpp}_0}\right) - \log x \, ,~~
\end{equation}
where $x^{\,}_0$ is a characteristic wealth scale that separates the lower and upper ends of the in distribution, so that 
%The scale $a$ corresponds to the crossover between the low and high ends of the income distribution. Assuming multiplicative Gaussian noise, 
\begin{equation}
\label{eq:money chempo}
    \mu \, = \, \pdv{\Phi}{x} \, = \, \frac{(1+\alpha)\, x^{\vpp}_0}{x+x^{\vpp}_0} - \frac{1}{x} \, ,~~
\end{equation}
is the chemical potential corresponding to $\Phi$ \eqref{eq:money stationary}. We assume multiplicative Gaussian noise with
\begin{equation}
    \label{eq:money noise}
    \expval{\eta(x,t) \eta(x,t')} \, = \, Q(x) \, \DiracDelta{t-t'} \, , ~~
\end{equation}
in which case the Fokker-Planck generator $W$ \eqref{eq:W form} takes the form
\begin{equation}
\label{eq:money generator}
    W \, = \, -\frac{1}{2} \, \pd{x}\,  Q(x) \left(\pd{x} + \mu(x)\right) \, ,~~
\end{equation}
via Eqs.~\eqref{eq:W nice} and \eqref{eq:nice Q def}, since there are no dissipationless terms required. 

We note that, at both small and large wealth $x$, the rate of income growth (or decay) should be proportional to $x$. Since $\mu(x)$ \eqref{eq:money chempo} scales as $1/x$, this is achieved if the noise variance $Q(x)$ \eqref{eq:money noise} is of the form
\begin{equation}
    Q(x) \, = \,  Q^{\vpp}_0 \, x^2 \, ~~
\end{equation}
for some constant $Q^{\,}_0$, so that the term in $W$ proportional to $\pd*{x}$ is linear in $x$. 

%From our perspective, we have essentially guessed the equations of motion based on plausible assumptions about the income distribution.  Let us now see what equations of motion we have predicted.  To compare with the literature, it is helpful to work with an It\^o SDE, and we find that  

From the perspective of our EFT formalism, we have essentially derived (or guessed) the equations of motion governing wealth based on plausible assumptions about (\emph{i}) the late-time distribution of wealth and (\emph{ii}) a rough idea of how wealth should grow or shrink on short time scales. From the Fokker-Planck generator $W$ \eqref{eq:money generator}, it is straightforward to work out the stochastic equation of motion for $x$. Using the It\^o regularization, we find
\begin{equation}
\label{eq:money SDE}
    \frac{\dot{x}}{x} \, = \, -Q^{\vpp}_0 \frac{[(1+\alpha)\, x^{\vpp}_0-3] \, x-3 \, x^{\vpp}_0}{x+x^{\vpp}_0}+ \sqrt{2 \, Q^{\vpp}_0} \, \xi \, ,~~
\end{equation}
where $\xi$ is an additive Gaussian noise source% in the It\^o prescription
, such that $\eta \propto \sqrt{x} \, \xi$. Note that Eq.~\eqref{eq:money SDE} agrees with the  literature: When $x \gg x^{\,}_0$ is large, we  find a contribution corresponding to multiplicative noise $\dot{x}/x \sim \xi$, as well as a deterministic (and  
approximately constant) contribution $\dot{x}/x \sim c$. This reproduces cartoon models of the ``Mathieu effect,'' which realize (approximate) power-law tails in a range of distribution functions that appear throughout the social sciences \cite{Newman_2005}. 

We comment that the standard approach in the literature---in the social and physical sciences---is to start by writing down an equation of motion, and then work out its properties, including the existence and form of a stationary distribution. We reiterate that the opposite approach---in which one starts by positing a stationary distribution $\Phi$ \eqref{eq:FPE stationary distribution} and works out all compatible dynamics---is just as valuable. The latter perspective, captured by our EFT formalism, unsurprisingly leads to similar models as have been identified in the literature. While this perspective may lead to more complicated (i.e., uglier) equations of motion, their derivation is more systematic, and the symmetries, late-time behavior, and possible phases are more readily understood. In many cases---including the dynamics of wealth considered above---these results can be recovered more expediently in the EFT approach. 

%It is more common in the literature to posit a plausible looking equation of motion, and then try to deduce the stationary distribution.   We reiterate that it is just as valuable to take the opposite approach: start with $\Phi$ and deduce the equations to study!   It is unsurprising that the methods lead to the same kinds of models.  As we have seen above, generally our strategy leads to uglier equations of motion, but whose symmetries are more readily understood, and which can be derived in fewer pages!

\section{Spontaneous breaking of generalized time-reversal symmetry}\label{sec:classify SSB}

The nonreciprocal Kuramoto model in Sec.~\ref{sec:NR Kuramoto SSB} provides a simple realization of ``activity,'' a crucial consequence of which is the \emph{spontaneous breaking} of generalized time-reversal symmetry gT. Importantly, the breaking of gT is %generally 
forbidden under thermal dynamics, and is thus a fingerprint of activity. We now systematically explore the mechanism(s) by which gT can be spontaneously broken from the perspective of the EFT framework of Sec.~\ref{sec:EFT}. This construction also %highlights a particularly 
provides an elegant application of the results of Sec.~\ref{sec:noether} regarding the incorporation of symmetries.

\subsection{General framework}\label{sec:general_T_breaking}
\label{sec:groupflow}

We first develop the mathematical machinery to investigate how generalized time-reversal can be broken spontaneously. For clarity, we consider the example of overdamped dynamics of rigid bodies, and comment on more general results.

\subsubsection{Example: rigid-body rotation}

Here we consider the  rotational dynamics of a rigid body---in which the constituent degrees of freedom rotate together---in $d=3$\footnote{We consider EFTs for rigid-body dynamics in further detail in Sec.~\ref{sec:dissipativerigid}}. For convenience, we identify the fixed center of mass with the origin $\vb{x}_{\rm COM}=0$. At the microscopic level, the rigid body is comprised of an enormous number of atoms or molecules labeled $a =1,\dots,N$. The microscopic configuration (phase) space $M$ is \emph{a priori} extremely large, with $M=\Reals^{3N}$ spanned by configurations $\vb{x} \in M$ with $\vb{x} = \{ x^{a}_i \}$, where $i = 1,2,3$ is a spatial index. Still, the configuration of a rigid body is highly constrained within $M = \Reals^{3N}$; at late times, the system relaxes to a configuration on $M$ that minimizes the stationary distribution $\Phi$ \eqref{eq:FPE stationary distribution}, as such configurations are the most probable. At sufficiently late times, the particles ``condense'' into a rigid body, at which point the only \emph{guaranteed} degeneracies of $\Phi$ are those related by global rotations of all particle positions,
\begin{equation}
\label{eq:rigid Phi rotational minima}
    \Phi\left( \{ x^a_i \}\right) \, = \,  \Phi\left( \{ R^{\vpd}_{ij} x^a_j \} \right) \, ,~~
\end{equation}
for some rotation matrix $R \in \SOrth{3}=G$, where all positions are defined with respect to the center of mass. We emphasize that $R$ does not depend on $a$---meaning that all particles rotate together under $R \in G$\footnote{The case in which $R$ depends on $a$ (or varies spatially) is treated in Ref.~\citenum{elipaper}.}. However, the configuration $\overline{\vb{x}} \in M$ that minimizes $\Phi$ does not have the symmetry $\vb{x}^* = R \, \vb{x}^*$ for $R \in G$ of $\Phi$ itself; hence, such configurations spontaneously break isotropy (the rotational symmetry $G$). For example, one could imagine a $\Phi$ whose minimum $\vb{x}^*$ corresponds to the $N$ particles realizing a cubic body; while $\Phi$ remains invariant under all rotations in $G$ via Eq.~\eqref{eq:rigid Phi rotational minima}, the cubic body is only invariant under the \emph{point group} $H=\mathrm{S}_4$ (the permutation group of four elements).

In the context of rigid bodies, it is standard practice to refer to a ``space frame'' and a ``body frame.'' The space frame is defined by a fixed coordinate axis in $\Reals^3$; different space-frame configurations correspond to different points in $M$. However, a particular configuration---such as the configuration $\vb{x}^*$ that minimizes $\Phi$---also identifies a particular rigid body, in the sense of defining a particular, spontaneous breaking of the symmetry group $G$, where different configurations $\vb{x}$ may realize distinct point groups $H \subset G$. Of course, rotations of a rigid body (under $R \in G$) do not change the \emph{type} of rigid body (i.e., they do not deform the body), and hence correspond to the same body. This implies an \emph{equivalence class} on $M$, corresponding to distinct rigid bodies, which are labeled by distinct body-frame configurations in $M/G$. Thus, points in the \emph{coset space} $M/G$ correspond to distinct ways of breaking $G$ spontaneously, and all points $\vb{x} \in M$ related by rotations $R \in G$ correspond to the same body-frame coordinate $\overline{\vb{x}}$. 

More importantly, at the level of EFT, the coarse-grained dynamics of the rigid body are entirely captured by rotations $R \in G$. The space-frame configurations $\vb{x} \in M$ corresponding to the same body-frame coordinate $\overline{\vb{x}} \in M/G$ not only realize rotations of the same body, but also have the same late-time probability due to Eq.~\eqref{eq:rigid Phi rotational minima}. Note that $\Phi$ is well defined on $M/G$ by Eq.~\eqref{eq:rigid Phi rotational minima}. Now, while deformations of the rigid body are possible in principle, they map ``low-energy'' configurations of $\Phi$ to higher-energy (i.e., less probable) configurations, which rapidly decay back to the low-energy manifold of $M$, corresponding to the body-frame coordinate $\overline{\vb{x}}^*$ that minimizes $\Phi$. Hence, the coarse-grained dynamics are captured by rotations $R \in G$, so that the space-frame coordinates evolve in time via 
\begin{equation}
\label{eq:rigid space and body}
    x_i^a (t) \, = \, R^{\vpp}_{iI}(t) \, \overline{x}^{\, a}_I \, ,~~
\end{equation}
where $x^a_i (t)$ is the time-dependent position of the $a$th particle in the \emph{space} frame (an inertial reference frame), and $\overline{x}^a_I$ is the position of the $a$th particle in a statistically favorable configuration  in the \emph{body} frame  (at $t=0$, for reference). The space-frame axes correspond to a fixed Cartesian frame, while the body-frame axes are fixed relative the rigid body; we use lowercase indices $i,j$ for the space frame and uppercase indices $I,J$ for the body frame. We generally assume that $\Phi$ has finitely many degenerate minima in $M/G$; for convenience, we often assume it has only one. The orthogonal matrix $R^{\,}_{iI}(t)$ effects a change of basis from the body frame to the space frame, as is standard to the treatment of rigid bodies~\cite{josesaletan}. The matrix $R(t)$ is dynamical degree of freedom for EFTs of rigid bodies.

We also stress that the notation of Eq.~\eqref{eq:rigid space and body} is useful not only in the context of rigid-body dynamics, but for continuous symmetries more generally. A dynamical system with configuration space $M$ and a continuous symmetry group $G$ generally realizes different dynamics in the space frame $M$ compared to the body frame $M/G$; moreover, from the EFT perspective, it is useful to distinguish these two. For convenience of presentation, we generally focus on the case of rigid bodies; we also use the terminology of ``space'' versus ``body'' frames in all contexts.

\subsubsection{Left and right symmetry groups}\label{sec:leftright}
Having argued that the matrix $R^{\,}_{iI}(t)$ is the correct coarse-grained degree of freedom for rigid bodies, we now comment on the additional symmetries constraining any EFT for $R^{\,}_{iI}(t)$, and how they transform $R$. As a reminder, a symmetry of the EFT is equivalent to a symmetry of its Fokker-Planck generator $W$ (see Sec.~\ref{sec:noether}).

We first observe that there is no ``preferred'' choice of coordinate axis for $\Reals^3$ in the space frame, or equivalently, of the origin for $G = \SOrth{3}$. Essentially, even when $G$ is spontaneously broken by a configuration $\overline{\vb{x}} \in M/G$, the same physics should recover for any rotation, captured by $\Lambda \in G$, which simply redefines the space-frame coordinates via
\begin{align}\label{eq: rigid redefinition}
    x^a_i \, \to \, y^\alpha_i \, &= \, \Lambda^{\vpp}_{ij} x^a_j \, ,~~
\end{align}
where $\Lambda^{\,}_{ij}$ is a space-frame rotation that acts on the left-hand side of Eq.~\eqref{eq:rigid space and body}. For the equality in Eq.~\eqref{eq:rigid space and body} to hold in this rotated basis for $M$, we also transform the element $R \in G$ according to
\begin{align}
\label{eq:rigid left G}
    R^{\vpp}_{iI}(t) \, \to \, \Lambda^{\vpp}_{ij} \, R^{\vpp}_{jI}(t) \, ,~~~~\Lambda \in G \, ,~~
\end{align}
which modifies $R$ \eqref{eq:rigid space and body} to reflect the rotated space-frame coordinates. Because the body-frame configurations $\overline{\vb{x}}$ belong to $M/G$ (they are fixed relative the body's principal axes), all points in $M$ related by rotations $\Lambda \in G$ are equivalent, and correspond to the same $\overline{\vb{x}}$. Hence, $\Lambda$ is not applied to the $I$ index of $R$ \eqref{eq:rigid space and body}.

This naturally leads to the distinction between left and right invariance of the dynamical variables $R(t)$ under (subgroups of) $G$. For a system with no symmetries whatsoever, the variables $R$ would have no invariance properties (this also results if a symmetry $G$ is \emph{explicitly} broken). If a system is invariant under a continuous symmetry $G$ that is respected not only by $\Phi$, but by the configurations that minimize $\Phi$ (i.e., there is neither explicit nor spontaneous symmetry breaking), then the variables $R$ would be invariant under $G$ from the left and right. In that case, $R$ is an element of $G$, and multiplying from the left or right by elements of $G$ also results in elements of $G$. The interesting scenario corresponds to \emph{spontaneous} breaking of $G$, as in the rigid-body example, in which case the space frame $M$ retains the full symmetry group $G$, but the body frame $M/G$ is symmetric only under a subgroup $H \subset G$ (e.g., $S_4 \subset \SOrth{3}$ in the case of the cube), corresponding to the part of $G$ that remains after $G$ is spontaneously broken. As a result,  $R^{\,}_{iI}(t)$ \eqref{eq:rigid space and body} is invariant under $G$ from the left (the space-frame index $i$) but only invariant under $H \subset G$ from the right (the body-frame index $I$). Essentially, the ``point group'' $H$ includes the transformations of the body itself that leave it invariant. Thus, for spontaneously broken continuous symmetries $G \to H$, the dynamical variables are left invariant under $G$ and right invariant under $H \subset G$.

%This is the first symmetry we will enforce on our $W$.  Symmetry of a theory under \eqref{eq:rigid left G} is referred to in the mathematical literature as \emph{left $G$-invariance}, and is a consequence of the fact that $G$ is spontaneously broken\footnote{since our theory is on the set of minima of $\Phi$, if we did not have $G$ symmetry, the minima we are describing would not be degenerate and therefore there would not have been SSB}.

%Indeed, one instructive way to think about left invariance is to consider the simplest object that could enter a theory for $R_{iI}(t)$ symmetric under Eq.~\eqref{eq:rigid left G}.  Since $R_{iI}R_{iJ} = \kron{IJ}$ is trivial, the simplest object is

We next consider how left and right invariance of $R$ manifest in the resulting theory. Using the notation convention of Eq.~\eqref{eq:rigid space and body}, left $G$ invariance of the Fokker-Planck generator $W(R^{\,}_{iI}, \pd*{iI})$ is ensured by demanding that lowercase (space-frame) indices $i,j,\dots$ of $R$ and/or $\p$ are only contracted with invariant tensors of $G$. For example, $\kron{i,j} \in G$ is an invariant tensor, and so a simple term that could appear in the rigid-body Fokker-Planck generator is 
\begin{align}
\label{eq:rigid body Omega}
    W = \partial_{iI}R_{iI }\Omega^{\vpp}_{JI} \, ,~~
\end{align}
where $\Omega$ is a matrix related to the components of the angular velocities $\vb*{\omega}$ in the body frame (see Sec.~\ref{sec:rigidbody}) via
\begin{align}
    \Omega_{IJ}=-1/2\epsilon_{IJK}\omega_K = x_I \dot x_J - x_J \dot x_I.
\end{align} 
Thus $\Omega$ encodes the physical degrees of freedom.  One can also check that $\Omega_{IJ}$ must be antisymmetric if $R_{iI}R_{iJ} = \delta_{IJ})$. 

Having established that we should enforce $G$ on the left (lower case) indices, let us turn now to the right (upper-case) indices.  We wish now to consider transformations where we rotate the body-frame axes with respect to the body itself.  However, a general rigid body will not itself be symmetric under every transformation in $G$!
If $\Omega_{JI}\neq 0$, then the body itself must explicitly break $G$---this follows simply from the fact that a general antisymmetric matrix $\Omega_{IJ}$ does not lead to covariance on the right-hand side of Eq.~\eqref{eq:rigid body Omega}. Let $H\subset G$ be the (possibly discrete) subgroup %which 
that represents the body-frame symmetry of the rigid object---for instance, a cylindrical body would have $H = \SOrth{2}$. Next, consider transforming the body-frame axes under $H$, so that $x_I \to V_{IJ} x_J$ for $V \in H$.  Then in Eq.~\eqref{eq:rigid space and body}, since the left-hand side is invariant under body-frame transformations, the matrix $R_{iI}$ on the right-hand side must transform with $V^{-1} = V^T$, i.e., 
\begin{align}\label{eq:rigid right H}
    R_{iI} \, \to \, R_{iJ} \, V_{JI}\, ~~~~ V \in H \, ,~~
\end{align}
and we further demand that this is a symmetry of the Fokker-Planck generator $W$ \eqref{eq:W nice}, which we refer to as \emph{right} $H$ \emph{invariance}. 
%More precisely, once we choose $R_{iJ}$ based on a particular reference configuration in the body-frame, it is not necessarily true that the body itself continues to be $G$ invariant.  
In the rigid-body example, one can show that $\Omega_{JI}$ itself respects right $H$ invariance, which forms the same group $\SOrth{3}$ if and only if the object is spherically symmetric!  To see this, recall first that the SDEs \eqref{eq:SDE main} of interest---at least in the overdamped limit---are linear in $\dot{R}$. Covariance of these equations of motion then corresponds to covariance of Eq.~\eqref{eq:rigid body Omega}, which in turn requires that
\begin{align}
\label{eq:rigid Omega V comm}
    \Omega_{IJ} \, = \, R_{iI} \, \dot{R}_{iJ} \, \to \,  R_{iK} \, R_{jL} \, \Omega_{KL} \, V_{KJ} \, V_{LJ} ~~~\implies ~~~~ \left[ \Omega, V \right] \, = \, 0 \, ,~~
\end{align}
meaning that $V$ and $\Omega$ commute, and thus,  $\Omega_{IJ}$ respects right $H$ invariance \eqref{eq:rigid right H}.

\subsubsection{Dynamics along \texorpdfstring{$G$}{G}}\label{subsec:leftright}
We are now ready to construct an effective theory for the low $\Phi$ dynamics in the case where the symmetry group $G$ is spontaneously broken.  Again, constructing an effective theory in our perspective amounts to building the symplest $W$ operator that is consistent with the desired symmetries. We should first note that the stationary distribution $\Phi$ is constant on $G$, since $G$ is %supposed to be a 
the manifold of degenerate minima of $\Phi$.  That means the ``chemical potentials'' vanish: $\chempo*{iI}\equiv \pdv*{\Phi}{R_{iI}} = 0$.  We further demand that the Fokker-Planck generator $W$ be compatible with the left $G$ invariance described in Sec.~\ref{subsec:leftright}.  Finally, if $R_{iI}(t) \in G$ for all time $t$, the time derivative $\dot{R}_{iI}$ must satisfy
\begin{align}
\label{eq:def_omega}
    \pd{t} R_{iI} \, = \, R_{iJ} \, \Omega_{JI}  \, ,~~
\end{align}
for an antisymmetric matrix $\Omega_{IJ} = -\Omega_{JI}$ that is an element of the Lie algebra of $G$.  The object $\Omega_{IJ}$, which encodes the angular velocity in the body frame, previously appeared in Eq.~\eqref{eq:rigid body Omega}.  Finally, if the right symmetry group $H$ is %not-trivial,
nontrivial, Eq.~\eqref{eq:rigid Omega V comm} must hold as well; this is satisfied by identifying the simplest $\Omega$ that commutes with all $V \in H$ \eqref{eq:rigid Omega V comm}.  The vanishing of $\chempo*{iI}$ (together with left and %write 
right symmetries) heavily constrains the dissipationless part of the Fokker-Planck operator $W$, giving
\begin{align}\label{eq:rigid noise free}
    W \, = \, \pdv{R_{iI}}\,R_{iJ}\Omega_{JI} \equiv \p_{iI}R_{iI}\Omega_{IJ} \, ,~~
\end{align}
where, in the rightmost equality above, we have introduced the convenient shorthand $\pd*{iI} \equiv \pdv*{R_{iI}}$.

However, in addition to noise-free dynamics, one can also consider adding a stochastic noise source $\xi_{iI}(t)$ to the equations of motion for $R_{iI}$, such that the Langevin SDE \eqref{eq:SDE main} is
\begin{align}\label{eq:rigid body SDE}
    \dot{R}_{iI} \, = \, R_{iJ} \, \Omega_{JI} + \xi_{iI}(t) \, , ~~
\end{align}
where, following Eqs.~ \eqref{eq:W nice} and \eqref{eq:nice Q def}, $\xi_{iI}$ should enter $W$ \eqref{eq:rigid noise free} at quadratic order in the gradients $\pd*{iI}$.  However, one must be careful that the addition of $\xi_{iI}(t)$ in Eq.~\eqref{eq:rigid body SDE} does not spoil the orthogonality of $R$, so that $R$ remains confined to  $G$. To maintain closure on $G$, the additive noise source $\vb*{\xi}$ must be replaced by a \emph{multiplicative} noise source $\eta_{iI}(R,t)$ of the form
\begin{align}
    \eta_{iI}(t) \, = \, R_{iJ} \, A_{IJ}(t)
\end{align}
where $A_{IJ}(t)$ is a stochastic antisymmetric matrix, again in the Lie algebra of $G$.  In the simple Gaussian case a $3 \times 3$ $A_{IJ}$ belongs to the Lie Algebra $\mathfrak{so}(3)$ and may therefore be expanded in terms of its basis matrices $\epsilon^\alpha_{IJ} = \epsilon_{\alpha I J}$ where $\epsilon_{\alpha I J}$ is the completely antisymemtric tensor , giving
\begin{align}
    A_{IJ(t)} = \epsilon^{\alpha}_{IJ}\, \xi^\alpha(t),
\end{align}
where $\epsilon^\alpha$ are the constant antisymmetric matrices that generate $G$ and $\xi^\alpha$ are stochastic variables with zero mean ($\expval{\eta^\alpha(t)} = 0$) and variance
\begin{align}
    \expval{\xi^{\alpha}(t) \xi^{\beta}(t^\prime} = Q^{\alpha \beta}\delta(t-t^\prime)
\end{align}
for a constant matrix $Q^{\alpha\beta}$.  Using our prescription \eqref{eq:W nice} for converting a stochastic noise in to a Fokker-Planck generator, we find that the $W$ describing noisy overdamped dynamics on $G$ is
\begin{align}
\label{eq:groupL}
    W = \p_{iI}R_{iJ}\Omega_{JI} -\frac12 \p_{iI} \left( R_{iK}\epsilon^\alpha_{KI}Q^{\alpha \beta} R_{jL}\epsilon^{\beta}_{LJ} \right)\p_{jJ}.
\end{align}
For completeness, we note that in the common simplifying case were the noise is isotropic and $Q^{\alpha \beta} = \Delta \delta^{\alpha\beta}$, one may check use properties of $\epsilon^\alpha_{IJ}$ to show that $W$ takes a simpler form
\begin{align}
     W = \p_{iI}R_{iJ}\Omega_{JI} -  \frac12 \Delta\, \p_{iI} \left( \qty(\delta_{ij} \delta_{IJ} - R_{iJ}R_{jI}) \right)\p_{jJ}.
\end{align}
In the following subsections, we will be mostly concerned with the deterministic part of $W$, but it is good to know that in principle we can include noise as well.

\subsubsection{Spontaneously breaking generalized time-reversal symmetry}\label{sec:gT}
Now that we have an EFT for dynamics on a general group $G$, we have a good platform to consider the spontaneous breaking of reversibility symmetry.
Spontaneous symmetry breaking (SSB) happens in a classical dynamical system when the individual stochastic trajectories (almost surely) do not sample the whole phase space, even restricted to the appropriate symmetry sector (e.g. at a given energy or conserved charge).
%Quantum mechanically, this is only possible in the thermodynamic limit since in any finite system, quantum tunneling allows superposition of the ground state (or steady state) that can be an invariant state. 
Strictly speaking, in the infinite time limit, SSB cannot occur if there are a finite number of degrees of freedom, so long as $\Phi$ is not infinite, or without some other fine tuning.  However, since we may often use a finite-dimensional theory (as in the Kuramoto model) to describe the mean-field limit of a sytsem in the thermodynamic limit, we can still start to classify possible patterns of SSB without immediately worrying about whether SSB is stable to such fluctuations in the $t\rightarrow\infty$ limit. 

%In a classical stochastic system, it is natural to think of thermal noise as making the steady state statistically invariant in any finite system, and only in the thermodynamic limit can SSB be well defined. Therefore, we claim 

%For example, the condition of long range order can be phrased as the existence of a nonzero charged operator, i.e. operators that transform nontrivially under the classical symmetry, on the steady state. With the above, let us consider the following construction of SSB.

In our framework, SSB must be a feature captured by the minima of $\Phi$, which are not invariant under the symmetry group of $\Phi$ as a whole.  This is only possible if $\Phi$ has more than one minimum in $M/G$, so we will now posit that this is indeed the case. For simplicity in the discussion below, we will consider the case where $\Phi$ has exactly two degenerate minima $m_\pm \in M/G$, which transform into each other under a $\Ints_2$ symmetry, which we will denote as $P$, which together with T forms the generalized time-reversal gT.  If $\Phi$ has a thermodynamically large barrier between $m_\pm$, then $P$ will be spontaneously broken---we will observe the system to be in either $m_+$ or $m_-$ with no transition between the two, and the configuration space of the resulting dynamics is effectively the corresponding disconnected sector of phase space parameterized only by $R$.

In this setting, let us now consider a Fokker-Planck generator $W$ which is gT-invariant.  This can be achieved by choosing $\Omega$, defined in (\ref{eq:def_omega}), to obey \begin{equation}\label{eq:T SSB}
    \Omega(m_\pm) = \pm \Omega_0;
\end{equation}
namely, $\Omega_{JI}$ has opposite sign at the two minima $m_\pm$.  Since $\Omega_{JI}$ flips sign under time-reversal, but also flips sign under $P$, $W$ (and thus the dynamical system) explicitly breaks both P and T, while being PT-symmetric.  

We emphasize that the \emph{dynamics} of the system explicitly break T, while the \emph{thermodynamics} of the system spontaneously break P.  Of course, since $W$ has exact PT-symmetry, and the symmetry breaking pattern of $\Phi$ is, by definition, insensitive to time-reversal symmetry, we can consider $\Phi$ to be PT-invariant as well.  Then, PT will be spontaneously broken as the system picks either $m_+$ or $m_-$ as the minimum on $M/G$.  In this way, our framework systematically describes a mechanism for spontaneously breaking a \emph{generalized} time-reversal symmetry.  We emphasize it is important that gT, and not just T, be broken spontaneously;  after all, the spontaneous breaking of P in the ``thermodynamic" $\Phi$ was crucial to the mechanism.

% In this setting if $\Omega \ne 0$, time-reversal symmetry is explicitly broken. To have \emph{spontaneous} ``time-reversal'' symmetry breaking, we need to instead consider theories where $\Phi$ has at least two-fold degeneracy on $M/G$.   
% %\footnote{If this is a thermal system, i.e. $\Phi = \beta H$, we can think of $P$ as SSB.}, 
% and one can imagine that by tuning some parameters the steady state goes from an state $\phi_0$ invariant under P to the two degenerate states $\phi_\pm$. In order to break the ``time-reversal'' symmetry spontaneously, we introduce the following driving force into $W$:
% \begin{align}
%     \Omega(\phi_\pm ) = \pm \Omega_0,\quad \Omega(\phi_0)=0.
% \end{align}
% Since $\Omega$ is T-odd, we immediately deduce that in the two degenerate steady state configuration, the theory is PT invariant, while in the state $\phi_0$ the theory is both T and P invariant. Since $\Omega$ only shows up in $W$ and not in $\Phi$, the T symmetry is \emph{dynamically} broken. The two degenerate states are, however, not invariant under PT symmetry:
% \begin{align}
%     \mathrm{PT}: (\phi_+,\Omega_0=\Omega(\phi_+)) \rightarrow (\phi_-, - \Omega_0=\Omega(\phi_-)).
% \end{align}
% Therefore, the PT symmetry is SSB, and it is not hard to see that the $\Omega$ is the order parameter.

A concrete example of this is found in the Kuramoto model of Section \ref{sec:kuramoto}.  Here, we saw that the two minima of $\Phi$ corresponded to $\mathrm{\Delta}\theta = \pm \theta_0$ in the spontaneous generalized time-reversal breaking phase; thus $ \lbrace \pm \theta_0\rbrace$ is the set of minima of $\Phi$ on $M/G$.  The residual symmetry group $G=\mathrm{U}(1)$ has a single generator corresponding to constant shifts in $\theta_A+\theta_B$.   There is a symmetry group P of $\Phi$, which corresponds to switching the species A and B of nonreciprocal rotors.  In the phase where PT-symmetry is spontaneously broken, applying the PT transformation to a typical \emph{trajectory} of the system will flip the signs of $\dot{\theta}_A + \dot\theta_B$ or $\Delta \theta$: we can think of either of these two quantities as order parameters for the SSB phase.\footnote{It would be interesting to connect these ideas to recent generalizations of the Landau paradigm described in \cite{McGreevy:2022oyu} to generalized symmetries.}

\subsection{Generalizations of the nonreciprocal Kuramoto model}
\label{subsec:Kuramoto general}
The group-theoretic construction outlined above gives a natural and straightforward way to find model systems with interesting dissipationless dynamics and spontaneous generalized time-reversal breaking. 
 One simply picks a $\Phi$ which has some discrete minima, and chooses an $\Omega$ compatible with the desired residual symmetries which also switches sign depending on the degenerate minimum into which the system condenses.  In this section, we will describe several examples of more complicated systems than the nonreciprocal Kuramoto model, especially models where $G$ is a non-Abelian group.  This section is \emph{not} an exhaustive account by any means but is instead meant to provide illustrative examples.  The methods of this section readily generalize to a large class of gT-breaking phase transitions, some of which are also described in Section \ref{sec:dissipativerigid}.  

\subsubsection{Kuramoto model with $N$ species}
To start, we consider a Kuramoto-like model with $N$ \emph{different} interacting two-dimensional spins. Intuitively, this model is just the mean-field limit of a Kuramoto model with $N$ different species. The problem is defined on the manifold $M = \mathrm S^1\times \mathrm S^1 \times\cdots \mathrm S^1 = \mathrm T^N$, since each copy of $\mathrm{S}^1$ denotes the angular coordinate of a single (species of) rotor. %Though this problem is trivial in terms of our general framework for time-reversal breaking, it illustrates the method and motivates further examples. We will see that imposing full permutation symmetry among the rotors prohibits and spontaneous $\mathrm T$-breaking, although a (parity-violating) model with less symmetry allows some.

For this problem, we take our continuous left symmetry to be $G = \mathrm{SO}(2)=\mathrm{U}(1)$.  We assume that $\Phi$ additionally has a discrete symmetry group $\mathrm S_N$ corresponding to permutations of the $N$ rotors.  To break gT spontaneously, however, we must further break $\mathrm{S}_N$ to a subgroup. $\mathrm \mathrm{S}_N$ has one natural $\mathbb Z_2$ subgroup: the sign of a permutation. Hence we will consider a theory that is only invariant under time-reversal together with reversing the sign of a permutation. We may write $\mathrm S_N = \mathrm{A}_N \rtimes \mathrm{P}$, where $\mathrm{P}$ (for `parity') is isomorphic to $\mathbb{Z}_2$ and is the sign of a permutation, while $\mathrm{A}_N$ is the \emph{alternating group} (of even permutations of $N$ objects). This model has a generalized time-reversal symmetry PT, and we will study models that spontaneously break PT while preserving $\mathrm{A}_N$.  
%In other words, we would like to have a $gT$ symmetry with $g = P$.

\begin{figure}[t]
    \centering
    \Large
    \resizebox{0.7\textwidth}{!}{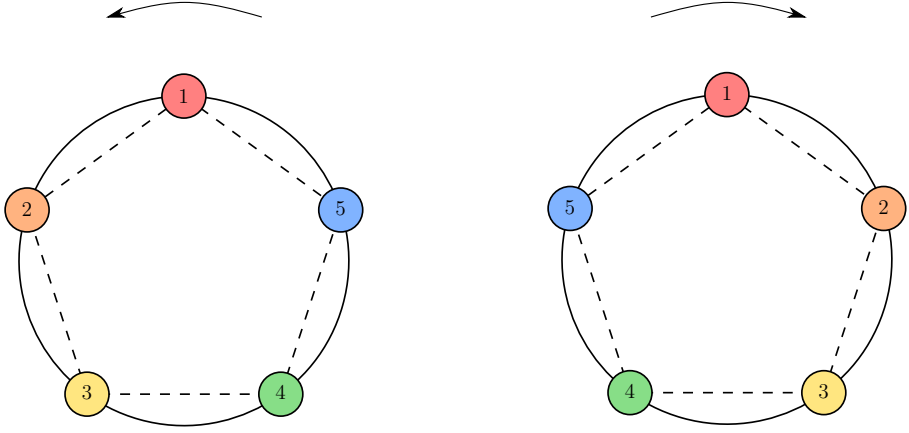}
    \normalsize
    \caption{Left: An \emph{even} permuation (read counterclockwise, starting from the top for sake of argument) of the spins on a circle rotates counterclockwise.  Right: An \emph{odd} permutation of spins rotates clockwise.  Generalized time-reversal is spontaneously broken; the direction of motion depends on the sign of the permutation of the minimum into which the system condenses.}
    \label{fig:circle}
\end{figure}

Although many of the statements made in this section do not depend on the explicit form of $\Phi$, it helps to have a simple example in mind; let us take a repulsive Kuramoto model, assuming that $N>2$:
\begin{align}
    \Phi = \frac12 \sum_{I\neq J} \cos(\theta_I-\theta_J) + \cdots. \label{eq:PhiKuramotoN}
\end{align}
The static equilibrium corresponds to the $N$ spins' being equally spaced around a circle as in \figref{fig:circle}, and this conclusion is agnostic to the actual form of the repulsive interactions.\footnote{This can be shown analytically by writing $\Phi = \frac{1}{4}|\e^{\ii\theta_1}+\cdots + \e^{\ii\theta_N}|^2 - \frac{N}{4}$.  When the $\theta$s are equally spaced on a ring, the first term (which is nonnegative) is exactly zero, thus implying it is a global minimum.  Without $\cdots$ in (\ref{eq:PhiKuramotoN}) there are many other global minima, which can be lifted by adding further terms such as $|\e^{\ii2\theta_1}+\cdots + \e^{\ii2\theta_N}|^2$, etc. }  We may then parameterize the orientation of the configuration using a rotation matrix $R_{iI} \in \mathrm{SO}(2)$.

Let us now turn to examining the T-breaking phases that could occur in such models.  To do so, with reference to \eqref{eq:rigid body SDE} and \eqref{eq:T SSB}, we must find an antisymmetric matrix $\Omega_{IJ}$ that is a body-frame ($\mathrm A_N$) invariant.  Such a matrix  must be proportional to $\epsilon_{IJ}$, the only antisymmetric $2 \times 2$ matrix, which is a body-group invariant.  We may therefore write
\begin{align}
    \Omega_{IJ} = \omega(\vb* \theta)\, \epsilon_{IJ} .
\end{align}

All that remains is to choose an $\omega$ that is invariant under even-signed permutations of the spins, and flips sign under odd ones.  For odd $N$,\footnote{If $N$ is even, then there will be particles separated by angle $\pi$ (equivalent to $-\pi$), meaning that $v(\pi)=0$ and thus $\omega=0$, and a more complicated choice is needed.} one option is to choose
\begin{align}
\label{eq:omega_Nspins_circle}
    \omega(\vb*\theta) = \prod_{I < J}v(\theta_I-\theta_J),
\end{align}
for a fixed odd function $v$, e.g., $v(\theta)=\sin(\theta)$.  This object changes sign under any single swap of two rotors, meaning in particular it is invariant under even permutations and differs by a sign for odd permutations.

In this case, the direction of motion depend on the sign of the permutation into which we condense---for example, even (odd) permutations could lead to anticlockwise (clockwise) motion, see \figref{fig:circle}.  The operation of time-reversal effectively takes us from a permutation read anticlockwise to the same one read \emph{clockwise}. 
%So if we define a parity-like operation $P$ that also reverses the permutation, changing its sign, $P$ also swaps between two minima, just like T. Accordingly, we have found the $\mathbb Z_2$ transformation that should be associated with $T$: it is just $P$.  
Hence, in the spinning phases, P and T are separately spontaneously broken, but PT is conserved.  

%We note finally that further reducing the residual symmetry group could allow other forms of spontaneous $T$-breaking, but it is not the focus of this subsection to catalogue each of these options.  

\subsubsection{Motion on a two-dimensional sphere}
\begin{figure}[t!]
    \centering
    \subfloat[]{\includegraphics[width=0.25\textwidth]{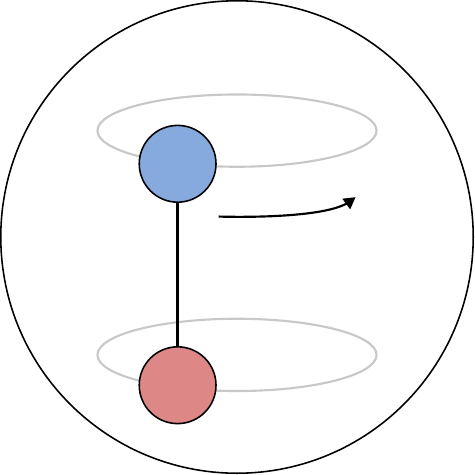}\label{fig:2s2}}\hspace{1em}
    \subfloat[]{\includegraphics[width=0.25\textwidth]{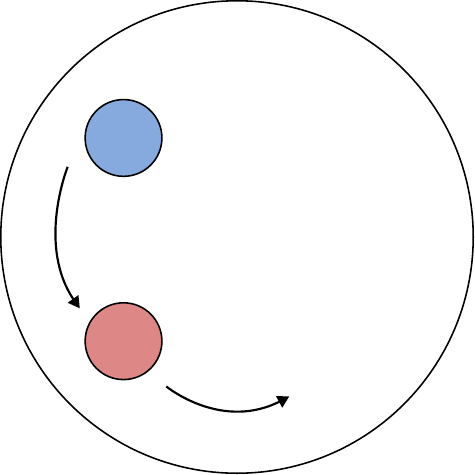}\label{fig:test}}\hspace{1em}
    \subfloat[]{\includegraphics[width=0.25\textwidth]{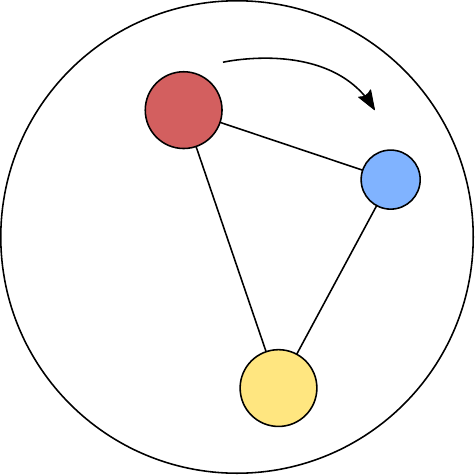}\label{fig:3s2}}
    \caption{Examples of Spontaneous $T$-breaking on $\mathrm{S}^2$.  (a) Two particles form a line and rotate together about an axis through the center of the sphere and parallel to the line joining the two particles.  In terms of equations, this is the situation of \eqref{eq:2s2}.  (b) Two particles lie on a circle and rotate around it.  This is the situation of \eqref{eq:2s22}. (c) Three particles form a triangle and rotate about an axis perpendicular to the triangle.  This is the situation of \eqref{eq:3s2}.  It is important to note that more exotic $T-$breaking could arise by demanding even less symmetry in the body frame.}
    %Four particles could, for example, form a tetrahedron.  In this case, no spontaneous $T$-breaking is possible, as discussed in \secref{sec:4s2}.  Geometrically, the lack of natural $T$-breaking corresponds to the high symmetry of the tetrahedron-there is no preferred rotation axis.  }
    \label{fig:sphere}
\end{figure}

One natural generalization of this nonreciprocal $N$-species Kuramoto model (in the mean-field limit) is to consider $N$ particles moving on the $M$-dimensional sphere $\mathrm{S}^M$.  We will illustrate a few such cases with $M=2$. As a minimal example, let us take $M=N=2$, with:  
\begin{align}\label{eq:2s22}
    \Omega_{IJ} = \epsilon_{IJK}\epsilon_{KLM}\,n^{(1)}_L n^{(2)}_{M}.
\end{align}
This term leads to nontrivial dynamics, for example, in a situation where $\Phi$ enforces the rotors to be separated by a fixed angle and the two species of particles $\vu n^{(1)}, \vu n^{(2)}$ have nonreciprocal interactions. This case may be visualized according to \figref{fig:test}. Here, the particles spin around an ``equator" as in the nonreciprocal Kuramoto model of Section \ref{sec:kuramoto} (with $M=1$).  An alternative model, with the same gT symmetry, corresponds to \begin{equation}
    \Omega_{IJ} = \epsilon_{IJK}\left(n^{(1)}_K - n^{(2)}_K\right);\label{eq:2s2}
\end{equation}
this case is depicted in \figref{fig:2s2}.

For something slightly more nontrivial, consider the case $N=3$ and $M=2$, with:
\begin{align}
\label{eq:3s2}
    \omega_{IJ} &= \epsilon_{IJK}\, \epsilon_{KLM}\, \qty(n^{(1)}_L -n^{(2)}_L)\qty(n^{(2)}_M-n^{(3)}_M)= \epsilon_{IJK}\, \epsilon_{KLM}\, \qty(n^{(1)}_Ln^{(2)}_M + n^{(2)}_Ln^{(3)}_M + n^{(3)}_Ln^{(1)}_M) \notag \\
    &\propto \sum_{\sigma \in \mathrm{S}_3} \mathrm{sign}(\sigma)\, \epsilon_{IJK}\, \epsilon_{KLM}\, n^{(\sigma(1))}_Ln^{(\sigma(2))}_M.
\end{align}
In this situation, the particles may form a triangle and rotate about an axis perpendicular to aligned with the triangle's normal vector.   Note that this normal vector is sensitive to the ``chirality" of the triangle on the sphere.  The motion in this case may be visualized according to \figref{fig:3s2}. 

If we try to go to $N=4$ and $M=2$, the situation becomes more complicated; this case, a systematic analysis (which relies on group theory) is discussed in Appendix \ref{app:S4}.

\section{Active and dissipative rigid-body dynamics} 
\label{sec:rigidbody}
Here we explore an interesting example that displays nontrivial dissipative and active phenomena: the dynamics of a free rigid body about a fixed rotation point. We will begin by casting ordinary rigid-body mechanics in a convenient Hamiltonian formalism that closely parallels the approach of \secref{sec:groupflow}.  Next, following the framework of this paper, we will modify the Hamiltonian dynamics by deriving the appropriate active (time-reversal breaking) and dissipative corrections that may appear in the presence of microscopic activity or noise.  Our constraints on the allowed active terms could be relevant for studying active systems like bacteria which can sometimes be approximated as rigid bodies~\cite{berg1993random}. Our minimal models of rotating bodies with angular-momentum conserving noise could provide toy models of (simple aspects of) galaxy formation~\cite{binney2011galactic,gunn1972galaxyformation}, tidal friction~\cite{hut1980tidal}, or other dissipative phenomena involving rotating objects in astrophysics, geophysics, and otherwise.  Another possible application is to the simpler modeling of rotating molecules interacting with atomic gases \cite{Raston2013}.

\subsection{Hamiltonian formalism of rigid-body dynamics}

Let us begin by studying ordinary rigid-body dynamics (in our universe's three spatial dimensions) in a Hamiltonian formalism.  To do so, we must first specify a set of coordinates for the rigid body, their conjugate momenta, and the form of the Poisson bracket.  With respect to fixed `space frame' axes, the configuration of such a rigid body may be specified by a rotation matrix $R_{iI}(t) \in \mathrm{SO}(3)$ from the body frame to the space frame, as described in Section \ref{sec:leftright}.
%To be explicit about this point, we have chosen to label the body frame coordinates with capital indices and the space frame coordinates with lower-case indices---exactly the same distinction as the left vs. right invariance discussed in \secref{sec:leftright}.  
Of course, being elements of $\mathrm{SO}(3)$, these matrices must be orthogonal:
\begin{align}
    \label{eq:orthogonal}
    R_{iI}R_{jI} = \delta_{ij},\ R_{iI}R_{iJ} = \delta_{IJ}.
\end{align}

We would now like to write down ordinary Hamiltonian mechanics for the problem that has configuration space SO(3) \cite{khesin}. In Hamiltonian mechanics, this corresponds to dynamics on $\mathrm{T}^*\mathrm{SO}(3)$, the cotangent bundle of the Lagrangian configuration space.  
% Next, define canonical momenta conjugate to $R_{iI}(t)$:
% \begin{align}
%     P_{iI} = \pdv{L}{\dot R_{iI}}.
% \end{align}
To do Hamiltonian mechanics, we also need a symplectic form $\omega$, related to the Poisson brackets.  The standard choice is to find an exact symplectic $2$-form $\omega$:
\begin{align}
    \omega = \dd{\alpha}
\end{align}
for a 1-form $\alpha$.  Hamiltonian mechanics comes from a particular choice of $\alpha$ commonly called the \emph{Liouville form} or \emph{canonical 1-form}~\cite{josesaletan}: 
\begin{align}
    \alpha = \frac12 P_{iI} \dd{R_{iI}}.
\end{align}
The odd-looking factor of $1/2$ will become clear later---essentially, it avoids double counting the dynamical variables.  With this choice, 
\begin{align}
    \omega = \dd{\alpha} = \frac12 \dd{P_{iI}} \wedge \dd{R_{iI}}.
\end{align}
This approach, however, poses two main complications.  First, the $R_{iI}$ variables are subject to the constraint \eqref{eq:orthogonal}, requiring the introduction of a Lagrange multiplier.  We would like to find clever coordinates where this can be neglected because the dynamics inherently respects the orthogonality constraint to begin with. Secondly, the momenta $P_{iI}$ have one body frame and one space frame index, so it is difficult to interpret these objects as physical momenta or angular momenta. But there is a trick that will allow us to obtain nontrivial Hamiltonian evolution for $\dot P$ in this situation while elegantly imposing the constraint.  Instead of $P_{iI}$, it helps to consider different variables $P_{IJ}$ (two body-frame indices) defined through:
\begin{align}
    P_{iI} = R_{iJ}P_{JI}.
\end{align}
This is simply a change of variables, but the body-frame object $P_{IJ}$ will end up encoding the angular momentum in the body frame coordinates, and this is the degree of freedom in which the standard Euler equations of rigid-body motion are expressed. But $R$ and $P_{IJ}$ are no longer canonical, so we must calculate their Poisson brackets.  Making the substitution to the $P_{IJ}$ coordinates, the Liouville form $\alpha$ becomes
\begin{align}
    \alpha = \frac{1}{2} R_{iJ} P_{JI} \dd{R_{iI}}.
\end{align}
Now is a good time to see how to efficiently enforce the orthogonality constraint (\ref{eq:orthogonal}).  Applying $\dd$ to the constraint, we find:
\begin{align}
    R_{iJ} \dd{R_{iI}} = - R_{iI} \dd{R_{iJ}}.
\end{align}
If this constraint held, observe that only the \emph{antisymmetric} part of the matrix $P_{IJ}$ contributes to $\alpha$---the symmetric part vanishes due to contraction with the antisymmetric matrix $R_{iJ}\dd{R_{iI}}$.  Since the symmetric part of $P_{IJ}$ does not appear in $\alpha$, it will also be absent from $\omega$, meaning it will have trivial dynamics.  It is natural, then, to work with only the antisymmetric part of $P_{IJ}$ from the start---we obtain the same $\omega$ as if we had begun with the full unconstrained $P$.  Moreover, we now have 3 momentum coordinates (the number of independent antisymmetric $3\times 3$ matrices) that can be ``conjugate" to the 3 degrees of freedom in the orthogonal matrix $R_{iI}$.  For clarity, let us define
\begin{align}
    L_{IJ} = \frac12(P_{IJ}-P_{JI})
\end{align}
as the momentum variables for this problem.  The key point is: we can take \textit{a priori} the antisymmetric matrices $L_{IJ}$ to be our degrees of freedom, and this choice automatically imposes the constraint on $R_{iI}$, \eqref{eq:orthogonal}. 
The $R_{iI}$ variables are now formally \emph{unconstrained}---so long as we choose initial conditions where $R_{iI}(0)$ is orthogonal, it will stay that way for all times if we can simply enforce a \emph{linear constraint} that $L_{IJ} = -L_{JI}$, which is not difficult and can be done at the outset.\footnote{Abstractly, one can think about this coordinate choice as follows. The cotangent bundle can be decomposed into position coordinates on the group SO(3), and Lie algebra $\mathfrak{so}(3)$-valued momentum coordinates $L_{A}$, given in (\ref{eq:Loneindex}).}
\begin{align}
    \alpha = \frac12 R_{iJ}\, L_{JI}\,\dd{R_{iI}}.
\end{align}

Having implemented the constraint and thus established that antisymmetric matrices $L_{IJ}$ are good momentum variables for the rigid-body problem, we must now calculate the symplectic form $\omega$. Carrying out the derivative of the Liouville form, we find
\begin{align}
    \omega = \frac12(L_{JI} \dd{R_{iJ}} \wedge \dd{R_{iI}} + R_{iJ}\dd{L_{JI}} \wedge \dd{R_{iI}}).
\end{align}
The unusual coupling between $R$ and $R$ in this form is responsible for the nontrivial Hamiltonian dynamics in this problem.  To explore it, let us first convert to the physical variables $L_A$ defined by
\begin{align}
    L_{IJ} = \epsilon_{IJA}L_A. \label{eq:Loneindex}
\end{align}
With this definition, we find the symplectic form
\begin{align}
\label{eq:physical_omega}
    \omega = \frac12(\epsilon_{JIK}L_K \dd{R_{iJ}} \wedge \dd{R_{iI}} + \epsilon_{JIK} R_{iJ}\dd{L_K} \wedge \dd{R_{iI}}).
\end{align}
It corresponds to a matrix according to:
\begin{align}
\label{eq:omega-rigid-body}
    \omega = \frac12 \omega_{\alpha \beta}\, \dd \xi_\alpha \wedge \dd \xi_\beta,
\end{align} 
Our goal now is to use $\omega$ to find the \emph{Poisson brackets} between the dynamical variables. The matrix elements of the \emph{inverse} of $\omega_{\alpha \beta}$ matrix yield the Poisson bracket according to 
\begin{align}
    \{f,g\} = \pdv{f}{\xi_\alpha}\, \omega^{-1}_{\alpha\beta}\pdv{g}{\xi_\beta}.
\end{align}
To calculate $\omega^{-1}$, however, we must be careful to restrict $R_{iI}$ to the three physical degrees of freedom.  To do so, it is convenient to momentarily consider a local patch of $\mathrm{SO}(3)$ parameterized by the Lie algebra-valued variables $\phi_{A}$ defined by:
\begin{align}
    R_{iI} = R^0_{iA}(\delta_{AI} + \epsilon_{AIB}\phi_B),
\end{align}
where $R^0$ is a constant element of $\mathrm{SO}(3)$.  A short calculation verifies that $\omega$ is given as 
\begin{align}
    \omega = -\frac12 \epsilon_{IJK} L_K \dd{\phi_I} \wedge \dd{\phi_J} + \dd{L_K}\wedge \dd{\phi_K}.
\end{align}
Here, we have expanded each term to lowest nontrivial order in $\phi$, which will ultimately lead to Poisson Brackets accurate to $O(\phi^0)$.  We may now write the matrix $\omega$, in $3\times 3$ blocks (with ordering $(\phi, L)$, as:
\begin{align}
    \omega_{IJ} = \begin{pmatrix}
        -\epsilon_{IJK}L_K & -\delta_{IJ}\\
        \delta_{IJ} & 0 
    \end{pmatrix}
\end{align}
Its inverse is readily found to be:
\begin{align}\label{eq:rigid-omega-inverse}
    \omega^{IJ} = 
    \begin{pmatrix}
        0 & \delta_{IJ} \\
        -\delta_{IJ} & -\epsilon_{IJK}L_K
        \end{pmatrix}
\end{align}
This result allows us to read off Poisson brackets between $\phi, L$, which in turn allow us to calculate the Poisson brackets between $R_{iI}, L_K$ as:
\begin{subequations}\begin{align}
    \{R_{iI},L_K\} &= R^0_{iJ}\epsilon_{JIB}\qty{\phi_B, L_K} = R^0_{iJ}\epsilon_{JIK} \rightarrow R_{iJ}\epsilon_{JIK}\label{eq:R-poisson}\\
    \{R_{iI},R_{jJ}\} &= 0, \\
    \{L_I,L_J\} &= -\epsilon_{IJK}L_K. \label{eq:ang-poisson}
\end{align}\end{subequations}
For \eqref{eq:R-poisson}, because our Poisson brackets are accurate only to order $\phi$, it is appropriate to replace $R^0_{iJ} = R_{iJ}$.  Additionally, the standard Poisson brackets for angular momentum \eqref{eq:ang-poisson} are read immediately from \eqref{eq:rigid-omega-inverse}, albeit with a relative sign, which is a consequence of $L_I$ being attached to the body frame rather than the space frame.

Now that the coordinates, momenta, and Poisson brackets are known, the next step is to actually specify a Hamiltonian for the problem.  Left SO(3)-invariance, together with the orthogonality constraint, implies that $H$ can only depend on $L_A$. The natural minimal choice is:
\begin{align}\label{eq:H rigid}
    H = \frac12 L_I I^{-1}_{IJ} L_J,
\end{align}
where $I^{-1}$ is a symmetric constant matrix compatible with the symmetries of the rigid body.  Defining $\vb*\omega = I^{-1} \vb L$ and $\Omega_{IJ} = \epsilon_{IJK}\omega_K$, we may then compute Hamilton's equations ($\dot{f} = \{f,H\}$) as
\begin{subequations}\begin{align}
    \dot{R_{iI}} &= R_{iL}\epsilon_{LIK} (I^{-1} \vb L)_K=R_{iL}\Omega_{LI},\\
    \dot{\vb L} &= \vb L \times \vb* \omega = -\vb*\omega \times (I\vb* \omega).
\end{align}\end{subequations}
The first line  verifies that the object $I^{-1} \vb L$ corresponds to the angular velocity in the rotating reference frame, and the second line is called \emph{Euler's equations}, giving the evolution of the angular momentum in the rotating frame~\cite{josesaletan}.  Evolution of $R(t)$ preserves the orthogonality of $R$ automatically, as it should, because $\Omega$ is antisymmetric. 

Finally, before turning to derive active and dissipative corrections to Hamiltonian mechanics in this context, it helps to put our results into the language of Sec.~\ref{sec:EFT}. To that end, we define the ``chemical potentials'' corresponding to $R_{iI}$ and $L_{I}$ as
\begin{subequations}
    \begin{align}
        \mu_{iI} &= \pdv{H}{R_{iI}} = 0,\\
        \nu_{I} &= \pdv{H}{L_I} = \omega_I\ \ (\nu_{IJ} = \epsilon_{IJK}\nu_K).
    \end{align}
\end{subequations}
For brevity, let us also introduce a shorthand notation for derivatives with respect to $R_{iI}$ and $L_a$:
\begin{subequations}\begin{align}
    \partial_{iI} &\equiv \pdv{R_{iI}}, \\
    D_{K} &\equiv \pdv{L_K}. 
\end{align}\end{subequations}
With these definitions, the Fokker-Planck operator for dissipationless Hamiltonian rigid-body dynamics may be written as
\begin{align}
\label{eq:rigidbodyhamiltonian}
    W = \p_{iJ} R_{iK}\epsilon_{KJA}\nu_A+ D_{A}\epsilon_{ABC}L_B\nu_C.
\end{align}
To be explicit, let us show directly that this $W$ is invariant under T.  For the first term, notice that $(\p_{iJ}R_{iK}) \nu_{KJ}=3\delta_{JK}{\nu_{KJ}} = 0$, due to $\nu_{JK}= -\nu_{KJ}$. Then under time-reversal, using that momenta $L_I$ are odd:
\begin{align}
    \p_{iJ} R_{iK}\nu_{KJ} \rightarrow R_{iK}(-\nu_{KJ})(-\p_{iJ}) = \p_{iJ} R_{iK}\nu_{KJ}.
\end{align}
Similarly, for the second term $(D_A\epsilon_{ABC}L_B \nu_C= 0)$, meaning that 
\begin{align}
D_A\epsilon_{ABC}L_B\nu_C\rightarrow \epsilon_{ABC}L_B\nu_C (D_A + \nu_A) = D_A\epsilon_{ABC}L_B\nu_C - \epsilon_{AAC}\nu_C - \epsilon_{ABC}L_B \frac{\partial^2 H}{\partial L_A\partial L_C} = D_A\epsilon_{ABC}L_B\nu_C.
\end{align}
Accordingly, we find that \eqref{eq:rigidbodyhamiltonian} is time-reversal symmetric. With the Hamiltonian mechanics derived in this subsection as a starting point, we are now ready to study modifications to $L$ (and therefore $W$) corresponding to activity or dissipation.

\subsection{Dissipative rigid-body dynamics}\label{sec:dissipativerigid}
Having cast rigid-body dynamics in a convenient Hamiltonian formalism, we may now apply the methods of this work to explore the natural dissipative terms that can complement the Hamiltonian dynamics of rigid bodies.  We will be especially interested in dissipative terms that violate energy conservation (associated with the macroscopic rotation of the collective body) while \emph{preserving} angular momentum conservation. This is certainly the situation for some systems of physical interest including stars and galaxies~\cite{binney2011galactic}, although neither may strictly be modelled as a rigid body.  

In order to impose angular momentum conservation, we will have to again use the Noether theorem of \secref{sec:noether}, so the first step is to write down the conserved quantity (angular momentum) interms of our variables $R_{iI}$ and $L_A$.  To obtain the angular momentum in the space frame in terms of the body frame objects $L_A$, we must pass to the space frame:
\begin{align}
    L_{a} = R_{aA} L_A
\end{align}
We would like all three components $L_a$ to be conserved. 
According to the Noether theorem \eqref{eq:noether}, conserving $L_a$ corresponds to demanding $W$ is invariant under the following shifts:
\begin{subequations}
\label{eq:angmomentshift}
    \begin{align}
    \p_{iJ} &\rightarrow \partial_{iJ} + \pdv{L_a}{R_{iJ}} = \p_{iJ} + \delta_{ai}L_J, \\
    D_{B} &\rightarrow D_{B} + \pdv{L_a}{L_B} = D_{B} + R_{aB}.
    \end{align}
\end{subequations}
The question addressed in this section is: what form of \emph{dissipative} terms are allowed, if the above shift symmetries are to be demanded?

Before tackling the question of dissipative terms, it is instructive to check that the ordinary Hamiltonian mechanics we have derived for the rigid body conserves angular momentum.  This amounts to checking that \eqref{eq:rigidbodyhamiltonian} is invariant under the shift symmetries \eqref{eq:angmomentshift}.  The change in the $W$ from \eqref{eq:rigidbodyhamiltonian} under \eqref{eq:angmomentshift} is then
\begin{align}
    \delta W &= \delta_{ai}L_I  R_{iJ}\epsilon_{IJB}\nu_B + R_{aA}\epsilon_{AJK}L_J \nu_K = 0.
\end{align}
As a result, we conclude that the Hamiltonian evolution of a rigid body conserves angular momentum in the space frame.

To find a dissipative term that conserves angular momentum, we should start by looking for an motif involving $\partial$ and $D$ that is invariant under \eqref{eq:angmomentshift}.  It is natural to simply look for such a term that is a linear combination of $\partial$ and $D$.  Following the same logic as above, one may show that the following object is invariant:
\begin{align}
    M_A = \p_{iJ} R_{iK} \epsilon_{AKJ}+ D_{I}\epsilon_{AIJ} L_J.
\end{align}
where $\epsilon_{AKJ} = \varepsilon^A_{KJ}$, for fixed $A$, are the generators of SO(3) in the adjoint representation.  We may combine these motifs into a dissipative term entering the Lagrangian, giving a general dissipative term at lowest order in the fields $R$ and $L$:
\begin{align}
      W^\prime = -(\p_{iJ} R_{iK} \varepsilon^{\alpha}_{KJ}+ D_A\varepsilon^\alpha_AB L_B)Q^{\alpha\beta}( R_{iK} \varepsilon^{\beta}_{KJ}\p_{iJ}+ \varepsilon^\beta_{AB} L_B(D_A+\nu_A)).
\end{align}
Here, the matrix $Q^{\alpha \beta}$ is positive semidefinite, and related to the microscopic noise variance in a way that we will make explicit shortly. The symmetric $Q$ matrix must be chosen to be invariant under the body group---the `right group invariance' of \secref{sec:classify SSB}. For the purposes of this work: we will focus mostly on the case of isotropic noise, $Q^{\alpha \beta} = \Gamma \delta^{\alpha\beta}$---in this simple case, the dissipative part of the Lagrangian becomes:
\begin{align}
    W^\prime = -\Gamma(\p_{iJ} R_{iK} \epsilon_{AKJ}+ D_{I}\varepsilon^\alpha_{IB} L_B)(\p_{iJ} R_{iK} \epsilon_{AKJ}+ (D_{I}+\nu_I)\varepsilon^\alpha_{IB} L_B).
\end{align}
Let us now explore the consequences of $W^\prime$ for the equations of motion.  The part of $W^\prime$ featuring only one derivative is:
\begin{align}
   W^\prime \sim -\Gamma (\p_{iJ} R_{iK} \epsilon_{AKJ}+ D_{I}\epsilon_{AIB} L_B)(\nu_I \epsilon_{AIB}L_B)
\end{align}
The dissipative part of the Lagrangian modifies both the equations of motion for $R_{iJ}$ and $L_{A}$. Beginning with the modification of $\dot R$ (including the contribution both of $W$ and $W^\prime$:
\begin{align}
    \dot{R}_{iI} = R_{iJ}\epsilon_{JIL}(\nu_L + \Gamma(L\times \vb*\nu)_L).
\end{align}
Importantly, the dissipative term modifies the relationship between angular velocity $\vb*\omega$ and angular momentum $\vb L$:  we find
\begin{align}\label{eq:dissipative-angular-velocity}
    \vb*\omega = \vb*\nu + \Gamma \vb L\times \vb*\nu.
\end{align}
Secondly, for $\dot{\vb L}$, we find a modification of Euler's equation:
\begin{align}\label{eq:dissipative-rigid-body}
    \dot{\vb L} = \vb L \times \vb*\nu + \Gamma\, \vb L \times (\vb L \times\vb*\nu) = \vb L \times \vb* \omega.
\end{align}
 Already, we can see a striking similarity to the Landau-Lifshitz-Gilbert damping term that arises in the Heisenberg spin chain (cf. \eqref{eq:llg}). This similarity is nontrivial: in the Heisenberg spin chain, we demanded only that the total angular momentum $S^2$ is conserved, while in the rigid-body case we conserve each component of $\vb L$ separately. The reason for the similarity is that the length of total angular momentum $\mathbf{L}\cdot \mathbf{L}$ is the same in both the space and body frame; however, it was not trivial that we can always choose dynamics for $R_{iI}$ that preserves the remaining two components of angular momentum!  Further, we have seen that the damping must break the simple relation between angular velocity and angular momentum, in order to conserve angular momentum in the presence of noise.  This is a subtle result that manifestly follows from our ``cookbook" protocol for developing theories of dissipative dynamics based on symmetry.
\begin{figure}
    \centering
    \includegraphics{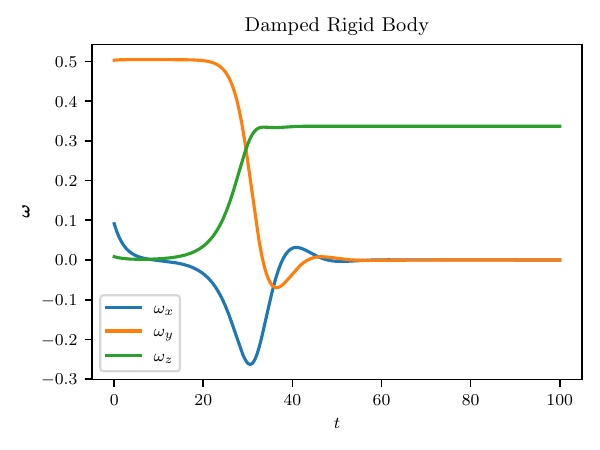}
    \caption{Simulation of the dynamics of a rigid body under \eqref{eq:dissipative-rigid-body}.  For the simulation, we use dimensionless units and work in the principal axis basis, where $I$ is dimensionless with $I_{xx}=1, I_{yy} = 2, I_{zz} = 3$.  We choose $\Gamma= 0.5$. The initial conditions are chosen such that the body rotates about the unstable $\vu y$ axis. As $t \rightarrow \infty$, the body's angular velocity $\vb* \omega$ (calculated according to \eqref{eq:dissipative-angular-velocity}) aligns with the body frame $\vu z$ axis, that is, the axis with largest moment of inertia.}
    \label{fig:enter-label}
\end{figure}

It is also worth considering the situation of a rigid body whose angular momentum is dissipating in addition to its energy. This would be the case, for example, for rotons in contact with a bath that carries angular momentum~\cite{degroot1962}. In this case where the equation of motion for $\vb L$ is subject to isotropic noise, we simply add the usual quadratic disspative term to $W$:
\begin{align}
    W' = -\Gamma D_I (D_I + \nu_I)
\end{align}
for $\Gamma > 0$ . This piece adds a damping term to Euler's equations:
\begin{align}\label{eq:eom rigid diss}
    \dot{\vb L} = \vb L \times \vb* \omega - \Gamma \vb*\omega.
\end{align}
A naive relaxation term $\dot{\vb L} \sim -\Gamma \vb L$ does not satisfy time-reversibility.

\subsection{Active rigid-body dynamics}
In this final subsection, we will explore the T-odd terms that may appear for rigid-body motion. What other dissipationless terms may be added in the absence of time-reversal symmetry?

Let us first try to guess modifications of Euler's equations that violate angular momentum conservation.  In the limit of weak noise, a general dissipationless term that is odd under time-reversal symmetry is:
\begin{align}\label{eq:W rigid active}
    W_{\mathrm{T-breaking}}=\p_{iJ} R_{iK}A_{KJ} + D_I B_{IJ}(D_J+\nu_J)
\end{align}
for an antisymmetric matrices $A$ and $B$ that may depend on even powers of the coordinates $L$ (due to left-SO(3) symmetry).  These matrices $A$ and $B$, having two body-frame indices, must also be an invariant tensor of the body group.  In the case where there exists a constant, antisymmetric matrix that is a body group invariant, we could then include a terms with constant $A$ or $B$, corresponding to driven rotation and precession about the axes chosen by $A$ and $B$.  In cases where such a constant antisymmetric invariant tensor does not exist, creating a body group-invariant object requires $A$ and $B$ to have dependence on the coordinates.  For instance, choosing $A = \nu$, $B = L$ corresponds to Hamiltonian mechanics, which is dissipationless but time-reversal invariant. To find a $T$-odd dissipationless term, one should choose $A$ and $B$ to be $T$-even.

We may, however, also be interested in active systems that still conserve angular momentum.  In that case, Noether's theorem constrains the form of $A$ and $B$ further.  If we wish to conserve all three components of $L_i$ (lab frame angular momentum), then \eqref{eq:angmomentshift} demands $A = c(\vb L)\nu$ and $B = c(\vb L) L$. In particular, the choice $c(\vb L) = 1$ corresponds to plain Hamiltonian mechanics.  In the absence of a $T$-odd scalar $c(\vb L)$ (a time-reversal pseudoscalar), angular-momentum conserving active terms are not possible.

If angular momentum is \emph{not} conserved however, $A$ and $B$ may be any antisymmetric matrices compatible with body frame symmetries. Let us write down a simple example where $B = 0$ (maintaining the form of Euler's equation) and:
\begin{align}
    A_{IJ} = \lambda \epsilon_{IJK}\epsilon_{KLM} L_N \nu_M
\end{align}
This modifies the relationship between angular velocity $\vb*\omega$ and angular momentum $\vb L$:
\begin{align}
    \vb*\omega = \vb* \nu + \lambda (\vb*\nu\times \vb L).
\end{align}
The equation of motion for $\vb L$ becomes
\begin{align}\label{eq:eom rigid active}
    \dot{\vb L} = \vb L \times \vb*\omega - \lambda (\vb* \nu \times \vb L).
\end{align}
Of course, in situations with even less symmetry, other active terms are possible.  If the body has a preferred axis, say $\vu z$, then one may choose $A_{IJ} = \epsilon_{IJK} \hat z_K$, corresponding to driven rotation about this axis.  Choosing $B_{IJ} = \epsilon_{IJK} \hat z_K$ instead corresponds to driven precession about the $\vu z$ axis.

\section{Field theory}\label{sec:continuum}
We now discuss genuine field theories using the formalism developed above.  Both to make contact with recent literature \cite{eft1,eft2}, and also to keep the discussion a little cleaner in theories where the number of fields can proliferate quickly, in this section we will use the notation $\pi_a = -\ii \partial_a$, and $H=-\ii W$.  Since we will work almost exclusively in linear response and in the limit of weak noise, it is simple to convert between the path integral approach described in Sec.~\ref{sec:MSR}, and the operator formalism used in the rest of the paper.   However, we always present formulas for $\mathcal{L}$ in such a way that they can be easily converted into operator expressions if desired.
 
The focus of this section will be on active matter whose dynamics break a $\mathbb{Z}_2$ symmetry (usually spatial parity in two dimensions) and time-reversal symmetry (but not their combination). Microscopic origins of such systems can include includes spinning particles, granular gases, and Brownian rollers \cite{CAM_epl,VV_2017_odd,VV_2021_fluctuating}. Although these microscopic models make particle-based simulations possible, a field theory can elucidate the aspects of the physics which are universal and model-independent.  Of course, our Wilsonian approach to stochastic dynamics is naturally designed for such effective theories.  

As we will see, our EFT approach allows a more streamlined derivation of a number of recent results in the literature on active solids and fluids, and in some cases also highlights some subtle or overlooked phenomena.

\subsection{Odd diffusitivity}\label{sec:odd diffusive}
To warm-up to the study of hydrodynamics, let us begin by considering a minimal model.  Firstly, we remind the reader that hydrodynamics is the effective theory that describes how a thermalizing system reaches equilibrium \cite{}.  The degrees of freedom within hydrodynamics are associated with the densities of conserved quantities, and/or the Goldstone bosons associated with any spontaneously broken continuous symmetries.  We will see examples of each in this paper, but we will leave the development of an exhaustive EFT framework for active fluids to another paper.

Now, we turn to our minimal model.  Suppose that we have a single conserved scalar field $n(\vb x,t)$---e.g., the number density of conserved particles---defined on a $d$-dimensional region coupled to a thermal bath of temperature $T_0=\beta_0^{-1}$.  We consider expanding the fluid around static equilibrium $n(\vb x,t) = n_0$: perturbations are denoted with $\delta n \equiv n - n_0$.  By the central limit theorem (and Wilsonian philosophy), $\Phi$ can be taken to be quadratic, at leading order:
\begin{align}
\label{eq:diffusionPhi}
    \Phi = \beta_0 \int \dd[d]{x} \frac{1}{2\chi}(\delta n)^2.
\end{align}
The constant $\chi$ is known as the charge/number susceptibility.  We will now derive the minimal Lagrangian in terms of $\delta n(\vb x)$ and its conjugate noise field $\pi(\vb x)$.  For sake of illustration, let us briefly deal first with the thermal case in which $T$-symmetry is present, and we will demand parity $P$ as well.  Additionally, since the total charge $\delta Q = \int \dd[3]{x} \delta n$ is conserved, the Noether theorem demands that the Lagrangian is invariant under  shifting $\pi \to \pi + 1$.  Then the simplest Lagrangian is 
\begin{align}
    \label{eq:normal_diffusion}
    \mathcal L = \pi \pd{t} \delta n + \ii \beta_0^{-1} \chi D_{ij}\pd{i} \pi \pd{j} \, (\pi - \ii \mu)
\end{align}
where $D_{ij}$ (positive semidefinite) are the diffusion constants, with prefactor $\beta^{-1}\chi$ chosen for convenience.  The chemical potential 
\begin{align}
    \mu(x) = \fdv{\Phi}{(\delta n(x))} = \frac{\beta_0\delta n}{\chi}.
\end{align}In the It\^o scheme, \eqref{eq:normal_diffusion} describes, through varying w.r.t the noise field $\pi$, a fluctuating diffusion equation
\begin{align}\label{eq:diffusion eom}
    \pd{t} \delta n = D_{ij} \p_i \p_j \delta n + \xi,
\end{align}
where the white noise $\xi$ satisfies
\begin{align}
\label{eq:whitederivativenoise}
    \expval{\xi(\vb x, t)\xi(\vb x^\prime, t^\prime)} = 2\beta_0^{-1} \chi D_{ij} \,\p_i\p_j \delta(\vb x - \vb x^\prime)\, \delta(t - t^\prime).
\end{align}
This is how the usual Fick's law arises from fluctuating hydrodynamics with derivative white noise. For normal isotropic achiral systems, $D_{ij}=D\delta_{ij}$ is the most general form possible, with $D\geq 0$ required for stability.  

For chiral systems, one can instead consider $D_{ij}=D^{\odd}\epsilon_{ij}$.  $D^{\mathrm{odd}}$ is unsurprisingly called the odd diffusivity \cite{odd_diff} . Note that $D^{\odd}$ does not contribute to the equation of motion:  this can be seen either  by noting that $D^{\mathrm{odd}}$ leads to a total derivative term in $\Phi$, or by evaluating the equation of motion to be $\partial_t n = D^{\mathrm{odd}}\epsilon_{ij}\partial_i\partial_j n = 0$.

Nevertheless, it is instructive to see how such a term can arise in the action, because, e.g., at the nonlinear level odd diffusivity can contribute to the equations of motion.\footnote{For example, imagine there are two species of charge $n^a,a=1,2$ satisfying diffusion equation $\p_t\delta n^a = \p_i(D_{ij}(n^b)\p_j\delta n^a)$. Taking $D_{ij}(n^b) = D^{\mathrm{odd}}\epsilon_{ij} n^b n^b$ as T-odd, we have $\p_t \delta n^a = D^{\mathrm{odd}}\epsilon^{ij} n^b \p_i \delta n^b \p_j\delta n^a$.} For the sake of simplicity, let us assume an isotropic system and work in $d=2$ so that there exists a rank-2 isotropic antisymmetric tensor.  Assume the stationary distribution \eqref{eq:diffusionPhi}.

Then way may write down the effective Lagrangian that satisfies $gT$ symmetry with $g = P$.  The minimal effective Lagrangian is given by
\begin{align}\label{eq:L diffusion active}
    \mL(\delta n,\pi) = \pi\p_t \delta n -\beta_0^{-1}   (\p_i \pi) \chi D^{\odd} \epsilon_{ij} \p_j \ii (\pi - \ii \mu)+\ii \beta_0^{-1} \chi D \p_i \pi \p_i(\pi-\ii \mu).
\end{align}
The antisymmetric of $D^{\antisym}_{ij} = D^{\odd}\epsilon_{ij}$ allows this coefficient to enter at linear order in $\pi$, meaning it is not related to noise---the statistics of the noise are still \eqref{eq:whitederivativenoise}. Upon variation of $\pi$, we find \eqref{eq:diffusion eom} with $D$ being the dissipative diffusion constant; $D^{\odd}$ is now a T-odd force in the effective field theory.  Although $D^{\odd}$ does not enter the equation of motion, it may nonetheless affect $\delta n$ when the boundary conditions depend only on the flux $J_i \equiv D_{ij}\pd*{j}\delta n$.  

\subsection{Odd elasticity}
We now consider an example of an active continuous medium: the two-dimensional odd elastic solid \cite{VV_odd_review}.  Following much of the literature \cite{odd_elasticity_2020}, we will focus on an overdamped elastic solid, such that the only degrees of freedom are the displacement fields of the solid $u_i(x)$.  These are Goldstone bosons of a spontaneously broken translation symmetry $\mathrm{U}(1)^2$, and for simplicity, following \cite{odd_elasticity_2020}, we assume that these are the only hydrodynamic degrees of freedom.

As usual in this framework, the first thing to do is to postulate the form of $\Phi$.  For the odd elastic solid, there are two plausible choices, and we will explore them both in turn.

\subsubsection{Gyroscopic potential}
The first plausible choice for $\Phi$ is
\begin{equation}\label{eq:odd_elasticity_phi1}
\Phi = \int \mathrm{d}^2x\,\left[c_1(\partial_i u_j)^2 + c_2(\partial_n u_n)^2\right]
\end{equation}
The reason this choice is natural is that as Goldstone bosons, we may expect that $u_i \rightarrow u_i+a_i$ should be a shift symmetry for any constant vector $a_i$.  This seemingly mundane choice will have physical consequences which we will soon discuss.  Notice that $\mu$ now has derivatives: treating $c_{1,2}$ as constants, we find that \begin{equation}
    \mu_i(x) = \frac{\delta \Phi}{\delta u_i(x)} = -c_1 \partial_j \partial_j u_i - c_2 \partial_i \partial_j u_j
\end{equation}
already contains two derivatives.

Next we turn to the Lagrangian $\mathcal{L}$.  Assuming PT-symmetry, and studying \emph{overdamped} Goldstone dynamics where $u_i$ (but not momenta) are the only degrees of freedom, we find
\begin{equation}
\label{eq:odd_elasticity_lagrangian}
    \mL(u,\pi) = \pi_i\partial_t u_i + \ii \pi_i\left(Q_1\delta_{ij}+Q_2\epsilon_{ij}\right)\left(\pi_j - \ii\mu_j\right).
\end{equation}
$Q_1$ corresponds to T-even effects, while $Q_2$ corresponds to T-odd effects.  The equations of motion are then:
\begin{equation}
\label{eq:EOM_first_oddelastic_pot}
    \partial_t u_i = -\left(Q_1\delta_{ij} + Q_2\epsilon_{ij}\right)\mu_j
\end{equation}
To understand what these mean, it is helpful to consider plane waves of the form $u_i \sim \e^{\ii(kx-\omega t)}$:
\begin{align}
    \partial_t u_i &= \left(Q_1\delta_{ij} + Q_2\epsilon_{ij}\right)\left(c_1k^2u_j + c_2 k_jk_iu_i\right) \notag \\
    &= \left(Q_1c_1k^2\delta_{ij}u_j + Q_2c_1k^2\epsilon_{ij}u_j +Q_1c_2\delta_{ij}k_jk_iu_i + Q_2c_2\epsilon_{ij}k_jk_lu_l\right).
\end{align}
Because we choose $k_i$ to lie in the $x$-direction, we find (neglecting noise contributions) 
\begin{subequations}
    \begin{align}
    \partial_t u_x &= -k^2\left(Q_1(c_1+c_2)u_x +Q_2c_1u_y\right) = -k^2 \left((B+\mu)u_x + K^\circ u_y\right) \\
    \partial_t u_y &= -k^2\left(Q_1c_1u_y - Q_2(c_1+c_2)u_x\right) = -k^2 \left(\mu u_y - (A+K^\circ) u_x\right).
\end{align}
\end{subequations}
where we have introduced the bulk $B$ and shear $\mu$ elastic moduli, along with their odd counterparts $A$ and $K^\circ$ respectively \cite{odd_elasticity_2020}. 
Observe that the following equality holds:
\begin{equation}
\label{eq:gyroscopic_ratio}
    \frac{A}{K^{\circ}}=\frac{B}{\mu}.
\end{equation}

An example of a system that obeys this ``gyroscopic condition" is found in \cite{gyroscopic_metamaterial}.\footnote{Mechanistically, this new force law is just a rotation of the passive force one would expect of Hookean springs. Since each spring has the same odd spring constant, this system rotates each force by the same angle. In other words, the net force is just a rotation of the passive force, which is exactly what a gyroscope does.} Consider a triangular lattice consisting of springs that have an additional odd spring constant: they obey the Hookean spring force law, but with a chiral transverse force proportional to the odd spring constant: \begin{equation}
    F_i = \left(k \delta_{ij} + k^\circ \epsilon_{ij}\right) \delta r_j.
\end{equation}
Here $\delta r_i$ is the displacement of each spring relative to its rest displacement.  The force-displacement matrix for this system is not symmetric which implies the linear response is nonconservative and thus time-reversal symmetry is broken. The coefficient $k^\circ$ multiplies the parity-odd $\epsilon_{ij}$, but the dynamics is overdamped and thus this dissipationless force also is odd under time-reversal; hence the system is overall PT-symmetric and it must obey (\ref{eq:gyroscopic_ratio}).
\begin{figure}[t!]
    \centering
    \subfloat[]{\includegraphics[width=0.25\textwidth]{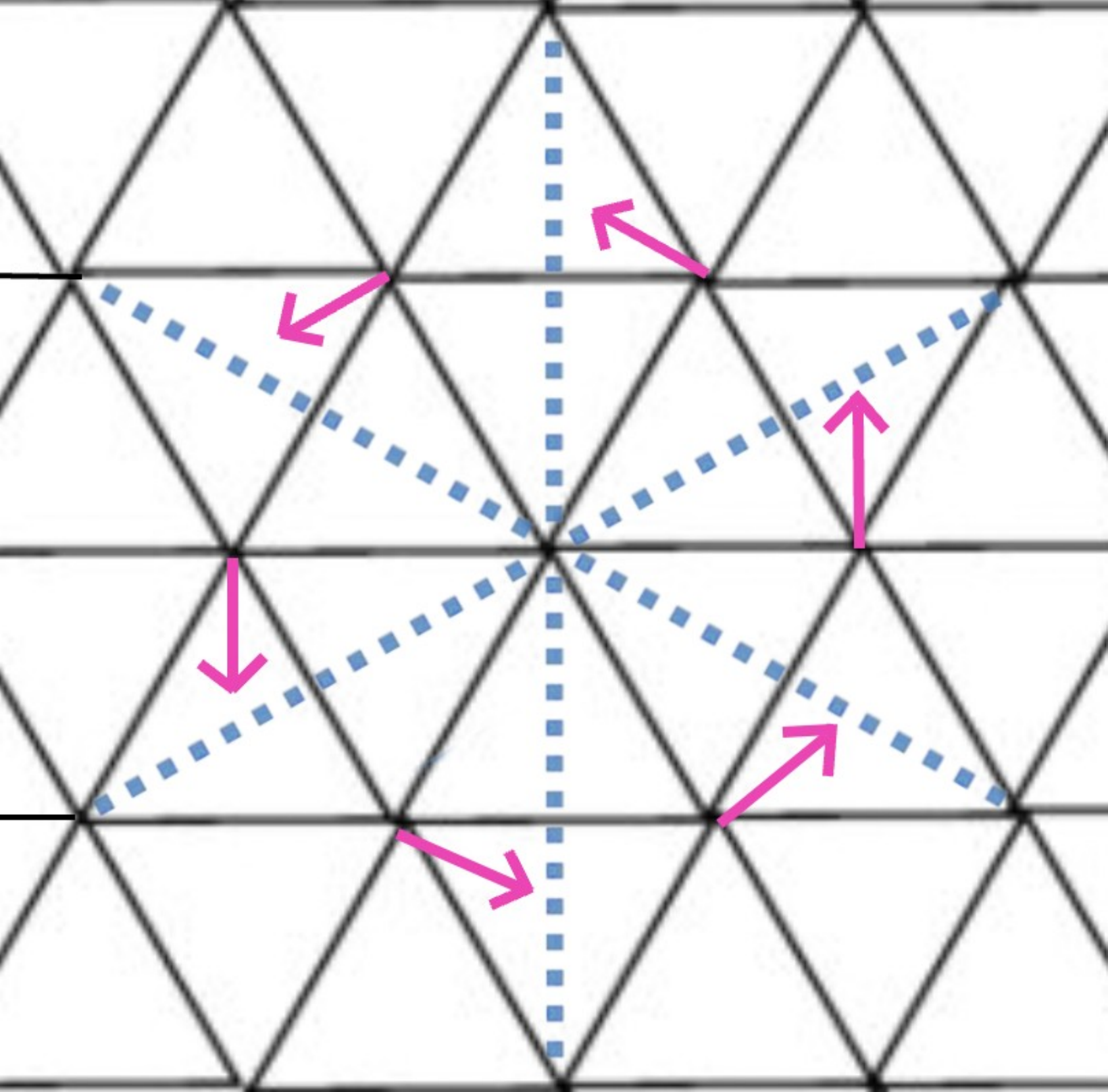}}\hspace{1em}
    \subfloat[]{\includegraphics[width=0.25\textwidth]{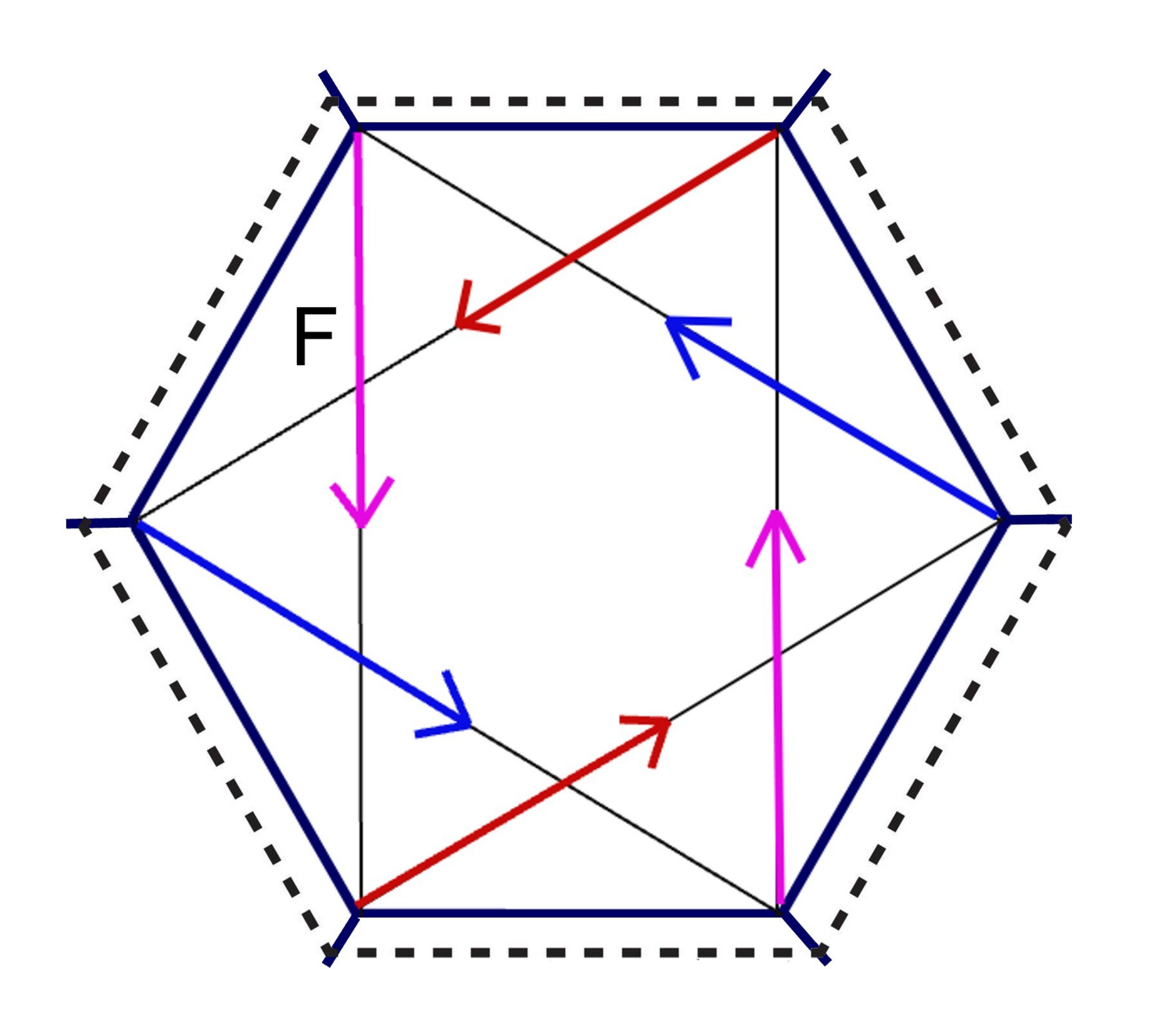}\label{fig:test}}
    \caption{{\bf (a)} Triangular lattice made of odd Hookean springs which obey the transverse force law. The pink arrows show that the pointwise force is just a passive rotation of the passive force one would expect of Hookean springs. Notice that this rotation of the effective force is what a gyroscope does.  {\bf (b)} A plaquette of the honeycomb lattice with next-nearest-neighbor bonds. Angular momentum is conserved since the tension along each bond is cancelled by its corresponding color.}
    %Four particles could, for example, form a tetrahedron.  In this case, no spontaneous $T$-breaking is possible, as discussed in \secref{sec:4s2}.  Geometrically, the lack of natural $T$-breaking corresponds to the high symmetry of the tetrahedron-there is no preferred rotation axis.  }
\end{figure}

To show that working with a microscopic system directly can lead to the explicit breaking of additional symmetries, let us add a next-nearest-neighbor odd spring constant with distinct parity-even and parity-odd coefficients ($k^{\,}_2$ and $k_2^\circ$, respectively).  Na\"ively, an identical argument to the above goes through, and the system should still be PT-symmetric. However, an explicit microscopic calculation \cite{odd_elasticity_2020} shows that theories with nearest- and next-nearest-neighbor odd springs does not obey \eqref{eq:gyroscopic_ratio}.  What happens is  that the stationary distribution, captured by $\Phi$, \emph{explicitly breaks parity}:
\begin{equation}
\label{eq:odd_elasticity_phi2}
    \Phi = \frac{1}{2}\int \thed^2x \left[ c^{\vpp}_1(\nabla \times u)^2 + (c_1+c_2)(\nabla \cdot u)^2+c^{\vpp}_3(\nabla \cdot u)(\nabla \times u) \right] \, ,~~
\end{equation}
where the P-odd coefficient $c_3$ depends on the details of the microscopic spring constants.  Repeating the calculation above, we find that 
\begin{subequations}
    \begin{align}
        B &= Q_1c_2 + 2Q_2c_3, \\
        \mu &= Q_1c_1-Q_2c_3,\\
        K^{\circ} &= Q_1c_3+Q_2c_1, \\
        A &= Q_2c_2-2Q_1c_3.
    \end{align}
\end{subequations}
 Thus the ratio (\ref{eq:gyroscopic_ratio}) is broken and we get four distinct elastic coefficients, due to the explicit breaking of parity. Observe that it is possible for $\mu<0$ or $B<0$ if $Q_2$ is large (which is not incompatible with the stability of the theory); however, $B+2\mu \ge 0$ since $Q_1, c_1, c_2 \ge 0$ are all required by stability of the fixed point.
 
 Interestingly, we can explicitly break parity, meaning that $c_3\ne 0$, but restore time-reversal symmetry, meaning that $Q_2 = 0$.  We still in this case find that the two new odd elastic moduli do not vanish, implying that time-reversal symmetry breaking was not essential to see odd elastic coefficients \cite{lier}.  In this case, we do see that $B,\mu>0$ directly, while $A=-2K^\circ$. 
 %The odd elastic coefficients for T-symmetry but P-breaking are; $B=Q_1c_1$, $\mu = Q_1c_2$, and $A=-2K^{\circ} = -2Q_1c_3$. 

Lastly, let us return to a point remarked at the beginning of this subsection:  the choice that $\Phi$ had a shift symmetry has led to direct physical consequences. Notice that the Lagrangian (\ref{eq:odd_elasticity_lagrangian}), or more formally the generator of the Fokker-Planck equation, has no derivatives acting on $\pi$.  This means that when we write down the noisy equations of motion:  $\partial_t u_i = \cdots + \xi_i$, the variance of the noise will be \begin{equation}
    \langle \xi_i(x,t)\xi_j(x',t')\rangle \sim \delta(x-x')\delta(t-t').
\end{equation}
In other words, this noise will \emph{translate} the solid in space.   This theory therefore corresponds to an odd solid that is pinned to some substrate, and therefore can slide around over time.  In most experiments, this is indeed likely to be the case.

\subsubsection{Translationally pinned potential}
However, it is also reasonable to postulate---at least theoretically---the existence of an intrinsically odd solid, which itself forms a closed system, and where the stochastic dynamics has only arisen due to integrating out short wavelength fluctuations of the solid itself.   Along the lines of the above discussion, let us now turn to discuss the case where the solid itself is taken to be an isolated system, that cannot slide itself spontaneously in the plane. Then, the quantities 
\begin{subequations}
    \begin{align}
        X_i &= \int \thed^2x \, u_x, \\
        Y_i &= \int \thed^2x \, u_y \, ,
    \end{align}
\end{subequations}
are conserved. Noether's theorem then implies that the Lagrangian must be invariant under $\pi_{x,y} \rightarrow \pi_{x,y}+1$, and therefore (\ref{eq:odd_elasticity_lagrangian}) must be modified to \begin{equation}\label{eq:odd_elasticity_lagrangian2}
        \mathcal{L} = \pi_i \partial_t u_i + \ii\partial_i \pi_j Q_{ijkl}\partial_k \left(\pi_l - \ii\mu _l\right).
\end{equation}
Due to the addition of two extra derivatives on $\pi_i$, if we take the same choice of $\Phi$ as before, evidently the elastic equations of motion will be fourth order.  On the other hand, since $X_i$ and $Y_i$ are conserved, it need not be the case that $\Phi$ depend on $u_i$ only via derivatives: we are \emph{not} just as likely to find the solid in all configurations translated by a constant.   So now, let us instead postulate \begin{equation}\label{eq:odd_elasticity_phi3}
    \Phi = \frac{1}{2}\int \mathrm{d}^2x \; u_i u_i
\end{equation}
For simplicity we have chosen units so there is no constant prefactor in $\Phi$ above.  In this case $\mu_i = u_i$,
and the equation of motion is
\begin{equation}
\label{eq:odd_elasticity_EOM_pot2}
    \partial_t u_i = Q_{ijkl}\partial_j\partial_k u_l + \xi_i,
\end{equation}
where \begin{equation}
    \langle \xi_j(x,t) \xi_l(x^\prime,t^\prime)\rangle = -2Q_{ijkl}\partial_k \partial_l \delta^{(2}(x-x^\prime) \delta(t-t^\prime).
\end{equation}
Now we want to know what components are allowed in the fourth rank tensor $Q_{ijkl}$. We consider the basis of $2\times 2$ matrices,  $\{\omega^i\}$ such that
\begin{equation}
    \omega^0 = \delta_{ij}, \quad \omega^1 = \epsilon_{ij}, \quad \omega^2 = \sigma_{ij}^z, \quad \omega^3 = \sigma_{ij}^x
\end{equation}
In this basis, we express the fourth rank tensor $Q_{ijkl}$ as a $4\times 4$ matrix;
\begin{equation}
    Q^{\alpha \beta} = \left(\omega^{\alpha}\right)_{ij}Q_{ijkl}\left(\omega^{\beta}\right)_{kl}
\end{equation}
First, we assume that the solid is isotropic for simplicity, as before.  This means that $\omega^{2,3}$ lie in the same real-valued representation of the rotation group distinct from $\omega^{0,1}$.  In order to be invariant under rotations, 
% $Q^{\alpha\beta}$ 
% \begin{equation}
%     Q^{\alpha \beta} = R^{\alpha \gamma}(\theta) Q^{\gamma \sigma} R^{\beta \sigma}(\theta)
% \end{equation}
% where
% \begin{equation}
%     R^{\alpha \gamma}(\theta) = 
%     \begin{pmatrix}
%         1&0&0&0\\
%         0&1&0&0\\
%         0&0&\cos(2\theta) & \sin(2\theta) \\
%         0 & 0 & -\sin(2\theta) & \cos(2\theta)
%     \end{pmatrix}
% \end{equation}
% In order for the matrix to be invariant with respect to this transformation, the matrix must be of the form
\begin{equation}
    Q^{\alpha \beta} =
    \begin{pmatrix}
        Q^{00} & Q^{01} & 0& 0\\
        Q^{10} & Q^{11} & 0& 0\\
        0 & 0 &  Q^{22} & Q^{23}\\
        0 & 0 &  -Q^{23} & Q^{22}
    \end{pmatrix}
\end{equation}
%\subsubsection{Parity-time transformation}
\begin{comment}
Now consider the parity transformation where 

\begin{equation}
    P =
\begin{pmatrix}
    1 & 0 \\
    0 & -1
\end{pmatrix}
\end{equation}
We will now consider how the basis elements transform under parity;
\begin{equation}
    \begin{split}
        P\omega^0P^{-1} &= \omega^0 \\
        P\omega^1P^{-1} &= -\omega^1 \\
        P\omega^2P^{-1} &= \omega^2 \\
        P\omega^3P^{-1} &= -\omega^3
    \end{split}
\end{equation}
\end{comment}
Table \ref{table:odd} describes how the components of $Q^{\alpha \beta}$ transform under parity, inversion, and time-reversal symmetry.
\newcommand{\otoprule}{\midrule[\heavyrulewidth]}
\begin{table}[t!]
\centering
%\begin{tabularx}{0.8\textwidth} { 
%  | >{\centering\arraybackslash}X 
%  | >{\centering\arraybackslash}X 
%  | >{\centering\arraybackslash}X
%  | >{\centering\arraybackslash}X | }
  %\hline
\begin{tabular}{c@{\hskip 3em}ccc}
 \toprule
 Component & Parity & Inversion & Time \\
 %\hline
 \otoprule
$\delta \otimes \delta $  & Even  & Even  & Even \\
%\hline
$\delta \otimes \epsilon + \epsilon \otimes \delta$  & Odd  & Even & Even  \\
%\hline
$\delta \otimes \epsilon - \epsilon \otimes \delta$  & Odd  & Even & Odd\\
%\hline
$\epsilon \otimes \epsilon$  & Even & Even & Even \\
%\hline
$\sigma_z \otimes \sigma_z$ & Even & Even & Even\\
%\hline
$\sigma_z \otimes \sigma_x - \sigma_x \otimes \sigma_z$  & Odd & Even & Odd\\
%\hline
$\sigma_x \otimes \sigma_x$  & Even & Even & Even\\
%\hline
\bottomrule
\end{tabular}
\caption{Tensor transformations for $Q_{ijkl}$ under parity, inversion, and time-reversal.}
\label{table:odd}
\end{table}
While we will not consider all possible combinations exhaustively, let us note that if we wish to have PT-symmetry only, the only further constraint is that $Q^{10}=-Q^{01}$.

As above, we can calculate the elastic moduli.  Assuming PT-symmetry, we find: \begin{subequations}
    \begin{align}
    B &= Q^{00}+ Q^{22}, \\
    \mu &= Q^{11} + Q^{22}, \\
    K^\circ &= Q^{01} + \frac{1}{2}Q^{23}, \\
        A &= 0.
    \end{align}
\end{subequations}  Note that because $\Phi$ is explicitly parity invariant, $A$ and $K^\circ$ are now necessarily T-odd coefficients.   

%ANDY: note that Q11 or Q10 can lead to angular orientation nonconservation, where \int epsilon_{ij}x_i X_j is not constant in time!

% The fact that $A=0$ implies that the solid is both insensitive to global rotations, \emph{and} that the dynamics will generically conserve the angular orientation of the body, since \begin{equation}
%     \frac{\mathrm{d}}{\mathrm{d}t} \int \mathrm{d}^2x \epsilon_{ij} X_i x_j = -\epsilon_{ij}Q_{ijkl} \int \mathrm{d}^2 x \; \partial_k u_l
% \end{equation}

\begin{comment}
Now if we Fourier transform \ref{eq:odd_elasticity_EOM_pot2}, the equation of motion becomes,
\begin{equation}
    i\omega \tilde{u}_i(k) = Q_{ijkl}k_kk_j\tilde{u}_l(k)
\end{equation}
and if we once again assume the material is isotropic then using the expression for $Q_{ijkl}$ above we get,
\begin{equation}
    -i\omega 
\begin{pmatrix}
    \dot{u_x} \\
    \dot{u_y}
\end{pmatrix} 
= k^2
\begin{pmatrix}
    Q^{00} + Q^{22} & Q^{01}+\frac{1}{2}Q^{23} \\
    -(Q^{01}+\frac{1}{2}Q^{23}) & Q^{22} - Q^{11} 
\end{pmatrix}
\begin{pmatrix}
    u_x \\
    u_y
\end{pmatrix}
\end{equation}
Comparing this matrix with \ref{odd_elasticity_matrix} we find that $A=0$, $K^{\circ} = Q^{01}+\frac{1}{2}Q^{23}$, $B = Q^{00}+Q^{11}$, and $\mu = Q^{22}-Q^{11}$. So there are three independent coefficients. This shows that in order to preserve PT symmetry, the solid must conserve angular momentum and thus have $A=0$.
\end{comment}

A system with $A=0$ and $K^{\circ}, B, \mu \ne 0$ arises in an active solid on a two-dimensional honeycomb lattice.  At vertices of the lattice, we place hinges between the edges, which are active components that deform according to the deformation of its nearest clockwise neighbor \cite{VV_odd_review}. We further assume that $B$ and $\mu$ are nonzero so that the solid resists changes in volume and shape (this can be realized by adding next-nearest-neighbor (NNN) Hookean springs). Each NNN bond responds to the strain of its counterclockwise neighbor in the unit cell. Reversing time would imply that each NNN bond responds to the tension of its clockwise neighbor which is tantamount to a parity (P) transformation. We therefore expect this system to have PT symmetry. Our effective theory shows that angular momentum must be conserved so long as $Q^{11}=0$, which is always the case when (as in this microscopic model) the system is invariant under a global rotation $(u_i=\epsilon_{ij}x_j$). In this microscopic model we can indeed confirm that angular momentum is conserved, and therefore $A=0$: all forces from Hookean springs are radial so momentum is conserved pair by pair, while the active hinges exert equal and opposite torques on the two bonds they connect.

\subsection{Chiral fluid}

Now, let us turn to the most complex example of fluid we will consider in this paper: a two-dimensional chiral fluid with both density and momentum conservation.  There are three possible ``odd" transport coefficients we should expect: odd diffusivity \cite{odd_diff} (which we have already discussed), odd viscosity \cite{VV_2017_odd}, and odd stress \cite{VV_2021_fluctuating}. While the former two have a close analogy with dissipationless Hall transport in equilibrium systems, the last one appears to us unique to active phases: in particular, it violates the thermal KMS condition in the presence of a generic background geometry \cite{huang_2022_discrete}.

\subsubsection{Odd viscosity}
Let us begin by discussing odd viscosity.  For simplicity, we'll momentarily neglect fluctuations of density, and assume that the most likely fluid configuration has zero velocity.  Then for small fluctuations around equilibrium, the stationary distribution is given by the nonrelativistic kinetic energy (with background mass density / momentum susceptibility~\cite{hartnoll_book} $\rho$):
\begin{align}\label{eq:Phi viscosity}
    \Phi = \beta_0 \int \dd[d]{x} \frac{\rho}{2} u_i u_i.
\end{align}
As usual, we introduce `chemical potentials' as
\begin{align}
    \mu_i = \fdv{\Phi}{u_i} = \rho u_i.
\end{align}
In the usual nonactive case, the effective Lagrangian should satisfy $T$ and $P$ and also be consistent with conservation of momentum, $\vb P = \int \dd[d]{x} \rho \vb u$.  The minimal Lagrangian therefore features two gradients of $\vb u$:
\begin{align}
\label{eq:normalviscL}
    \mL(u,\pi) = \pi_i\p_t u_i +\ii \beta_0^{-1}\rho^{-2} \eta_{ijkl} \p_i \pi_j \p_k (\pi_l-\ii \mu_l).
\end{align}
The constant object $\eta_{ijkl}$ is known as the viscosity tensor.  In the It\^{o} scheme of the MSR path integral, integrating out $\pi_i$ leads to a fluctuating diffusion equation for the velocity:
\begin{align}\label{eq:viseom}
    \rho \p_t u_i = \eta_{ijkl}\p_j\p_k u_l + \xi_i,
\end{align}
where the statistics of the white noise $\xi_i(\vb x, t)$ are such that
\begin{align}
\label{eq:viscnoise}
    \langle \xi_i(\vb{x},t)\xi_j(\vb{x}^\prime,t^\prime) \rangle = \beta_0^{-1}(\eta_{ikjl} + \eta_{jlik})\p_k\p_l \delta(\vb{x}-\vb{x}^\prime) \delta(t-t^\prime).
\end{align}
Thus viscous hydrodynamics naturally emerges as the simplest $P$, $T$-invariant effective field theory for conserved momentum.  We will discuss the pressure and stress, so far omitted, in the next subsection.

However, in a chiral fluid composed, for example, of spinning particles, P and Tmay be violated such that only PT is conserved.  In that case, we may decompose 
\begin{align}
    \eta_{ijkl} = \eta^{\even}_{ijkl}+\eta^{\odd}_{ijkl},
\end{align}
where $\eta^{\even}_{ijkl} = \eta^{\even}_{klij}$, $\eta^{\odd}_{ijkl} = -\eta^{\odd}_{klij}$.   The latter is called the odd viscosity \cite{VV_odd_review}.  Note that $\eta^{\mathrm{even}}$, but not $\eta^{\mathrm{odd}}$, contribute to the variance of the statistical fluctuations; hence $\eta^{\mathrm{even}}$ must a positive semidefinite tensor.

% To see explicitly that the odd viscosity is dissipationless, we may calculate its contribution to viscous heat generation per unit area as
% \begin{align}
%     Q = \eta^{\odd}_{ijkl}\p_i u_j\p_k u_l = 0.
% \end{align}
% Thus the odd viscoisty is dissipationless in the usual sense, and we will presently see that it is also described by a dissipationless force in our EFT in the sense of \secref{sec:EFT}.

To find odd viscosity in our effective field theory, we simply need to enforce PT instead of T and P separately.  The minimal Lagrangian becomes:
\begin{align}\label{eq:L viscosity active}
    \mL(u,\pi) = \pi_i\p_t u_i +\beta_0^{-1}\rho^{-2} \eta^{\odd}_{ijkl}  \p_i \pi_j \p_k \mu_l+\ii \beta_0^{-1}\rho^{-2} \eta^{\even}_{ijkl} \p_i \pi_j \p_k (\pi_l-\ii \mu_l),
\end{align}
where, as in the case of odd diffusivity, the antisymmetry of $\eta^{\odd}_{ijkl}$ means this coefficient enters only at linear order in $\pi_i$. Accordingly, $\eta^{\odd}_{ijkl}$ is not related to noise---the statistics of the noise remain \eqref{eq:viscnoise}. The resulting equation of motion is in agreement with \eqref{eq:viseom}. 

Our approach differs from \cite{VV_2021_fluctuating}; in this work, the authors attributed the odd viscosity to an ``antisymmetric" non-Markovian noise where $\langle \xi(t) \xi(t^\prime) \sim f(t-t^\prime)$ with $f$ an odd, sharply localized function.  In our discussion, odd viscosity is allowed due to the gT invariance of the effective Lagrangian, and it is independent of noise as a dissipationless force in the effective theory.  Odd viscosity, like odd diffusivity, therefore emerges naturally from the simplest PT-invariant effective field theory of a dissipative incompressible fluid.  The fact that we obtain the same hydrodynamic universality class as \cite{VV_2021_fluctuating} therefore explains the emergence of an FDT and a second law of thermodynamics in more microscopically active models as well.  Like in the discussion of odd elasticity, our approach does not give an unambiguous definition of the stress tensor, so certain global correlation functions of stress will not be universally captured by our EFT.

\subsubsection{Fluid with spinning particles}\label{sec:fluid spin}
Let us now work in a regime where both $n$ and $\vb u$ are nonzero. We also consider an additional spin density $s$, as slowly varying fields.  The spin density $s$ could correspond to a local angular momentum density of spinning particles, and we will first assume that $s$ is a conserved quantity. The minimal steady-state distribution is given by
\begin{align}\label{eq:Phi spin}
    \Phi = \beta_0\int \ud^d x~\left[\frac{\rho}{2}(u_i)^2+\frac{1}{2\chi}(\delta n)^2+\frac{1}{2\chi_s}(\delta s)^2 \right],
\end{align}
where $\chi_s$ is the spin susceptibility. Spinning motions are P-odd and T-odd, but PT-even.

Under PT symmetry, the leading Lagrangian is given by
\begin{align}\label{eq:pt L spin}
    \mL =&  \pi_i \p_t u_i+\pi^n\p_t \delta n+\pi^s \p_t \delta s - n_0 u_i \p_i \pi^n  -\frac{n_0}{\chi\rho}\p_i\pi_i\delta n - s_0 u_i\p_i \pi^s - \frac{s_0}{\chi_s\rho}\p_i\pi_i\delta s\nonumber\\
    &+\ii \beta_0^{-1} \rho^{-2} \eta_{ijkl}^{\mathrm{even}} \p_i\pi_j \p_k (\pi_l - \ii \mu_l)+\sum_{\alpha,\beta=n,s}\ii \beta_0^{-1} D^{\alpha\beta,\even}\p_i\pi^\alpha \p_i(\pi^\beta-\ii\mu^\beta)\nonumber\\
    &+\beta_0^{-1}\rho^{-2} \eta^{\odd}_{ijkl}  \p_i \pi_j \p_k \mu_l+\sum_{\alpha,\beta=n,s}\beta_0^{-1}  D^{\alpha\beta,\odd} \epsilon_{ij} \p_i \pi^\alpha \p_j \mu^\beta,
\end{align}
where $D^{ns}=D^{sn}$. Importantly, we do not include the following invariant term
\begin{align}\label{eq:K term spin}
    \mL\sim\sum_{\alpha=n,s}\frac{\ii}{2} \beta_0^{-1} K^\alpha \epsilon^{ij}(\p_i\pi_j(\pi^\alpha-\ii\mu^\alpha)+\p_i(\pi_j-\ii\mu_j)\pi^\alpha).
\end{align}
This is because to fulfill the positivity constraint of the noise invariance matrix, one has to also include terms with zero derivative order:
\begin{align}\label{eq:gamma spin}
    \mL\sim \sum_{\alpha=n,s} \ii \beta_0^{-1}\Gamma^{\alpha}\pi^\alpha(\pi^\alpha
    -\ii\mu^\alpha).
\end{align}
Moreover, this zeroth order term is forbidden by the shift symmetry $\pi^\alpha\to \pi^\alpha+1$ corresponding to the conservation of charge and spin. Indeed, \eqref{eq:gamma spin} will lead to equations of motion: $\p_t n\sim -\Gamma^{n} n$ and $\p_t s\sim -\Gamma^{s} s$ that decay to zero. Without them, the stochastic equations of motion corresponding to \eqref{eq:pt L spin} are given by
\begin{subequations}\label{eq:eom spin}
    \begin{align}
        \p_t \delta n+ n_0 \p_i u_i  - \chi^{-1}D^{n,\even} \p_i^2 \delta n - \chi_s^{-1}D^{ns,\even} \p_i^2 \delta s
        % +\frac{1}{2}K^n \rho \epsilon^{ij}\p_i u_j 
        &=\xi^n,\\
        \p_t\delta s +s_0\p_i u_i - \chi_s^{-1}D^{s,\even} \p_i^2 \delta s -\chi^{-1} D^{ns,\even} \p_i^2 \delta n 
        % +\frac{1}{2}K^s \rho \epsilon^{ij}\p_i u_j
        &=\xi^s,\\
        \rho \p_t u_i +\chi^{-1} n_0 \p_i\delta n+\chi_s^{-1} s_0 \p_i\delta s - \eta_{jikl} \p_j\p_k u_l 
        % +\frac{1}{2} \rho \epsilon^{ij}(K^n \chi^{-1}\p_j \delta n+ K^s \chi_s^{-1}\p_j \delta s)
        &=\xi_i,\label{eq:momentum conservation spin}
    \end{align}
\end{subequations}
with the noise variance
\begin{subequations}\label{eq:noise spinning PT}
\begin{align}
    \expval{\xi^\alpha(\vb x, t)\xi^\beta(\vb x^\prime, t^\prime)} &= 2\beta_0^{-1}  D^{\alpha\beta,\even} \,\p_i^2 \delta(\vb x - \vb x^\prime)\, \delta(t - t^\prime),\\
    \expval{\xi_i(\vb x, t)\xi_j(\vb x^\prime, t^\prime)} &= 2\beta_0^{-1}  \eta_{ikjl}^{\mathrm{even}} \,\p_k\p_l \delta(\vb x - \vb x^\prime)\, \delta(t - t^\prime),
    % \expval{\xi_j(\vb x, t)\xi^\alpha(\vb x^\prime, t^\prime)} &= 2\beta_0^{-1} \rho K^\alpha \epsilon^{ij}\p_i \delta(\vb x - \vb x^\prime)\, \delta(t - t^\prime),
\end{align}
\end{subequations}
for $\alpha,\beta=n,s$. Therefore, we find that it describes the odd fluid dynamics, including the odd diffusivity and the odd viscosity that we discussed above, with two species of charge. In particular, to the leading order, the equation of state is given by
\begin{align}
    \ud p \approx \chi^{-1} n_0\ud n + \chi_s^{-1}s_0\ud s+\cdots,
\end{align}
where $p$ is the thermodynamic pressure, and there exist no odd stress.

However, if we were allowed to include \eqref{eq:K term spin}, it will actually lead to the desirable odd stress. To see it, we vary \eqref{eq:K term spin} with respect to $\pi_i$, which gives 
\begin{align}
    \sigma_{ij}^{\mathrm{P-odd}} = \sum_{\alpha=n,s} \frac{1}{2}\beta_0^{-1} \rho K^\alpha \epsilon^{ij} \mu^\alpha,
\end{align}
where $\sigma_{ij}^{\mathrm{P- odd}}$ is the odd stress that is defined by $\rho\p_t u_i +\p_j \sigma_{ij} =\cdots $ and antisymmetric.
To avoid that $\mu^\alpha\to 0$ as we argued above, we can introduce an active torque $\tau$ into the equations of motion for $\delta s$: \cite{VV_2017_odd}
\begin{align}
    \p_t\delta s \sim \tau - \Gamma^s s,
\end{align}
such that a spinning steady state with $s_0 \equiv \tau/\Gamma^s$ can exist. Now, the odd stress is given by
\begin{align}
    \sigma_{ij}^{\mathrm{P-odd}} =\frac{1}{2}\rho \chi_s^{-1} K^s \epsilon^{ij} s,
\end{align}
in agreement with \cite{VV_2017_odd}. Several remarks follow. First, the odd stress strength $K^s$ is constrained by the positivity of the noise invariance: $(K^s)^2\leq \eta \Gamma^s$. Second, the torque $\tau$ is not an ingredient that can be captured in the EFT  -- $\mL\sim \pi^s\tau$ is not invariant under PT -- but the resulting steady state with $s_0\neq 0$ is a valid starting point for our EFT and we have seen that it leads to the odd stress. Let us turn to the relevant changes of the equations of motion due to this odd stress. We have, to the leading order, 
\begin{subequations}\label{eq:eom spin active}
    \begin{align}
        \p_t \delta n+n_0\p_i u_i &=0,\\
        \p_t \delta s+s_0\p_i u_i + \frac{1}{2} \rho K^s \epsilon^{ij}\p_i u_j&=-\Gamma^s \delta s,\\
        \rho\p_t u_i +\chi^{-1} n_0 \p_i \delta n+ \chi_s^{-1} s_0 \p_i \delta s + \frac{1}{2} \rho \chi_s^{-1}K^s \epsilon^{ij}\p_j\delta s&=0.
    \end{align}
\end{subequations}
Solving the above equations to the linear order in $O(k)$, we find, schematically, a longitudinal sound mode $\omega \approx \pm v_0 k$ with $v_0 = n_0/\sqrt{\chi\rho}$, a decaying spin density mode $\omega\approx -\ii \Gamma^s$, and a transverse diffusion mode $\omega\sim  -\ii k^2$.

There is an intrinsic spin density $\delta s= \epsilon^{ij}\p_i u_j$ corresponding to the angular momentum of the fluid. It also contributes to the odd stress if the rotational symmetry is broken explicitly. Specifically, there exists a rotational viscosity \cite{caleb1,caleb2,bradlyn}
\begin{align} \label{eq:rotvisc}
    \eta^{\even}_{ijkl} \sim \eta_\circ \epsilon^{ij}\epsilon^{kl},
\end{align}
and plugging it in \eqref{eq:momentum conservation spin}, we obtain
\begin{align}
    \sigma_{ij}^{\mathrm{odd}} = - \eta_\circ \epsilon^{ij} \epsilon^{kl}\p_k u_l.
\end{align}
Notice that this is typically a higher derivative contribution to the odd stress, and that this contribution is P-even and T-even separately.\footnote{The ``odd" stress refers to the fact that $\sigma_{ij}^{\mathrm{odd}}\sim \epsilon^{ij}$, but the tensor structure of (\ref{eq:rotvisc}) renders it even under P.}  Recently, this relaxation of intrinsic angular momentum has been argued to relax the momentum along the helical axis of a fluid with helical symmetry in thermal equilibrium \cite{Farrell:2022xnf}.

\subsubsection{Odd stress and PT-odd charge}
Lastly, suppose that we add to our fluid a P-odd but T-even charge, which we will denote $w$.   A possible candidate for this P-odd and T-even charge is to consider a fluid consisting of two kinds of chiral particles (which cannot be created or destroyed), with $n_{\mathrm{L,R}}$ the densities of left- and right-handed particles respectively.  Since parity will exchange left/right-handed particles, we would then obtain both a P-even and P-odd charge: $n=n_{\mathrm{L}}+n_{\mathrm{R}}$ and $w=n_{\mathrm{L}}-n_{\mathrm{R}}$.   Both of them, however, remain T-even.
%we have two anti-spiral T-even charges $n_+,n_-$, their difference $n_5\equiv n_+-n_-$ is P-odd since P will flips the orientation of the helix.

We take similarly a minimal steady-state distribution
\begin{align}\label{eq:Phi odd stress}
    \Phi = \beta_0\int \ud^d x~\left[\frac{\rho}{2}(u_i)^2+\frac{1}{2\chi}(\delta n)^2+\frac{1}{2\chi_w}(\delta w)^2 \right],
\end{align}
where $\chi_w$ is the susceptibility associated with $w$. Then, the leading order Lagrangian that is invariant under PT symmetry is given by
\begin{align}\label{eq:pt L odd stress}
    \mL =&  \pi_i \p_t u_i+\pi^n\p_t \delta n+\pi^w \p_t \delta w - n_0 u_i \p_i \pi^n  -\frac{n_0}{\chi\rho}\p_i\pi_i\delta n - w_0 \epsilon^{ij}u_i\p_j \pi^w +\frac{w_0}{\chi_w\rho}\epsilon^{ij}\p_i\pi_j\delta w\nonumber\\
    &+\ii \beta_0^{-1} \rho^{-2} \eta_{ijkl}^{\mathrm{even}} \p_i\pi_j \p_k (\pi_l - \ii \mu_l)+ \sum_{\alpha=n,w}\ii \beta_0^{-1} \chi_\alpha D^{\alpha,\even}\p_i\pi^\alpha \p_i(\pi^\alpha-\ii\mu^\alpha)\nonumber\\
    &+\beta_0^{-1}\rho^{-2} \eta^{\odd}_{ijkl}  \p_i \pi_j \p_k \mu_l+\sum_{\alpha=n,w}\beta_0^{-1} \chi_\alpha D^{\alpha,\odd} \epsilon_{ij} \p_i \pi^\alpha \p_j \mu^\alpha \nonumber\\
    &+  \beta_0^{-1} D_{\mathrm{mix},\odd} \left(\p_i\pi^n\p_i\mu^w -\p_i\pi^w\p_i\mu^n  \right),
\end{align}
where the subscript ``odd'' indicates the term is T-odd.
In the above equation, we did not include
\begin{align}
    \mL\sim  \ii \frac{1}{2} \beta_0^{-1} D_{\mathrm{mix},\even}\epsilon^{ij}\left(\p_i\pi^n\p_j(\pi^w-\ii\mu^w) +\p_i(\pi^n-\ii\mu^n)\p_j \pi^w \right),
\end{align}
which is a total derivative,
and the invariant terms
\begin{align}
    \mL\sim\frac{\ii}{2} \beta_0^{-1} K^n \epsilon^{ij}(\p_i\pi_j(\pi^n-\ii\mu^n)+\p_i(\pi_j-\ii\mu_j)\pi^n) + \frac{\ii}{2} \beta_0^{-1} K^w(\p_i\pi_i(\pi^w-\ii\mu^w)+\p_i(\pi_i-\ii\mu_i)\pi^w)
\end{align}
for the same reason as in \secref{sec:fluid spin}. Hence, the stochastic equations of motion is
\begin{subequations}\label{eq:pt eom odd stress}
    \begin{align}
        \p_t \delta n+ n_0 \p_i u_i  - D^{n,\even} \p_i^2 \delta n - \chi_s^{-1}D^{\mathrm{mix},\odd} \p_i^2 \delta w&=\xi^n,\\
        \p_t\delta w -w_0\epsilon^{ij}\p_i u_j - D^{w,\even} \p_i^2 \delta w +\chi^{-1} D^{\mathrm{mix},\odd} \p_i^2 \delta n &=\xi^w,\\
        \rho \p_t u_i +\chi^{-1} n_0 \p_i\delta n+\chi_w^{-1} w_0 \epsilon^{ij} \p_j\delta w - \eta_{jikl} \p_j\p_k u_l &=\xi_i,
    \end{align}
\end{subequations}
with the noise variance
\begin{subequations}\label{eq:noise spinning PT}
\begin{align}
    \expval{\xi^\alpha(\vb x, t)\xi^\alpha(\vb x^\prime, t^\prime)} &= 2\beta_0^{-1} \chi_\alpha D^{\alpha,\even} \,\p_i^2 \delta(\vb x - \vb x^\prime)\, \delta(t - t^\prime),\\
    \expval{\xi_i(\vb x, t)\xi_j(\vb x^\prime, t^\prime)} &= 2\beta_0^{-1}  \eta_{ikjl}^{\mathrm{even}} \,\p_k\p_l \delta(\vb x - \vb x^\prime)\, \delta(t - t^\prime),
\end{align}
\end{subequations}
for $\alpha=n,w$. A crucial difference from \secref{sec:fluid spin} is that there exists an odd stress (pressure) obeying
\begin{align}
    \ud p_{\mathrm{P-odd}} \approx \chi_w^{-1} w_0 \ud w.
\end{align}
The odd stress results in a transversal sound mode, which is absent in the normal hydrodynamics. To see it, we compute the quasinormal modes corresponding to \eqref{eq:pt eom odd stress} to the leading (ideal) order. This leads to
\begin{align}
    \omega = \pm \frac{n_0}{\sqrt{\chi\rho}} k ,\quad \omega = \pm \frac{w_0}{\sqrt{\chi_w\rho}} k,
\end{align}
where the former is the usual longitudinal sound mode, and the latter is the transversal sound mode due to the odd stress.  It would be interesting if this effect can be discovered in active chiral matter.

\subsection{Fracton hydrodynamics}
\label{sec:fracton}

``Fracton hydrodynamics" is the name for the universality classes of hydrodynamics associated to theories with multipolar conservation laws (e.g. charge and dipole conservation together) \cite{gromovfrachydro,knap2020,morningstar,hart2021hidden,Glorioso:2021bif,Grosvenor:2021rrt,Burchards:2022lqr,Guo2022}.  These exotic conservation laws require a careful formulation of hydrodynamic EFTs \cite{Guo2022}, including in nonthermal systems.  As mentioned in Sec.~\ref{sec:challenge} for example, certian terms naively allowed in equations of motion can be forbidden by an EFT.  Using the weak noise limit formalism described in Section \ref{sec:MSR}, \cite{Guo2022} has studied theories of fracton hydrodynamics without time-reversal symmetry. To quickly re-cap why this formalism is very helpful, notice that if \begin{equation}
    0=\frac{\mathrm{d}}{\mathrm{d}t}\int \mathrm{d}^dx \; (a+b_ix_i)n 
\end{equation}
for any $a$ and $b_i$, and $n(x)$ is a conserved charge density, we must write down a theory invariant under $\pi \rightarrow \pi + a + b_i x_i$.  Hence, the lowest derivative motif of $\pi$ that can show up in $\mL$, outside of $\pi \partial_t n$, becomes $\partial_i\partial_j \pi$.  So long as the time-reversed theory remains local, one can show that in generic dimensions\footnote{In one dimension only, one can also add a third-derivative term which contributes a total derivative after applying time-reversal.}, the most general Lagrangian at lowest order in derivatives is \begin{equation}
    \mL = \pi \pd{t} n +  \ii D_{ijkl}(n)\pd{i}\pd{j} \pi \pd{k} \pd{l} \left(\pi-\ii \mu\right). \label{eq:dipole L}
\end{equation}
If we take 
\begin{equation}
    \Phi \, = \, \int \thed^dx \, \frac{n^2}{2\chi} \label{eq:Phirhoneg} \, ,~~
\end{equation}
for constant $\chi$, then Eq.~\eqref{eq:dipole L} predicts subdiffusive with $\pd*{t} n \sim \pdp{x}{4} n$, in agreement with Ref.~\citenum{gromovfrachydro,Guo2022}, and resolving the puzzle of Sec.~\ref{sec:challenge}.

Note that the above choice of $\Phi$ \eqref{eq:Phirhoneg} is only reasonable if $n<0$ makes sense---e.g., there are both positive and negative charges in the theory.  If we wish to consider only positive charges, and we wish to consider nonlinear dynamics close to $n=0$, an appropriate, alternative choice of $\Phi$ is 
\begin{align}\label{eq:dipolePhi classical}
    \Phi = \int \ud x~ n\log n ,
\end{align}
which is inspired by the formula for Shannon entropy. Keeping the Lagrangian \eqref{eq:dipole L}, and working in $d=1$ spatial dimensions for simplicity, a short calculation reveals the equation of motion 
\begin{align}\label{eq:dipole eom}
    \p_t\delta n + \p_x^2\left( D(n)\p_x^2\mu\right)=0.
\end{align}
Using (\ref{eq:dipolePhi classical}) we find \begin{align}
    \mu = \frac{\p \Phi}{\p \delta n} =  \log n-1.
\end{align}
In order for the equation of motion to be regular, we demand \begin{equation}
    D(n) = D_0 n^2 + \cdots
\end{equation}
as $n\rightarrow 0$.  The resulting nonlinear equation of motion reproduces the predictions of a recent paper \cite{Han:2023ici} in the absence of noise. This effective theory approach conveniently streamlines what is otherwise a somewhat complicated microscopic calculation of the equations of motion. Dynamics conserving higher multipole moments can be similarly derived following \cite{Guo2022}.
%significantly streamlined the derivation of this effective theory.

% More recently, another recent work has proposed a natural nonlinear generalization of the a different (nonlinear) diffusion process.   We work in $d=1$ for simplicity. Following \cite{Guo2022}, the Noether theorem demands the Lagrangian is invariant under $\pi\to \pi+ x$, so the simplest choice is
% \begin{align}
%     \mL = \pi \p_t \delta n +\ii  D(n) \p_x^2\pi \p_x^2(\pi-\ii\mu),
% \end{align}
% where we take $\beta_0=1$.
% Various variables follow the notations in \secref{sec:odd diffusive}. The equation of motion is given by

% If we assume $\Phi$ to be \eqref{eq:diffusionPhi} and take $D(n)=D$, then we obtain the subdiffusive dynamics studied in \cite{Guo2022}. To obtain the equation of motion in \cite{Han:2023ici}, we take $\Phi$ to be

% where $S$ is the thermodynamic entropy. This form of $\Phi$ is equivalent to finding a steady state that extremizes the entropy.
% % A priori, however, $\Phi$ is not subject to the dipole conservation, which is instead the constraint for the Lagrangian.
% Using \eqref{eq:dipolePhi classical}, we find

% Plugging it in \eqref{eq:dipole eom} and taking $D(n)= D n^2$, we have $\p_t \delta n+ \p_x^2 J=0$ with $J = D n^2\p_x^2 \log n$. This is precisely the dipole current obtained in \cite{Han:2023ici}. We note that $D(n)\sim n^2$ is the smallest scaling as $n\rightarrow 0$ for which the equation of motion is not singular as $n\rightarrow 0$, so in this sense it is the one that should be chosen within the EFT. 

One further generalization considered in \cite{Han:2023ici}, which again immediately follows from our effective field theory, is to consider quantum particles (bosons or fermions) subject to the same dipole-conserving dynamics.  In our EFT, we can take the \emph{same} Lagrangian \eqref{eq:dipole L}, but simply modify $\Phi$ to 
\begin{align}\label{eq:dipolePhi quantum}
    \Phi_s= \int \ud x\left( n \log n - s(1+s n)\log(1+s n) \right)
\end{align}
where we take $s=1$ for bosons and $s=-1$ for fermions.
Calculating
\begin{align}
    \mu_s = \frac{\p \Phi_s}{\p \delta n} = \log \frac{n}{1+s n}
\end{align}
along with $D_s(n)=D n^2(1+s n)^2$, which is the minimal scaling in which the equation of motion is not singular, we find the same equations of motion as \cite{Han:2023ici}. 

Although this theory is nonlinear, it is straightforward to convert the discussion above into the operator formalism, where we maintain an exact implementation of time-reversal symmetry using regularizations analogous to App.~\ref{app:diffusion}.

\section{Lattice model for active phase separation}\label{sec:phase separation}
In this final section, we describe a lattice discretization of active phase separation  between two interacting species.   Such models have been of recent interest \cite{saha2020prx,You_2020}.  Here, we show how to readily build such a model consistently coupled to noise---as usual, in our construction, we will have explicit control over the thermodynamic (time-reversal even part of the) phase diagram of the model.   While we leave an exhaustive analysis of this model to future work, we include the construction here to emphasize the versatility of our methods.

Let us begin by overviewing how an active Cahn-Hilliard model for phase separation has been derived in the literature \cite{saha2020prx}.  Let $\phi_{\alpha,i}$ denote the concentration of species $\alpha\in \lbrace 1,2\rbrace$ at site $i$.  We assume that the net concentration of each species is conserved in the dynamics as a whole.  A nonactive dynamical system can be obtained on a discretized lattice by considering the noisy stochastic equation \begin{equation}
    \dot \phi_{\alpha,i} = - \Gamma \sum_{j} \Lambda_{ij} \frac{\partial \Phi}{\partial \phi_{\alpha,j}} + \text{noise} \label{eq:sec8passive}
\end{equation}
where, as usual in this paper, $\Phi$ denotes the logarithm of the stationary distribution:\footnote{Note that the $\lambda$ term is different here vs. e.g. Ref. \cite{saha2020prx}; however our construction is not too sensitive to the choice of $\Phi$.}
\begin{equation}\label{eq:Phi separation}
    \Phi = \beta \left[ \sum_{\alpha=1}^2 \left[ \sum_{i\sim j}  \frac{1}{2}(\phi_{\alpha,i}- \phi_{\alpha,j})^2 + \sum_i \left(\frac{\phi_{\alpha,i}^4}{4}-\frac{\phi_{\alpha,i}^2}{2}\right)  \right] + \lambda \sum_i (\phi_{1,i}+ \phi_{2,i})^2 \right],
\end{equation}
where $i\sim j$ denotes nearest-neighbor sites on a two-dimensional square lattice; the parameter $\lambda$ denotes an interaction between the two species.  Here we have also defined $\Lambda_{ij}$ to be the graph Laplacian: \begin{equation}
    \Lambda_{ij} = \left\lbrace \begin{array}{ll} -1 &\ i,j\text{ adjacent} \\ 4 &\ i=j \end{array}\right..
\end{equation}  
The noise in (\ref{eq:sec8passive}) conserves the net concentration of each species, and will be explicitly defined in the Fokker-Planck formalism shortly.  Note that $(\nabla^2)_{ij}\approx -\Lambda_{ij}$. 
If $\beta$ is very large, we expect to see $\phi_{\alpha,i}\approx \pm 1$, so that the first term in $\Phi$ is small.   When $\lambda \gg 1$, we would anticipate phase separation, where $\phi_{1,i}\approx -\phi_{2,i}$.     To add activity to this passive model, Ref. \cite{saha2020prx} added a nonreciprocal diffusion term to the equation of motion: \begin{equation}
     \dot \phi_{\alpha,i} = - \Gamma \sum_j \Lambda_{ij}\frac{\partial \Phi}{\partial \phi_{\alpha,j}} + \kappa \sum_{\beta ,j} \epsilon_{\alpha\beta} \Lambda_{ij} \phi_{\beta j} +  \text{noise}. \label{eq:activekappa}
\end{equation}

It is deliberately the point of Ref. \cite{saha2020prx} that the $\kappa$-dependent term above does not obviously follow from considering derivatives of $\Phi$ with respect to $\phi_{\alpha i}$ -- this term will thus break detailed balance.  The perspective of our paper is that a minimal model for active phase separation can instead demand that $\Phi$ remains the stationary distribution \emph{even out of equilibrium}; therefore, we elect to incorporate the nonreciprocal diffusion ($\kappa$ term) consistent with stationarity. We can achieve this as follows. With Gaussian noise and detailed balance, the passive model we would write down becomes a lattice discretization of the passive model of \cite{saha2020prx}: \begin{equation}
    W = \sum_{\alpha, i,j} \Gamma \Lambda_{ij} \partial_{\alpha, i} \left(\partial_{\alpha, j}  + \mu_{\alpha,j}\right).
\end{equation}
where $\Gamma$ denotes the noise strength. To write down active terms consistent with this stationary distribution, we can invoke \eqref{eq:v_anti} and look for an antisymmetric matrix $V_{\alpha i, \beta j}$ that leads to the desired terms.  Our strategy will be to look for $V$ that give us a desired ``active term", up to presumably irrelevant nonlinear corrections (with the caveat that we will not perform an exhaustive renormalization group analysis in this paper).  The key observation is that even in equilibrium configurations where $\mu_{\alpha i}=0$, we can still find active terms in the equations of motion of the form $[\partial_{\beta j}V_{\alpha i, \beta j}]$, and so we will choose a $V$ so that this divergence gives us a discretization of the terms of interest.   Consider the following choice of $V$: \begin{equation}\label{eq:W separation active}
    V_{\alpha i,\beta j} = -\kappa \epsilon_{\alpha\beta} \Lambda_{ij} \left(\frac{\phi_{\alpha i}^2 + \phi_{\beta j}^2}{2} -1 \right)
\end{equation}
where $\epsilon_{12}=-\epsilon_{21}=1$ is the two-dimensional Levi-Civita tensor. 
This choice of $V$ is chosen to be small near the global minima of $\Phi$ (where $\phi^2 =1$); hence, it will make the ``unwanted" terms added to the equations of motion to enforce stationarity as ``small" as possible.  Explicit calculation gives us \begin{subequations}
    \begin{align}
        \sum_{\beta,j}\frac{\partial V_{\alpha i, \beta j}}{\partial \phi_{\beta j}} &= -\sum_{\beta,j}\kappa \epsilon_{\alpha \beta} \Lambda_{ij} \phi_{\beta j}  \\
        \sum_{\beta,j}V_{\alpha i, \beta j} \mu_{\beta j} &= -\sum_{\beta,j}\kappa \beta \epsilon_{\alpha \beta} \left(\frac{\phi_{\alpha i}^2 + \phi_{\beta j}^2}{2}-1\right)  \Lambda_{ij} \left[\sum_k \Lambda_{jk}\phi_{\beta k} - \phi_{\beta j} + \phi_{\beta j}^3 + 2\lambda (\phi_{1,j}+\phi_{2,j})\right]
    \end{align}
\end{subequations}
Notice that the continuum limit of the first line is precisely the active term considered in \cite{saha2020prx}, and which appeared in (\ref{eq:activekappa}) -- $\kappa \epsilon_{\alpha\beta} \nabla^2 \phi_\beta$ -- while the second line is an irrelevant correction to the standard passive dynamics, if we expand around a static solution $\phi_{1i}=-\phi_{2i}=1$: the $V\mu$ term is \emph{quadratic} in fluctuations while the $\partial V$ term is \emph{linear} in fluctuations, with each term having at least two spatial derivatives in the continuum limit. It is thus reasonable to speculate that this lattice model ought to lie in the same universality class as the active scalar mixture model, in which we simply do not include the $V\mu$ term in the equations of motion.

In our construction, we know precisely that the stationary distribution of the dynamics is independent of $\kappa$.  Thus while $\kappa$ may lead to nonreciprocal phases, as we saw in the Kuramoto model, these nonreciprocal phases will provably obey all the laws of equilibrium statistical mechanics -- in particular, in one dimension, this model cannot exhibit any long-range order.

This construction is curious in that it uses precisely the regularization subtlety which App.~\ref{app:diffusion} aims to avoid, in order to construct equations of motion with desired active terms (at leading order in a long wavelength expansion).  This means that the construction of this section cannot be reproduced by any MSR field theory in the continuum in which time-reversal takes the simple form (\ref{eq:MSR T pi relation}).  We leave a more detailed investigation of such methods to future work \cite{elipaper}.

\section{Outlook}
In this paper, we have proposed an effective theory formalism for nonequilibrium systems that extends the approaches recently pioneered for thermal systems \cite{harder2015thermal, eft1, eft2, haehl2016fluid, jensen2018dissipative}.  Numerous examples and applications demonstrate the broad applicability of our approach, along with its relative ease of use; see Tab.~\ref{tab:summary application} for a summary.  Many of the theories described in this manuscript were first derived and studied elsewhere, but our effective theory approach provides an instructive derivation of these theories, and can provide broader context into how one particular model fits into a broader paradigm of nonequilibrium matter.

Our methods can elucidate subtleties that are easily overlooked when only considering equations of motion, such as the subtle nature of parity and generalized time-reversal symmetries (and their breaking) in recently developed theories of active odd elastic solids.  In other cases, our methods give a clear framework within which existing models can be generalized: e.g., Kuramoto models of chiral rotors on higher-dimensional spheres.

Perhaps most importantly, our conceptual starting point for building models of active matter is not to begin by breaking detailed balance or a second law of thermodynamics. Instead, we begin by understanding the typical states in the stationary distribution (i.e., we deduce $\Phi$), and in turn, understand how it is possible for the system to ``flow" around phase space.  From our perspective, a second law can often emerge, albeit with an entropy that may differ from equilibrium analogues (see App.~\ref{app:entropies}).

\begin{table}[t!]
    \centering
    
    \begin{tabular}{l@{\hskip 4em}c@{\hskip 1em}c@{\hskip 1em}c@{\hskip 1em}c}
    \toprule

    Theory & Equation of motion & $\Phi$ & $V_{ab}$ & MSR-compatible \\
   \otoprule
     Overdamped harmonic oscillator & \eqref{eq:HOlangevin} & \eqref{eq:H0Over Phi}& N/A & const. noise\\
     %\hline
     Finite-mass harmonic oscillator & \eqref{eq:finite-mass SHO EOM} & \eqref{eq:HO phi}& symplectic & const. noise\\
      %
     %\hline
     Vortex in superfluid thin films & \eqref{eq:vortexdiss} & \eqref{eq:vortex Ham}& symplectic & const. noise\\
     %
     %\hline
     Landau-Lifshitz-Gilbert spin & \eqref{eq:LLG H} & \eqref{eq:llg}& symplectic & const. noise\\
     %
     %\hline
     \midrule
     Odd harmonic oscillator & \eqref{eq:eomoddHO} & \eqref{eq:HO phi}& \eqref{eq:W SHO active terms} & const. noise\\
     %
    % \hline
     Non-reciprocal Kuramoto model & \eqref{eq:NRK fun EOM} & \eqref{eq:weirdPhi} & \eqref{eq:nonrecipWkuramoto} & const. noise\\
     %
     %\hline
     Wealth & \eqref{eq:money SDE} & \eqref{eq:money stationary} & N/A & const. noise\\
     \midrule
     %\hline
     Rigid body & \eqref{eq:eom rigid diss}\eqref{eq:eom rigid active} & \eqref{eq:H rigid} & \eqref{eq:W rigid active} & const. noise\\
     %
     %\hline
     \midrule
     Odd diffusion & \eqref{eq:diffusion eom} & \eqref{eq:diffusionPhi} & \eqref{eq:L diffusion active} & yes\\
     %
     %\hline
     Odd elasticity & \eqref{eq:EOM_first_oddelastic_pot}\eqref{eq:odd_elasticity_EOM_pot2} & \eqref{eq:odd_elasticity_phi1}\eqref{eq:odd_elasticity_phi2}\eqref{eq:odd_elasticity_phi3} & \eqref{eq:odd_elasticity_lagrangian}\eqref{eq:odd_elasticity_lagrangian2} & yes\\
     %
     %\hline
     Odd viscosity & \eqref{eq:viseom} & \eqref{eq:Phi viscosity} & \eqref{eq:L viscosity active} & yes\\
     %
     %\hline
     Odd stress & \eqref{eq:eom spin active} \eqref{eq:pt eom odd stress} & \eqref{eq:Phi spin} \eqref{eq:Phi odd stress} & \eqref{eq:K term spin} \eqref{eq:pt L odd stress} & yes\\
     %
     %\hline
     %\midrule
     Fracton hydrodynamics & \eqref{eq:dipole eom} & \eqref{eq:dipolePhi classical} \eqref{eq:dipolePhi quantum} & N/A & yes\\
     %
     %\hline
     \midrule
     Phase separation & \eqref{eq:activekappa} & \eqref{eq:Phi separation} & \eqref{eq:W separation active}  & break It\^o\\
    \bottomrule
    \end{tabular}
    \caption{Summary of the various example systems considered in this paper. If the stationary state is given by some Hamiltonian $H$, then $\Phi = \beta H$ corresponds to thermal equilibrium. 
    The symplectic $V_{ab}$ refers to the Hamiltonian dynamics in \secref{subsec:Hamiltonian setup}.
    The last column indicates whether the theory is compatible with MSR for different regularizations: ``const. noise'' means to be compatible with \secref{sec:MSR} the noise must not be multiplicative; ``yes'' means that MSR path integrals are exact upon applying the spatial regularization of \appref{app:diffusion}, MSR is applicable; ``break It\^o'' means the MSR formalism of Sec.~\ref{sec:MSR} is not applicable. A formal resolution to the case of ``break It\^o'' is provided in App.~\ref{app:PI}. } 
    \label{tab:summary application}
\end{table}

\subsection{Generalizations and comparisons to previous literature}

Looking forward, there are a number of open directions for extending this work, which we now enumerate. First, this paper has focused on active \emph{classical} matter; some of us will propose a notion of active \emph{quantum} matter in a follow-up paper \cite{aqm}.  We also plan to develop a more systematic classification for active systems with spontaneously broken symmetries, including spacetime symmetries, which may lead to new universality classes with ``flocking" phenomena \cite{elipaper}.

It is also important to understand the extent to which our framework does \emph{not} crisply capture known phases of active matter.  The key challenge is that, in our formalism, it is important to know the stationary distribution $\exp[-\Phi]$ of the dynamics; in many currently studied systems of active matter, this is not easily determined.  Moreover, it is possible that $\Phi$ is not expressible within a Wilsonian (i.e., derivative, few-body) expansion.  For example, the Toner-Tu model violates the equilibrium Mermin-Wagner theorem \cite{TonerTu}, meaning that $\Phi$ must be nonlocal.  It would be interesting to understand if there is a systematic means by which to construct a nonlocal $\Phi$, while maintaining strict locality in the equations of motion and noise correlations.

On an optimisitc note, it is sometimes the case that \emph{equilibrium} quantum phase transitions once believed to be beyond Landau's paradigm are now argued to be compatible therewith \cite{McGreevy:2022oyu} (once one considers, e.g., higher-form symmetries).   So it might be the case that our EFT approach to active matter can classify all known phases, once suitable generalized symmetries and/or constructions of $\Phi$ are identified.  We leave these important open questions to future work.

The thermal EFTs for fluids with conserved charges and energy/momentum  neatly couple to background gauge fields and geometry, respectively.  In \emph{quantum} field theory, it seems crucial to have these background couplings to ensure that a theory is not anomalous; however, we have not found that such couplings are necessarily natural within our formalism.  We anticipate that it is not possible, in general, to couple to background geometry in the same way as in a thermal fluid---a simple obstruction we have already encountered is the existence of odd stress in certain active chiral fluids, since this effect is forbidden by KMS symmetry \cite{huang_2022_discrete}.  The lattice model of \cite{Qi:2022vyu} demonstrates that such anomalous effects, forbidden in thermal systems, are readily possible in microscopic classical models of active systems.  However, our formalism additionally allows for time-reversal-breaking terms are possible that are not compatible with any known anomaly in quantum field theory.  A simple example is a biased diffusion equation $\pd*{t} \rho + \pd*{x} v(\rho) - D\pdp{x}{2} \rho + \cdots = 0 $---for generic $v(\rho)$ and $\Phi = \int \thed x \, \rho^2$, one can show that this theory is consistent with stationarity, and analyze it using the EFT framework we develop.  However, the apparent anomaly coefficient is nonlinear and field dependent, rendering it beyond the existing classification of anomalies in quantum field theory.  As far as we can tell, one must use our MSR framework to describe such a theory, rather than a ``geometrically inspired" approach.

\subsection{Towards classifying active phases of matter}\label{sec:classify active}

Finally, conclude by posing the question, \emph{when can a phase only be realized out of equilibrium}?  Certain phases of matter may arise in active matter, where the microscopic degrees of freedom are driven by, e.g., internal batteries, and yet the long-wavelength description of such phases may reproduce a thermal theory (such as a diffusion equation).  In what phases can we show that \emph{no thermal system} can realize an active phase of matter?
%or whether it has a counterpart in an effectively thermal fluid which obeys a generalized KMS symmetry (e.g. incorporating spatial parity together with time-reversal).

A first proposal is that any system that breaks detailed balance (time-reversal) is active.  For example, such fluids can have odd viscosity and diffusion.  However, such odd transport coefficients can be reproduced by PT-symmetric thermal systems \cite{huang_2022_discrete}.  An experimental ``realization" of such a system---at least at the level of linear response---is an electron fluid in a background magnetic field \cite{Berdyugin_2019}, albeit strictly speaking, momenta is only conserved if the fluid is overall charge neutral, as can be realized in graphene \cite{Crossno_2016, Lucas:2017idv}. Since such fluids can have odd viscosity, this time-reversal-odd transport coefficient is not unique to active fluids.

An example of a system that appears to be truly nonequilibrium---yet also describable fully in our framework---is the chiral phase of the nonreciprocal Kuramoto model \cite{nonreciprocalPhase}.  To understand why, notice that if we have a Hamiltonian system that spontaneously breaks time-reversal symmetry, it must be the case that the minima of $\Phi$ (where the system is most likely to be found) are \emph{not} extrema of $H$ (as there is no deterministic dynamics if the gradient of $H$ vanishes).  This means that $\Phi \ne \beta H$ is required to have any mathematical ``Hamiltonian description" of a nonreciprocal phase with spontaneous T-breaking.  At best, therefore, one relaxes to a strongly nonthermal stationary distribution, so it makes sense to us to call this phase active.

That active matter can both realize genuinely new phases, or simply reproduce physics possible in thermal systems, may have interesting implications for frameworks in which the amount of entropy production in the steady-state is used as a diagnostic for ``activity"   \cite{schnakenberg,oliveira}.  Such definitions can lead to entropy production from a single particle spinning in a cyclotron orbit in a magnetic field (a system which has PT, but not T, symmetry). On the other hand, at least in hydrodynamics, there is  a natural notion of entropy production in hydrodynamics that vanishes for a fluid in thermal equilibrium, even when placed in a background magnetic field \cite{hartnoll,surowka}.

Our proposal that nonreciprocal phases where a generalized time-reversal symmetry is spontaneously broken can only be realized in active matter could be a first step towards identifying phases unique to active matter, yet surely there are other possibilities as well.  We expect that the tools, and simply the perspective, advocated in this manuscript can play an important role in comprehensively understanding what is possible in active systems constrained not only by thermal equilibrium, but mere statistical stationarity.

\section*{Acknowledgments}
We thank Jay Armas, Michel Fruchart, Jinkang Guo, Oliver Hart, Akash Jain, Marvin Qi,  Leo Radzihovsky, Yi-Zhuang You, and especially Vincenzo Vitelli, for useful discussions.  This work was supported in part by the Alfred P. Sloan Foundation through Grant FG-2020-13795 (AL), the National Science Foundation through CAREER Grant DMR-2145544 (XH, IZ, AL) and CAREER Award DMR-2045181 (PG), through the Gordon and Betty Moore Foundation's EPiQS Initiative via Grant GBMF10279 (XH, JHF, AL),  by
the Laboratory for Physical Sciences through the Condensed Matter Theory Center of the University of Maryland, College Park (PG), by the Department of Energy through Award DE-SC0019380 (PG), and by the Simons Foundation through Award No. 620869 (PG) and a Simons Investigator Award via Leo Radzihovsky (AJF).  AL thanks NORDITA and  KITP for hospitality while this manuscript was written. KITP is supported by the National Science Foundation through Grant PHY-1748958.

\begin{appendix}

\renewcommand{\thesubsection}{\thesection.\arabic{subsection}}
\renewcommand{\theequation}{\thesection.\arabic{equation}}

\section{Topological exceptions to the form of dissipationless terms}\label{sec:topoappendix}
Are expressions of the form \eqref{eq:v_anti} are the \emph{only} solutions to the condition \eqref{eq:dissipationless_terms}? In fact, there exists a single, special case that is not captured by constructions of the form \eqref{eq:v_anti}. Suppose that the (phase-space) coordinates $\{ q^{\,}_a \}$ belong to an $N$-dimensional manifold $M$; for simplicity suppose that $M$ is compact. In this case, one can use de  Rham cohomology to classify all possible solutions to (\ref{eq:v_anti}) by expressing the equation in the language of differential forms:
\begin{equation}
\label{eq:dissipationless de Rham}
    0 \, = \, \pd{a} \left(v_a \e^{-\Phi}\right) = \mathrm{d} * \left( v \e^{-\Phi}\right) \, . ~~
\end{equation}
In the right equation above, we wrote $v=v_a\mathrm{d}x_a$ as a 1-form; the Hodge star $*$ converts $*\mathrm{d}x_1 = \mathrm{d}x_2\wedge \cdots \wedge \mathrm{d}x_N $, e.g., and allows us to express divergences in terms of the exterior derivative $\mathrm{d}$ acting on forms. The solutions captured by \eqref{eq:v_anti} in terms of the antisymmetric matrix $V^{\,}_{ab}$ in \eqref{eq:nice Q def} are those in which $*v$ is an exact form: \begin{equation}
    *v = \mathrm{d}(*V)
\end{equation}
where $V$ is a 2-form, which can be thought of as an antisymmetric matrix. The de Rham cohomology ensures that these are \emph{all} solutions to \eqref{eq:dissipationless de Rham} up to certain topological solutions that can exist when the manifold $M$ has noncontractible cycles (a nontrivial $\mathrm{H}^1(M)$ cohomology group).  For example, if we look at $M=\mathrm{T}^N$, where $\mathrm{T}^N$ is the $N$-dimensional torus with coordinates $(x_1,\ldots, x_N)$, each of which obeys $x_i \sim x_i + 1$, we find $N$ linearly independent topological solutions to \eqref{eq:dissipationless de Rham} corresponding to 
\begin{equation}
    v \e^{-\Phi} = \thed x^{\vpp}_i
\end{equation}
for any $i$.  These solutions correspond to flows that cycle around one direction of the torus persistently. In Sec.~\ref{sec:general_T_breaking}, we will be particularly interested in the existence of such cycles on nontrivial manifolds $M$. 

\section{Second law of thermodynamics} \label{app:entropies}
In this appendix, we describe settings in which the second law of thermodynamics arises naturally in our formalism, even in nonequilibrium matter.

\subsection{Linear response}\label{app:entropy}
In Sec. \ref{sec:MSR}, we discussed an analogy between our formalism and recent EFTs for dissipative thermal systems.  In thermal field theories for dissipative systems, there is an elegant way to construct an entropy current whose divergence is nonnegative \cite{eft2,Glorioso:2016gsa,Liulec}.  The existence of this entropy current underpins the Landau formulation of hydrodynamics.  Let us now show that -- subject to two key assumptions -- the prescription of Sec. \ref{sec:2ndlaw} will reproduce the thermal EFT derivation of an entropy current and extend it to active fluids.

We proceed by using the Lagrangian formalism, closely following \cite{eft2,Glorioso:2016gsa,Liulec} to derive an entropy current.  Then, we will connect to the formalism in Sec. \ref{sec:2ndlaw}. Before starting the derivation, let us emphasize that the two key assumptions we will be using.  Firstly, we will use that we work only in the linear response regime (we hope to generalize this elsewhere).  Secondly, we will assume Gaussian noise; recall that the construction of Sec. \ref{sec:2ndlaw} is also restricted to this case.  With these assumptions, gT is equivalent to the KMS-like transformation \eqref{eq:MSR T pi relation}, which simplifies the resulting analysis greatly.

Let us begin by considering the general (and most interesting) limit of a field theory (Section \ref{sec:continuum}).  Under gT symmetry, $\mL$ will transform to $\tilde \mL$ where tilde represents gT transformations, and we have
\begin{align}
    \tilde \mL - \mL = \p_t V^t + \partial_i V^i = \partial_\mu V^\mu. \label{eq:dmuVmu}
\end{align} 
For this appendix only, we will use an Einstein summation convention on spacetime $\mu$ indices. 
Expanding $V^\mu = \ii V^\mu_{(0)} + V^\mu_{(1)}+\ldots$, where $V^\mu_{(k)} \sim O(\pi^k)$, we find that if using the manifestly gT/KMS-invariant $\mL$ in \secref{sec:MSR}, there may be only one contribution $V^\mu_{(0)}\neq 0$ and $V^\mu_{(k)}=0,\; k>0$.  
To build the entropy current explicitly, it is therefore convenient to integrate by parts and write $\mL$ as, schematically,
\begin{align}
    \mL = E_a[q] \pi_a+\frac{\ii}{2} \pi_a F_{ab}[q,\overleftrightarrow{\p}] \pi_b,
\end{align}
where $E_a[q],F_{ab}[q,\overleftrightarrow{\p}]$ are local functions except that $F_{ab}[q,\overleftrightarrow{\p}] = F_{ba}$ contains derivatives acting symmetrically on the $\pi_a$, while $E_a\pi_a$ has no derivatives acting on $\pi_a$. 
 For example, for the theory of charge diffusion, we would write \begin{equation}
    \mL = \pi \partial_t \rho + \ii \sigma \partial_i \pi \partial_i \left(\pi - \ii \mu\right) \rightarrow  \pi \partial_t \rho - \pi \sigma\partial_i\partial_i \mu  + \ii \sigma \partial_i \pi \partial_i \pi.
\end{equation} Using this Lagrangian, we can have nontrivial $V^\mu_{(1)}$ as well, which is important to have entropy production out of equilibrium. Defining a diagonal matrix $\eta_{ab}$ whose $\pm 1$ entries encode whether the variable $q_a$ is even or odd under T, we can explicitly use the definition (\ref{eq:dmuVmu}) to obtain
\begin{subequations}
    \begin{align}
        \p_\mu V^\mu_{(0)} &=\eta_{ab} \tilde E_a \mu_b - \frac{1}{2} \mu_a \tilde F_{ab} \mu_{b},\\
        \p_\mu V^\mu_{(1)} &= -\eta_{ab}\tilde E_{a} \pi_{b} - E_{a} \pi_{a} +\frac{1}{2}\mu_{a} \tilde F_{ab} \pi_{b} + \frac{1}{2} \pi_{a} \tilde F_{ab} \mu_{b},
    \end{align}
\end{subequations}
where we used $\pi_{a} \tilde F_{ab} \pi_{b} =  \pi_{a}  F_{ab}\pi_{b}$.
Let us define $\hat V^\mu_{(1)}$ by replacing the $\pi$ by its corresponding conjugate chemical potential $\mu$ in $V^\mu_{(1)}$, and then define the entropy current as
\begin{align}
    s^\mu \equiv V^\mu_{(0)} + \hat V^\mu_{(1)}.
\end{align}
Hence, we find
\begin{align}\label{eq:entropy production}
    \p_\mu s^\mu & = -E_{a} \mu_{a} + \frac{1}{2} \mu_{a} \tilde F_{ab} \mu_{b} = \frac{1}{2} \mu_a  F_{ab} \mu_b \geq 0,
\end{align}
where we used $E_a=0$ as the equation of motion, and %the semi definite positivity 
positive semidefiniteness comes from the fact that distribution of fluctuations is bounded, and thus $\im \mL \geq 0$. When there is no dissipation, we find a conserved entropy current $\p_\mu s^\mu =0$, and, from the construction in \secref{sec:MSR}, we can immediately see that the entropy is equal to the steady-state distribution
\begin{align}\label{eq:total entropy}
    \Phi =  - \int \ud^d x  V^0_{(0)} = - \int \ud^d x s^0.
\end{align}

As an illustrative example, consider a simple chiral fluid with degree of freedom $u_i$ and $n$. From \eqref{eq:pt L spin} with the spin turned off, we can extract the $V^\mu$ as
\begin{subequations}
    \begin{align}
        \p_\mu V^\mu_0 &= -\p_t u_i \mu_i - \p_t \delta n \mu^n - n_0\p_i(u_i \mu^n) -\beta_0^{-1} \rho^{-2} \eta^{\even}_{jikl}\p_j (\mu_i \p_k \mu_l) - \beta_0^{-1} D^\even_{ij}\p_i(\mu^n \p_j \mu^n)\nonumber\\
        &+\beta_0^{-1} \rho^{-2}\eta^{\odd}_{jikl}\mu_i \p_j \p_k \mu_l +\beta_0^{-1} D^{\odd}_{ij}\mu^n \p_i \p_j \mu^n,\\
        \p_\mu \hat{V}^\mu_1 &= 2\beta_0^{-1} \rho^{-2} \eta^{\even}_{jikl}\p_j (\mu_i \p_k \mu_l) +2\beta_0^{-1} D^\even_{ij}\p_i(\mu^n \p_j \mu^n).
    \end{align}
\end{subequations}
This leads to the entropy current
\begin{subequations}
    \begin{align}
        s^0 &= - \beta_0 \frac{\rho}{2} u_i^2 - \beta_0 \frac{1}{2\chi} \delta n^2 ,\\
        s^i &= - \beta_0 \frac{n_0}{\chi} \delta n u_i + \beta_0 \frac{1}{\chi} D_{ij} \delta n\p_j \delta n  + \beta_0 \eta_{ijkl}u_j \p_k u_l,
    \end{align}
\end{subequations}
where $D_{ij}$ and $\eta_{ijkl}$ include both T-even and T-odd terms.
Notice that \eqref{eq:total entropy} is satisfied.
Upon using equations of motion, we arrive at
\begin{align}
    \p_\mu s^\mu = \beta_0 \frac{1}{\chi} D^{\even}_{ij} \p_i \delta n \p_j \delta n + \beta_0  \eta^{\even}_{ijkl} \p_i u_j \p_k u_l \geq 0.
\end{align}
While this goes against the common expectations in the literature as far as we can tell, we showed how this arises naturally at least in simple settings.  We leave a careful generalization of this argument to arbitrary nonlinear systems, using our operator formalism, to future work.

Next, let us connect back to the discussion of Sec. \ref{sec:2ndlaw}. For simplicity of notation in the following two paragraphs, we describe systems with a finite-dimensional phase space, but the conclusions are general.  As before, we are focused in this appendix on the linear response limit. In this limit, we can take $P(q_a,t)$ to be a perturbed Gaussian distribution with variance $\sigma_a^2$ and mean zero:
\begin{align}\label{eq:Pt app}
    P(q_a,t)\propto \exp\left[ - \sum_a \frac{(q_a - \hat q_a(t))^2}{2\sigma_a^2}\right],
\end{align}
where $\hat q_a(t)$ represent the degrees of freedom of hydrodynamics (e.g., in a nonfluctuating limit we would have above equation of motion $E_a(\hat q_a)=0$) -- i.e. they are expectation values of $q_a$ instantaneously in $P(t)$. We have, without loss of generality in the linear response regime, further rotated the variables $q_a$ so that the genuine stationary state has \begin{equation}
    \Phi = \sum_a \frac{q_a^2}{2\sigma_a^2}.
\end{equation} Plugging in \eqref{eq:EFT entropy def}, with $P^{\rm ss} = P(q_a;\hat q_a=0)$, we find
\begin{align}
    D(\hat q_a) = - \sum_a \frac{\hat q_a^2}{2 \sigma_a^2} = \Phi(\hat q_a).
\end{align}
We see that the entropy \eqref{eq:total entropy} agrees with the Kullback-Leibler divergence \eqref{eq:EFT entropy def}. 

As a further sanity check, it is helpful to explicitly confirm that the FPE of \eqref{eq:Pt app} precisely reproduces the naive hydrodynamic equation for $\hat q_a$ alone.  First we evaluate the FPE directly:
\begin{align}\label{eq:FPE Pt app}
    \frac{\ud }{\ud t}P = -WP =  \p_a Q_{ab} (\p_b+\mu_b) P = - (\mu_a - \hat \mu_a) Q_{ab}\hat{\mu}_b P.
\end{align}
where $\hat \mu_a = \hat q_a/\sigma_a^2$.  Then, we can evaluate the time-derivative by simply assuming the form (\ref{eq:Pt app}), and checking when the two answers are identical: \begin{align}\label{eq:FPE Pt app 2}
     \frac{\ud }{\ud t}P = \left(\mu_a - \hat \mu_a\right) \frac{\mathrm{d}\hat \mu_a}{\mathrm{d}t} P.
\end{align} So the FPE \eqref{eq:FPE Pt app} agrees with \eqref{eq:FPE Pt app 2} so long as we enforce the noise-free equation of motion
\begin{align}
    E_a(\hat \mu_a) = - \p_t \hat \mu_a +  Q_{ab}\hat \mu_b=0.
\end{align}
This short calculation explains why our notion of second law always exists within linear response: the full FPE is mathematically equivalent to an equation for the averages $\hat q_a$ alone \cite{bankslucas}.

\subsection{A nonequilibrium system}

Lastly, let us describe how such an entropy current arises even in a driven nonequilibrium system that breaks time-reversal symmetry: a diffusing charge between two reservoirs held at different densities.   The degrees of freedom are $\rho(x)$ for $|x| \le a$, and we take the boundary conditions to be fixed: \begin{equation}
    \rho(\pm a) = \pm a \alpha,
\end{equation}
for some constant $\alpha$.   The steady-state solution to the linear diffusion equation $\partial_t \rho  =D \partial_x^2 \rho$ is \begin{equation}
    \rho_{\mathrm{ss}}(x) = \alpha x.
\end{equation}
In a thermal fluid which relaxes to a uniform density, the entropy production rate is locally proportional to $D(\partial_x \rho)^2$, so one could compute the entropy production in this diffusive theory as \begin{equation}
    \dot{S}_{\mathrm{ss}} \sim  \int\limits_{-a}^a \mathrm{d}x D (\partial_x \rho_{\mathrm{ss}})^2 = 2Da \alpha^2 > 0, 
\end{equation}
and thus deduce that this system is out of equilibrium as entropy is locally produced for all times (and leaks through the boundaries).   

In our framework, this situation is understood a bit differently (and we hope that the contrast clarifies the sense in which nonequilibrium systems nonetheless can obey a second law).  The steady state of the dynamics is readily found by making the elementary observation that if we write \begin{equation}
    \rho(x,t) = \rho_{\mathrm{ss}}(x) + \tilde\rho(x,t),
\end{equation}
then (fixing the charge susceptibility to be 1, i.e. $\mu=\tilde\rho$) we see that 
\begin{equation}
    \Phi \sim \frac{1}{2}\int\mathrm{d}x \; \tilde\rho^2
\end{equation}
is the same $\Phi$ that would arise for a theory of conserved charge with ordinary ``equilibrium boundary conditions" ($\tilde\rho(\pm a)=0$).  From our perspective, the entropy current becomes \begin{subequations}
    \begin{align}
        s^0 &= -\Phi =  -\frac{1}{2}\tilde \rho^2 = -\frac{1}{2}(\rho - \alpha x)^2 , \label{eq:s0us} \\
        s^x &= -D\tilde \rho \partial_x \tilde \rho = -D (\rho-\alpha x)(\partial_x \rho - \alpha).
    \end{align}
\end{subequations}
We have written the equations in terms of both the original $\rho$, and the ``nicer" $\tilde \rho$, to emphasize that the entropy current naturally defined by our formalism becomes space-dependent in the ``physical" coordinates due to the nonequilibrium boundary conditions.  One can straightforwardly check that this entropy current obeys the second law of thermodynamics for the nonequilibrium steady state.  We thus put forth the point of view that restoring the second law by \emph{changing the definition of entropy} is an alternative perspective to the common notion of defining nonequilibrium systems as those in which entropy is produced.  While our perspective then loses a physically motivated definition for nonequilibrium, we gain the power of the second law and the fluctuation-dissipation theorem in return.  

Finally, we emphasize that in this simple example,  our proposal that entropy production vanishes in the nonequilibrium steady state is similar to previous hydrodynamic theories in background fields \cite{hartnoll,surowka}.  In this setting, one may elect to define the entropy \emph{density} differently as
\begin{equation}
    s^0 = -\frac{1}{2}\rho^2,
\end{equation}
instead of (\ref{eq:s0us}), although a quick calculation confirms that $\partial_t \int \mathrm{d}x s^0$ is the same with both definitions in this theory.  Beyond this convention, and the normalization of the entropy current, our results then agree with the hydrodynamic theories of \cite{hartnoll,surowka}.

\section{Derivation of the MSR path integral}\label{app:PI}

To compute the Jacobian in \eqref{eq:zxi}, we adopt the $\alpha$-regularization \cite{lubensky} as a generalization of the It\^o ($\alpha=0$) and Stratonovinch ($\alpha=1/2$) regularizations. The discretization of the path integral depends on the choice of $\alpha$ as follows. By dividing the time to $N+1$ intervals $\Delta_t$, the matrix elements of the Jacobian become $\mJ^{(ij)}_{ab} = \p \hat{N}^{(j)}_{b} / \p q^{(i)}_{a}$, with $q^{(i)}_{a} \equiv q_a(t_i)$ and
\begin{align}
    \hat{N}^{(j)}_{b} = q_{b}^{(j)}-q_{b}^{(j-1)} - \Delta_t \left[ f_{b}\left(\alpha q^{(j)}+(1-\alpha)q^{(j-1)}\right)  +b_{bc}\left(\alpha q^{(j)}+(1-\alpha)q^{(j-1)}\right) \xi_{c}^{(j-1)}  \right] \, ,~~
\end{align}
and we see that $\mJ_{ab}^{(ij)}$ is a lower-triangular matrix in $(ij)$ with diagonal elements
\begin{align}
    \mJ_{ab}^{(ii)} = 
    \delta_{ab}-\alpha\Delta_t \p_a f_b-\alpha\Delta_t \p_a b_{bc}\xi_{c}^{(i-1)}.
\end{align}
Using $\det A = \exp \Tr \ln A$, we obtain the Jacobian $\mJ = \prod_i \mJ^{(ii)}$, where
\begin{align}\label{eq:jacobiandiag}
    \mJ^{(ii)}[\rho;\xi] = 1-\alpha\Delta_t \p_a f_a-\alpha\Delta_t \p_a b_{ab}\xi_{b}^{(i-1)}+\frac{(\alpha\Delta_t)^2}{2}\left(\p_a b_{ac}\p_b b_{bc^\prime} - \p_a b_{bc}\p_b b_{ac^\prime} \right)\xi_{c}^{(i-1)}\xi_{c^\prime}^{(i-1)}+\ldots.
\end{align}
In the above equation, we have only kept terms at leading order $\mO(\Delta_t)$ and used $\xi^{(i)}\sim (\Delta_t)^{-1/2}$.
Since the third term scales as $(\Delta_t)^{1/2}$, direct exponentiation does not work. Therefore, we first replace every noise term in $\mJ^{(ii)}$ by $ \xi_{a}^{(i-1)} \Delta_t=b^{-1}_{ab}(q_{b}^{(i)}-q_{b}^{(i-1)}-\Delta_t f_{b})$ according to the equation of motion, then perform the noise averaging. With the help of an auxiliary field $\pi_{\alpha}$, we obtain
\begin{align}
    Z &= \int \theD \xi ~ Z[\xi] \e^{-\frac{1}{2}\int \ud t~\xi^2} = \int \theD \rho \prod_i \mJ^{(ii)}[\rho]  \int \theD \pi ~ \e^{\ii \int\ud t \left[\pi_a \p_t q_a - f_a\pi_a+\frac{\ii}{2}\pi_a Q_{ab}\pi_b\right] }.
\end{align}
After applying the trick 
\begin{align}
    \left(q_{b}^{(i)}-q_{b}^{(i-1)}-\Delta_t f_{b} \right)\e^{\ii \int\ud t \left[\pi_a \p_t q_a - f_a\pi_a\right] } = -\ii \frac{\p}{\p \pi_{b}^{(i)}}\e^{\ii \int\ud t \left[\pi_a \p_t q_a - f_a\pi_a \right] },
\end{align}
and integrating by parts, we arrive at
\begin{align}\label{eq: L alpha}
    Z = \int \theD q \theD \pi \prod_i \left(1-\alpha\Delta_t \p_a f_a + \ii \alpha\Delta_t\p_a b_{ab} b_{cb}\pi_c + \frac{\alpha^2\Delta t}{2}\left(\p_a b_{ac}\p_b b_{bc} - \p_a b_{bc}\p_b b_{ac} \right)  \right)  ~ \e^{\ii \int\ud t \left[\pi_a \p_t q_a - f_a\pi_a+\frac{\ii}{2}\pi_a Q_{ab}\pi_b\right] }.
\end{align}
Now we can exponentiate the Jacobian and obtain $Z = \int \theD q\theD \pi \e^{\ii \int \ud t L}$ with the MSR Lagrangian given by
\begin{align}\label{eq:effL_old}
    L(q,\pi) = \pi_{a}\p_t q_{a} - \left(f_a - \alpha b_{ac}\p_b b_{bc}\right)\pi_{a}+\ii\alpha\p_a f_a-\frac{\ii\alpha^2}{2}\left(\p_a b_{ac}\p_b b_{bc} - \p_a b_{bc}\p_b b_{ac} \right)+\frac{\ii}{2} \pi_{a} Q_{ab}\pi_{b}.
\end{align}
There are two problems to address
The first problem is the existence of the $O(\pi^0)$ terms. In thermal equilibrium, one can use causality or retardness of response function to argue that these $O(\pi^0)$ terms must vanish \cite{ghostbusters}, but this is not known for systems out-of-equilibrium. We will leave this problem open, although we find it might be possible to argue that the loop diagrams that violate the causality would exactly cancel. In fact, these $O(\pi^0)$ terms does not show up once we convert \eqref{eq:effL_old} into the FPE, which is given by \cite{lubensky}
\begin{align}\label{eq:MSR FPE}
    \p_t P = - \p_a\left[ (f_a +\alpha \p_b b_{ac} b_{bc})P - \frac{1}{2}\p_b(Q_{ab}P)\right].
\end{align}
The second problem is how to implement the T symmetry in \eqref{eq:effL_old}. We wish to have the minimal transformation: $\pi\to -\pi+\ii\mu$ as in \eqref{eq:MSR T pi relation}. Let $f_a =G_a+v_a $ with $G_a/v_a$ T-even/odd. The minimal transformation requires the T-even force to be
\begin{align}\label{eq:G noise eqn1}
    G_a - \alpha b_{ac}\p_b b_{bc}=-\frac{1}{2} Q_{ab}\mu_b.
\end{align}
On the other hand, in order for \eqref{eq:MSR FPE} to support a steady-state distribution up to a T-odd force, we find
\begin{align}\label{eq:G noise eqn2}
    G_a + (\alpha-\frac{1}{2})\p_b b_{ac}b_{bc} - \frac{1}{2} b_{ac}\p_b b_{bc} = -\frac{1}{2}Q_{ab}\mu_b.
\end{align}
To satisfy both \eqref{eq:G noise eqn1} and \eqref{eq:G noise eqn2}, we must choose $\alpha=1/2$. Therefore, to implement \eqref{eq:MSR T pi relation} as our T symmetry, we must take $\alpha=1/2$ in the Lagrangian \eqref{eq:effL_old}. Note that in the main text, we can deal with the It\^o scheme $\alpha=0$ when we assume the weak noise limit, such that $G_a\approx -\frac{1}{2}Q_{ab}\mu_b$ for both \eqref{eq:G noise eqn1} and \eqref{eq:G noise eqn2}, or we use the regularization of App.~\ref{app:diffusion} in the continuum limit.

Stratonovinch regularization $\alpha=1/2$ is also the most natural choice to make the T-odd force invariant under the generalized T symmetry (gT). Taking $\alpha=1/2$, the T-odd terms in \eqref{eq:effL_old} are
\begin{align}
    H^{\odd} = \pi_a v_a - \ii \frac{1}{2} \p_a v_a.
\end{align}
One can show that only under the constraint \eqref{eq:dissipationless_terms} can $H^{\odd}$ be gT-invariant. This generalizes the weak noise limit in the main text. Moreover, if the system only contains T-odd terms, i.e. dissipationless, we can also let the gT symmetry to act on an arbitrary regularization $\alpha$ but, at the same time, shift $\alpha\to 1-\alpha$. For example, $H^{\odd}_{\alpha=0} = \pi_a v_a$ is invariant under gT by transforming to $H^{\odd}_{\alpha=1} = \pi_a v_a - \ii \p_a v_a$. This thus explains why the naive KMS symmetry discussed at the end of \secref{subsec:Hamiltonian setup} for Hamiltonian dynamics with a state-dependent symplectic form will not know the shift in $\Phi$ \eqref{eq:Phiham} -- the gT symmetry includes both the KMS transformation and the shift in $\alpha$.

Lastly, we note that it is challenging to straightforwardly construct higher-order terms in (\ref{eq:effL_old}), whereas such a generalization would be straightforward in the It\^o path integral, so long as time-reversal is implemented by (\ref{eq:MSR T pi relation}).  For this reason, we believe that these MSR path integrals will be most valuable in models where the method of App.~\ref{app:diffusion} can be employed to neglect subtleties with regularization.

\subsection{Correlation functions}
An advantage of the above Lagrangian is that, although reversibility is not manifest, it does allows us to compute response/correlation functions using a more familiar diagrammatic approach.  As one example, let us show how the FDT \eqref{eq:corr fdt} is derived using the T symmetry.  We perturb the steady-state distribution by $\Phi \to \Phi - h_a(t) q_a(t)$. Under such perturbation, the forces will shift in the Lagrangian, so we obtain (note that we take $\alpha=1/2$)
\begin{align}\label{eq:perturb ss}
    -\frac{\ii}{2}\langle q_a(t)Q_{bc}\pi_c(t^\prime) \rangle_0 -\ii \langle q_a(t) \frac{\delta v_c}{\delta h_b} \pi_c(t') \rangle_0 - \frac{1}{2} \langle q_a(t) \p_c\frac{\delta v_c}{\delta h_b} (t') \rangle_0 =\frac{\delta \langle q_a(t)\rangle}{\delta h_b(t^\prime)}|_{h_a=0},\quad \mathrm{for}\;\; t>t^\prime.
\end{align}
According to \eqref{eq:dissipationless_terms}, $\delta v_a$ satisfies
\begin{align}\label{eq:deltav}
    \mu_a \frac{\delta v_a}{\delta h_b} - \p_a \frac{\delta v_a}{\delta h_b} = v_b.
\end{align}
Consider acting T on the pieces in \eqref{eq:perturb ss} that are proportional to $\delta v_a$:
\begin{align*}
    \mathrm{T:} \;\; -\ii \langle q_a(t) \frac{\delta v_c}{\delta h_b} \pi_c(t') \rangle_0 - \frac{1}{2} \langle q_a(t) \p_c\frac{\delta v_c}{\delta h_b} (t') \rangle_0 & \to -\ii \langle q_a(t') \frac{\delta v_c}{\delta h_b} \pi_c(t) \rangle_0 - \frac{1}{2} \langle q_a(t') \p_c\frac{\delta v_c}{\delta h_b} (t) \rangle_0\nonumber\\
    &\;\;\;\;\;- \langle q_a(t') \frac{\delta v_c}{\delta h_b} \mu_c(t) \rangle_0+\langle q_a(t') \p_c\frac{\delta v_c}{\delta h_b} (t) \rangle_0.
\end{align*}
Using the T-invariance of it and \eqref{eq:deltav}, we obtain, for $t>t'$,
\begin{align}\label{eq:correlation v shift}
    -\ii \langle q_a(t) \frac{\delta v_c}{\delta h_b} \pi_c(t') \rangle_0 - \frac{1}{2} \langle q_a(t) \p_c\frac{\delta v_c}{\delta h_b} (t') \rangle_0 = \langle q_a(t) v_b(t^\prime)\rangle_0. 
\end{align}
Plugging \eqref{eq:correlation v shift} in \eqref{eq:perturb ss} and using \eqref{eq:MSR T pi relation}, we obtain
\begin{align}
    \chi_{ab}(t,t^\prime)\equiv\frac{\delta \langle q_a(t)\rangle}{\delta h_b(t^\prime)}|_{h=0} =  \frac{1}{2}\langle q_a(t)Q_{bc}\mu_c(t^\prime) \rangle_0 + \langle q_a(t) v_b(t^\prime)\rangle_0,\quad \mathrm{for}\;\; t>t^\prime.
\end{align}
From \eqref{eq:perturb ss} and employing the time-translation symmetry, $\p_t\langle q_a(t)q_b(t^\prime) \rangle_0 = -\p_{t^\prime}\langle q_a(t)q_b(t^\prime) \rangle_0$, we arrive at our generalized FDT:
\begin{align}\label{eq:fdt2}
    \p_t\langle q_a(t)q_b(t^\prime) \rangle_0 = -\chi_{ab}(t,t^\prime)\Theta(t-t^\prime).
\end{align}
Remarkably, both the T-odd force $v_a$ and the multiplicative noise are encoded in the response function $\chi_{ab}$ which can be probed by perturbing the steady-state distribution. We note that \eqref{eq:fdt2} generalizes the result in \cite{lubensky} to $v_a\neq 0$.

\section{Convenient regularization of a nonlinear diffusion equation}
\label{app:diffusion}
In this appendix\footnote{We thank Akash Jain for discussions that led to the calculation of this appendix.} we further elaborate on subtleties for continuum, hydrodynamic field theories, associated with the apparent mismatch between the KMS transformation and time-reversal or detailed balance in our stochastic framework.  We will see that (at least in this simple example) it is possible to choose a discretization in which, \emph{in the long wavelength limit}, time-reversal symmetry and the KMS transformation exactly agree.

We consider a theory of nonlinear diffusion which, in the continuum, takes the form \begin{equation}
    \partial_t \rho = \partial_x \left(D(\rho)\partial_x \rho + \text{noise}\right).
\end{equation}To apply our Fokker-Planck formalism precisely, it is necessary to discretize this equation \emph{in space}: $\rho_n(t)$ (for $n\in\mathbb{Z}$) become our degrees of freedom, which model $\rho(x_n,t)$ for some discretization $\lbrace x_n\rbrace$ of the real line.  For convenience in what follows, we take the lattice spacing to be 1, so $x_n=n$, so that in the long-wavelength limit, \begin{equation}
    (\partial_x \rho)_n \approx \rho_{n+1}-\rho_n \approx \rho_n - \rho_{n-1}\approx \frac{\rho_{n+1}-\rho_{n-1}}{2}\approx\cdots .
\end{equation} 
We will use this relative freedom to our advantage, in what follows.  In order to discretize the continuum nonlinear diffusion theory, we wish to build a Fokker-Planck generator $W  = -\partial_a Q_{ab}(\partial_b + \mu_b)$ (with $\mu_b = \rho_b$ for simplicity).  One choice is:
\begin{equation}
    Q^{(1)}_{n,m} = \displaystyle \left\lbrace \begin{array}{ll} \displaystyle D\left(\dfrac{\rho_n+\rho_{n-1}}{2}\right) + D\left(\dfrac{\rho_n+\rho_{n+1}}{2}\right) &\ n=m \\ \displaystyle -D \left(\dfrac{\rho_n+\rho_{m}}{2}\right) &\ |n-m| = 1\\ 0 &\ \text{otherwise}\end{array}\right..
\end{equation}
Note that $Q^{(1)}$ has a null vector $\rho_b=1$, compatible with Noether's theorem from Sec.~\ref{sec:noether}.  Another logical choice is 
\begin{equation}
    Q^{(2)}_{n,m} = \displaystyle \left\lbrace \begin{array}{ll} \displaystyle \dfrac{2D(\rho_n)+D(\rho_{n+1})+D(\rho_{n-1})}{2} &\ n=m \\ \displaystyle -\dfrac{D(\rho_n)+D(\rho_{m})}{2} &\ |n-m| = 1\\ 0 &\ \text{otherwise}\end{array}\right.
\end{equation}
which also has the desired null vector encoding charge conservation.  A simple calculation shows that \begin{equation}
    \partial_m Q^{(1)}_{nm} = 0
\end{equation}
while \begin{equation}
    \partial_m Q^{(2)}_{nm} \ne 0.
\end{equation}
Therefore, so long as we have in mind the continuum KMS-invariant theory as being discretized according to $Q^{(1)}$, the MSR Lagrangian exactly encodes time-reversal symmetry, and the regularization subtleties are not important.

A similar method can be employed to study more complicated corrections, such as \begin{equation}
    \partial_t \rho = \partial_x \left(D(\rho) \partial_x \rho + F(\rho) (\partial_x \rho)^3 + \cdots + \text{noise} \right).
\end{equation}
In this example, one could choose to add (this construction is not unique):
\begin{equation}
    Q^{(F)}_{n,m} = \displaystyle \left\lbrace \begin{array}{ll} \displaystyle F\left(\dfrac{\rho_{n+2}+\rho_{n-1}}{2}\right)\left(\dfrac{\rho_{n+2}-\rho_{n-1}}{3}\right)^2 + F\left(\dfrac{\rho_{n-2}+\rho_{n+1}}{2}\right)\left(\dfrac{\rho_{n-1}-\rho_{n}}{3}\right)^2 &\ n=m \\ \displaystyle -F\left(\dfrac{\rho_{n\pm2}+\rho_{n\mp1}}{2}\right)\left(\dfrac{\rho_{n\pm2}-\rho_{n\mp 1}}{3}\right)^2 &\ m=n\pm 1\\ 0 &\ \text{otherwise}\end{array}\right..
\end{equation}
This theory does, on long wavelengths, reduce to the same continuum theory as desired, while obeying $\partial_m Q^{(F)}_{nm}=0$.

The deviations between KMS and time-reversal can always be removed up to an arbitrary order in derivatives by choosing a suitable discretized regularization of $Q$, as long as the derivative expansion is treated perturbatively, as stated in the main text.  This strategy will thus work so long as we can approximate $\Phi$ to be a strictly local integral in space (e.g. there are no Goldstone bosons), as in this case one can always choose a ``point splitting" of the continuum equations of motion that ensures that $\partial_b Q_{ab}=0$, while $Q_{ab}\mu_b$ matches the continuum theory to arbitrary orders in derivatives (by adding ``counter terms" to cancel the discrepancies that arise at lower orders).  Thus one can safely use a KMS-invariant path integral to describe a T-invariant continuum theory.  

An interesting exception to the validity of this regularization, which we will not explore in detil in this paper, is when the steady state of the model possesses an intrinsic length scale that is shorter or comparable to the hydrodynamic scales of interest, such as in the presence of a domain wall -- see e.g. Section \ref{sec:phase separation}. In these situations, the derivative expansion is not strictly perturbative, and thus the above regularization is not directly applicable.

\section{Breaking gT in a four-species Kuramoto model on the 2-sphere}
\label{app:S4}
We describe the case of $N=4$ species of particles on the 2-sphere $\mathrm{S}^2$, and look for generalized Kuramoto models %that could exhibit
compatible with spontaneous gT breaking.   We begin by looking for dynamics with an $\mathrm{S}_4$-invariant $\Phi$, where $\mathrm{S}_4$ is the permutation group acting on four objects.   There is no obvious physical application for this dynamics; however, the symmetry group of $\Phi$ is %complex enough 
sufficiently complicated that one needs to think in group theoretic terms %in order 
to make progress.  Hence, the %group theoretical 
group-theoretic arguments %made 
below are  helpful in determining whether gT symmetry can be spontaneously broken in models whose stationary state $\Phi$ has a non-Abelian symmetry, and generalize beyond this particular example. 

First, we recall a few group theory facts about $\mathrm{S}_4$, following the notation of \cite{zee}.  $\mathrm{S}_4$ has a normal subgroup isomorphic to $\mathbb{Z}_2$, which corresponds to the sign of a permutation, with alternating group $\mathrm{A}_4=\mathrm{S}_4/\mathbb{Z}_2$.  There are 5 irreducible representations of $\mathrm{S}_4$, which we denote using the high energy physics notation with $\mathbf{1}$, $\mathbf{1}^\prime$, $\mathbf{2}$, $\mathbf{3}$, $\mathbf{3}^\prime$.  

\begin{figure}[t]
    \centering
    \subfloat[]{\includegraphics[width=0.25\textwidth]{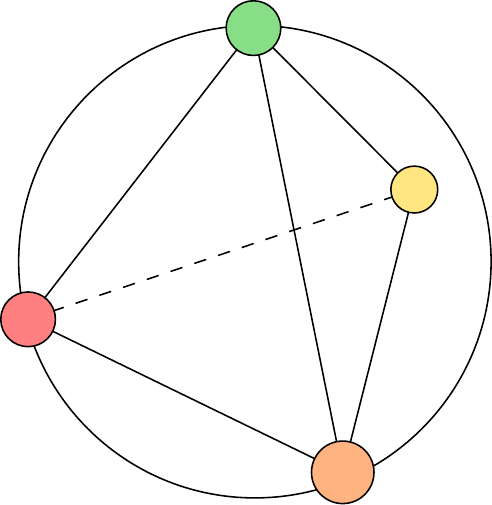}\label{fig:4s2}}\hspace{1em}
    \subfloat[]{\includegraphics[width=0.25\textwidth]{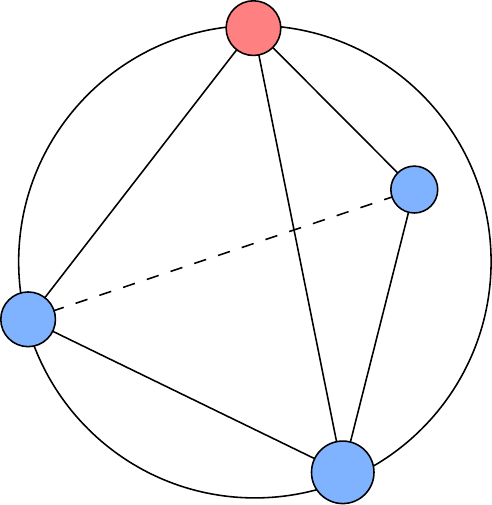}}\label{fig:nospin}\hspace{1em}
    \subfloat[]{\includegraphics[width=0.25\textwidth]{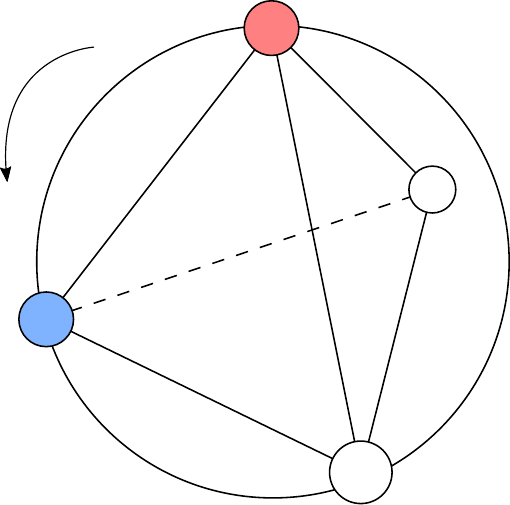}}\label{fig:spin}
    \caption{Positive and negative examples of spontaneous gT-breaking of 4 species of particles on $\mathrm{S}^2$, which coalesce into   (a) $\mathrm{A}_4$ symmetry: in this case, no spontaneous gT-breaking is possible, as explained in the appendix.  Geometrically, the lack of natural gT-breaking corresponds to the high symmetry of the tetrahedron---there is no preferred rotation axis.  (b) Three blue rotors and one red with nonreciprocal interactions (red pursues blue, blue avoids red).  There is still no spontaneous gT-breaking since there is no natural axis around which to rotate.  If the blue species have $\mathrm{S}_3$ symmetry, the only natural choice of $\Omega_{IJ}$ is oriented in the direction of the red species.  (c) Two ``inert'' rotors an two others with nonreciprocal interactions. This time, spontaneous gT-breaking is allowed since there is a clear axis around which to rotate, given by (\ref{eq:2s22}).}
    %  }
    \label{fig:sphere2}
\end{figure}

With these facts in mind, we look for a theory in which the $\mathbb{Z}_2$ normal subgroup of $\mathrm{S}_4$ is spontaneously broken, while also playing the role of the ``g" in gT-symmetry.  In other words, we wish to find an $\Omega_{IJ}$ which flips sign if we apply a sign-changing permutation, but is otherwise invariant.  In order to build such an $\Omega_{IJ}$ in an $\mathrm{S}_4$-covariant manner without explicitly breaking the SO(3) rotational symmetry of $\mathrm{S}^2$, we need to ensure that the $IJ$ indices on $\Omega$ come from $n_I^{(a)}$ and $n_J^{(b)}$.  Moreover, there cannot be any dangling $(a)$ or $(b)$ indices, namely we must write $\Omega_{IJ}$ in terms of an $\mathrm{S}_4$-covariant expression that flips sign under permutation.  To deduce a possible construction, we use the following group theoretical facts.   A ``defining" representation of $\mathrm{S}_4$ is the four-dimensional representation $\mathbf{4} = \mathbf{1}\oplus \mathbf{3}$, on which the permutation group elements are represented by permutation matrices with 1s and 0s only.  Notice that with respect to the $(a)$ index, $n^{(a)}_I$ transforms in the $\mathbf{4}$. The antisymmetric $IJ$ indices on $\Omega$ must come from an antisymmetric tensor product which contains representations 
\begin{equation}
\mathbf{4}\otimes_{\mathrm{A}}\mathbf{4} = \mathbf{3}\oplus \mathbf{3}^\prime.
\end{equation}
Under $\mathrm{S}_4$, we wish for $\Omega_{IJ}$ to be an element of the signed $\mathbf{1}^\prime$ irreducible representation.  Since $\mathbf{3}\otimes \mathbf{1}^\prime = \mathbf{3}^\prime$, we wish to find a scalar object with respect to SO(3) (no dangling $IJ$ indices) that transforms in either the $\mathbf{3}$ or $\mathbf{3}^\prime$.  An appropriate inner product between these two motifs in three-dimensional representations of $\mathrm{S}_4$ will give us the desired $\Omega_{IJ}$.   

Actually, it is a little easier to look for objects in the $\mathbf{4}^\prime = \mathbf{1}^\prime \oplus \mathbf{3}^\prime$ and $\mathbf{4}$ representations, since permutations act more simply on these representations.  We have found that the following transforms in the $\mathbf{4}^\prime$: \begin{align}
    G^{(1)}_{IJ} = n^{(2)}_{[I}n^{(3)}_{J]} + n^{(3)}_{[I}n^{(4)}_{J]} + n^{(4)}_{[I}n^{(2)}_{J]}, \;\;\;\; -G^{(2)}_{IJ} = n^{(3)}_{[I}n^{(4)}_{J]} + n^{(4)}_{[I}n^{(1)}_{J]} + n^{(1)}_{[I}n^{(3)}_{J]}, \notag \\
     G^{(3)}_{IJ} = n^{(4)}_{[I}n^{(1)}_{J]} + n^{(1)}_{[I}n^{(2)}_{J]} + n^{(2)}_{[I}n^{(4)}_{J]}, \;\;\;\; -G^{(4)}_{IJ} = n^{(1)}_{[I}n^{(2)}_{J]} + n^{(2)}_{[I}n^{(3)}_{J]} + n^{(3)}_{[I}n^{(1)}_{J]},
\end{align}
while in the $\mathbf{4}$ we have \begin{equation}
    F^{(a)} = \left|- n^{(a)}_I  + \sum_{b=1}^4 n^{(b)}_I \right|^2 .
\end{equation}
Hence, \begin{equation}\label{eq:C4}
    \Omega_{IJ} = \sum_{a=1}^4 F^{(a)}G^{(a)}_{IJ}
\end{equation}
transforms in the $\mathbf{1}^\prime $ of $\mathrm{S}_4$, implying that the resulting $W$ is gT-invariant.  The resulting theory will spontaneously break gT symmetry so long as $\Omega_{IJ}$ does not vanish in the minima of $\Phi$.  These minima of $\Phi$ must also spontaneously break $\mathrm{S}_4$ in a generic configuration where $\Omega_{IJ} \ne 0$.  We have numerically confirmed that for generic choices of $n^{(a)}_I$, $\Omega_{IJ}$ given in (\ref{eq:C4}) is indeed nonvanishing while transforming in the $\mathbf{1}^\prime$.

In the simplest theory, we may attempt to choose $\Phi$ such that its minima correspond to the $n^{(a)}_I$ equidistant on $\mathrm{S}^2$.  The unique such configuration is a tetrahedron, as shown in \figref{fig:sphere2}.  Unfortunately, one can check that (\ref{eq:C4}) vanishes in such a configuration; intuitively this is because there is no preferred axis to rotate around, when in such a highly symmetric configuration we can perform an even permutation on the configuration by a global rotation (which would in turn rotate $\Omega_{IJ}$).  A preferred axis can arise as one deviates from the tetrahedral configuration.

If the four particles form a tetrahedral configuration, but the interactions between them are not $\mathrm{A}_4$-symmetric, then it is possible to have nontrivial dynamics: see, e.g., \figref{fig:sphere2} for a few different possibilities.  

Finally, if the four particles form instead a square configuration on $\mathrm{S}^2$---in this case, there are two sensible equilibrium configurations with unique T-breaking dynamics.  The first reasonable equilibrium configuration would feature the 4 particles forming a square in some planar slice of $\mathrm{S}^2$, for example, around the equator. Take the axis perpendicular to the square to be the $3$ direction.  In that case, none of the permutations in $\mathrm{A}_4$ correspond to rotations outside of the $1-2$ plane! Therefore the matrix $\epsilon_{IJ3}$ is invariant, and we may include a term $\Omega_{IJ} \sim \epsilon_{IJ3}$, 
which is similar to the model studied in \eqref{eq:3s2}.

\end{appendix}

\let\oldaddcontentsline\addcontentsline%
\renewcommand{\addcontentsline}[3]{}%
\bibliography{nonthermalEFT}
\let\addcontentsline\oldaddcontentsline%

\end{document}